

\hfuzz=50pt
\vsize=7.5in
\hsize=5.6in
\magnification=1200
\tolerance 10000
\input revmacs.tex

\def\qvecsq{\vec q^{\mkern2mu\raise1pt\hbox{$\scriptstyle2$}}}

\def\xiva{{\xi_\sst{V}^{(a)}}}

\def\evec{{\vec e}}
\def\sigpin{{\Sigma_{\pi\sst{N}}}}

\def\ref#1{$^{#1}$}

%
%

\pageno=0

\baselineskip 12pt plus 1pt minus 1pt

\stylephrep
\idraft=0

\secnum=1

\baselineskip 12pt plus 1pt minus 1pt
\nopagenumbers
\pageno=0

\centerline{\bf INTERMEDIATE--ENERGY SEMILEPTONIC PROBES}
\centerline{{\bf OF THE HADRONIC NEUTRAL CURRENT}\footnote{*}{\eightrm
This work is supported in part through funds provided
by the U.~S.~Department of Energy (D.O.E.) under contracts
\#DE--AC02--76ER03069 and \#DE--AC05--84ER40150
and grant \#DE--FG02--88ER40415 and by the National
Science Foundation under grant \#PHY91-15574.
}}
\vskip .35truein
\centerline{M.~J.~Musolf$^1$,
T.~W.~Donnelly$^2$,
J.~Dubach$^3$, S.~J.~Pollock$^4$,}
\vskip .18truein
\centerline{S.~Kowalski$^5$, and E.~J.~Beise$^6$}
\vskip 0.9truein
\item{$^1$} Center for Theoretical Physics,
Laboratory for Nuclear Science and Department of Physics,
Massachusetts Institute of Technology,
Cambridge, Massachusetts 02139 USA;
Department of Physics, Old Dominion University, Norfolk, Virginia 23529
USA\footnote{$^\dagger$}{\eightrm Present Address} and CEBAF Theory Group,
MS-12H, 12000 Jefferson Ave., Newport News, Virginia 23606 USA$^\dagger$.
\vskip 0.15truein
\item{$^2$} Center for Theoretical Physics,
Laboratory for Nuclear Science and Department of Physics,
Massachusetts Institute of Technology,
Cambridge, Massachusetts 02139 USA.
\vskip 0.15truein

\item{$^3$} Department of Physics and Astronomy,
University of Massachusetts,
Amherst, Massachusetts 01003 USA.
\vskip 0.15truein

\item{$^4$}
Institute for Nuclear Theory,
University of Washington, HN-12,
Seattle, Washington 98195 USA; and NIKHEF-K Theory, Postbus 41882, 1009DB
Amsterdam, The Netherlands$^\dagger$.
\vskip 0.15truein

\item{$^5$} Laboratory for Nuclear Science and Department of Physics,
Massachusetts Institute of Technology,
Cambridge, Massachusetts 02139 USA.
\vskip 0.15truein

\item{$^6$} W.~K.~Kellogg Radiation Laboratory,
California Institute of Technology,
Pasadena, California 91125 USA; and
Department of Physics, University of Maryland, College
Park, Maryland 20742 USA.
\vskip .05truein
\parskip 6pt plus 1pt
\parindent 5ex
\baselineskip 12pt plus 1pt
\vfil
\eject
\pagenumbers
\noindent {\bf Table of Contents:}
\def\leaderfill{\leaders\hbox to 1em{\hss.\hss}\hfill} 

\noindent\hangindent=60pt\hangafter=1
\bigskip
\bigskip
\line{I. INTRODUCTION \leaderfill 3}
\bigskip
\line{II. PHYSICS ISSUES \leaderfill 6}
\medskip
\line{\qquad II.A. Strangeness Content of the Nucleon \leaderfill 6}
\smallskip
\line{\qquad II.B. Additional Aspects of PV Electron Scattering Studies \hfill}
\line{\qquad\qquad\qquad of Hadronic Structure \leaderfill 15}
\smallskip
\line{\qquad II.C. Electroweak Tests  \leaderfill 20}
\bigskip
\line{III. FORMALISM  \leaderfill 30}
\medskip
\line{\qquad III.A. Currents and Couplings  \leaderfill 30}
\smallskip
\line{\qquad III.B. Higher--order Processes and Renormalization  \leaderfill
36}
\smallskip
\line{\qquad III.C. Single--Nucleon Matrix Elements of the
Electroweak Current \leaderfill 46}
\smallskip
\line{\qquad III.D. Nuclear Matrix Elements of the Electroweak
Current  \leaderfill 51}
\smallskip
\line{\qquad III.E. Lepton Scattering from Nucleons and Nuclei  \leaderfill
69}
\bigskip
\line{IV. SPECIFIC CASES  \leaderfill 87}
\medskip
\line{\qquad IV.A. Elastic Scattering from the Proton  \leaderfill 87}
\smallskip
\line{\qquad IV.B. Elastic Scattering from Spin--0 Nuclei  \leaderfill 95}
\smallskip
\line{\qquad IV.C. Elastic Scattering from the Deuteron  \leaderfill 105}
\smallskip
\line{\qquad IV.D. Other Discrete Nuclear Transitions  \leaderfill 111}
\smallskip
\line{\qquad IV.E. Axial--Vector Hadronic Response  \leaderfill 118}
\smallskip
\line{\qquad IV.F. Quasielastic Scattering  \leaderfill 123}
\smallskip
\line{\qquad IV.G. The Nucleon--to--Delta Transition  \leaderfill 132}
\smallskip
\line{\qquad IV.H. Parity--Violating Deep Inelastic Scattering
\leaderfill 137}
\smallskip
\line{\qquad IV.I. Atomic Parity Violation  \leaderfill 142}
\smallskip
\line{\qquad IV.J. Neutrino Scattering from Nucleons and Nuclei
\leaderfill 147}
\vfill\eject
\line{V. EXPERIMENTAL CONSIDERATIONS \leaderfill 164}
\medskip
\line{\qquad V.A. General Considerations  \leaderfill 164}
\smallskip
\line{\qquad V.B. Past Studies of Electron and Neutrino Scattering  \leaderfill
172}
\smallskip
\line{\qquad V.C. Present and Proposed Experiments --- Future
Prospectives \leaderfill 187}
\bigskip
\line{VI. SUMMARY AND CONCLUSIONS \leaderfill 196}
\bigskip
\line{REFERENCES \leaderfill 200}

\vskip 1truein
\noindent {\bf Abstract:}
\bigskip
\bigskip
{\narrower
        The present status and future prospects of intermediate--energy
semileptonic neutral current studies are reviewed. Possibilities for
using parity--violating electron scattering from nucleons and nuclei
to study hadron structure and nuclear dynamics are
emphasized, with particular attention paid to probes of strangeness
content in the nucleon. Connections are drawn between such studies and tests
of the electroweak gauge theory using electron or neutrino scattering.
Outstanding theoretical issues in the interpretation of semileptonic
neutral current measurements are highlighted and the prospects for
undertaking parity--violating electron or neutrino scattering
experiments in the near future are surveyed.
\smallskip}
\vfill\eject

\noindent{\bf  I.\quad  INTRODUCTION}
\medskip

        Nearly a generation has passed since the minimal, SU(2$)_L\times
$U(1$)_Y$ model of electroweak interactions was first proposed by Weinberg,
Salam, and Glashow [Wei67, Sal68, Gla70].
During that time, a wide range of experiments,
varying in energy scales from a few eV in atomic parity--violation tests
up to $\sim 100$ GeV in $e^+e^-$ annihilation, have been performed with the
objective of testing the tree--level predictions of the theory. Of particular
interest have been experiments exploring the structure of the weak neutral
current (designated in this work by \WNC), whose form is
governed by the degree of mixing between the SU(2$)_L$
and U(1$)_Y$ sectors:
$$
J_\lambda^\sst{NC}\>=\> J_\lambda^{W^0}-
        4 Q_f \sstw J_\lambda^\sst{EM} \ \ .\eqno\nexteq\nameeq\Ejnc
$$
Here $J_\lambda^{W^0}$ is the neutral, weak isospin partner of the
charge--changing weak currents, $Q_f$ is the electric charge of an
elementary fermion (lepton or quark), $J_\lambda^\sst{EM}$ is
the electromagnetic current, and $\theta_\sst{W}$, the Weinberg or
weak--mixing angle, characterizes the degree of mixing between the gauge
sectors. In the limit that $\sstw\to 0$, the two sectors
decouple, and the neutral current takes on the same structure as the
charge--changing weak current, $J_\lambda^\sst{W^\pm}$. Results from the \lq\lq
first generation" of neutral current studies showed remarkable agreement with
predictions based on Eq.~(\Ejnc), provided $\sstw\approx 1/4$. In fact,
global analyses of neutral current data from all energy scales are consistent
with a common value of $\sstw$ to roughly 1.5\% accuracy
[Ell90, Alt90, Ken90, Lan90]\footnote{$^\dagger$}{We
follow the \lq\lq mass--squared" definition of the weak mixing angle:
$\sstw\equiv 1-\mws/\mzs$; alternate definitions are discussed below.}.

        The advent of the $Z^0$--resonance physics at SLC and LEP, along with
recent improvements in both experimental and theoretical \lq\lq technology"
relevant to atomic parity measurements, presages a new era of higher precision
neutral current tests, pushing the level of precision well beyond the 1\%
level. Indeed, determinations of $\sstw$ from polarization asymmetries in
$e^+e^-$ annihilation could reduce the present uncertainty by an order of
magnitude [Ell90, Alt90]. At this level of precision, such experiments
would be sensitive to both higher--order predictions of the minimal Standard
Model, thereby testing the quantum field character of the theory
as well as \lq\lq new physics" beyond the Standard Model, such as additional
neutral vector bosons or technicolor [Mar90, Pes90, Ken90, Gol90].

        In this review, we consider semileptonic neutral current
studies at intermediate--energy scales (up to a few GeV) as might be performed
at facilities such as CEBAF, MIT/Bates, Mainz,
and LAMPF. The interpretation of experiments
at these energy scales involves an additional level of complexity not
encountered in deep inelastic scattering or high--energy $e^+e^-$
annihilation. In the latter cases, the effects of strong interactions are
sufficiently small to permit a perturbative treatment of semileptonic
processes
in terms of quarks and gluons. Thus, in such experiments, the interpretation
of results in terms of electroweak interactions among elementary particles
is relatively straightforward and unambiguous. At the lower--energy scales of
interest here, however, QCD becomes a strong--coupling theory and
such a perturbative treatment is no longer
justified. Instead, one must work with explicitly collective hadronic degrees
of freedom such as mesons, nucleons, and nuclei. Since low-- and
intermediate--energy properties of hadrons cannot
at present be reliably calculated from
first principles, the interpretation of intermediate--energy electroweak
studies necessarily involves some degree of theoretical uncertainty.
One then faces the question as to what limitations this uncertainty imposes
on the precision with which the Standard Model may be confidently tested using
intermediate--energy semileptonic probes.

On the other hand, studies of hadron structure in the strong--coupling
regime are of central importance to particle and nuclear physics.
The diversity of ways in which the hadronic neutral current is manifested in
intermediate--energy measurements of semileptonic scattering from nucleons
and nuclei makes this a potentially very fruitful direction to follow in
the near future. In particular, the subject which provides the main focus
of the present work, parity--violating electron scattering, presents a
rich new array of
hadronic quantities, such as nucleon form factors and nuclear response
functions, which cannot be observed in higher--energy experiments. Indeed,
since a determination of many of these observables is not possible using
purely QED--type processes, intermediate--energy neutral current studies
potentially offer
a unique window on aspects of hadronic structure which are of fundamental
interest in their own right.
The realization of
this potential presents both experimentalists and theorists a variety
of challenges. Once some of these challenges have been met and the hadronic
structure better understood, experiments in
this sector could become more relevant in the search for \lq\lq new" or
\lq\lq non--standard" physics.

        In the remainder of this article, we attempt to clarify both the
precision to which one may reliably test the Standard Model using
intermediate--energy, semileptonic processes as well as the kind of hadronic
structure information which might realistically be extracted from such
experiments. We focus
primarily on parity--violating (PV)
polarized electron scattering and make contact with
neutrino scattering as needed.
Our analysis includes both theoretical as well
as experimental treatments of past, present, and prospective experiments of
these types. We also discuss atomic PV and deep--inelastic PV
electron scattering for purposes of comparison.

Our objective is to provide both a general overview of the field as
well as a \lq\lq roadmap" to assist in determining what types
of precision neutral current scattering measurements might be undertaken in
the future. As a consequence, an important theme running through our
discussion will be considerations of experimental
\lq\lq do--ability". Indeed,
the relative importance of various hadronic quantities appearing in the
observables of interest generally depends both on the choice of kinematic
conditions as well as on experimental capabilities ({\it e.g.},
luminosities, beam polarizations, solid angles, {\it etc.\/}).  Since the
achievable statistical precision also depends on all of these factors,
issues of theoretical interpretability are correlated with
considerations of experimental do--ability. In analyzing this
do--ability/interpretability correlation for specific cases, we try to make
reasonable assumptions for experimental conditions,
based on what one expects to be achievable in the next decade at
various accelerators. Hence, our projections of constraints on
quantities of interest from different prospective experiments are
somewhat time--dependent and should be taken more as benchmark
indicators than as definitive statements.

        The remainder of this article is organized as follows.
In Sect.~II we review  the central physics issues,
including a brief summary of the status of
Standard Model tests, a discussion
of the hadronic form factors and response functions which might be probed
with neutral current scattering, and a brief overview of the theoretical
 hadronic
uncertainties which enter the interpretation of these experiments.
In Sect.~III we introduce our formalism and, in
Sect.~IV, give a case--by--case theoretical analysis of various classes of
neutral current scattering measurements.  Section V contains a discussion of
a variety of experimental considerations. Section VI
summarizes our conclusions,
particularly in regard to a prospective program of future experiments and
the attendant theoretical and experimental challenges. More detailed
treatments of the subjects reviewed in this paper may be found
elsewhere [Mus92a, Don92].

\vfil\eject
\secnum=2
\neweq

\noindent{\bf  II.\quad  PHYSICS ISSUES}
\medskip

        The observables measured in low-- and intermediate--energy
neutral current processes depend on two classes of physics: (a)
the structure of the underlying electroweak gauge theory and (b)
hadronic matrix elements of the quark currents.
For {\it high--energy} neutral current observables hadronic contributions
are in general reliably
calculable, whereas at the energy scales of interest here, they are
more difficult to calculate but are more selectively obtained from
experiment. Thus, if one considers
the underlying gauge theory to be well--tested from experiments in other
sectors, intermediate--energy semileptonic neutral current scattering
provides an opportunity to study a variety of interesting hadronic
physics issues not accessible at other energy scales.  We begin this
section by discussing the strangeness content of the nucleon and then
continue in Sect.~II.B with a summary of some additional aspects involved
in probing hadronic structure using neutral current scattering; both subjects
will be treated in more detail in the rest of the article. Following
these discussions we turn in Sect.~II.C to the issue of undertaking
Standard Model tests using semileptonic processes.

\bigskip
\noindent {\bf II.A.\quad  Strangeness Content of the Nucleon}

        The simplest \lq\lq valence--quark" picture of hadrons depicts the
nucleon as consisting solely of $u$-- and $d$--quarks. Many low--energy
properties of the nucleon can be understood in this picture without explicitly
accounting for the presence of heavier quarks in the \lq\lq sea". However,
there exists evidence that not {\it all}\ low--energy nucleon properties can
be understood within this framework. In particular,
analyses of $\sigpin$, the so called \lq\lq sigma term" extracted from
$\pi$--N scattering, suggests a non--negligible value for proton matrix
elements
of the $\sbar s$ operator. Defining the ratio $R_s$ as
$$
R_s\>\equiv\> {\langle p\vert \sbar s \vert p\rangle\over \langle p\vert
        \ubar u+\dbar d+\sbar s\vert p\rangle}\ \ ,  \eqno\nexteq\nameeq\ERsubs
$$
one has from these analyses that $R_s\approx$ 0.1--0.2 [Che76, Che71a, Don86a,
Gas91] --- see Table~2.1. Neglect of $\sbar s$ pairs in the sea would, in
this framework,
imply a value for the nucleon mass of $\approx 600$ MeV. Although the dynamical
origin of this surprising result is not definitively understood, Donoghue
and Nappi [Don86a] have shown that a large value of $R_s$ is plausible
within the SU(3) Skyrme model and MIT bag model. The more recent
calculation of Ref.~[Mus93a] suggests that strange hadronic components of
the nucleon wave function ({\it e.g.}, a $K\Lambda$ intermediate state)
may account for a substantial portion of the extracted value of $R_s$.

\midinsert
$$\hbox{\vbox{\offinterlineskip
\def\strut{\hbox{\vrule height 15pt depth 10pt width 0pt}}
\hrule
\halign{
\strut\vrule#\tabskip 0.2cm&
\hfil$#$\hfil&
\vrule#&
\hfil$#$\hfil&
\vrule#&
\hfil$#$\hfil&
\vrule#\tabskip 0.0in\cr
& \multispan5{\hfil\bf TABLE 2.1\hfil} & \cr\noalign{\hrule}
& \hbox{Source } && R_s && \hbox{Reference}
& \cr\noalign{\hrule}
& \Sigma_\sst{\pi N} \hbox{(A)}&& 0.09 &&\hbox{[Gas91]}
& \cr
& \Sigma_\sst{\pi N} \hbox{(B)}&& 0.21\pm 0.03  && \hbox{[Don86a]} 
& \cr
& \hbox{Skyrme} && 0.23 && \hbox{[Don86a]} 
& \cr
& \hbox{MIT Bag} && 0\to 0.29 && \hbox{[Don86a]} 
& \cr
& \hbox{Kaon Loops} && -0.007\to 0.47&& \hbox{[Mus93a]} 
& \cr\noalign{\hrule}}}}$$
\smallskip
\baselineskip 10pt
{\ninerm
\noindent\narrower {\bf Table 2.1.} \quad
Strange quark scalar density of the nucleon, $R_s$, as defined in
Eq.~\ERsubs.
First two rows give
extractions from $\pi N$ scattering data;
the remaining rows give predictions from various theoretical models.
\smallskip}
\endinsert

\baselineskip 12pt plus 1pt minus 1pt

        These results suggest that nucleon matrix elements of other
strange--quark operators could differ non--negligibly from zero. Of particular
interest are $s$--quark vector and axial--vector currents, parameterized in
terms of strangeness form factors $\GES(Q^2)$, $\GMS(Q^2)$, and $\GAS(Q^2)$
(see Eqs.~(\FGtilde), (\Fgalstr), and (\Fgalstra)). Of these three
form factors, experimental
constraints have been reported only for $\GAS$. Its value at $Q^2=0$ has
been derived from deep inelastic ${\vec\mu}\vec p$ scattering [Ash89]
and from the BNL neutrino--proton
scattering determination of the isoscalar axial--vector form factor, $\GATEZ$
[Ahr87] (see Eq.~(\FGAtilde b)). Results
taken from these two determinations
appear in Table~2.2, along with a variety of theoretical predictions.
The value extracted from the EMC data assumes an SU(3) parameterization
of the baryon octet matrix elements of the axial--vector current.
Breaking of SU(3) symmetry could be large [Jen91, Par91],
resulting in a reduction of the EMC value for $\GAS(0)$ by a factor of
three or more. The BNL measurements were performed at nonzero $Q^2$,
and the $Q^2$--dependence of $\GATEZ$ was fit to a dipole form. The
experimenters report a non--negligible correlation between $\GATEZ(0)$
and the dipole mass parameter, $M_\sst{A}^{T=0}$.
The results in Table~2.2 were derived
assuming that $M_\sst{A}^{T=0}$ is the same as the mass parameter appearing
in the
isovector axial--vector form factor, and does not take into account the strong
$M_\sst{A}^{T=0}$ correlation.
Additional uncertainties in the BNL
results arise from nuclear physics sources. Since 80\% of the events
in the experiment were generated by quasielastic (QE) scattering from protons
in $^{12}$C nuclei, knowledge of the $^{12}$C QE response is necessary in
the interpretation of the measured cross section (see Sect.~IV.J).
Evidently, a more precise determination of $\GAS(Q^2)$ is warranted.

\midinsert
$$\hbox{\vbox{\offinterlineskip
\def\strut{\hbox{\vrule height 15pt depth 10pt width 0pt}}
\hrule
\halign{
\strut\vrule#\tabskip 0.2cm&
\hfil$#$\hfil&
\vrule#&
\hfil$#$\hfil&
\vrule#&
\hfil$#$\hfil&
\vrule#\tabskip 0.0in\cr
& \multispan5{\hfil\bf TABLE 2.2\hfil} & \cr\noalign{\hrule}
& \hbox{Source } && \GATEZ(0) && \hbox{Reference}
& \cr\noalign{\hrule}
& \hbox{EMC} && -0.194 \pm 0.055 &&\hbox{[Ash89]} &\cr
& \hbox{BNL} && -0.15 \pm 0.09 &&\hbox{[Ahr87]} & \cr
& \hbox{ET$+$QM} && -0.075 &&\hbox{[Col78, Wol79]} & \cr
& \hbox{ET$+$OZI} && -0.046 &&\hbox{[Col78, Wol79]} & \cr
& \hbox{Kaon Loops (a)} && -0.037\to-0.041 &&\hbox{[Mus93a]} & \cr
& \hbox{Kaon Loops (b)} &&  -0.004 &&\hbox{[Koe92]} & \cr
& \hbox{Kaon Loops (c)} &&  \pm 0.20 &&\hbox{[Hol90]} & \cr
& \hbox{SU(3) Skyrme (B)} && -0.10 &&\hbox{[Par91]} &
\cr\noalign{\hrule}}}}$$
\smallskip
\baselineskip 10pt
{\ninerm
\noindent\narrower {\bf Table 2.2.} \quad
Experimental results and theoretical predictions for low--\absQ2\
behavior of the isoscalar axial--vector form factor. At tree--level
in the Standard Model, one has $\GATEZ=\GAS$.
First two rows give experimental values from EMC and BNL measurements.
Statistical and systematic errors in the EMC value have been combined
in quadrature. Third and fourth lines give predictions using effective
theory (ET) approach of Ref.~[Col78], together with
quark model (QM)
and SU(3)/Zweig Rule (OZI)
estimates of nucleon matrix elements given in
Ref.~[Wol79]. Following three rows give kaon--strange baryon loop predictions
under different assumptions about the meson--baryon form factor. The
prediction of Ref.~[Hol90] gives magnitude only. Final row gives
broken (B) SU(3) Skyrme model prediction.
\smallskip}
\endinsert

\baselineskip 12pt plus 1pt minus 1pt

{}From a theoretical point of view, elastic $\nu$N scattering offers some
advantage over parity violating $\vec eN$ in terms of measuring $\GAS$.
The latter always introduces a small coefficient, $\gve  = -(1-4\sstw)$
(see Eq.~(\Fermc a) and Table~3.1)
multiplying the nucleon axial--vector form factor,
thus generically reducing the sensitivity to this term by an
order of magnitude.
Perhaps more significantly for theoretical {\it interpretation},  there exist
large and theoretically uncertain radiative
corrections arising in the $V(e)\times A(\hbox{N})$ amplitude which do not
enter the amplitude for neutrino probes of the nucleon axial--vector
current [Mus90].
The LSND experiment planned for LAMPF, intended primarily as a search
for $\nu_\mu\to\nu_e$ and $\nubar_\mu\to\nubar_e$ oscillations,
will also detect a significant number of
recoil protons from elastic $\nu p$ scattering [Lou89].
The  effective $|Q^2|$ for these scatterings is  lower by approximately
an order of magnitude from
the BNL experiment, so that the LSND results could permit an extraction
of $\GAS(0)$ with significantly lower sensitivity to $M_\sst{A}^{T=0}$.
The impact of nuclear physics uncertainties in this determination remains
to be analyzed. An additional source of uncertainty arises from lack
of knowledge of the neutrino flux [Ahr87]. A measurement of the ratio
$$
R_\nu^{pn} = {d\sigma(\nu , \nu' p)/d Q^2\over d\sigma(\nu , \nu' n)/d Q^2}
 \eqno\nexteq\nameeq\ERnupn
$$
could be quite sensitive to $\GAS(0)$ while being independent of
uncertainty in the neutrino flux [Gar92].

To date, no experimental constraints have been published
for $\GES$ and $\GMS$. For this reason, one must turn to theoretical
predictions
to obtain some suggestion for the scale of these form factors. While such
predictions are model--dependent, there exists one rigorous theoretical
constraint on $\GES$.
Since the nucleon has no net strangeness,
$\GES$ must vanish at $Q^2=0$. Consequently, it is convenient to
characterize the low--\absQ2\ behavior of this form factor by a dimensionless
mean square \lq\lq strangeness radius", $\rhostr$ (see Eqs.~(\Frhomus a)
and (\Fgalstr a)).
The strange magnetic form factor is not constrained by any symmetry, and
one characterizes its low--\absQ2 scale (see
Eqs.~(\Frhomus a) and (\Fgalstr b)) by a
strange magnetic moment, $\mustr \equiv \GMS(0)$. Theoretical predictions
for $\rhostr$ and $\mustr$ are given in Table~2.3; the non--leading
$Q^2$--dependence of $\GES$ and $\GMS$ is discussed more fully in
Sect.~III and Ref.~[Jaf89].

\midinsert
$$\hbox{\vbox{\offinterlineskip
\def\strut{\hbox{\vrule height 15pt depth 10pt width 0pt}}
\hrule
\halign{
\strut\vrule#\tabskip 0.2cm&
\hfil$#$\hfil&
\vrule#&
\hfil$#$\hfil&
\vrule#&
\hfil$#$\hfil&
\vrule#&
\hfil$#$\hfil&
\vrule#\tabskip 0.0in\cr
& \multispan7{\hfil\bf TABLE 2.3\hfil} & \cr\noalign{\hrule}
& \hbox{Source } && \rhostr && \mustr && \hbox{Reference}
& \cr\noalign{\hrule}
& \hbox{Poles} && -2.12 \pm 1.0 && -0.31 \pm 0.009 && \hbox{[Jaf89]} & \cr
& \hbox{Kaon Loops (a)} && 0.41\to 0.49 && -0.31\to -0.40 &&
\hbox{[Mus93a]}&\cr
& \hbox{Kaon Loops (b)} && 0.173 && -0.026 && \hbox{[Koe92]} &\cr
& \hbox{Kaon Loops (c)} && - && \pm 0.8 && \hbox{[Hol90]} &\cr
& \hbox{SU(3) Skyrme (B)} && 1.65 && -0.13 && \hbox{[Par91]} & \cr
& \hbox{SU(3) Skyrme (S)} && 3.21 && -0.33 && \hbox{[Par91]} &
\cr\noalign{\hrule}}}}$$
\smallskip
\baselineskip 10pt
{\ninerm
\noindent\narrower {\bf Table 2.3.} \quad Theoretical predictions for
low--\absQ2\ behavior of $s$--quark vector current form factors.
First row gives average of all fits in Ref.~[Jaf89]. Following
three rows give predictions of various loop calculations (see also
Table~2.2). The estimate of Ref.~[Hol90] predicts magnitude only
and not sign.
Final two rows are taken from Ref.~[Par91] for broken (B) and
symmetric (S) SU(3) Skyrme models.
To set the scale, note that the dimensionless mean--square EM charge
radius of the neutron is $\rho_n\approx 1.91$
(see Eqs.~(\Fradneu b) and (\Frhomus a)), while
the isoscalar nucleon magnetic moment is $\mu^\sst{T=0} = 0.44$.
\smallskip}
\endinsert

\baselineskip 12pt plus 1pt minus 1pt

In principle, one might hope to extract information on the strange
vector current form factors by re--analyzing existing data from the BNL
experiment as well as the elastic PV $^{12}\hbox{C}(\vec e,e)$ measurement
at MIT/Bates [Sou90a] and the Mainz $^9\hbox{Be}(\vec e,e')$ QE
experiment [Hei89]. However,
for reasons particular to each of these experiments, the form factor
constraints extracted would be unreliable.  While nonzero values of
$\GES$ and $\GMS$ would contribute a significant percentage to the
total $\sigma_{\nu\sst{N}}$ and $\sigma_{\nubar\sst{N}}$ in a BNL--type
experiment [Bei91a], extraction of limits on the vector current
strangeness form factors from the BNL experiment would require detailed
knowledge of the $\nu$ energy spectra, normalizations, correlations and
other systematics. While such an analysis lies beyond the scope of the
present review, one may arrive at an idea of the level of constraints
possible by modeling a BNL--type experiment. Assuming measured cross
sections at the BNL data points are given by their Standard Model
predictions with no strangeness, and assuming only 10\% statistical
uncertainty in the data, one would obtain a simultaneous extraction of
the vector strangeness form factors with correlated uncertainties on
the order of  $\delta\mustr\approx \pm 0.3$, $\delta\rhostr\approx \pm
5.0$ (67\% confidence). More detailed discussions of these constraints
from a hypothetical BNL--type experiment appear in Sect.~IV.J.

     In the case of the elastic scattering $^{12}\hbox{C}({\svec e},e)$
experiment, only
$\GES$ contributes to $\alr$ since the target is spin--0. The uncertainty
in the experimental value of  $\alr(^{12}\hbox{C})$ is significantly larger,
at the small $|Q^2|$ of this experiment, than the size of the $\GES$
contribution one would expect from the pole--model prediction [Jaf89]
(see Table~2.3). Assuming the low--$|Q^2|$ behavior of $\GES$ were given
by the pole--model value, neglect of this
form factor would introduce roughly a three--percent theoretical
error in $\alr$, significantly lower than $\approx$ 25\% experimental error.
Backward--angle quasielastic electron scattering experiments of the Mainz
variety are quite insensitive to isoscalar form factors, as discussed in
Sect.~IV.F. Again, the scale of model predictions
for the strangeness contribution to $\alr(^9\hbox{Be, QE})$ is
significantly smaller than the
20\% uncertainty in the measured asymmetry [Bei91a, Hei89].

        In the absence of reliable experimental bounds on $\GES$ and $\GMS$,
the theoretical predictions of Table~2.3 take on added interest. These
calculations are highly model--dependent and should be viewed more
as indications of the possible overall scale and sign of the form factors
than as firm and rigorous predictions. Nevertheless,
the approaches followed are instructive as to the different types
of physics which could give rise to strangeness contributions of varying
magnitudes. In general, two  pictures have been used to represent
the presence of non--valence $s\sbar $ pairs in the nucleon: (a) a
perturbative, or \lq\lq high energy", picture making explicit use of quarks
(Fig.~2.1), and (b) a phenomenological, or \lq\lq low
energy", representation employing hadronic degrees of freedom (Fig.~2.2).
In the former approach, one assumes the dominant effect of heavy $q\bar q $
pairs is to \lq\lq dress", or renormalize, operators involving the light
valence quarks via gluon exchange. By integrating out the heavy
quarks, one derives an effective theory involving only the lighter degrees
of freedom appropriate for physics at momentum scales below the heavy--quark
masses. In the case of the nucleon, for which typical internal momenta are
on the order of $\Lambda_\sst{QCD}\sim$ few hundred MeV$/c$, it is sensible
to integrate out the $c$, $b$, and $t$ quarks, leaving the $u$, $d$, and $s$
quarks as effective degrees of freedom. Following this procedure, Kaplan
and Manohar [Kap88] find that the heavy--quark
renormalization of the light--quark vector currents
is small enough to be ignored in the analysis of the
NC observables of interest here. The induced isoscalar
axial--vector current, however, is non--negligible, especially since
this current vanishes at tree--level in the Standard Model. Prior to the
work of Ref.~[Kap88], Collins {\it et al.} [Col78] carried the integration
down through the $s$--quarks, and derived a somewhat larger effective isoscalar
axial--vector current. While integrating out the $s$--quarks is questionable in
this case, given that $m_s\sim\Lambda_\sst{QCD}$, the estimate of Ref.~[Col78]
gives some indication of the magnitude of $s$--quark effects in the
axial--vector current. Nucleon matrix elements of the isoscalar axial--vector
current have been estimated in two ways by Wolfenstein [Wol79]. Combining these
results with the estimates of Ref.~[Col78] for the coefficient of
this operator leads to the two entries in Table~2.2. One observes that
these estimates agree in sign but are smaller in magnitude than the EMC and
BNL results.

        The second approach to modeling strangeness focuses on the
low--momentum ($p\lapp\Lambda_\sst{QCD}$) part of the virtual $s\sbar$ loop.
At these momentum scales, the heavy--quark pair interacts with the valence
quarks for a sufficiently long time to permit the formation
of virtual strange hadronic intermediate
states. To estimate contributions of this type, one must rely on
phenomenological hadronic models, since a first--principles approach involving
quark degrees of freedom directly is at present intractable. The pole--model
approach followed in Ref.~[Jaf89] represents one estimate of this type.
In this analysis, $\mustr$ and $\rhostr$ were obtained from a three--pole
($\phi$, $\omega$, and one higher--mass vector meson)
fit to the isoscalar nucleon EM form factors under {\it ad hoc} assumptions
about the asymptotic (large--$|Q^2|$) behavior of the strangeness form factors.
Alternatively, several authors have performed loop
estimates [Hol90, Koe92, Mus93a], in which the virtual $s\sbar$ pair appear as
a
kaon--strange baryon intermediate state. As indicated in Table~2.3, the
loop contributions to the strangeness radius are typically much smaller in
magnitude than the pole contribution, whereas, depending on the cut--off
procedure employed, the loop and pole contributions to $\mustr$ may be
comparable. The extent to which these two types of contributions are
either independent and ought to be added or represent the same physics
is not entirely clear. Thus, one ought to take them as indicative of the
magnitude and sign of $s$--quark contributions rather than as definitive
predictions. One also ought to note that the calculations of Refs.~[Hol90,
Koe92] did not satisfy gauge invariance at the level of the Ward--Takahashi
identity. A phenomenological approach in a somewhat different spirit is
represented by the Skyrme model predictions, which yield values of $\rhostr$
of roughly the same magnitude as the pole estimate, but having the opposite
sign.

        While it is of interest to test these model predictions
experimentally, the presence of $s$--quark form factors also impacts on the
feasibility of future semileptonic scattering studies as electroweak tests.
For example, an {\it uncertainty} in $\GES$ on the order of the prediction
of Ref.~[Jaf89] would seriously hamper one's ability to constrain
extensions of the Standard Model with PV electron scattering. Consequently,
it is important not only to constrain the strangeness form factors at a level
needed to distinguish among model predictions, but ultimately also to reduce
the uncertainty in these constraints to a level below what is tolerable in
electroweak tests. With these objectives in mind,
a number of new experiments have been proposed or discussed which could
provide constraints on $\GES$, $\GMS$, and $\GAS$. The so--called \lq\lq
SAMPLE"
experiment underway at  MIT/Bates [McK89] seeks to extract
information on $\mustr$ by measuring the backward--angle asymmetry in
$\evec p$ scattering at an incident energy of $\epsilon\sim 200$ MeV.
It is expected that this measurement will
constrain $\mustr$ with an uncertainty of $\delta\mustr=\pm 0.22$
[Bei92]. The \lq\lq $G^0$" experiment, conditionally approved for CEBAF,
seeks to determine the $Q^2$--dependence of $\GMS$ over the range $0.1\leq
|Q^2|\leq 0.5$ $(\hbox{GeV}/c)^2$, with an uncertainty having the same scale
as the expected SAMPLE error in $\mustr$ at low--$|Q^2|$ and decreasing for
larger $|Q^2|$ [Bec91]. These measurements
could rule out several of the predictions listed in Table~2.3.
A fundamental limitation on further tightening of these constraints using
SAMPLE--type experiments is imposed by hadronic uncertainties in the radiative
corrections entering the axial--vector contribution to the
backward--angle asymmetry.
The extent to which these uncertainties would limit one's ability to
determine $\mustr$ from PV $\evec p$ scattering is discussed in detail in
Sect.~IV.A. The uncertainty in the isovector component of these radiative
corrections could be experimentally reduced by a measurement of the
backward--angle QE PV asymmetry. Alternatively, a measurement of the
backward--angle
elastic $\evec D$ asymmetry, which is significantly less sensitive to
axial--vector uncertainties than is $\alr(\evec p)$, might permit an
improvement
over the SAMPLE $\mustr$ constraints by a factor of two or more. These
possibilities are discussed more fully in Sect.~IV.

        Proposals to determine $\GES$ with forward--angle PV elastic
$\evec p$ scattering and PV electron scattering from $^4$He have
been made at CEBAF [Fin91, Bec91, Bei91b] and have been conditionally
approved. A low--$|Q^2|$ forward--angle measurement of
$\alr(\evec p)$ with 10\% experimental error could ultimately determine the
strangeness radius to a precision of $\delta\rhostr=\pm 2$.
A large fraction of the uncertainty actually arises from a lack of knowledge
of $\mustr$, which also
contributes to the forward--angle asymmetry. An improvement in the
SAMPLE $\mustr$ bounds would correspondingly allow a tighter determination
of $\rhostr$.

        Alternatively, a determination of $\GES$ with elastic
scattering from $^4$He would be independent of uncertainties in $\mustr$.
To the extent that $\GES$ can be extracted from the $^4$He asymmetry
in a nuclear--physics--independent manner, a series of low-- and
moderate--$|Q^2|$
measurements of $\alr(^4\hbox{He})$ could produce constraints on
$\GES$ significantly more stringent than would be attainable from a series
of measurements on the proton alone (see Fig.~4.7). To achieve such a
result, experimental uncertainties at roughly the 1\% (low--$|Q^2|$) and
10\% (moderate--$|Q^2|$) levels would be necessary. From the standpoint of
theoretical interpretability, scattering from a $(J^\pi  T)=(0^+ 0)$
nucleus such as $^4$He is an attractive case.
At the simplest level, one has
$$\alr(0^+0) \propto
{ {  \rbra{\hbox{g.s.}} \rho^\sst{NC}(T=0)\rket{\hbox{g.s.}} }
  \over
  {  \rbra{\hbox{g.s.}} \rho^\sst{EM}(T=0)\rket{\hbox{g.s.}} }
 }\ \ , \eqno\nexteq\nameeq\EAlhe
$$
where $\rho^\sst{NC (EM)}(T=0)$ is the isoscalar weak neutral current
(electromagnetic) charge operator and
$\rbra{\hbox{g.s.}}\ \ \rket{\hbox{g.s.}}$ is a reduced ground state
matrix element (see Eqs.~(\Felzera) and (\Felzerb)).
To the extent that transition meson--exchange currents ({\it e.g.}, $\pi-\rho$)
can be neglected, one expects the
nuclear many--body contributions to these matrix elements to cancel from
the ratio, leaving only a dependence on the electroweak gauge theory
parameters and single nucleon charge form factors.

        Corrections to this na\"\i ve result arise from a number of
effects. The $^4$He ground state is not an eigenstate of strong isospin,
and mixing of higher--lying $T=1$ states into the nominally $T=0$ ground
state introduces a nuclear--structure--dependent correction into the
asymmetry [Don89].
For sufficiently small momentum transfer, one expects this correction, and
the associated theoretical uncertainty, to fall below a problematic level
for a determination of $\GES$. Indeed, at low momentum transfer both the
isospin--mixing correction and the term containing $\GES$ are proportional
to $Q^2$.  From the estimates of the former in Ref.~[Don89] it appears for
favorable cases such as $^4$He and $^{12}$C that the isospin--mixing
correction implies an uncertainty in $\rho_s$ at low--$|Q^2|$ of approximately
$|\Delta\rho_s|\sim 0.02$, which is considerably smaller than the
level of precision at which $\GES$ is likely to be determined in the
foreseeable future.  At moderate--$|Q^2|$ it is more difficult to be
certain of the scale of the isospin--mixing corrections, although the
modeling performed in Ref.~[Don89] suggests that they continue to
remain small compared to the uncertainties in $\rho_s$ that arise from
other sources, such as those mentioned below, at least for the favorable
cases of helium and carbon.  In contrast, as discussed in Ref.~[Don89],
heavier nuclei (in the $s$--$d$ shell and beyond) are expected to have
important isospin--mixing corrections and consequently are likely to
prove unsuitable for $\GES$ determinations.  We shall return to discuss
this point in more detail in Sect.~IV.B.

        A potentially more serious issue in the interpretation of this
asymmetry is the contribution from multi--boson--exchange \lq\lq dispersion"
corrections (see Fig.~3.7). Data for elastic parity--conserving (PC)
scattering from another spin--0, isospin--0 nucleus --- $^{12}$C ---
suggests that these corrections are significantly larger than one
expects based on analyses of dispersion corrections in $ep$
scattering [Kal89],
and that at low--$|Q^2|$ they could enter the denominator
of $\alr(0^+0)$
in Eq.~(\EAlhe)\
at a potentially problematic level.
Furthermore, theoretical calculations [Fri74]
of the $^{12}$C PC dispersion corrections are thus far in rather poor
agreement with the data.
It is unlikely that the PV dispersion corrections
to the neutral current amplitude entering the numerator of the asymmetry
will be measured directly, so that a nuclear--model--dependent estimate
of these corrections will be needed. In short, a better understanding of
dispersion corrections constitutes an interesting challenge for nuclear theory
posed by the interpretation of the $^4$He asymmetry.

        Going beyond elastic semileptonic scattering, there exist several
other possibilities for probing nucleon strangeness matrix elements
with lepton probes: PV QE electron scattering,
inelastic neutrino scattering and PV in heavy muonic atoms. In the
case of QE electron scattering, as discussed in Sect.~IV.F, contributions
to the asymmetry
from $\GMS$ and $\GAS$ are suppressed, since they enter with a multiplicative
factor for the isoscalar magnetic moment. In the case of the only
QE PV experiment thus far completed, a backward--angle measurement $\alr
(^9\hbox{Be})$, the largest strangeness form factor sensitivity is to
$\GES$. Under the prediction of Ref.~[Jaf89], the latter would generate a
3\% contribution to the asymmetry [Bei91a]
--- a value significantly smaller than
the 20\% experimental error. A future backward--angle measurement of
$\alr(\hbox{QE})$ is more suited to a determination of $\GATEO$, which enters
multiplied by the isovector magnetic moment. A determination of this
term, with its large and theoretically uncertain radiative correction,
could reduce the associated uncertainty in a backward--angle $\evec p$
determination
of $\GMS$. The forward--angle QE asymmetry is dominated by the transverse
vector current response, since the longitudinal PV response is
fortuitously suppressed (see Eq.~(\Fsmallqe) and following). However, this
asymmetry is potentially quite sensitive to $\GES$, and a series of
1\% measurements of $\alr(\hbox{QE})$ at low-- and moderate--momentum transfer
could produce constraints on $\GES$ nearly equivalent to those
ultimately attainable from a series of elastic $\alr(^4\hbox{He})$
measurements (see Fig.~4.7). As in the case of elastic scattering,
the interpretation of a forward--angle $\alr(\hbox{QE})$ measurement
would require the resolution of a number of theoretical issues,
including contributions from final--state interactions, meson--exchange
currents, and nuclear correlations. In fact, the forward--angle
asymmetry appears to be quite sensitive to isospin--dependent nuclear
correlations at momentum transfers roughly below one GeV$/c$ [Don92, Had92,
Alb93a].

In principle, inelastic excitation of discrete states by PV electron
scattering could be used to study specific aspects of the weak neutral
current.  This general approach of using nuclear transitions with
carefully chosen spins and parities to ``filter out'' pieces of the
electroweak currents was developed in the 1970's (see, for example, the
review of the ideas involved in Ref.~[Don79a]).  For neutrino scattering
we shall see that this way of proceeding is a fruitful one when we
return to that subject in Sect.~IV.J.  Unfortunately, as discussed in
Sect.~IV.D, very few practical cases of discrete--state inelastic
scattering emerge in the case of PV electron scattering due to the
poor figures--of--merit for such reactions which stem from the small
cross sections that occur ({\it i.e.,\/} much smaller that the
coherent elastic scattering cross sections --- see Figs.~3.10--3.12).

        In contrast to semileptonic scattering measurements, atomic PV
experiments are generally much less sensitive to nucleon form factors
due to the very small effective momentum transfer associated with the
interaction of an atomic electron with the nucleus. In particular, a PV
$^{4}$He$(\evec, e)$ would have to be carried out at $q \approx
30$ MeV/$c$ to be as insensitive to $\rhostr$ as is a measurement of
the weak charge in $^{133}$Cs atomic PV. For this reason, atomic PV
is apparently more suitable as a low--energy testing ground of the
Standard Model than is semileptonic scattering, up to uncertainties
associated with atomic and nuclear structure. However, in the case of
muonic atoms, the muon is much more tightly bound (for a given set of
orbital quantum numbers) than a corresponding atomic electron, making
it sensitive to physics at hadronic length scales.
In fact, a 1\% measurement of
the weak charge in a heavy muonic atom could constrain $\rhostr$ as
tightly as would a forward--angle measurement of $\alr(\evec p)$, if
the nuclear neutron distribution were also known to sufficiently
high precision.

\vfil\eject
\noindent {\bf II.B.\quad Additional Aspects of PV Electron Scattering Studies}
\smallskip
\noindent {\bf\qquad\quad\  of Hadronic Structure}
\medskip
We now touch upon several additional issues where PV electron scattering may
provide information on hadronic structure with the focus on issues
other than the specific study of strangeness in the nucleon.  Similar
discussions occur for lepton scattering in general and neutrino scattering
in particular (see Ref.~[Don79a]) --- in the present section we restrict our
attention in this regard to PV electron scattering.
\smallskip

\noindent\undertext{Nuclear Parity Violation}

        Achieving a full understanding of the strong N--N interaction (as well
as three-- and higher--body forces) has been a long--standing problem in
nuclear
physics. Of more recent interest have been attempts to study the weak
interaction between nucleons. Since the PC part of this interaction has
strength of $\hcal{O}(\Gmu\mns)\sim 10^{-5}$ and is masked by the much stronger
electromagnetic [$\hcal{O}(\alpha)\sim 10^{-2}$] and strong
[$\hcal{O}(\alpha_s)\sim 1$] interactions, one must isolate the weak N--N
interaction by measuring PV observables, which provide a window on the PV
part of the weak N--N force. Direct $W^\pm$-- and $Z^0$--exchange between
nucleons has a range of $\sim 0.002$ fm and is suppressed by the short--range
repulsion between the nucleons. Consequently,
one conventionally models a longer--range PV N--N interaction as being
mediated by the exchange of light pseudoscalar and vector
mesons (Fig.~3.9),
where the virtual meson is emitted from one nucleon through a PC strong
interaction and absorbed by a second nucleon via a PV weak interaction.
The resultant two--body N--N PV potential can
be parameterized in terms of seven PV meson--nucleon couplings, $\hnnm$,
corresponding to different exchanged mesons ($M$) and isospin
channels (see Eq.~(\Fhpv)).\footnote{$^\dagger$}{It is conventional
in the literature to
denote the PV pion--nucleon coupling by $\fpi$; this unfortunate choice
of notation is not to be confused with the pion--decay constant, also
denoted by $\fpi$.}

        Theoretical predictions, based on quark--model calculations, give a
rather wide range of values for the $\hnnm$ [Des80]. Assuming
that these couplings can be extracted from nuclear PV experiments with
manageable nuclear physics uncertainties, such experiments test particle
physics methods for estimating low--energy hadronic matrix elements of
four--quark
operators. At the nuclear level, one may treat the $\hnnm$ as
experimentally determined parameters, and to the extent that a variety of
nuclear PV measurements may be fit with the same values for these parameters,
one has confirmation of this conventional picture of nuclear PV. In fact,
nuclear PV experiments completed to date have placed only rather modest
constraints on the $\hnnm$, all of which are
consistent with the theoretical
\lq\lq reasonable ranges" of Ref.~[Des80] (for a summary
of these experiments and associated theoretical issues, see, {\it e.g.,\/}
Ref.~[Ade85]). However,
further tightening of these constraints is necessary if estimates of
the four--quark hadronic matrix elements are to be tested in this way.
At the nuclear level, PV experiments have produced a self--consistent
set of values for the $\hnnm$, with one glaring exception: the value of
$h_\sst{\pi NN}$ extracted from the PV $\gamma$--decay of $^{18}$F is
about a factor of three smaller than suggested by all other experiments
(see Fig.~2.3). Proposed resolutions for this discrepancy include the need
to account for nucleon strangeness in estimating the $\hnnm$ [Dai91]
and possible contributions generated by a PV $NN\pi\pi$ vertex [Kap92].
Clearly, additional experimental information would be
desirable as one attempts to solve this puzzle.

        Several authors have suggested the possibility of using PV electron
scattering as a new probe of nuclear PV [Hen73, Hen79, Ser79, Hen82].
The contribution of nuclear PV to $\alr$ would
be generated by the processes illustrated in Fig.~3.9.
For spin $\geq 1/2$ targets, the resultant $\gamma$--nucleus
interaction will effectively include an axial--vector
component due to mixing of
opposite parity nuclear levels by the PV N--N force (Fig.~3.9a,b) and PV
meson--exchange currents (Fig.~3.9c).
For {\it elastic} scattering, this vertex induces
the so--called \lq\lq anapole moment" (AM) [Zel57, Zel60] which, unlike
other electromagnetic moments, couples only to virtual photons (for a
review of its properties, see Ref.~[Mus91]). At low--$|Q^2|$, the corresponding
amplitude has the same form as low--$|Q^2|$ $Z^0$--exchange with magnitude
growing as $A^{2/3}$. Thus, for heavy nuclei
the AM contribution to $\alr$ can be as large
as the leading axial--vector neutral current term [Fla84, Hax89, Bou91].
In such cases, one could reasonably expect to separate the contribution of
nuclear PV to $\alr$ from that of the NC, thereby providing a new means for
constraining models of nuclear PV. Alternatively, Flambaum, Khriplovich, and
others have observed that measurements of nuclear spin--dependent observables
in atomic PV experiments would also be sensitive to the nuclear AM [Fla80].
In fact,
evidence for a large AM term, consistent with theoretical predictions, was
reported from a recent atomic PV experiment on $^{133}\hbox{Cs}$ performed
by the Boulder group [Noe88].

        Theoretical evaluation of the PV contribution to elastic
scattering from $^{13}$C has been completed by Serot [Ser79], who found
that the effect of nuclear PV should be comparable to the NC contribution
at low electron energies. In the same work, it was found that nuclear
PV should dominate the forward--angle asymmetry for excitation of the
$(J^\pi T)=(1^+ 1)$ level in $^{12}$C. Hwang, Henley, and Miller [Hwa81]
considered PV electrodisintegration of the deuteron, and found comparable
contributions from nuclear PV and from the NC at low incident energies.
These studies are suggestive, and further analysis of PV electron
scattering as a probe of nuclear PV --- particularly for heavy targets ---
is warranted.

\bigskip
\goodbreak
\noindent\undertext{Determination of $\GEn$ and the Ground--State
Neutron Distribution}

        Present information on the
spatial distribution of charge in the neutron and
of neutrons in nuclei is much more limited than the corresponding information
regarding protons. In the case of the neutron charge distribution --- embodied
in momentum--space form $\GEn(Q^2)$ --- only the mean--square
radius is known to high precision [Sim80, Dum83].
Away from $Q^2=0$, the uncertainty
in $\GEn(Q^2)$ can be as much as 50\%.
The situation is even more
uncertain for $\rho_n(r)$, the ground--state neutron distribution in nuclei.
{}From a theoretical perspective, the most interpretable data on the
ground--state rms neutron radius, $R_n$, comes from moderate--energy elastic
$p$--nucleus  scattering.
Different analyses of the data on $^{208}$Pb allow a
difference between $R_n$ and
the proton rms radius, $R_p$, of $\sim 0.3$ fm (see the discussion in
Ref.~[Don89]) and it is not certain which of $R_n$ and $R_p$
is larger. Knowledge of $R_n$ has particular importance in the
interpretation of atomic PV experiments, where uncertainty in $R_n$ could
limit the future use of such measurements as Standard Model tests [For90,
Pol92a].  In addition, reactions involving the scattering of hadrons other
than single protons are used in extracting information on the
nuclear density distributions and specifically on the neutron distribution
in the nuclear ground state.  Examples where nuclei far from the valley of
stability can be explored include pion double charge exchange reactions
and the scattering of radioactive beams.  Having a few ``benchmark''
values of $R_n$ for selected nuclei might prove valuable to help
calibrate those hadronic analyses.

Both the issue of the role played by the charge distribution of the
neutron and that of determining the rms radius of the ground--state neutron
distribution have been addressed in past work [Don88, Don89] and will be
discussed further in Sect.~IV.B. In summary, it was shown that it would be
possible to obtain more precise information on $\GEn(Q^2)$ and
$\rho_n(r)$ than presently exists by using a combination of purely
electromagnetic and PV electron scattering.  This possibility relies on
the fact that isoscalar and isovector densities enter the
electromagnetic and neutral current response functions in different
linear combinations.  In effect the idea relies on isospin being a good
quantum number: PC and PV electron scattering in principle yield two
independent responses (for given kinematics) which can be recast in
terms of two form factors, one isoscalar and the other isovector.
Using the fact that the proton and neutron are (approximate)
eigenstates of isospin with $T=1/2$ (see the brief discussion above in
Sect.~II.A and Sects.~IV.A,B) the results may
then be rewritten in terms of
proton and neutron form factors.  The particular case of elastic
scattering from the proton is an important example.  There only PC and
PV scattering from the proton is involved and yet the charge form
factor of the neutron $\GEn$ plays an important role.  For typical
forward--angle scattering kinematics the effect of ignoring $\GEn$ would
amount to an error of
between 10 and 25\% in the asymmetry, as discussed in Ref.~[Don88]. Given
this rather high sensitivity to $\GEn$ and the poor knowledge presently
available about its magnitude away from $|Q^2|\approx 0$, it was
suggested in Ref.~[Don88] that PV asymmetry measurements be used together
with the usual PC determinations of $\GEp$, $\GMp$ and $\GMn$ to
extract information on $\GEn$.  Specifically, in that work it was estimated
that a 10\%
determination of $\GEn$ at $|Q^2|=0.4\ (\hbox{GeV}/c)^2$,  near the
maximum in the estimated figure--of--merit (FOM), could be obtained in a
$^1$H$(\evec, e)^1$H experiment using 4 GeV electrons at $\sim  9^o$
in 350 hours of running time.
A determination of $\GEn$ near the
beginning of the region where the FOM begins to fall--off
($|Q^2|\approx 1.0\ (\hbox{GeV}/c)^2$) would also appear to be feasible,
though with less precision due to uncertainties in $\GEp$ and $\GMn$
which also enter the asymmetry.  As discussed in more detail in Sect.~IV.A,
what actually enters in the asymmetry is not just $\GEn$, but the
combination $\GEn + \GES$. Thus, to extract values for the individual
form factors will require additional measurements such as using elastic
scattering from $0^+0$ nuclei to determine $\GES$ as outlined above or
using PC polarized electron scattering with hadronic polarizations to
determine $\GEn$.

        As for a determination of $R_n$, a 1\% measurement for
$^{208}$Pb could be obtained using PV electron scattering at
$\epsilon =300$ MeV in the region below the first diffraction minimum in the
charge form factor ($q\approx 0.6\ \hbox{fm}^{-1}$), with $\sim$ 10 days
of running time [Don89]. A similar determination of the neutron radius
for $^{133}$Cs also appears feasible [Sic91]. In addition to reducing
$R_n$ uncertainties below a problematic level for the interpretation of
atomic PV measurements using a single isotope, such measurements would
also serve to calibrate theoretical estimates of $\rho_n(r)$ as needed
in the analysis of atomic PV measurements with a series of isotopes.
Atomic theory uncertainties would be eliminated in the latter approach,
and lack of knowledge of ground--state neutron distributions would
appear to introduce the dominant theoretical error.

\bigskip
\noindent\undertext{Isospin Decomposition of Response Functions}

        One of the outstanding, unsolved puzzles in electron scattering
is the apparent failure of the Coulomb sum rule in QE scattering (see, for
example, Ref.~[Ber91]). Among the various explanations proffered for this
failure [Alb93a], and one which could be
further explored with semileptonic scattering, is the role played by
isospin--dependent nuclear correlations. In the absence of such
correlations, the longitudinal PV QE response is suppressed (see
Eq.~(\Fsmallqe)).
Consequently, the sensitivity of this response function to differences between
the isovector and isoscalar nuclear correlations is significant, so that
a measurement of this response could provide a useful window on such
correlations. A comparison of such a measurement with determinations
of the QE response in purely hadronic scattering would also be of interest
[Alb88].  We return in Sect.~IV.F to discuss PV QE electron scattering
in more detail.

\bigskip
\noindent\undertext{The Nucleon--to--Delta Transition}

The parity--violating asymmetry in the region of the $\Delta(1232)$ is
insensitive to many details of nucleon and $\Delta$ structure, and thus
may provide an interesting electroweak test. This possibility is
discussed in some detail in Sect.~IV.G. There are also several additional
contributions to $\alr(N\to\Delta)$ which are of
interest in themselves: (a) contributions from the axial--vector NC, and
(b) effects of non--resonant backgrounds. Knowledge of these contributions
is also relevant for the interpretation of NC QE electron and
neutrino scattering experiments, such as the QE $^9$Be$(\evec,
e')$ measurement at Mainz mentioned above, which integrate into the
\lq\lq dip" region
and tail of the $\Delta$ resonance. For the contribution from the
axial--vector current, which is pure isovector, knowledge of $\bra{\Delta}
J_{\mu 5} \ket{N}$ may be obtained from charge--changing $\nu$N
reactions by performing an isospin rotation, although information obtained by
this means is at present rather uncertain, with model predictions
varying by as much as 50\% at low--$|Q^2|$ [Sch73]. Recent experimental
results [Jon89, Kit90] have been fit with a specific (highly model--dependent)
form [Adl68] containing only one remaining free parameter
for $|Q^2|\lapp 1$ GeV$^2$, with a 10\% or more
uncertainty in the single fit parameter. At larger--$|Q^2|$, the fit is
even less satisfying.
The low--energy limit of $\alr(N\to\Delta)$ would
be the appropriate kinematic regime in which to probe $\bra{\Delta}
J_{\mu 5} \ket{N}$ directly, since in this case it is suppressed only by
the leptonic NC coupling $\gve$ and is dominated by a single
multipole.

 In the case of non--resonant background vector current contributions,
one requires electroproduction data for both proton and neutron targets
in order to obtain the isospin amplitudes needed in the NC transition, and
while such data exist for the proton, much less is known for the
neutron [Vap88]. There is independent interest in further
$\Delta(1232)$ electroproduction experiments at CEBAF [Bur89], as well as
at medium--energy accelerators, and consequently it is expected that
considerably better information will be available in due course.
In particular, future CEBAF experiments
will include measurements of non--resonant background multipoles
and a complete isospin decomposition.  In the absence of the latter,
combining $\alr(N\to\Delta)$ with existing electromagnetic data could
itself be used to measure this isospin decomposition of the backgrounds ---
a quantitative analysis of the latter possibility is in progress [Pol93].

\vfil\eject
\noindent {\bf II.C.\quad  Electroweak Tests}

        The motivation for much of the earliest discussions of
low-- and intermediate--energy semileptonic NC experiments was that
of testing the Standard Model of electroweak interactions [Fei75, Wal77,
Don79a].
Indeed, the goal of the MIT/Bates
$\alr(^{12}\hbox{C})$ experiment [Sou90a] was a determination of the isoscalar
hadronic NC vector coupling $\tilde\gamma$ (defined in Table~3.2),
while the Mainz QE
$\alr(^9\hbox{Be})$ measurement [Hei89]
sought to complement the SLAC deep--inelastic
$\alr(^2\hbox{H})$ measurement [Pre78, Pre79]
by constraining the isovector and
isoscalar hadronic NC axial--vector couplings, $\tilde\beta$ and $\tilde
\delta$, respectively. In a similar vein, the Brookhaven $\nu_\mu p\,
(\nubar_\mu p)$ experiment [Ahr87] provided constraints on the proton neutral
current couplings. Since one may predict the values of these
couplings using the Standard Model,
the aforementioned intermediate--energy scattering
experiments could be used as tests of the standard electroweak theory.
With the advent of very high precision $e^+e^-$
measurements at the $Z^0$ pole and of precision atomic PV experiments, the
role of intermediate--energy scattering in testing the Standard Model seems
less clear than previously thought. At present, these experiments seem
more suited as probes of hadron structure, and as such, they occupy a
unique position in the broad context of NC studies. Nevertheless, it is
still of interest to ask how these experiments might contribute in the
search for physics beyond the Standard Model. To that end, we
review briefly both the present status of electroweak tests and the
issues involved in the use of semileptonic scattering for this purpose.
More extensive reviews of electroweak tests may be found elsewhere
[Ell90, Alt91, Lan90].

\bigskip
\noindent II.C.1.\quad  STATUS OF ELECTROWEAK TESTS

        The minimal SU(2$)_L\times$U(1$)_Y$ Standard Model (one Higgs
doublet) with three generations of fermions (including massless neutrinos)
depends on 17 arbitrary parameters: the nine fermion masses, the Higgs
mass $\mh$, the three angles and phase which parameterize the
Kobayashi--Maskawa
(K--M) matrix and three parameters in the gauge sector. Since the K--M
parameters do not enter in tree--level neutral current amplitudes, we will not
discuss them further. Their values have been determined or constrained by
charge--changing processes, such as semileptonic decays and deep--inelastic
neutrino scattering
[RPP92]. Of the masses, only two remain largely undetermined:
$\mh$ and the top--quark mass, $\mt$. In the gauge sector, one has a choice
about which three parameters to treat as independent
inputs.\footnote{$^\dagger$}{This
point, as well as the choice of renormalization scheme, is discussed in
more detail in Sect.~III.} It is conventional to take the fine--structure
constant, $\alpha$, the Fermi constant measured in muon decay, $G_\mu$,
and the $Z^0$ mass, $\mz$, as these three parameters, since they are the three
most accurately know quantities in the gauge
sector.\footnote{$^\ddagger$}{
Although $G_\mu$ does not enter the Standard Model lagrangian directly,
it is simply related to parameters which do through Eq.~(\FGmu). }\
To date, they have been
determined to the precision indicated in Table~2.4 [RPP92].

\midinsert
$$\hbox{\vbox{\offinterlineskip
\def\strut{\hbox{\vrule height 15pt depth 10pt width 0pt}}
\hrule
\halign{
\strut\vrule#\tabskip 0.2cm&
\hfil$#$\hfil&
\vrule#&
\hfil$#$\hfil&
\vrule#\tabskip 0.0in\cr
& \multispan3{\hfil\bf TABLE 2.4\hfil} & \cr\noalign{\hrule}
& \hbox{Parameter} && \hbox{value}
& \cr\noalign{\hrule}
& \alpha&& 1/137.0359895(61)&\cr
& G_\mu&& 1.16639(2)\times 10^{-5}\ \ \hbox{GeV}^{-2}& \cr
& \mz&& 91.173\pm 0.020\ \ \hbox{GeV}&
\cr\noalign{\hrule}}}}$$
\smallskip
\baselineskip 10pt
{\ninerm
\noindent\narrower {\bf Table 2.4.} \quad
Gauge sector input parameters in the Standard Model.
\smallskip}
\endinsert

\baselineskip 12pt plus 1pt minus 1pt

\noindent In the on--shell renormalization scheme, where the weak mixing angle
is defined via
$$
        \sstw\>\equiv\> 1-{\mws\over\mzs}\ \ ,\eqno\nexteq\nameeq\Esinstw
$$
both $\mw$ and $\sstw$ are determined as functions of $(\alpha, \Gmu, \mz)$
through the relation [Sir80]
$$
        \sstw\cstw\>=\> {\pi\alpha\over\sqrt{2}\Gmu}{1\over\mzs}(1-\Delta
        r)^{-1}\eqno\nexteqp\nameeq\Esctw
$$
or
$$
\mws={1\over 2}\mzs\Bigl[1+\sqrt{1-A_o}\Bigr]\ \ ,\eqno\sameeq
$$
where
$$
A_o={2\sqrt{2}\pi\alpha\over\Gmu\mzs}{1\over 1-\Delta r}\eqno\sameeq
$$
and where $\Delta r$ is a radiative correction to muon--decay [Sir80]. In this
scheme, $\sstw$ and $\mw$ depend as well on $\mh$ and $\mt$ through the
dependence of $\Delta r$ on these masses. Hence, $\sstw$ cannot be determined
with the same precision as $(\alpha, \Gmu, \mz)$ until both $\mt$ and $\mh$
are known.

        Although $\sstw$ is not rigorously an independent parameter within the
context of the minimal
Standard Model,\footnote{$^*$}{This statement assumes that ($\alpha, \mz,
\Gmu$)
are taken as inputs.} one may test the theory by treating it
{\it as if\/} it were independent and comparing
experimental determinations based on this
assumption with values of $\sstw$ obtained from Eqs.~(\Esinstw) or
(\Esctw a). Such an
analysis, when performed without considering higher--order process
({\it e.g.,\/}
$\Delta r\to 0$), would test the consistency of the tree--level theory with
neutral current data. However, present (and prospective) experiments
are sensitive to $\hcal{O}(\alpha\Gmu)$  effects, so one must account for
second--order electroweak corrections to tree--level amplitudes when
interpreting
precision neutral current results. On the one hand, consideration of these
radiative corrections complicates the analysis of the structure of the theory:
virtual Higgs boson and top--quark loops introduce an $(\mh, \mt)$--dependence
into processes not involving these particles explicitly, thereby adding
ambiguity to comparisons of $\sstw$ derived from Eqs.~(\Esinstw),
(\Esctw a), and directly
from experiments.\footnote{$^\dagger$}{Some of this ambiguity may be eliminated
through an alternate choice of renormalization scheme, such as
$\overline{\hbox{MS}}$.} On the other hand, sensitivity to radiative
corrections allows one to test the quantum field theory nature of
the Standard
Model, much as measurements of the anomalous magnetic moment of
the electron --- a quantity arising solely
from higher--order processes --- allow one to test QED as a quantum field
theory.

Were $\mt$ and $\mh$ known precisely, the weak mixing angle would be
determined from Eq.~(\Esctw a) to much better than 1\% accuracy. For
example, taking $\mt=100$ GeV and $\mh=250$ GeV one finds [Lan90]
$$
        \sstw\vert_\sst{\mz}\>=\>0.2316\pm 0.0002\pm 0.0004\ \ ,
$$
where the first uncertainty is experimental, the second is theoretical,
and where a somewhat older value of $\mz$ has been used. The $\mh$--dependence
of $\Delta r$ is rather weak, so that once $\mt$ is known, $\sstw$ will be
fixed by the theory to much better than 1\% .
For comparison, the most precise determinations of $\sstw$
via Eq.~(\Esinstw) are obtained from
recent $p\bar p$ collider measurements of the ratio $\mw/\mz$
[Ali92, Abe91]. The
associated uncertainty in $\sstw$ is roughly 2--4\%.
When combined with the average LEP value for $\mz$, these results
also yield the most precise value of $\mw$ obtained to date, with a
corresponding error in this mass of about 0.5\% .

        Determinations of $\sstw$, treated as an independent, experimentally
measured quantity, come from a variety of experiments. In the purely leptonic
sector, early measurements of $\sigma(\nubar_e e)$, $\sigma(\nu_\mu e)$, and
$\sigma(\nubar_\mu e)$, taken together with forward--backward (FB) asymmetry
measurements in $e^+e^-$ annihilation, confirmed predictions based on
Eq.~(\Ejnc)
provided $\sstw\approx 0.22$ [Che84--Ch.~12]. More recent results from
CHARM, BNL, and LAMPF experiments have yielded values for $\sstw$ with
uncertainties ranging from 5--20\% [All90, Abe89, Dor89, Gei89, Abe87].

In the neutrino--quark sector, the first deep inelastic $\nu$--nucleon
experiments with isoscalar targets showed consistency with quark neutral
currents having the form in Eq.~(\Ejnc) with $\sstw=0.23\pm 0.023$
[Com83--Ch.~8].
The value of $\sstw$ obtained using isoscalar targets is relatively
independent of hadronic structure models [Pas73].
Determinations derived from deep inelastic $\nu(\nubar)$--nucleon scattering
with $T_3\not=0$ targets depend more strongly on details of quark distributions
in the nucleon [Com83, Che84, Ama87].
In both cases, a value of $\sstw$ is determined from ratios of
neutral current and charged current cross sections, such as
$$
R_{\nu(\nubar)}^\sst{DIS}\>=\> {\sigma(\nu N\to\nu X)\over\sigma(\nu N\to
\mu X)}\ \ .
$$
\smallskip\noindent
The average of the latest CDHS and CHARM  data
[All87, Abr86, Blo89, Gei89] gives a $\nu N\to \nu X$
value for
$\sstw$ with roughly 3\% combined experimental and theoretical
error. The dominant theoretical error arises from uncertainty in the
value for $m_c$ used in modeling the charm production threshold [Ama87].
In contrast with other direct determinations of $\sstw$, those
derived from deep inelastic $\nu$--N
scattering are quite insensitive to $\mt$, owing to an accidental cancellation
in radiative corrections to $R_{\nu(\nubar)}^\sst{DIS}$ [Alt91].
A much less precise value for $\sstw$ has been derived from the
Brookhaven (quasi)--elastic $\nu_\mu p/\nubar_\mu p$ experiment [Ahr87]
where a value of $\sstw$ with roughly 14\% error is reported.
We discuss these results in more detail in Sect.~IV.J.

        Standard Model tests using charged lepton probes fall into two classes.
The most precise results have been obtained from atomic PV
experiments with heavy atoms. Atomic PV observables result from the mixing
of opposite parity atomic states, induced by the PV NC interaction of the
atomic electrons with the nucleus.
The quantity of interest is the so--called \lq\lq weak
charge", $Q_\sst{W}$ (see Eqs.~(\Fdqw)),
 which depends, at the simplest level,
on the vector NC couplings to the proton and neutron.
Extraction of $Q_\sst{W}$ from the experimental observable depends on details
of atomic structure,
so that the uncertainty in the quoted value of
$\sstw$ depends on atomic theory uncertainties as well as on experimental
errors and theoretical uncertainties associated with higher--order effects.
Recent improvements in atomic theory
techniques have reduced this error in $Q_\sst{W}$
for $^{133}\hbox{Cs}$ to roughly the 1\% level [Blu90], whereas the
experimental error is roughly 2\% from the latest experiment by Noecker {\it et
al.} [Noe88]. Additional theoretical uncertainty enters the extraction of
$\sstw$ from $Q_\sst{W}$ due to the dependence of the weak charge on
the ground state neutron distribution [For90, Pol92a]. A 10\% uncertainty in
$R_n$ for $^{133}$Cs
would result in a 1\% error in $\sstw$. The corresponding errors in $\sstw$
from the experiment and from atomic theory are roughly 3\% and 2\%,
respectively.

Less precise Standard Model tests
have been obtained from PV electron--hadron
scattering. In addition to the pioneering deep inelastic $\evec D$
measurement at SLAC [Pre78, Pre79], which yielded a 9\% determination of the
weak mixing angle, determinations have also been carried out using
quasielastic and elastic PV electron scattering
on $^9$Be [Hei89] and $^{12}$C, respectively [Sou90a].
The former yielded a 7\% determination, whereas the error from the
elastic $^{12}$C measurement is nearly 25\% .

        A measure of the consistency of the minimal Standard Model with
all neutral current data can be obtained by performing \lq\lq global"
fits to the data. Several authors have recently reported on such fits.
Given the $(\mt, \mh)$--dependence of $\Delta r$ in Eq.~(\Esctw a) as well as
in
the
radiative corrections to other neutral current observables, such fits
produce ranges for $\sstw$ correlated with ranges for $\mt$ and $\mh$.
Representative results from different fits are summarized in Table~2.5.

\midinsert
$$\hbox{\vbox{\offinterlineskip
\def\strut{\hbox{\vrule height 15pt depth 10pt width 0pt}}
\hrule
\halign{
\strut\vrule#\tabskip 0.2cm&
\hfil$#$\hfil&
\vrule#&
\hfil$#$\hfil&
\vrule#&
\hfil$#$\hfil&
\vrule#\tabskip 0.0in\cr
& \multispan5{\hfil\bf TABLE 2.5\hfil} & \cr\noalign{\hrule}
& \hbox{Fit} && \sstw && \mt
& \cr\noalign{\hrule}
& \hbox{[Lan90]} && 0.2272\pm 0.004 && 139 {+33\atop -39} \pm 16
        &\cr
& \hbox{[Alt91] }&& 0.228\pm 0.005 && 140 \pm 45 & \cr
& \hbox{[Ell90] }&& 0.2273\pm 0.003 && 127 {+24\atop -30} &
\cr\noalign{\hrule}}}}$$
\smallskip
\baselineskip 10pt
{\ninerm
\noindent\narrower {\bf Table 2.5.} \quad
Global fits of $\sstw$, $\mt$, and $\mh$ to all electroweak data. Fit of
Ref.~[Alt91] includes only data from LEP, $p\bar p$ determination of $\mw/\mz$,
and $\nu N\to \nu X$. The $\sstw$ and $\mt$ values of Ref.~[Ell90] assume
$\mh=\mz$.  The authors of Ref.~[Ell90] also performed a fit in which
$\mh$ was allowed to vary and found $\mh > 1.8$ GeV at a 68\%
confidence level.
\smallskip}
\endinsert

\baselineskip 12pt plus 1pt minus 1pt
\noindent  These fits all indicate a central value for $\sstw$ near 0.23 with
uncertainties of approximately 1--2\%.
Most of this uncertainty arises
from uncertainty in $\mt$. Once $\mt$ is known, precision improves by an
order of magnitude, signalling consistency of the Standard Model with all
neutral current data at much better than the 1\% level. Even the present
1--2\% level of consistency is impressive, given the wide range in
energy scales which the data encompass.

        With present and future NC measurements approaching the
1\% level of precision or better, tests of this sort could be sensitive to
physics beyond the minimal SU(2$)_L\times$U(1$)_Y$ Weinberg--Salam theory.
While an in--depth discussion of this \lq\lq non--standard" physics lies beyond
the scope of this review and can be found elsewhere [Ama87, Alt91, Mar90,
Pes90, Gol90] we highlight
a few aspects relevant to intermediate--energy experiments. Deviations of the
so--called $\rho$--parameter from its Standard Model value could indicate
the presence of extra, non--doublet Higgs bosons, the presence of additional,
very heavy $W^\pm$ and $Z^{0\prime}$ bosons, or heavy fermion loops associated
with super--symmetric (SUSY) extensions of the Standard Model [Alt91, Ama87].
The natural parameter
for discussing grand unified theories (GUT's), $\sstwh$ (defined in
Sect.~III),  seems to rule
out non--SUSY GUT's such as SU(5) while providing some
consistency with SUSY
grand unification [Ama87].

        Recently, a new framework has been introduced
for discussing certain types of non--standard physics, such as SUSY or
technicolor, which would enter neutral current observables through gauge--boson
self energies [Ken90, Pes90, Mar90].
While some extensions of the Standard Model --- such as those
associated with tree--level exchange of extra $Z^{0\prime}$ bosons --- are not
described by this framework, it nonetheless constitutes a useful means of
comparing different observables for purposes of electroweak tests.
In one version of this
parameterization, new physics is described by two parameters: $T$, which is
particularly sensitive to mass splittings in boson or fermion
isomultiplets, and $S$, which characterizes degenerate heavy physics
contributions. In the treatment of Ref.~[Mar90] which we
follow here, a nonzero value for $T$ would also signal a value of
$\mt$ different from 140 GeV. Measurements of
$Z^0$ widths are dominantly sensitive to $T$ (at $\approx$ 1\% level), while
$Z^0$ asymmetries and neutrino scattering measurements are roughly equally
sensitive to both $S$ and $T$. Atomic PV and elastic PV electron scattering, on
the other hand, would probe primarily for nonzero values of $S$. In fact,
results from the Cs atomic PV experiment [Noe88] constrain this parameter
to the level $|\delta S|\approx \pm 2.3$, where the dominant error is
experimental. A reduction in this uncertainty
by a factor of ten would render one sensitive to contributions from either
additional generations of heavy fermions or the minimal one--doublet
technicolor model [Mar90]. Anticipated future improvements in the atomic PV
experimental uncertainty would leave theoretical atomic structure
uncertainty as the dominant source of error. It is expected that the
overall atomic PV uncertainty in $S$ will decrease by a factor of three or
more in the future [Lan91].

The completion of
one or more PV electron scattering experiments with sufficient precision
could complement constraints from atomic PV and reduce the uncertainty in
$S$ to an interesting level. For purposes of illustration, we plot in Fig.~2.4
the present constraints on $S$ and $T$ from atomic PV as well as potential
constraints from future, high--precision measurements of $\alr$. We assume
all low-- and intermediate--energy experiments agree on common central values
for these parameters, so that the axes give deviations from these values
allowed by experimental and theoretical uncertainty. From Fig.~2.4 one
observes both the rationale for exploring $\alr$ measurements as electroweak
tests as well as the precision needed to make such measurements relevant.
Indeed, a 10\% measurement of the elastic $\evec p$ asymmetry or 1\%
measurements of either the elastic $^{12}\hbox{C}(\evec, e)$ or
$N\to\Delta$ asymmetries could nicely complement atomic PV and constrain
$S$ to an interesting level.

\bigskip
\goodbreak
\noindent II.C.2.\quad  PRECISION TESTS WITH INTERMEDIATE--ENERGY SCATTERING
\medskip

\noindent \undertext{Experimental Considerations.}

        From the above discussion, it should be clear that any new
electroweak tests using neutrino or PV electron scattering must attain
$\sim$ 1\% precision (or 10\% for $\evec p$ scattering)
in order to be both meaningful and competitive with
tests in other sectors. This requirement presents both experimental as well
as theoretical challenges. Let us begin by discussing the former. Since the
typical asymmetries are very small (ranging from about $10^{-6}$ to a few
times $10^{-5}$ for electrons of a few 100 MeV to a few GeV), attaining
1\% precision requires a large number of scattering events with specified
electron helicity.  For instance, considering just the statistical precision
that can be reached, this range of asymmetries implies $10^{14}$ to
$10^{16}$ 100\% polarized electron scattering events to determine the
asymmetries to 1\%. Given that accelerators in the medium-- and high--energy
regime will deliver up to 100--200 $\mu$A ({\it e.g.,\/} CEBAF is
designed for a maximum current of 200 $\mu$A) and that there are limitations
on how thick practical targets can be, as discussed in Sect.~V, there is a
limit to how high the attainable luminosities can be for the foreseeable
future.  Typically the best that can be achieved is
${\cal L}\sim$ few $\times 10^{38}$
cm$^{-2}$ s$^{-1}$.  For typical asymmetries, cross sections and the
resulting FOM (see Sect.~III.E.2) even these extreme values
of luminosity still imply experimental running times of 100's to 1000's of
hours.  It should be obvious that if luminosities on the order of a few
$\times 10^{38}$ cm$^{-2}$ s$^{-1}$ or more cannot be reached, then most
high precision PV electron scattering experiments cannot be attempted.
As we shall discuss in detail for specific cases of interest (Sect.~IV),
for a few carefully selected hadronic transitions, the FOM is
large enough to allow us to contemplate reaching the 1\% precision level.
However, in many cases, such as for typical inelastic
excitation of discrete nuclear states, the FOM's are projected to be so
small that even getting to the 100\% precision level may be impractical.

In addition to these considerations, there exist
additional factors which only increase the level of experimental difficulty.
For example, the incident electron beams are not 100\% polarized.
While considerable progress has been made in recent years, and there is
hope that advances will continue in the next decade, the present
state--of--the--art restricts high--current (100--200 $\mu A$ average)
polarized electron sources to below 50\% polarization.
Since the experimental FOM is proportional to the square of the
measured asymmetry, this amounts to a factor of four reduction in the
effective do--ability of polarized electron experiments.  Higher polarizations
have been attained, but with lower average currents.

Other factors also enter into the final evaluation of the feasibility of
PV electron scattering experiments.  For instance, specific
measurements may require
very forward--angle scattering and the issue
of limited detector solid angle must be faced.  Alternately,
certain measurements may require detectors with sufficiently high energy
resolution to restrict the asymmetry determination to a single specific
transition. However, the available solid angle --- and the corresponding
achievable luminosity --- is generally limited for such high--resolution
detectors. Of course, if one does not require such high resolution, then
rather different poor--resolution, large solid angle detectors can be employed.
An example of the latter is a measurement of the elastic $\evec p$ asymmetry,
for which one only needs a resolution of a few hundred MeV  in order to
exclude contributions from excitation of the $\Delta(1232)$ resonance.

Beyond these considerations of statistical precision and resolution, there
are issues relating to systematic errors that are
discussed in some detail in Sect.~V.  As a distillation of past experience in
performing PV electron scattering experiments at SLAC, Mainz and
MIT/Bates one arrives at the expectation that systematic errors can be
controlled sufficiently well to attain the goal of 2\%--3\% asymmetry
determinations.  It should be realized that this is no mean feat: for an
asymmetry of $10^{-6}$ this implies keeping the sum of all systematic effects
below $10^{-8}$.

\bigskip
\goodbreak
\noindent \undertext{Theoretical Interpretability.}

        Even if these challenges could be surmounted, the
{\it interpretability} of such high precision results would be, at this
point, somewhat questionable, owing to theoretical uncertainties involved
in calculating strong interaction effects at low and intermediate
energies. While some of these uncertainties would be much less
problematic for the interpretation of precision neutrino scattering
measurements, experimental considerations suggest that such measurements will
not approach the 1\% level in the foreseeable future. The issue is more
relevant, however, for PV electron scattering, where some hope does exist for
attaining the requisite level of experimental precision. Over time, theoretical
progress, coupled with the completion of measurements constraining some
of the present sources of uncertainty in the PV asymmetry, may make electroweak
tests at the level suggested in Fig.~2.4 possible with PV electron scattering.
In what follows, we outline what would be required to realize this possibility.

        By a suitable choice of target and/or experimental kinematics, much
of the hadronic physics content of the ratio $\alr$ can be eliminated,
leaving only the dependence on the electroweak couplings, which would be
determined
from the Standard Model and its possible extensions.
Two cases of particular interest are low--$|Q^2|$,
forward--angle elastic scattering from the proton and elastic scattering
from $(J^\pi T)=(0^+ 0)$ nuclei. The recent $^{12}\hbox{C}({\svec e},e)$
experiment at Bates,
from which a 25\% measurement of $\sstw$ was extracted, falls into the
latter category. As suggested by Fig.~2.4, however, a future experiment
of this type would need to achieve $\sim$ 1\% precision in order to be
competitive with atomic PV as a probe of non--standard physics.
Forward--angle experiments on the proton have been
discussed as possibilities for CEBAF [Nap90]. In both of
these cases, the leading term in $\alr$ is nominally independent of
hadronic physics associated with the target (see
Eqs.~(\Fepstanda) and (\Fzerostd)).
Target--dependent corrections to the leading terms arise from a number of
sources: hadronic form factors, hadronic contributions to radiative
corrections, isospin impurities in the hadronic ground state,
and, in the case of an $A>1$ target, many--body currents.

        In the case of forward--angle scattering from the proton, the
most serious uncertainties appear to be introduced by nucleon form
factors. The contribution from these form factors to $\alr(\theta\to
0)$ vanish as $Q^2\to 0$, and na\"\i vely one would expect to minimize
their impact by working at sufficiently small momentum--transfer.
However, the FOM for such a measurement also
decreases with $|Q^2|$. Consequently, one cannot go to arbitrarily low
$|Q^2|$ without sacrificing the statistical precision needed to make a
meaningful electroweak test. In this regard, lack of knowledge of
$\GES$ is particularly problematic. Assuming that the strangeness form
factors were to be determined by a series of $\alr(\evec p)$ measurements
alone,
the remaining uncertainty in $\GES$ would still be larger than needed to
permit one to constrain $S$ and $T$ with PV electron scattering at the
level indicated in Fig.~2.4. Additional, although somewhat smaller
uncertainties
are introduced by lack of knowledge of other form factors.
This difficulty might be overcome either by
reducing the uncertainty in the strangeness form factors with an appropriate
combination of $\alr$ measurements with $A>1$ targets, or by building a
detector with sufficient solid angle at more forward angles than
are contemplated
for existing or planned detectors. Alternatively, if it were possible to
achieve very high currents ($\rapp$ few hundred mA) of polarized beams
at low--energy accelerators ($\epsilon\sim$ few hundred MeV), this
form factor issue could be surmounted.

For scattering from
$(0^+ 0)$ targets, the primary form factor ambiguity is also associated
with a term in the asymmetry containing $\GES$. Although this form
factor vanishes at the photon point, one cannot reduce its contribution by
going to arbitrarily
low $|Q^2|$ without increasing the statistical error in $\alr$ beyond
the 1\% level. Fortunately, one has some hope of constraining this
term to the precision needed for a Standard Model test. We show in
Sect.~IV.B how a series of two measurements of $\alr(0^+ 0)$ would be
sufficient for this purpose. In the absence of such a determination,
the uncertainty in $\GES$ remaining after its
determination with $\evec p$ scattering would, given some model assumptions,
severely restrict the kinematics at which a meaningful
$(0^+ 0)$ electroweak test would need to be performed.
For purposes of comparison, we note in passing
that given the very small effective momentum transfer in the interaction
of an atomic electron with the nucleus, the impact of nucleon form
factors in atomic PV electroweak tests can largely be ignored. Apart
from the uncertainties associated with the neutron distribution and
atomic structure, then,  atomic PV offers a distinct advantage over
semileptonic
scattering as far as testing the Standard Model is concerned.

        For both $\alr(\evec p)$ and $\alr(0^+ 0)$, additional uncertainties
are generated by hadronic contributions to higher--order electroweak
amplitudes. Hadronic contributions to the $Z^0$--$\gamma$ mixing tensor
have been estimated using a dispersion analysis, and the associated uncertainty
appears significantly below a problematic level [Mar84, Deg89].
Of more concern are the dispersion corrections discussed above in connection
with strangeness measurements. While the scale of hadronic uncertainties
in the one--body (single nucleon) dispersion amplitudes is likely to be
no greater than 1\% of the tree--level amplitude [Dre59, Gre69, Mar84]
data from elastic scattering on $^{12}$C [Kal89]
and $^{208}$Pb [Bre90] suggest a significant many--body enhancement, at least
in
the case of purely electromagnetic scattering.
Since the EM and NC dispersion corrections are unlikely to cancel from
the PV asymmetry, a better understanding of this contribution appears
necessary in the interpretation of $\alr(0^+ 0)$ as an electroweak test.

        One expects the effect of isospin--mixing in the nucleon to be
suppressed, given the scale of mass splittings (and, hence, energy
denominators) in the baryon spectrum. In particular, the first isospin--3/2
components that might isospin--mix with an isospin--1/2 ``proto--nucleon''
to form the physical nucleon occur at the opening of the $\pi N$ channel,
{\it viz.,\/} at 140 MeV. In the case of nuclei, however, the
typical energy splitting between low--lying states is roughly an order
of magnitude smaller, so that one might anticipate the presence of
non--trivial isospin impurities in the nuclear ground state. In the case of
$(0^+ 0)$ $s$--$p$ shell nuclei, the effect of isospin--mixing on $\alr$
at low momentum transfer has been estimated to be below 1\% [Don89].
This result follows from the difficulty in supporting an isovector
monopole matrix element in the relevant nuclear model space. Thus,
isospin--mixing is unlikely to present a problematic uncertainty in
a $(0^+ 0)$ electroweak test.

        Once one goes beyond the special cases of elastic scattering from
$^1$H and $(0^+ 0)$ targets,
additional uncertainties enter the use of $\alr$ measurements to test the
Standard Model. In the case of QE scattering, for example, one
encounters nuclear physics uncertainties associated with contributions
from the pion--production cross section, \lq\lq dip region", and radiative
tail as well as from the QE peak. The present theoretical understanding
of these regions is somewhat limited, particularly with regard to the dip
region, and the magnitude of these uncertainties is difficult to quantify.
For PV QE electron scattering, one might anticipate some cancellation
of nuclear physics uncertainties from $\alr(\hbox{QE})$, since the asymmetry
depends on a {\it ratio} of nuclear response functions. Investigations
using a relativistic Fermi Gas model for the QE response in fact indicate
that $\alr(\hbox{QE})$ is significantly less sensitive to various nuclear
model parameters than are the individual response functions appearing in
the PC and PV cross sections [Don92].
To the extent that future $\alr(\hbox{QE})$
measurements are carried out at kinematics for which non--QE contributions
are negligible, this result gives one hope that nuclear physics uncertainties
might be minimized.

Even so, uncertainties associated with single--nucleon
form factors, which plague potential $\evec p$ electroweak tests, also
enter $\alr(\hbox{QE})$ at a non--negligible level. In the recent
Mainz QE $^9\hbox{Be}(\evec , e')$ experiment, carried out at backward--angles,
uncertainties in the radiative corrections to the axial--vector term
induce a 2--3\% uncertainty in the extracted value of $\sstw$ [Mus92a].
Similarly, a forward--angle determination of $\sstw$
is likely to be highly sensitive to uncertainty in $\GES$. Consequently,
PV QE electron scattering appears more suited to the determination of
nucleon form factors than to tests of electroweak theory. As far as
QE neutrino scattering is concerned, as in the BNL and
LSND experiments,  an electroweak test based on a measurement
of the cross section alone incurs both form factor and nuclear physics
uncertainties.  It is likely that LSND will be limited by statistical
accuracy: the systematic errors on the cross section are projected to be
less than 5\%, corresponding to an error in $\eta_s$ of about 0.03.
Quasielastic neutrino scattering does present at least one
theoretical advantage over electron scattering in that
the large and uncertain radiative correction
to the axial--vector response in electron scattering does not arise in
processes involving neutrinos.

The inelastic $N \rightarrow \Delta(1232)$ transition, being purely
isovector, provides direct information on vector and axial--vector isovector
couplings.  In contrast with the $\vec e p$ elastic cases discussed
above, isoscalar nucleon structure plays a small role, and will not
dominate the resulting uncertainty in extracted electroweak
parameters. More problematic are contributions from
the axial--vector NC matrix elements. As discussed above,
these matrix elements are at present insufficiently well known for the
purposes of $N\to\Delta$ electroweak tests at
CEBAF energies or lower.  The axial--vector contribution is, however,
explicitly suppressed by the inverse of the incident beam energy, and
thus the forward angle, high--energy limit should be the appropriate
kinematic regime for such tests.

 Non--resonant background contributions to $\alr(N\to\Delta)$ are also a
potential source of serious uncertainty.  A full isospin decomposition
of $\Delta$ electroproduction amplitudes at the same $Q^2$ as where a
PV asymmetry measurement could be attempted would be
sufficient to eliminate this problem, and may be undertaken at
CEBAF [Bur89].  In the absence of such data, one must turn to
predictions from theoretical models (see, for example, Ref.~[Li82])
to estimate non--resonant isoscalar backgrounds which
complicate the interpretation of the electroweak tests.  Estimates of
the resulting uncertainty are still needed, and some work in this
direction is in progress.

        From the foregoing discussion, it should be clear to the reader that
semileptonic, NC scattering observables are sensitive to a variety of
of physics issues (hadron structure, extensions of the Standard Model, nuclear
dynamics) at potentially significant levels. At the same time, the tasks
of choosing the appropriate combination of measurements and reducing the
sources of theoretical uncertainty in order to draw meaningful conclusions
from such measurements are non--trivial. In the near term, it appears that
intermediate--energy, semileptonic NC scattering is best suited as a probe of
nucleon and nuclear structure --- particularly nucleon strangeness. Its use
to search for physics beyond the Standard Model awaits both the completion
of the \lq\lq first generation" of NC studies and progress on the theoretical
issues outlined above. In the remainder of this article, we provide the
rationale for these conclusions in greater detail.

\vfil\eject

\secnum=3
\neweq

%

\noindent{\bf III.\quad FORMALISM}
\medskip

     The formalism for treating semileptonic electroweak interactions is
straightforward and largely parallels the treatment of the
electromagnetic interaction between charged leptons and hadrons based
on quantum electrodynamics.  Our goal here is to use this formalism in a
way that emphasizes the physics issues summarized in the previous
section.  The first step will be to construct the relevant hadronic
(both nucleon and nuclear) currents from the underlying fundamental
electroweak interactions between leptons and quarks.  Since the physics
of interest generally requires one
to consider effects at the 1\% level, it will be
necessary to take the theory beyond tree--level to include
contributions from radiative corrections and other higher--order processes.
Upon taking nucleon matrix elements of the appropriately renormalized
current operators, one encounters the interplay of strong and electroweak
dynamics which complicates the interpretation of semileptonic scattering
at low-- and intermediate--energies.
With the hadronic currents in hand, we turn to the specifics of
parity--violating (PV) polarized electron scattering and neutrino scattering.
We subsequently discuss the complications generated by
the additional level of hadronic structure associated with
nuclear ($A>1$) targets.

\bigskip
\noindent{\bf III.A.\quad Currents and Couplings}

        We begin with the fundamental couplings of an elementary
fermion (lepton or
quark) to the photon and $Z^0$ which we write, respectively, as
$$ieQ_f\gamma_\mu\eqno\nexteqp$$
and
$$i{gM_Z\over 4M_W}\gamma^\mu\left(\gvf +\gaf\gamma_5\right)\
.\eqno\sameeq$$
Here $e$ and $g$ are the electromagnetic and weak coupling strengths,
respectively, $Q_f$ is the electromagnetic charge of the fermion,
and the vector and
axial--vector ``charges'', $\gvf$ and $\gaf$, respectively, are given by
$$\gvf = 2T_3^f - 4Q_f\sin^2\theta_W\eqno\nexteqp\nameeq\Fermc$$
and
$$\gaf = -2T_3^f\ ,\eqno\sameeq$$
where $T_3^f$ is the third component of a {\it weak}
isospin vector operator; acting on
the  weak isodoublets $\pmatrix{\nu_e\cr e^-}$, $\pmatrix{u\cr d}$,
$\ldots$, $T_3^f$ gives $+{1\over2}$
for the upper component and $-{1\over2}$ for the lower. Thus, one obtains
the values given in Table~3.1.

\midinsert
$$\hbox{\vbox{\offinterlineskip
\def\strut{\hbox{\vrule height 15pt depth 10pt width 0pt}}
\hrule
\halign{
\strut\vrule#\tabskip 0.2cm&
\hfil$#$\hfil&
\vrule#&
\hfil$#$\hfil&
\vrule#&
\hfil$#$\hfil&
\vrule#\tabskip 0.0in\cr
& \multispan5{\hfil\bf TABLE 3.1\hfil} & \cr\noalign{\hrule}
& \hbox{fermion} && \gvf && \gaf
& \cr\noalign{\hrule}
& \nu_e,\nu_\mu && 1  && -1  &\cr
& e^-,\mu^- && -1+4\sstw  && 1  &\cr
& u,c,t && 1-\coeff{8}{3}\sstw  && -1  &\cr
& d,s,b && -1+\coeff{4}{3}\sstw  && 1  &\cr
\noalign{\hrule}}}}$$
\smallskip
\baselineskip 10pt
{\ninerm
\noindent\narrower {\bf Table 3.1.} \quad
Standard Model values (columns two and three) for the
hadronic neutral current couplings of
elementary fermions
(first column). Note that the nomenclature introduced in Ref.~[Don79a]
differs from that used in the present work: there the leptonic neutral
current coupling were defined with respect to the quantities in the table
by $a_\sst{V}=\gvf$ and $a_\sst{A}=-\gaf$.
\smallskip}
\endinsert

\baselineskip 12pt plus 1pt minus 1pt

     With these couplings, it is straightforward to compute
the one--boson--exchange (photon or $Z^0$) amplitudes for scattering
of a lepton from a hadronic electromagnetic (EM) current and weak neutral
current (NC) associated with Figs.~3.1a and b, respectively:
$$
M^\sst{EM} =  {4\pi\alpha\over Q^2}Q_\ell \ell^\mu
J_\mu^\sst{EM}\eqno\nexteq\nameeq\Emem
$$
and
$$
M^\sst{NC} = -{G_\mu\over 2\sqrt{2}}\left\lbrack \gvl\ell^\mu +\gal\ell^{\mu 5}
\right\rbrack\left\lbrack J_\mu^\sst{NC} +
J_{\mu 5}^\sst{NC}\right\rbrack\ .\eqno\nexteq\nameeq\Emnc
$$
In Eq.~(\Emnc) we have neglected the $Q^2$--dependence of the $Z^0$ propagator
since $\vert Q^2\vert <\!\!< M_Z^2$
for all of the kinematics we consider in the present work.
The PV component of the neutral current amplitude is
$$M^{PV} = -{G_\mu\over 2\sqrt{2}}\left\lbrack
\gvl \ell^\mu J_{\mu 5}^\sst{NC} +
\gal\ell^{\mu 5}J_\mu^\sst{NC}\right\rbrack\ .\eqno\nexteq\nameeq\Empv$$
In these expressions, $Q_\ell$, $\gvl$, and $\gal$ are the leptonic
 electromagnetic,
vector, and axial--vector charges, respectively, and $Q_\mu \equiv K_\mu
-K^\prime_\mu$ is the four--momentum transfer with $K(K^\prime)$ the
initial (final) lepton four--momentum.  Here, for reasons discussed in
Sect.~II.C.1, we have chosen to write $M^\sst{NC}$ in terms of the Fermi
constant for muon decay, $G_\mu$ instead of the weak coupling strength,
$g$.  At tree--level, these constants are related by
$$
G_\mu = {g^2\over 4\sqrt{2}\mws}={\pi\alpha\over\sqrt{2}\mws\sstw}
\ \ \  .\eqno\nexteq\nameeq\EGmu
$$
Once one works beyond tree--level,  Eq.~(\EGmu) must be modified to
account for electroweak radiative corrections to muon decay, $\Delta r$
[Sir80]:
$$
G_\mu={\pi\alpha\over\sqrt{2}\mws\sstw}{1\over 1-\Delta r}\ \ \ .\eqno\nexteq
$$
\par
The lepton vector and axial--vector currents, $\ell^\mu$ and $\ell^{\mu 5}$,
respectively, are just the Dirac currents
$$\ell^\mu \equiv \bar u_\ell\gamma^\mu u_\ell\eqno\nexteqp$$
$$\ell^{\mu 5} \equiv \bar u_\ell\gamma^\mu\gamma^5 u_\ell\ .\eqno\sameeq$$
where $u_\ell$ is the lepton spinor.
Since the hadrons are composed of quarks, the currents $J_\mu^\sst{EM}$,
$J_\mu^\sst{NC}$, and $J_{\mu 5}^\sst{NC}$ are the hadronic matrix
elements of the electromagnetic, vector, and axial--vector quark current
operators:
$$J_\mu^\sst{EM} \equiv \langle H \vert \hat J_\mu^\sst{EM} \vert H\rangle
\eqno\nexteqp$$
$$ J_\mu^\sst{NC} \equiv \langle H \vert \hat J_\mu^\sst{NC} \vert  H\rangle
 \eqno\sameeq$$
$$ J_{\mu 5}^\sst{NC} \equiv \langle H \vert \hat J_{\mu 5}^\sst{NC}
\vert  H\rangle \ \ ,\eqno\sameeq$$
where $\vert H\rangle$ is any hadronic state (in the present context
a nucleon or a nucleus) and,
$$\hat J_\mu^\sst{EM} \equiv \sum_q Q_q\bar u_q\gamma_\mu u_q
\eqno\nexteqp\nameeq\Eqexpand$$
$$\hat J_\mu^\sst{NC} \equiv \sum_q \gvq\bar u_q\gamma_\mu u_q \eqno\sameeq$$
$$\hat J_{\mu 5}^\sst{NC} \equiv \sum_q \gaq\bar u_q\gamma_\mu\gamma_5 u_q
\ \ ,\eqno\sameeq$$
where the sums are over all quark flavors, $u$, $d$, $c$, $s$, $\ldots$.
\par

     In what follows, we assume the structure of the hadronic states
is dominated by the lighter quarks and limit the sums in Eqs.~(\Eqexpand)
to $q$ =
$u$, $d$, and $s$. The error introduced by neglect of the heavier quarks
is expected to be
of order $10^{-4}$ ($10^{-2}$) for the vector (axial--vector) currents
(see Eqs.~(\Frvacbt)).
With this truncation, it is convenient to decompose the current
operators in terms of the SU(3) octet and singlet currents:
$${\hat V}_\mu^{(a)} \equiv \bar q {\lambda^a\over 2}
\gamma_\mu q \eqno\nexteqp$$
$${\hat A}_\mu^{(a)} \equiv \bar q {\lambda^a\over 2} \gamma_\mu \gamma_5 q
\ \ ,\eqno\sameeq$$
where $q$ represents the triplet of quarks, $q \equiv \pmatrix{u\cr d\cr s}$,
$\lambda^0=\coeff{2}{3}\hbox{\bf 1}$,
and the $\lambda^a$, $a=1,\ldots, 8$ are the Gell--Mann SU(3) matrices,
normalized to $Tr(\lambda^a\lambda^b)=2\delta^{ab}$.  (Note that no assumption
of SU(3) symmetry is implied in the use of the SU(3) decomposition of the
{\it operators}. Such an assumption enters only when SU(3) symmetry is
employed to determine {\it matrix elements} of these operators.)  In the
case of the electromagnetic and weak neutral currents, one requires only
the six diagonal terms
$${\hat V}_\mu^{(0)} = {1\over 3} \left\lbrack \bar u\gamma_\mu u + \bar
d\gamma_\mu d + \bar s\gamma_\mu s\right\rbrack \eqno\nexteqp$$
$${\hat V}_\mu^{(3)} = {1\over 2} \left\lbrack \bar u\gamma_\mu u - \bar
d\gamma_\mu d \right\rbrack \eqno\sameeq$$
$${\hat V}_\mu^{(8)} = {1\over 2\sqrt{3}} \left\lbrack \bar u\gamma_\mu u +
\bar
d\gamma_\mu d -2 \bar s\gamma_\mu s\right\rbrack \eqno\sameeq$$
$${\hat A}_\mu^{(0)} = {1\over 3} \left\lbrack \bar u\gamma_\mu \gamma_5 u
+ \bar
d\gamma_\mu \gamma_5 d + \bar s\gamma_\mu \gamma_5
s\right\rbrack \eqno\sameeq$$
$${\hat A}_\mu^{(3)} = {1\over 2} \left\lbrack \bar u\gamma_\mu \gamma_5 u
- \bar
d\gamma_\mu \gamma_5 d \right\rbrack \eqno\sameeq$$
$${\hat A}_\mu^{(8)} = {1\over 2\sqrt{3}}
\left\lbrack \bar u\gamma_\mu \gamma_5 u + \bar
d\gamma_\mu \gamma_5 d -2 \bar s\gamma_\mu \gamma_5
s\right\rbrack \ . \eqno\sameeq$$
     At the level of {\it strong} isospin, both the 0$^{\rm th}$
(singlet) and 8$^{\rm th}$
(octet) SU(3)
components are ``isoscalar'' operators, while the 3$^{\rm rd}$  octet
component is ``isovector''.  The 8$^{\rm th}$
and 3$^{\rm th}$  components
of the vector current are related to the isoscalar and
isovector electromagnetic currents, respectively,  by
$${\hat J}_\mu^\sst{EM} (T=0) = {1\over\sqrt{3}} {\hat V}_\mu^{(8)}
\eqno\nexteqp\nameeq\Ejemmu$$
$${\hat J}_\mu^\sst{EM} (T=1) = {\hat V}_\mu^{(3)} \ \ .\eqno\sameeq$$
Since we wish to emphasize the strange--quark contribution to
various processes, it is useful to note the following relationship between the
two isoscalar currents:
$${\hat V}_\mu^{(0)} = {2\over\sqrt{3}}{\hat V}_\mu^{(8)} + {\hat V}_\mu^{(s)}
\eqno\nexteqp$$
$${\hat A}_\mu^{(0)} = {2\over\sqrt{3}}{\hat A}_\mu^{(8)} +
{\hat A}_\mu^{(s)}  \ \ ,\eqno\sameeq$$
where
$${\hat V}_\mu^{(s)} \equiv \bar s\gamma_\mu s \eqno\nexteqp$$
$${\hat A}_\mu^{(s)} \equiv \bar s\gamma_\mu\gamma_5 s \ .\eqno\sameeq$$
With these relations we can re--write the
weak neutral currents of Eqs.~(\Eqexpand) in
the following convenient forms:
$$\hat J_\mu^\sst{NC}
=\xi_V^{T=1}{\hat J}_\mu^\sst{EM}(T=1) +
\sqrt{3}\xi_V^{T=0} {\hat J}_\mu^\sst{EM}(T=0) +\xi_V^{(0)} {\hat V}_\mu^{(s)}
\eqno\nexteq\nameeq\Ejncmu$$
and
$$\hat J_{\mu 5}^\sst{NC}
=\xi_A^{T=1}{\hat A}_\mu(T=1) +
\xi_A^{T=0} {\hat A}_\mu^{(8)} +\xi_A^{(0)} {\hat A}_\mu^{(s)} \ \ ,
\eqno\nexteq\nameeq\Ejncmuf$$
where
$$\xi_V^{T=1} = \gvu-\gvd \eqno\nexteqp\nameeq\Exig$$
$$\xi_V^{T=0} = \sqrt{3}(\gvu+\gvd) \eqno\sameeq$$
$$\xi_V^{(0)} = \gvu+\gvd+\gvs \eqno\sameeq$$
$$\xi_A^{T=1} = \gau -\gad \eqno\sameeq$$
$$\xi_A^{T=0} = \sqrt{3}(\gau+\gad) \eqno\sameeq$$
$$\xi_A^{(0)} = \gau+\gad+\gas \ .\eqno\sameeq$$
The extra factor of $\sqrt{3}$ appearing in Eq.~(\Ejncmu) but not in
Eq.~(\Ejncmuf)
results from the use of the relation between the octet and isoscalar EM
currents
in Eq.~(\Ejemmu a).

     Writing the hadronic neutral current in the form of Eqs.~(\Ejncmu) and
(\Ejncmuf) delineates
between the physics associated with the underlying electroweak gauge theory
and the hadronic physics associated with the quark currents.  The content
of the former is contained in the couplings $\xi_{\sst V,A}^{(a)}$,
whereas hadronic
effects enter via matrix elements of the latter.  At tree--level, the couplings
$\xiva$ are determined by the underlying electroweak gauge theory via
Eqs.~(\Exig).
In Table~3.2 we give their values in the Standard Model as well as
their relation to other hadronic NC couplings defined in the literature.
Note that in earlier treatments, possible contributions from $s$--quarks
were ignored, so that no equivalents to $\xivz$ and $\xiaz$ are listed.

\midinsert
$$
\hbox{\vbox{\offinterlineskip
\def\strut{\hbox{\vrule height 15pt depth 10pt width 0pt}}
\hrule
\halign{
\strut\vrule#\tabskip 0.2cm&
\hfil$#$\hfil&
\vrule#&
\hfil$#$\hfil&
\vrule#&
\hfil$#$\hfil&
\vrule#&
\hfil$#$\hfil&
\vrule#\tabskip 0.0in\cr
& \multispan7{\hfil\bf TABLE 3.2\hfil} & \cr\noalign{\hrule}
& \hbox{Coupling} && \hbox{Hung and}
&& \hbox{Donnelly and} && \hbox{Standard}& \cr
& \ && \hbox{Sakurai}
&& \hbox{Peccei} && \hbox{Model}&
\cr\noalign{\hrule}
& \xi^{(0)}_V && \ && \  &&  -[1+\rvz]  &\cr
& \sqrt{3}\, \xi^{T=0}_V && -6\tilde\gamma/\gae &&
 2\beta^{(0)}_V && -4\sstw[1+\rvtez] & \cr
& \xi^{T=1}_V && -2\tilde \alpha/\gae && 2\beta^{(1)}_V &&
2(1-2\sstw)[1+\rvteo]&\cr
\noalign{\hrule}
& \xi^{(0)}_A && \ && \  && 1+\raz  & \cr
& \xi^{T=0}_A && -2\sqrt{3}\,\tilde\delta/\gve &&
- 2/ \sqrt{3}\,\beta^{(0)}_A && \sqrt{3}\ratez & \cr
& \xi^{T=1}_A && -2\tilde \beta/\gve && - 2\beta^{(1)}_A && -2[1+\rateo]&
\cr\noalign{\hrule}}}}
$$
\smallskip
\baselineskip 10pt
{\ninerm
\noindent\narrower {\bf Table 3.2.} \quad Hadronic weak neutral current
couplings. Columns two and three give equivalent couplings in the notation
of Refs.~[Hun76] and [Don79a], respectively. Column four
gives Standard Model values, including effects of higher--order contributions
($\rva$). At tree--level, one has $\rva=0$.
\smallskip}
\endinsert

\baselineskip 12pt plus 1pt minus 1pt

{}From Eqs.~(\Ejncmu) and (\Ejncmuf) one sees
that, given a measurement of $J_\mu^\sst{NC}$
($J_{\mu 5}^\sst{NC}$) and independent determinations of the vector
(strong) isoscalar EM, isovector EM , and $s$--quark
(axial--vector isoscalar, octet, and $s$--quark)
current matrix
elements, it is
in principle possible to extract the $\xiva$ ($\xi_\sst{A}^{(a)})$ and
thereby perform an
electroweak test. As discussed in Sect.~III.B below, these couplings
depend on higher--order electroweak processes in the Standard Model,
hadronic--structure effects, and possible contributions from extensions
of the Standard Model, such as those characterized by the $S$ and $T$
parameterization discussed in Sect.~II.C. The extent to which one
might place constraints on the latter, however, depends on both one's
ability to calculate the hadronic effects appearing in the NC couplings as well
as the precision with which one may determine the EM, octet, and strangeness
matrix elements associated with the right hand side of
Eqs.~(\Ejncmu) and (\Ejncmuf).

     Another
class of hadronic contributions to the hadronic NC couplings arises from
heavy--quark
($c$, $b$, $t$) renormalizations of the light quark current operators.
The scale of these contributions has been estimated in Ref.~[Kap88]
following the effective theory approach discussed in Sect.~II.A (see
especially Fig.~2.1). They may be included in the $\rva$ as follows:
$$
\eqalignno{\rvz&\longrightarrow\rvz(\hbox{ewk})-\Delta_\sst{V}&\nexteqp
     \nameeq\Ervacbt\cr
         \rvtez&\longrightarrow\rvtez(\hbox{ewk})-\Delta_\sst{V}&\sameeq\cr
         \raz&\longrightarrow\raz(\hbox{ewk})+2\Delta_\sst{A}&\sameeq\cr
         \ratez&\longrightarrow\ratez(\hbox{ewk})+\coeff{4}{3}\Delta_\sst{A}
\ \ ,&
         \sameeq\cr}
$$
where $\Delta_\sst{V}\lapp 10^{-4}$ and $\Delta_\sst{A}\sim 10^{-2}$ according
to Ref.~[Kap88] and where the $\rva(\hbox{ewk})$ are contributions
from higher--order electroweak processes.

Information on the matrix elements
of the currents entering the right--hand side of
Eqs.~(\Ejncmu) and (\Ejncmuf) can be obtained
from a variety of sources. Information on
the EM matrix elements is obtainable from parity--conserving (PC) electron
scattering, while a determination of the axial--vector octet matrix elements
may be performed using nuclear beta--decay and hyperon semileptonic decays.
As noted in Sect.~II.A, much less is presently know about the strangeness
matrix elements, and their determination constitutes one of the chief goals
of the experiments discussed in the present work. The strategy for doing
so follows from the form of Eqs.~(\Ejncmu) and (\Ejncmuf).
If the $\xi$'s as well as
EM and octet axial--vector matrix elements are all taken as inputs from other
experiments, then determinations of the NC matrix elements would allow
extraction of information on the $s$--quark currents. Discussion of different
targets and kinematic
conditions that might be used for this purpose makes up the bulk of
Sect.~IV below. Once the $s$--quark matrix elements are sufficiently
constrained it could be possible to perform independent determinations
of the $\xi$'s by going to the appropriate kinematic regime. As indicated
by the hypothetical constraints on ``non--standard" physics
of Fig.~2.4, one has reason to contemplate such a program. In what
follows,
we discuss in more detail how such new physics appears in the hadronic NC
couplings.

\vfil\eject

%

\def\sstwh{{\sin^2{\hat\theta}_\sst{W}}}
\def\xivp{{\xi_\sst{V}^p}}
\def\xivn{{\xi_\sst{V}^n}}

\noindent{\bf III.B.\quad Higher--order Processes and Renormalization}
\medskip

        As indicated by the foregoing discussion, the corrections
$\rva$ to the tree--level expressions for the couplings
$\xi_\sst{V,A}^{(a)}$ receive
a variety of contributions: electroweak radiative corrections within
the framework of the Standard Model, corrections generated by physics
beyond the Standard Model, and hadronic physics effects of various types.
We have already considered one contribution of the latter type, namely,
the correction generated by neglect of $c$, $b$, and $t$ quarks in an
effective theory approach. In this section, we discuss other
higher--order contributions germane to analysis of the experiments of interest.

\bigskip
\noindent\undertext{Standard Model Radiative Corrections}

     Na\"{\i}vely, one might assume higher--order processes involving multiple
photon and/or $Z^0$ exchanges as in Figs.~3.2
to be suppressed by at least $\alpha / 4\pi
\approx 10^{-3}$ relative to the simple single--photon or single--$Z^0$
exchanges of Fig.~3.1.  Thus, one might hope to be able to neglect
these  effects without introducing significant error in interpreting
semileptonic  processes.  However, such an assumption breaks down in at least
two ways.  First, calculations of one--loop electroweak corrections
may be significantly enhanced over their generic $\alpha/4\pi$ scale due
to the presence of large logarithms of the form $\ln{(M_W/ p_f)}$,
where $p_f$ is a momentum scale (or mass)
characteristic of one of the scattering leptons or quarks.
In addition, there exist cases where the tree--level contribution is suppressed
relative to the generic scale of tree--level NC amplitudes.
One case is obvious
from Table~3.2; $\xi_A^{T=0} =0$ at tree level in the Standard Model so that
the
existence of such a term depends on higher--order processes.  In two other
cases, PV observables associated with charged--lepton probes of the
$\bra{H}J_{\mu 5}^\sst{NC}\ket{H}$ and
processes which involve the neutral current charge form factor of the proton,
the tree--level
amplitudes are proportional to $1-4\sstw\approx 0.08$. Higher--order
contributions need not carry this small factor and, thus, may be relatively
more
important than one na\"{\i}vely expects.

        An in--depth treatment of theoretical estimates of the
higher--order terms can be found elsewhere [Sir80, Mar80, Mar81, Aok82,
Mar 83, Mar84, Ros90 and references therein]; we limit ourselves
to discussing briefly the main physics issues and quoting typical results.
The problem of calculating electroweak radiative corrections to tree--level
semileptonic amplitudes is considerably more complicated than in the case
of purely leptonic scattering. In the latter instance, the scattering
leptons interact only electroweakly, so that once a renormalization framework
is chosen and all parameters of the electroweak model determined, the theory
makes precise and unambiguous predictions for higher--order leptonic
processes. Complications
arise in semileptonic scattering from the interplay of the strong and
electroweak interactions involving complex intermediate states characterizing
the structure of the hadronic target. In high--energy processes, such as
$e^+e^-$ annihilation or deep inelastic scattering, one can
reliably estimate strong interaction structure effects by treating hadronic
quarks as quasi--free and using perturbative QCD. Such a first--principles
approach breaks down, however,
for medium-- and low--energy processes ($|Q^2| < 1$ \gevocsq)
where the strong coupling $\alpha_s(Q^2)$ becomes large. In this
regime, one is generally forced to rely on hadronic models in order to account
for strong interaction effects. In the case of the vector currents,
one may eliminate much of the hadronic uncertainty
by employing Eq.~(\Ejemmu) and experimentally determined matrix elements of the
EM currents. In other cases, however,
model--dependence introduces an intrinsic theoretical uncertainty into
estimates of the higher--order terms.

        In order to delineate between pure Standard Model electroweak
corrections and those containing hadronic physics uncertainties, it is
convenient to consider first corrections to scattering involving a
single quark at a time (Fig.~3.3). Calculations of these
``one--quark" amplitudes closely parallel the treatment of higher--order
corrections in purely leptonic processes. The effect of the one--quark
contributions is to renormalize the $\xi$'s appearing in the current
operators in a manner independent of the target's structure. However,
since the quarks from which the lepton scatters are confined to a
hadron ({\it e.g.}, the nucleon) with a radius of about 1 fm, one ought to
account for confinement where appropriate in renormalizing the quark
current operators. To this end, it is reasonable to suppress
contributions to loops from momenta $\lapp 1/R_{\rm nucleon}$
through the use of constituent, rather than
current, quark masses ($M_{u,d}\sim 330$ MeV) in the quark propagators.
In all other respects, the quark fields appearing in Eqs.~(\Eqexpand)
are those associated with the current quarks of the QCD lagrangian. The
procedure for treating internal quark loops ({\it e.g.}, Fig.~3.4), in
which the virtual $\bar qq$ pair need not be confined to the hadron
interior, is somewhat different, as discussed below.

        At this one--quark level, several theoretical issues
should be mentioned. First is the choice of renormalization scheme. Since the
use of the classical, or tree--level, theory to compute higher--order processes
results in amplitudes which are infinite, the starting theory must be
re--defined in such a way as to generate finite, higher--order amplitudes. The
prescription followed in performing this re--definition is the \lq\lq
renormalization scheme". Because calculations of electroweak amplitudes
are performed only to finite order in perturbation theory, estimates
obtained under different renormalization schemes will, in general, differ
slightly [Jeg89]. We follow
the so--called ``on--shell" renormalization
scheme (OSR), which represents a natural extension of the prescription
generally followed in renormalizing QED. In this scheme, poles in the
renormalized propagators occur at physical particle masses and elementary
vertices are renormalized with all particles on--shell. Moreover, $\sstw$ is
defined in terms of the vector boson masses (Eq.~(\Esinstw)). The independent
parameters in the theory are $(\alpha, \mz, G_\mu)$ in the gauge boson
sector and the fermion and Higgs masses in the remaining sectors. All
other parameters (including $\mw$ and $\sstw$) are determined as functions
of these parameters.

     Another widely used scheme is the so--called ``MS--bar"
($\overline{\hbox{MS}}$),
or modified minimal subtraction, scheme. In contrast to the OSR prescription,
the $\overline{\hbox{MS}}$ procedure  defines finite inverse propagators and
vertices by subtracting only the divergent parts of the loops, without placing
physical
requirements on the finite remainders. Moreover, the weak mixing angle in
this scheme carries a dependence on a renormalization mass scale $\mu$
(typically $\mz$ or $\mw$) and is no longer simply related to the vector
boson masses as in Eq.~(\Esinstw):
$$
\sstw=1-{\mws\over\mzs}\longrightarrow \sstwh(\mu)\ ,
$$
where $\sstwh(\mu)$ defines the $\overline{\hbox{MS}}$ weak mixing angle.
The $\overline{\hbox{MS}}$ prescription has the attraction that
$\sstwh(\mu)$ is straightforwardly related to the running EM and semi--weak
couplings $e(\mu)$ and $g(\mu)$ arising in grand--unified theories [Mar81].
Moreover, the $\rva(\overline{\hbox{MS}})$ display
a weaker dependence on the presently undetermined top--quark mass than do
the $\rva(\hbox{OSR})$. On the other hand, OSR has the advantages of
a simple, scale--independent definition of $\sstw$ and of being a
straightforward extension of the procedure conventionally followed in QED.
There exist at least two other prescriptions used in the literature:
the so--called ``$*$--scheme" of Refs.~[Ken89] and
the scheme followed in Ref.~[Con89]. In these latter schemes,
the weak--mixing angle depends on the value of $Q^2$ at which a given
process is studied, whereas in OSR, $\sstw$ is independent of momentum
transfer and the entire $Q^2$--dependence is carried by the $\rva$.

        A second significant issue alluded to above
is the dependence of electroweak amplitudes
on the unknown Higgs and top--quark masses. This dependence enters the
electroweak scattering amplitudes in two ways. First, $\sstw$--dependent
Born--level amplitudes vary with $\mh$ and $\mt$ when $\sstw$ is determined
using relation (\Esctw a). Top--quark and Higgs loops in muon decay amplitudes
induce an $(\mh, \mt)$ dependence
in $\Delta r$ appearing in Eq.~(\Esctw a) and, consequently, in $\sstw$.
Second, top--quark and Higgs loops in the neutral current amplitudes themselves
also introduce a dependence on  the unknown masses. The dependence on $\mt$ is
stronger than the dependence on $\mh$; the latter enters only logarithmically,
while top--quark loops introduce terms of order $(\mt/\mw)^2$ in the vector
boson propagators. For $\mt > \mw$, these terms become significant.

        Third, it is important to bear in mind that the electroweak corrections
$\rva(\hbox{ewk})$ depend on both the species of lepton probe as well as the
$Q^2$ of the process under consideration. In general, the corrections for
electron scattering differ from those for neutrino processes. This distinction
is particularly important in the interpretation of probes of the isoscalar,
axial--vector hadronic NC, as we discuss below. As for the dependence on
momentum transfer, the $\rva$ vary rather gently with $Q^2$. According
to Ref.~[Mar80], for example, the corrections for neutrino scattering
$\rvtez$, $\rvteo$, and $\RVn$ (defined in Eq.~\Frvpn b)
change by $\lapp$ 0.001 over the range
$0\leq |Q^2|\leq 20\ (\hbox{GeV}/c)^2$, while the variation in $\RVp$
(defined in Eq.~\Frvpn a) is
$\sim 0.01$ over the same $Q^2$--range (calculated
assuming $\mh=10\mz$ and $\mt=18$
GeV). For the medium--energy
experiments under consideration here, corrections of this order are
negligible. More significant for semileptonic experiments is the
much greater $Q^2$--dependence of the hadronic form factors
appearing in matrix elements of the quark current operators. In the case
of elastic $\evec p$ scattering, for example, one must know the $Q^2$ behavior
of the {\it neutron} electromagnetic charge form factor, $G_\sst{E}^n(Q^2)$, in
order to determine the longitudinal proton neutral current response (see
discussion of Sect.~IV.A  below). The uncertainty in $G_\sst{E}^n(Q^2)$ is
significantly larger than any error incurred by neglecting the
$Q^2$--dependence
of the lepton--quark amplitudes and represents a much more serious issue for
the interpretation of high--precision experiments.

        Fourth, we note that one class of diagrams not included in the
calculations cited below are those corresponding to bremsstrahlung from
either of the scattering fermions. Formally, the inclusion of
such diagrams is required to
cancel infrared divergences in the one--loop amplitudes [Kin62, Lee64].
The finite remainders, however, depend on details of the experimental
configuration, such as detector resolution. Hence, one requires detailed
knowledge of the specific experiment before arriving at an estimate of
bremsstrahlung contributions. In certain special cases, such as scattering
from $(0^+ 0)$ targets, bremsstrahlung cancels from the PV asymmetry,
thereby simplifying the theoretical interpretation.
We also point out that the interpretation of PV
electron scattering asymmetries requires knowledge of QED radiative corrections
as well those which renormalize the NC couplings of Table~3.2. The QED
corrections for PC electron scattering have been worked out in detail
elsewhere [Mo69].

        Fifth, in order to make contact with notation used elsewhere in the
literature, the radiative corrections for atomic PV are sometimes written
in the form (see also Eqs.~(\Fxiprne))
$$
\eqalign{\xivp=2C_{1p}&=\rho_\sst{PV}'[1-4\kappa_\sst{PV}'(0)\sstwh(\mzs)]
        \cr
           \xivn=2C_{1n}&=\rho_\sst{PV}'\ \ \ ,\cr}\eqno\nexteq
$$
where the $C_{1p}$ ($C_{1n}$) are the neutral vector current couplings
(normalized differently than in the present work) and where the radiative
corrections are contained in the parameters $\rho_\sst{PV}'$ and
$\kappa_\sst{PV}'(0)$. The corrections for $Q^2\not=0$ are obtained by
replacing $\kappa_\sst{PV}'(0)\to\kappa_\sst{PV}'(Q^2)$ (the $Q^2$--dependence
of $\rho_\sst{PV}'$ is negligible). This parameterization of the radiative
corrections is motivated by a \lq\lq factorization" of the corrections into
two general classes: (a) corrections associated with the SU(2)$_L$ components
of the weak currents, and (b) those associated with the mixing of the SU(2)$_L$
and U(1)$_Y$ sectors. The $\rho_\sst{PV}'$ and
$\kappa_\sst{PV}'$ differ from the $\rho$ and $\kappa$ terms appearing in
neutrino scattering amplitudes by process--dependent terms. Writing
$\rho_\sst{PV}'=1+\delta\rho$ and $\kappa_\sst{PV}'=1+\delta\kappa$, the
$R_\sst{V}$ can be written as
$$
\eqalignno{\rvtez&=\delta\rho+\delta\kappa&\cr
           & &\cr
           \rvteo&=\delta\rho-\biggl[{2\sstw\over(1-2\sstw)}\biggr]
           \delta\kappa &\nexteq\cr
          & &\cr
          \RVp&=\delta\rho-\biggl[{4\sstw\over(1-4\sstw)}\biggr]
          \delta\kappa&\cr
          & &\cr
          \RVn&=\delta\rho\ \ \ .&\cr}
$$
Our rationale for writing corrections in terms of the $R$ rather than
$\rho$ and $\kappa$ is to facilitate comparison of electroweak radiative
corrections with other contributions ({\it e.g.}, the $\Delta_\sst{V,A}$
of Eqs.~(\Ervacbt)) which
do not simply factorize into contributions associated with the two gauge
groups.

        With the aforementioned considerations in mind, we note the scale of
typical results and refer the reader to the literature for more detail. In the
OSR scheme, one has that the $R_\sst{V}^{(a)}(\hbox{1-quark})$
are on the order of 1--5\%
at $\mt=120$ GeV and $\mh$=100 GeV for both electron and neutrino scattering.
For the axial--vector corrections, on the other hand,
$R_\sst{A}^{(a)}(\hbox{1-quark})$ $\sim$ few percent
for neutrino scattering, significantly
smaller than for electron scattering where
$R_\sst{A}^{(a)}(\hbox{1-quark})$ $\sim$ 25--50\%, due to the
logarithmic enhancements and tree--level suppression factors mentioned above.
This result has important consequences for the interpretation of observables
containing $\bra{H}J_{\mu 5}^\sst{NC}\ket{H}$.
For comparison, one has in the $\overline{\hbox{MS}}$ scheme
$$
\eqalignno{\RVp&=-0.054\pm 0.033&\cr
           \RVn&=-0.0143\pm 0.0004&\nexteq\nameeq\Ervoneq\cr
           \rvteo&=-0.017\pm 0.002&\cr
           \rvtez&=-0.0113&\cr}
$$
for $\mh=100$ GeV and $\mt=140$ GeV, where the values have been extracted
from Ref.~[Mar90] and where the uncertainties include estimated
uncertainties in hadronic contributions (see below). Although the values
were computed for atomic PV, they should not differ appreciably for
PV electron scattering given the gentle $Q^2$--dependence of the $\rva$.

\bigskip
\goodbreak
\noindent\undertext{Hadronic Contributions}

     Higher--order processes which depend on both the electroweak and strong
interactions generally introduce some degree of theoretical uncertainty
into the $\rva$ due to the present incalculability of strong interaction
effects at low--momentum scales from first principles in QCD. Such effects
enter higher--order corrections in two ways: (a) via \lq\lq internal" quark
loops (Fig.~3.4), in which the quarks in the virtual $q\bar q$ pair
interact strongly
with each other (Fig.~3.4a) or with quarks in the target (Fig.~3.4b); and (b)
via strong interactions among the valence quarks of the target, thereby
introducing hadronic intermediate states into the higher--order electroweak
amplitudes (Fig.~3.5).
In the case of internal loops, the effects of strong interactions
of Fig.~3.4a may be estimated using a dispersion theory analysis of
$e^+ e^-$ annihilation. Such an approach has been followed by the authors
of Refs.~[Mar84, Deg89].
The scale of this uncertainty, which is included in the error appearing in
Eqs.~(\Ervoneq), is given by these authors to be $\delta R_V^{(T)} \approx
\pm 0.002$.
Since this uncertainty is target--independent,
one may associate it conceptually with the 1--quark corrections. In contrast,
contributions of the type in Fig.~3.4b depend on the target, and no
estimate of their scale has been performed to date.

        Whereas the 1--quark corrections effectively renormalize the
quark current {\it operators} of Eqs.~(\Eqexpand ff), contributions involving
strong interactions among the target quarks (Fig.~3.5) arise only
in hadronic {\it matrix elements} of these operators. We label the
latter ``many--quark" hadronic
corrections to distinguish them from corrections
requiring no knowledge of target structure. The present intractability of
a first--principles estimate of $\rva(\hbox{had})$ requires one to rely on
phenomenological models, which necessarily introduces considerable
theoretical uncertainty. While more theoretical work on this topic is
needed, estimates of $R_\sst{A}^{(T)}(\hbox{had})$ for PV electron
scattering have been performed by the authors of Ref.~[Mus90],
based on the loop and pole contributions of Fig.~3.6. Results are
summarized in Table~3.3.

\midinsert
\vbox{
$$
\vbox{\offinterlineskip
\hrule
\halign{\vrule\strut#&\strut\quad\hfil#\hfil\quad&\vrule\strut#
        &\strut\quad\hfil#\hfil\quad
        &\vrule\strut#&\strut\quad\hfil#\hfil\quad&\vrule\strut#
        &\strut\quad\hfil#\hfil
        \quad&\vrule\strut#&\strut\quad\hfil#\hfil\quad&\vrule\strut#\strut\cr
&\multispan9{\hfil\bf Table 3.3 \hfil}&\cr\noalign{\hrule}
&\omit&&\omit&&\omit&&\omit&&\omit&\cr
&Source&&$\sqrt{3}\ratez$&&$\rateo$
&&$\sqrt{3}\ratez$&&$\rateo$&\cr
&\omit &&$(\hbox{best})$&&$(\hbox{best})$
&&$(\hbox{range})$&&$(\hbox{range})$&\cr
&\omit&&\omit&&\omit&&\omit&&\omit&\cr
\noalign{\hrule}
&\omit&&\omit&&\omit&&\omit&&\omit&\cr
&Standard Model&& -0.43 && -0.47 && - && - &\cr
&Hadronic&& -0.19 && 0.13 && -0.43 $\to$ 0 && -0.07$\to$ 0.31 &\cr
&\omit&&\omit&&\omit&&\omit&&\omit&\cr
\noalign{\hrule}
&\omit&&\omit&&\omit&&\omit&&\omit&\cr
&total&& -0.62 && -0.34 && -0.86$\to$ -0.43 && -0.54$\to$ -0.16 &\cr
&\omit&&\omit&&\omit&&\omit&&\omit&\cr}
\hrule}
$$}
\smallskip
\baselineskip 10pt
{\ninerm
\noindent\narrower
{\bf Table 3.3}. Estimates of Standard Model (one-quark)
and hadronic
contributions to the $R_\sst{A}^{(T)}$ for PV electron
scattering.  \lq\lq Best" values in hadronic contributions correspond
to use of theoretical \lq\lq best" estimates of Ref.~[Des80] for
weak meson--nucleon vertices appearing in Fig.~3.6. Ranges (columns
4 and 5) correspond to theoretical ranges in these couplings.
One--quark contributions are evaluated at $\mt=120$ GeV and $\mh=100$ GeV
in the OSR scheme.
\smallskip}
\endinsert

\baselineskip 12pt plus 1pt minus 1pt

While the $R_\sst{A}^{(T)}(\hbox{had})$
receive contributions from a plethora of diagrams not considered in
Ref.~[Mus90], these results ought to indicate both the {\it scale}
of hadronic effects as well as the degree of theoretical uncertainty.
{}From Table~3.3, one observes that the resultant hadronic uncertainty may
be as large as the Standard Model contributions to
$R_\sst{A}^{(T)}(\hbox{1-quark})$. Note that column four gives hadronic
uncertainty in the
induced isoscalar axial--vector current for PV electron scattering.
These results have important consequences for the interpretation of
PV electron scattering experiments, as discussed in Sect.~IV. Moreover,
since it is unlikely that theoretical uncertainty will be
significantly reduced in the foreseeable future, one
would ideally like to determine these corrections experimentally. Possibilities
for doing so are also discussed in Sect.~IV.

     Another type of many--quark diagram, the ``dispersion'' correction,
involves multi--boson exchanges between lepton and hadron as shown in Fig.~3.7.
The difficulty in reliably computing this contribution derives from the
lack of knowledge of the structure of the hadronic intermediate states.  One
does not expect this difficulty to be relevant to diagrams involving the
exchange of two heavy vector bosons, since the corresponding amplitudes are
dominated by large loop momenta ($k_{\rm loop}\sim \mw$). Since such
diagrams generate the dominant dispersion corrections for neutrino scattering,
theoretical uncertainties in neutrino dispersion corrections should be
tolerably small. For amplitudes involving charged lepton probes
where at least one of the exchanged bosons is a photon, however, the loop
integrals no longer need be dominated by large momenta, so that contributions
from low--lying target structure may be important. At the one--quark level,
for example, the PV $e$--$q$ amplitudes arising form the $Z^0$--$\gamma$ box
graph contain large logarithms of the form $\ln |s|/\mws$ and
$\ln |u|/\mws$, where the Mandelstam variables $s$ and $u$ depend on
the momenta of the lepton probe and target quark. Such a logarithmic
dependence on widely different momentum scales indeed suggests the possible
importance of low--lying target structure. In the case of two--$\gamma$
exchange, theoretical estimates of $ep$ dispersion corrections suggest
that hadronic contributions do not differ significantly from the generic
$\hbox{\cal O}(\alpha/4\pi)$ scale [Dre59, Gre69]. Experimental determinations
of PC $ep$ dispersion corrections are
consistent with this conclusion, though the level of precision is not
sufficient to constrain theoretical predictions. The situation appears
rather different for scattering from nuclei, where data from recent
measurements on $^{12}$C suggests significant many--body enhancement of PC
dispersion corrections over their one--body ($ep$) scale. This situation
is discussed more fully in Sect.~III.D.

        Dispersion corrections to NC $eN$ amplitudes have received less
theoretical attention, and there exists no experimental information in this
case. The authors of Ref.~[Mar84]
have estimated hadronic contributions
to the $Z^0$--$\gamma$ box amplitude in atomic PV by including only nucleon
intermediate states. The resultant corrections to the tree--level PV vector
current amplitudes are $\RVp\approx 0.013$ and $\RVn\approx -0.0006$. The
corresponding contributions to $\rvtez$ and $\rvteo$ are on the order of
a few tenths of a percent. No study has been made of many--body contributions,
which might result in enhancements over the one--body scale similar to those
seen in PC scattering from carbon.

        With these remarks in mind, we note several general features of
dispersion corrections as they enter PV electron scattering. The relevant
quantity for the experiments of interest here is the difference
$$
R_\sst{V,A}^{\rm disp}\approx R_\sst{V,A}^{\gamma Z^0}-R_\sst{V}^{\gamma\gamma}
\ \ , \eqno\nexteq$$
where $R_\sst{V,A}^{\gamma Z^0}$ ($R_\sst{V}^{\gamma\gamma}$) are the dominant
dispersion corrections to the PV (PC) electron scattering amplitudes. The
reason one is interested in the difference is that the asymmetry measured
in PV electron scattering is governed by the ratio of NC to EM amplitudes
(see Sect.~III.D). Although one might hope for significant cancellation
between these two corrections, there exist several reasons why a
non--negligible difference may occur.
First, electromagnetic and
neutral current dispersion corrections display different $Q^2$--dependences at
low--$Q^2$. The one--photon exchange diagram has a pole at $Q^2=0$, whereas
the two--photon box--diagram cannot have such a pole, since it is one--particle
irreducible.
The neutral current dispersion correction, however, need not vanish at
$Q^2=0$ since the tree--level $Z^0$--exchange amplitude has a pole at
$Q^2=\mzs$
rather than $Q^2=0$.
Additional differences follow from the structure of the diagrams. The
$Z^0$--$\gamma$ box diagrams, for example, generate a different weighting of
loop
momentum than occurs in the 2--$\gamma$ exchange graphs, due to the presence of
a heavy vector boson propagator in the loop. Consequently, transitions to
different intermediate hadronic or leptonic states are weighted differently
in the two cases. Moreover, the isospin content of the neutral vector current
{\it transition} matrix elements to hadronic intermediate states is not the
same as for the electromagnetic current, due to the different isospin structure
of the two currents. Given these features, along with experimental results
for $R_\sst{V}^{\gamma\gamma}$ for $^{12}$C, one has reason to invest
addition theoretical effort in order better understand this higher--order
process. Such study is especially warranted if one hopes to extract
information on ``new" physics from the $\rva$.

\bigskip
\noindent\undertext{Non--standard Physics}

        As discussed in Sect.~II, there exists a variety of scenarios for
extending the Standard Model beyond the minimal SU(2$)_L\times$U(1$)_Y$
Weinberg--Salam theory. While space does not permit an extensive review
of different scenarios and useful discussions may be found in the literature
(see, {\it e.g.} [Ama87]), we return to the $S$, $T$ framework to illustrate
how
non--standard physics could appear in the observables of interest here, to
determine what level of experimental precision is needed in these observables
to provide for interesting electroweak tests, and to set the corresponding
scale of maximum tolerable theoretical uncertainty ({\it e.g.}, from hadronic
contributions) in their interpretation. To that end, we focus on the vector
current corrections $R_\sst{V}^{(a)}$ which, owing to current conservation
and one's knowledge of the $J_\mu^\sst{EM}$, appear to contain smaller
hadronic physics uncertainties than the $R_\sst{A}^{(a)}$. Since we
subsequently consider PV electron scattering from the proton in Sect.~IV.A,
we define the corresponding correction $\RVp$ as
$$
\xi_\sst{V}^p = {1\over 2}[\xivteo+\sqrt{3}\xivtez] =
(1-4\sstw)[1+\RVp]\ \ , \eqno\nexteqp\nameeq\Exiprne
$$
where $\xi_\sst{V}^p$ is the NC coupling to the proton at $Q^2=0$.
Likewise we have
$$
\xi_\sst{V}^n = {1\over 2}[-\xivteo+\sqrt{3}\xivtez] =
-[1+\RVn] \eqno\sameeq
$$
for neutrons. These quantities are related through the following:
$$
\eqalignno{\RVp &= {1\over{1-4\sstw}}\Bigl[ (1-2\sstw)\rvteo
-2\sstw\rvtez \Bigr] &\nexteqp\nameeq\Ervpn\cr
\RVn &= (1-2\sstw)\rvteo+2\sstw\rvtez\ \ . &\sameeq\cr}
$$
Following Ref.~[Mar90], it is straightforward to determine the dependence of
the $R_\sst{V}^{(a)}$ on $S$ and $T$ [Mus93c]:
$$
\eqalignno{\rvtez(\hbox{new})&\approx 0.016S -
0.003T&\nexteqp\nameeq\ERsubv\cr
         \rvteo(\hbox{new})&\approx -0.014S + 0.017T&\sameeq\cr
         \RVp(\hbox{new})&\approx -0.206S + 0.152T &\sameeq\cr
         \RVn(\hbox{new})&\approx 0.0078T \ \ . &\sameeq\cr}
$$
Note that these relations are arrived at within the
$\overline{\hbox{MS}}$ renormalization
scheme and that a nonzero value of $T$ would signal a top--quark mass different
from 140 GeV.  It is also worth noting that the relatively larger
sensitivity of $\RVp$ to $S$ and $T$ is a consequence of the small scale of
the leading--order proton coupling (see Eq.~(\Exiprne)).

        The prospective constraints on $S$ and $T$ shown in Fig.~2.4 are
derived from Eq.~(\ERsubv), assuming a 1\% determination of $\xivtez$ from
elastic PV C$(\evec, e')$ scattering, a 10\% determination of $\xi_\sst{V}^p$
from $\alr(\evec p)$, and a 10\% extraction of $\xivteo$ from
$\alr(N\to\Delta)$. While the first two of these prospective determinations
could nicely complement results from cesium atomic PV, roughly an order of
magnitude improvement in $\alr(N\to\Delta)$ would be needed to make the latter
competitive. We note that the impact of hadronic uncertainties in the
radiative corrections on the prospective low--energy $S$ and $T$ constraints
of Fig.~2.4 is significantly smaller than either the assumed experimental
error or the uncertainty associated with poorly constrained hadronic form
factors.

\bigskip
\noindent\undertext{Higher--order Contributions: Summary}

All of the
higher--order processes discussed above can be folded into corrections to
the couplings of Table~3.3.  Explicitly, we have
$$
\eqalignno{R_V^{T=0}& = R_V^{T=0}(\hbox{1-quark}) +R_V^{T=0}(\hbox{had})
+R_V^{T=0}(\hbox{new}) -\Delta_V&\nexteqp\cr
R_V^{T=1}& = R_V^{T=1}(\hbox{1-quark}) +R_V^{T=1}(\hbox{had})
+R_V^{T=1}(\hbox{new})&\sameeq \cr
R_V^{(0)}& = R_V^{(0)}(\hbox{1-quark}) +R_V^{(0)}(\hbox{had})
+R_V^{(0)}(\hbox{new}) -\Delta_V&\sameeq\cr
R_A^{T=0}& = R_A^{T=0}(\hbox{1-quark}) +R_A^{T=0}(\hbox{had})
+R_A^{T=0}(\hbox{new}) +2\Delta_A&\sameeq\cr
R_A^{T=1}& = R_A^{T=1}(\hbox{1-quark}) +R_A^{T=1}(\hbox{had})
+R_A^{T=1}(\hbox{new})&\sameeq \cr
R_A^{(0)}& = R_A^{(0)}(\hbox{1-quark}) +R_A^{(0)}(\hbox{had})
+R_A^{(0)}(\hbox{new}) +{4\over3}\Delta_A\ \ .&\sameeq\cr}
$$

\bigskip
\noindent\undertext{Isoscalar Axial--Vector Current}

Before leaving this section, we note that the higher--order processes are
particularly important for the
isoscalar axial--vector current:
$$
\jncf(T=0) = \xiatez A_\mu^{(8)} + \xiaz A_\mu^{(s)}\ \ .\eqno\nexteq
\nameeq\Eaxxter
$$
Since $\xiatez=0$ in lowest order in the Standard
Model, one might na\"{\i}vely conclude
that a measurement of the isoscalar axial--vector matrix element
would directly probe for strange quarks in a non--strange
hadron. It should be emphasized, however, that this na\"{\i}ve conclusion holds
only because the tree--level isoscalar
Standard Model coupling vanishes and {\it not} because
matrix elements of $A_\mu^{(8)}$, which contains the isoscalar combination of
$u$-- and $d$--quark axial--vector currents, are zero. Moreover, once
higher--order
processes renormalize the tree--level axial--vector couplings, $\xiatez$
becomes
nonzero. In this case, $\bra{H}\jncf(T=0)\ket{H}$ does not necessarily
vanish even in the absence of strange quarks.
A more accurate statement
is that
$\jncf(T=0)$ is {\it suppressed} in the absence of strange--quarks, since
it enters only at higher--order in electroweak couplings. In contrast,
$s$--quark
contributions to the isoscalar axial--vector current enter at leading order in
electroweak couplings ($\xiaz=1$ at tree--level). However, one expects
{\it matrix elements} of $A_\mu^{(s)}$ to be suppressed for non--strange
hadrons, so that strange--quark contributions to $\bra{H}\jncf(T=0)
\ket{H}$ should also be small.

 {\it A priori}, one has no way of knowing whether $\xiatez\bra{H}
A_\mu^{(8)}\ket{H}$ or $\xiaz\bra{H}A_\mu^{(s)}\ket{H}$ is larger. In
fact, estimates of higher--order electroweak amplitudes and an SU(3)
analysis of the EMC data suggest that the two contributions could be of
comparable magnitude for PV electron scattering (see Tables~2.2 and 2.3).
Furthermore, the theoretical uncertainty in
$\ratez(\hbox{had})$ discussed above introduces an uncertainty into
$\bra{N}\jncf(T=0)\ket{N}$ amplitudes having the same magnitude as
$s$--quark contributions.  The same conclusion holds for atomic PV
probes of the axial--vector hadronic current.  In short, the neglect of the
second term in the right--hand side of Eq.~(\Eaxxter) is potentially
misleading as
far as the interpretation of PV $(\vec e,e^\prime)$ measurements and
atomic PV is concerned. In the case of $\nu$--N scattering, on the other
hand, the strange--quark term appears to dominate, while theoretical
uncertainties in higher--order processes are small. This difference between
neutrino and electron scattering follows from the two factors mentioned
previously: (i) logarithmic enhancements appearing in one--loop amplitudes
involving photon--exchange, and (ii) the fortuitous suppression of tree--level
electron--hadron amplitudes by the $\gve=-1+4\sstw$ factor which does
not arise in all of the one--loop amplitudes.

\vfil\eject

%

\def\qvecsq{\vec q^{\mkern2mu\raise1pt\hbox{$\scriptstyle2$}}}

\def\xiva{{\xi_\sst{V}^{(a)}}}

\def\evec{{\vec e}}
\def\sigpin{{\Sigma_{\pi\sst{N}}}}

\noindent{\bf III.C.\quad Single--Nucleon Matrix Elements of the Electroweak
Current}
\medskip

     Single-nucleon matrix elements of the currents of
Eqs.~(\Ejemmu), (\Ejncmu), and (\Ejncmuf) are
restricted by Lorentz covariance, together
with parity and time--reversal invariance, to the general forms
$$
\bra{N(P')}\jem(0) \ket{N(P)}\>=\>\ubar(P')\Bigl[F_1\gamma_\mu+i{F_2\over
        2\mn}\sigma_{\mu\nu}Q^\nu\Bigr] u(P)\eqno\nexteqp
$$
$$
\bra{N(P')}\jnc(0) \ket{N(P)}\>=\>\ubar(P')\Bigl[\ftilo\gamma_\mu+i{\ftilt\over
        2\mn}\sigma_{\mu\nu}Q^\nu\Bigr] u(P)\eqno\sameeq
$$
$$
\bra{N(P')}\jncf(0)\ket{N(P)}\>=\>\ubar(P')\Bigl[\gtila\gamma_\mu+{\gtilp
 \over\mn}Q_\mu\Bigr]\gamma_5 u(P)\ \ ,\eqno\sameeq
$$
where $Q=P'-P$ is the four--momentum
transferred to the nucleon (see Fig.~3.1).  Denoting
the magnitude of the three--momentum transfer by $q=\vert\ \qv\ \vert$ and
the energy transfer by $\omega$, we have $ Q^2=\omega^2-q^2\le 0$.
Here $F_{1(2)}$ are the usual electromagnetic Dirac and Pauli form factors of
the nucleon. Throughout the remainder of this work, we usually use Sachs form
factors [Sac62] defined in terms of the Dirac and Pauli vector current form
factors as
$$
\eqalign{G_\sst{E}(Q^2)\>&=\>F_1(Q^2)-\tau F_2(Q^2)\cr
         G_\sst{M}(Q^2)\>&=\>F_1(Q^2)+F_2(Q^2)\ \ ,
         \cr}\eqno\nexteq$$
where $\tau\equiv |Q^2|/ 4\mns$ and $\mn$ denotes the nucleon mass (we
ignore the $n$--$p$ mass difference at this point).

{}From Eqs.~(\Ejncmu) and (\Ejncmuf) the form
factors entering the matrix elements of the neutral
currents may be written
$$
\gtil_{a}(Q^2)=\xivteo G_{a}^\sst{T=1}\tau_3+\sqrt{3}\xivtez
        G_{a}^\sst{T=0}+\xivz G_{a}^{(s)},\quad
        \quad a=E,M\eqno\nexteqp\nameeq\EGtilde
$$
for the vector current form factors and
$$
\gtil_{a}(Q^2)=\xiateo G_{a}^{(3)}\tau_3+\xiatez G_{a}^{(8)}+
        \xiaz G_{a}^{(s)},\quad\quad\> a=A,P\eqno\sameeq
$$
for the axial--vector form factors. Here,
the $G_{a}^\sst{T=0,1}$ denote the isoscalar and isovector
combinations of the electromagnetic Sachs form
factors of the nucleon and $\tau_3 = +1$($-1$) for the proton(neutron).
For use elsewhere we also define the combinations
$$
\gtil_{A}^{(T=1)}(Q^2)=\xiateo G_{A}^{(3)}\eqno\nexteqp\nameeq\EGAtilde
$$
and
$$
\gtil_{A}^{(T=0)}(Q^2)=\xiatez G_{A}^{(8)}+
        \xiaz G_{A}^{(s)} \ \ .\eqno\sameeq
$$
Assuming the nucleon to be an eigenstate of isospin, one may write\footnote{*}{
In other work, such as Refs.~[Don79a] and [Don89], the isoscalar and isovector
nucleon form factors are defined without the factor of $1/2$.}
$$
\eqalign{\GETEO\>&=\>\coeff{1}{2}[\GEp-\GEn]\cr
        \GETEZ\>&=\>\coeff{1}{2}[\GEp+\GEn]\cr}\eqno\nexteqp\nameeq\Esachsa
$$
$$
\eqalign{\GMTEO\>&=\>\coeff{1}{2}[\GMp-\GMn]\cr
        \GMTEZ\>&=\>\coeff{1}{2}[\GMp+\GMn]\ \ ,\cr}\eqno\sameeq
$$
where at $Q^2=0$ one has
$$
\GETEO(0)\>=\GETEZ(0)\>=\>\coeff{1}{2}
         \eqno\nexteqp
$$
$$
\eqalign{\GMTEO(0)\>&=\>\coeff{1}{2}(\mu_p-\mu_n)\equiv \mu^\sst{T=1}\cr
         \GMTEZ(0)\>&=\>\coeff{1}{2}(\mu_p+\mu_n)\equiv \mu^\sst{T=0}\ \ ,
         \cr}\eqno\sameeq
$$
where the magnetic moments of the proton and neutron are $\mu_p=$ 2.79 and
$\mu_n=$ -1.91, respectively.
The $Q^2$--dependence of these form factors is known, to varying degrees of
accuracy, from unpolarized electron scattering. The data may be summarized
using the so--called \lq\lq Galster parameterization" [Gal71]
$$
\eqalignno{\GEp&=\GdipV&\nexteqp\nameeq\Eradneu\cr
         \GEn&=-\mu_n\tau\GdipV\xi_n&\sameeq\cr
         &\cr
         \GMp&=\mu_p\GdipV&\nexteqp\cr
         \GMn&=\mu_n\GdipV\ \ , &\sameeq\cr}
$$
where
$$
\eqalignno{\GdipV&=(1-Q^2/M_\sst{V}^2)^{-2} =
         (1+\lamdV\tau)^{-2}&\nexteqp\nameeq\Egdip\cr
         \xi_n&=(1+\lamn\tau)^{-1}&\sameeq\cr}
$$
with $\lamdV=4.97$ and $\lamn=5.6$.  This parameterization of the form
factors agrees with measured proton form factors for $\vert Q^2\vert$ of
present interest to better that 5\% (the
proton magnetic form factor deviates from the dipole form by about 3\% at
low momentum transfer [Hoh76]).  The neutron magnetic form factor is known
to about 7\% at low $\vert Q^2\vert$ [Mar93] and is about 2 standard
deviations above the dipole at $\vert Q^2\vert =0.1$ GeV$^2$/$c^2$.  The
neutron electric form factor is presently constrained to the 50\% level
[Mad92, Pla90] below $\vert Q^2\vert = 1.0$ GeV$^2$/$c^2$.  At higher
momentum transfer, $G_E^n$ appears to be consistent with zero [Lun93].

Except for the value of $\GES$ at the photon point, the vector current
strangeness form factors are unknown.  Since the nucleon has no net
strangeness, one must have that $\GES(0)=0=\FOS(0)$. It is
conventional to describe the leading, non--trivial $Q^2$--behavior of
charge form factors in terms of mean--square radius, defined, for example, in
terms of $F_1$. Consequently, one defines an mean--square \lq\lq strangeness
radius" as [Jaf89]
$$
\rsstr\equiv 6{d\FOS\over dQ^2}\Bigr\vert_{Q^2=0}\ \
.\eqno\nexteq\nameeq\Ers
$$
Similarly, the strangeness magnetic moment is given by the form factor
at $Q^2=0$:
$$
\mustr\equiv\FTS(0)\ \ .\eqno\nexteq
$$
Working instead with the Sachs form factors and the dimensionless variable
$\tau$, as is conventional in studies of the nucleon electromagnetic
form factors, one has the analogous quantities
$$
\rhostr\equiv {d\GES\over d\tau}\Bigr\vert_{\tau=0}\>=\>-\coeff{2}{3}\mns
        \rsstr -\mustr\eqno\nexteqp\nameeq\Erhomus
$$
$$
\mustr\equiv\GMS(0)\ \ .\eqno\sameeq
$$
As discussed in Sect.~II, there presently exist no published experimental
constraints on $\rhostr$ and $\mustr$, though a careful re--analysis of
the BNL $\nu p/\nubar p$ data might provide interesting bounds on the
strangeness magnetic moment (see Sect.~IV.J). In the absence of such
experimental information, one must turn to theoretical model predictions
for these form factors (see Table~2.3). In Sect.~IV, we review several
experimental scenarios for constraining $\GES$ and $\GMS$.
For the purpose of analyzing
the sensitivity of prospective experiments to these form factors, it is
necessary to choose some parameterization for the non--leading
$Q^2$--dependence.
To that end, we work with a natural extension of the Galster parameterization
and take
$$
\eqalignno{\GES&=\rhostr\tau\GdipV\xies&\nexteqp\nameeq\Egalstr\cr
         \GMS&=\mustr\GdipV\xims\ \ , &\sameeq\cr}
$$
where $\GdipV$ is defined in Eq.~(\Egdip a) and
$$
\xi_\sst{E,M}^{(s)}=(1+\lambda_\sst{E,M}^{(s)}\tau)^{-1}\ \ \ .\eqno\nexteq
\nameeq\Egalstrxi
$$
In the remainder of this work, we set $\lambda_\sst{M}^{(s)}=0$ and
take $\lambda_\sst{E}^{(s)}$ as a free parameter to be determined by
experiment.
With this choice, the strangeness vector current form factors have the
same asymptotic behavior as the neutron EM form factors: $G_\sst{E,M}^{(s)}
\sim 1/\tau^2$ as $\tau\to\infty$. Given that these form factors arise
entirely from sea quarks, it would be reasonable to assume a more rapid
fall--off with $\tau$. Nevertheless, for purposes of comparing the constraints
which different intermediate--energy, low-to-moderate $\tau$
scattering experiments might derive on the form factors, this choice is
sufficient.

\par
        In contrast to the electromagnetic and weak neutral vector currents,
the
neutral axial--vector current is not conserved. Consequently, the $Q^2=0$
values of the axial--vector form factors are not constrained by any symmetry or
nucleon quantum numbers.  If, however, one assumes the nucleon to be a state
of good isospin, $G_A^{(3)}(0)$ may be determined from Gamow--Teller
$\beta$--decay rates. If, in addition, one takes the eight lowest--lying
baryons to constitute an exact SU(3) octet, one may also extract $G_A^{(8)}$
from hyperon $\beta$--decay measurements. Under these two assumptions, one
has
$$
G_A^{(3)}(0)=\coeff{1}{2}(D+F)\eqno\nexteqp\nameeq\EGafd
$$
$$
G_A^{(8)}(0)=\coeff{1}{2\sqrt{3}}(3F-D)\ \ ,\eqno\sameeq
$$
with $D+F\equiv \ga =1.262$ and $F/D \approx 0.64$
[Gai84, Dub90, Fre90, Bou83].
Na\"\i vely, one might expect corrections to Eq.~(\EGafd b)
to have the same scale as that of
SU(3)--breaking in the baryon octet, namely, $(M_\sst{\Sigma}-\mn)/\mn
\approx 0.27$. Estimates using the Skyrme model  [Par91] and
chiral perturbation theory [Jen91] suggest that the
scale of these corrections could be significantly larger, with the result
that the value of $G_\sst{A}^{(8)}$ could be uncertain by a factor of two
or more. The impact of this uncertainty would not be serious for the
interpretation of $\nu$ scattering experiments, since $\xiatez$ is small
in this case, but would introduce problematic uncertainty for the
analysis of PV asymmetries beyond that associated with $\ratez$ as discussed
above. As for the $Q^2$--dependence of these form factors, data can be
parameterized adequately with a dipole form
$$
\eqalignno{G_\sst{A}^{(3)}&=\coeff{1}{2}(D+F)\GdipA&\nexteqp\cr
         G_\sst{A}^{(8)}&=\coeff{1}{2\sqrt{3}}(3F-D)\GdipA\ \ , &\sameeq\cr}
$$
where
$$
\GdipA = (1-Q^2/M_\sst{A}^2)^{-2} = (1+\lamdA\tau)^{-2}\ \ . \eqno\nexteq
\nameeq\GDaxial
$$
{}From charged current quasielastic neutrino reactions on isoscalar targets,
one has the world average value for the axial--vector dipole mass parameter
$M\sst{A} = 1.032\pm 0.036$ or $\lamdA = 3.32^{+0.24}_{-0.22}$
[Ahr87 and references therein; see also Kit90].

        Values for $\GAS(0)$ have been extracted from the BNL
$\nu p/\nubar p$ experiment as well as from the EMC measurement of
$\int dx g_1(x)$. The latter determination makes use of SU(3) symmetry
and a QCD--corrected parton model
to write the $g_1$ sum rule for the proton as [Ash89]
$$
\int_0^1 dx g_1^p(x) =
{1\over 12}\biggl\{ \Bigl[1-{\alpha_s\over\pi}\Bigr]\Bigl[a_3+{1\over\sqrt{3}}
a_8\Bigr] + 2\sqrt{2\over 3}\Bigl[ 1- {33-8N_f\over 33-2N_f}{\alpha_s\over
\pi}\Bigr]a_0\biggr\}\ \ , \eqno\nexteq
$$
where $N_f$ is the number of flavors of quarks,
$a_{3,8}=2 G_\sst{A}^{(3,8)}(0)$, and $a_0$ may be re--written in
terms of $G_\sst{A}^{(8)}(0)$ and $\GAS(0)$ in order to extract a value
of the strangeness axial--vector form factor. Results of this analysis, as
well as that of the BNL data, are given in Table 2.2. Several theoretical
predictions agree, at least in rough order of magnitude, with these results.
The presence
of SU(3)--breaking of the scale suggested by Refs.~[Par91, Jen91]
could reduce the EMC value for $|\GAS(0)|$ by roughly a factor of three.
In the case of the BNL determination, the data were collected at sufficiently
large $|Q^2|$ to generate a significant correlation between $\GAS(0)$ and
the axial--vector dipole mass parameter of Eq.~(\GDaxial). It
is clearly of interest
to reduce the resulting uncertainty by performing a determination at much
lower momentum transfer and ways for doing so are discussed in Sect.~IV.J.
In analyzing these proposals, we make a choice for the $Q^2$--dependence
of $\GAS$ similar to that of the vector current strangeness
form factors:
$$
\GAS(Q^2)=\eta_s(D+F)\GdipA\xias\ \ , \eqno\nexteq\nameeq\Egalstra
$$
where
$$
\eqalignno{\xias&=(1+\lambda_\sst{A}^{(s)}\tau)^{-1}&\nexteqp\cr
         \eta_s&=\GAS(0)/g_\sst{A}\ \ . &\sameeq\cr}
$$
We will generally set $\lambda_\sst{A}^{(s)}=0$ in
what follows.

        Although less is known about the induced pseudoscalar form factor,
$\gtilp$, it does not contribute to observables measured in PV electron
scattering to leading order in electroweak couplings.
Its contribution to
elastic neutrino scattering vanishes for $m_\nu=0$ and is usually
neglected in this case.

\vfil\eject

%

\def\qvecsq{\vec q^{\mkern2mu\raise1pt\hbox{$\scriptstyle2$}}}

\def\xiva{{\xi_\sst{V}^{(a)}}}

\def\evec{{\vec e}}
\def\sigpin{{\Sigma_{\pi\sst{N}}}}

\noindent{\bf III.D.\quad Nuclear Matrix Elements of the Electroweak Current}
\medskip

In this section we discuss the nuclear matrix elements of the electroweak
current.  For scattering from nuclear targets, one must model both the
nucleon and nuclear structure in obtaining the matrix elements.
Accordingly, one focus in this section is to indicate how this interplay
is handled in terms of a hierarchy of one--body, two--body, \dots\ nuclear
operators. When considering initial and final nuclear states with good
angular momentum quantum numbers it is usually convenient
to expand the hadronic currents in terms of
multipole projections of the non--relativistic charge and three--current
operators.  This expansion leads naturally to definitions of basic vector and
axial--vector form factors, as discussed in Sect.~III.D.1.  The dominant
contributions to the nuclear matrix elements usually arise from the
one--body parts of the general current operators (see Sect.~III.D.2) with
typically small corrections from two--body
meson--exchange currents (see Sect.~III.D.3).  In many instances, as discussed
in more detail in Sect.~IV, assuming the nuclear states to be eigenstates
of parity and isospin, in addition to angular momentum, and truncating the
many--body hierarchy to one-- and two--body operators is sufficient in
treating the nuclear physics contributions to observables of interest.
However, the extraction of nucleon structure information from these observables
often requires that we understand the residual uncertainty in the nuclear
modeling, so we must also evaluate the role played by dynamical ingredients
which go beyond these basic assumptions.  Two specific problems of this
type are mentioned in the last two subsections: the role
of isospin--mixing in the nuclear states (Sect.~III.D.4) and
the general character of parity--mixing in the nuclear states, including the
the role of the anapole moment (Sect.~III.D.5). Although these additional
ingredients generally
produce negligible changes to the interpretation
of the hadronic matrix elements within the context of the basic set of
assumptions, there exist a few cases, discussed in detail in Sect.~IV,
for which it may be possible to
address these issues using PV electron scattering.

\bigskip
\goodbreak
\noindent III.D.1.\quad ELECTROWEAK FORM FACTORS

The multipole expansions of the EM charge and three current are given by
$$
\eqalignno{\hat\rho({\vec q})
\>&=\>\sum_{J\geq 0}\sqrt{4\pi(2J+1)}i^J \hat M_{J0}(q)
        &\nexteqp\nameeq\Erjmult\cr
        &\cr
\hat J_{\lambda}({\vec q})
\>&=\>-\cases{ \sum_{J\geq 0}\sqrt{4\pi(2J+1)}i^J \hat L_{J0}(q),
                &$\lambda=0$ \cr &\cr
\sum_{J\geq 1}\sqrt{2\pi(2J+1)}i^J
                [\hat T^{\rm el}_{J\lambda}(q)
      +\lambda \hat T^{\rm mag}_{J\lambda}(q)],
                &$\lambda=\pm 1$ \ \ ,\cr}&\sameeq\cr}
$$
where the Coulomb ($\hat M$), longitudinal ($\hat L$), and transverse electric
and magnetic ($\hat T^{\rm el,\>mag}$) multipole operators
are defined in Refs.~[deF66, Don75, Wal75, Don79a].  As usual, the
$z$--axis has been chosen to lie in the $\vec q$--direction.  Let us assume an
initial state with good quantum numbers $J_i^{\pi_i}T_i,\ M_{J_i},\ M_{T_i}$
and a final state with good quantum numbers $J_f^{\pi_f}T_f,\ M_{J_f},\
M_{T_f}$.
Since our main focus is PC/PV electron scattering (see Sect.~III.E) which
involves only neutral electroweak currents, we require matrix
elements between states in a given nucleus; hence,
$M_{T_i}=M_{T_f}=M_T$.  The following electromagnetic form factors may then
be defined:
$$
\eqalign{F_{CJ}(q)&\equiv{1\over{\sqrt{2J_i +1}}}\sum_{T=0,1}
(-)^{T_f-M_T}\left(\matrix{T_f &T &T_i \cr -M_T &0 &M_T \cr}\right)
<J_f;T_f\doubred {\hat M}_{J;T}(q)\doubred J_i;T_i> \cr
          F_{LJ}(q)&\equiv{1\over{\sqrt{2J_i +1}}}\sum_{T=0,1}
(-)^{T_f-M_T}\left(\matrix{T_f &T &T_i \cr -M_T &0 &M_T \cr}\right)
<J_f;T_f\doubred {\hat L}_{J;T}(q)\doubred J_i;T_i> \cr
          F_{EJ}(q)&\equiv{1\over{\sqrt{2J_i +1}}}\sum_{T=0,1}
(-)^{T_f-M_T}\left(\matrix{T_f &T &T_i \cr -M_T &0 &M_T \cr}\right)
<J_f;T_f\doubred {\hat T}_{J;T}^{\rm el}(q)\doubred J_i;T_i> \cr
          F_{MJ}(q)&\equiv{1\over{\sqrt{2J_i +1}}}\sum_{T=0,1}
(-)^{T_f-M_T}\left(\matrix{T_f &T &T_i \cr -M_T &0 &M_T \cr}\right)
<J_f;T_f\doubred i{\hat T}_{J;T}^{\rm mag}(q)\doubred J_i;T_i>\ \ , \cr}
\eqno\nexteq\nameeq\Exdvec
$$
where the symbols $\doubred$ are used to denote matrix elements reduced
both in angular momentum and in isospin.  With the factor $i$ in the
definition of the magnetic form factor it can be shown that a phase
convention can be adopted in which the form factors are all real.
For conserved vector currents (CVC) such as the electromagnetic current, one
may eliminate matrix elements of the longitudinal projection of the
the three--vector current, which involves only the divergence of the current,
using instead matrix elements of $[\Hhat,\rohat]$:
$$
<J_f;T_f\doubred {\hat L}_{J;T}\doubred J_i;T_i> ={\omega\over q}
<J_f;T_f\doubred {\hat M}_{J;T}\doubred J_i;T_i>
\ \ , \eqno\nexteq
$$
where $\omega=E_f-E_i$ with $E_i$($E_f$) the energy of the initial (final)
nuclear state.  Thus, in considering such CVC cases, one need discuss only
the Coulomb and transverse operators. The conventions used in defining the
form factors in Eqs.~(\Exdvec) are those of
Refs.~[deF66, Don75, Don79a, Don79b, Don80, Don84, Don89]. They
may be compared
with different choices of normalization by computing the elastic C0 form factor
in the long wavelength limit ($q\rightarrow 0$):
$$
\eqalign{<J_i;T_i\doubred {\hat M}_{0;0}\doubred J_i;T_i>
&\longrightarrow {1\over\sqrt{4\pi}}\sqrt{(2J_i+1)(2T_i+1)}{1\over 2}A \cr
<J_i;T_i\doubred {\hat M}_{0;1}\doubred J_i;T_i>
&\longrightarrow {1\over\sqrt{4\pi}}\sqrt{(2J_i+1)(2T_i+1)}
\sqrt{T_i(T_i+1)} \ \ , \cr}\eqno\nexteq\nameeq\Elwlim
$$
using the general arguments in Ref.~[Don84].  This implies that
$$
F_{C0} \longrightarrow Z/\sqrt{4\pi} \eqno\nexteq
$$
in the long wavelength limit.

The above definitions can easily be extended to
include charge--changing weak interaction processes such as
$\beta$--decay, charged lepton capture and $\nu$--reactions
as well [Don79a]. In that case only
isovector matrix elements occur
and the following vector current form factors appear
as natural generalizations of Eqs.~(\Exdvec):
$$
\eqalign{F_{CJ}^{\pm}(q)&\equiv{1\over{\sqrt{2J_i +1}}}(-)^{T_f-M_{T_f}}
\left(\matrix{T_f &1 &T_i \cr -M_{T_f} &\pm 1 &M_{T_i} \cr}\right)
<J_f;T_f\doubred {\hat M}_{J;1}(q)\doubred J_i;T_i> \cr
          F_{LJ}^{\pm}(q)&\equiv{1\over{\sqrt{2J_i +1}}}(-)^{T_f-M_{T_f}}
\left(\matrix{T_f &1 &T_i \cr -M_{T_f} &\pm 1 &M_{T_i} \cr}\right)
<J_f;T_f\doubred {\hat L}_{J;1}(q)\doubred J_i;T_i> \cr
          F_{EJ}^{\pm}(q)&\equiv{1\over{\sqrt{2J_i +1}}}(-)^{T_f-M_{T_f}}
\left(\matrix{T_f &1 &T_i \cr -M_{T_f} &\pm 1 &M_{T_i} \cr}\right)
<J_f;T_f\doubred {\hat T}_{J;1}^{\rm el}(q)\doubred J_i;T_i> \cr
          F_{MJ}^{\pm}(q)&\equiv{1\over{\sqrt{2J_i +1}}}(-)^{T_f-M_{T_f}}
\left(\matrix{T_f &1 &T_i \cr -M_{T_f} &\pm 1 &M_{T_i} \cr}\right)
<J_f;T_f\doubred i{\hat T}_{J;1}^{\rm mag}(q)\doubred J_i;T_i>\ \ ,
\cr}\eqno\nexteq\nameeq\Exdvcc
$$
where now $M_{T_f}=M_{T_i}\pm 1$.  In fact the isovector--vector reduced matrix
elements appearing in all electroweak processes are the same (the
``isovector triplet hypothesis'') except for the electroweak couplings,
1 or $\xivteo$ (see below). The isovector form factors in Eqs.~(\Exdvec)
and (\Exdvcc) differ only in that the appropriate 3--j symbols must be
used:
$$
F^{\lambda}(T=1) \sim
\left(\matrix{T_f &1 &T_i \cr -M_{T_f} &\lambda &M_{T_i} \cr}\right)\ \ ,
\eqno\nexteq
$$
where one has the isovector part of one of the electromagnetic form factors
in Eqs.~(\Exdvec) when $\lambda=0$ and one of the charge--changing weak
form factors in Eqs.~(\Exdvcc) when $\lambda=\pm 1$.
Of course, when discussing charge--changing weak interaction processes it is
also necessary to consider axial--vector currents as well as vector currents.
The analogous operators for multipole expansions of the
axial--vector currents will be denoted with a
\lq\lq 5" superscript:
$$
\eqalign{F_{CJ_5}^{\pm}(q)&\equiv{1\over{\sqrt{2J_i +1}}}(-)^{T_f-M_{T_f}}
\left(\matrix{T_f &1 &T_i \cr -M_{T_f} &\pm 1 &M_{T_i} \cr}\right)
<J_f;T_f\doubred i{\hat M}_{J;1}^5(q)\doubred J_i;T_i> \cr
          F_{LJ_5}^{\pm}(q)&\equiv{1\over{\sqrt{2J_i +1}}}(-)^{T_f-M_{T_f}}
\left(\matrix{T_f &1 &T_i \cr -M_{T_f} &\pm 1 &M_{T_i} \cr}\right)
<J_f;T_f\doubred i{\hat L}_{J;1}^5(q)\doubred J_i;T_i> \cr
          F_{EJ_5}^{\pm}(q)&\equiv{1\over{\sqrt{2J_i +1}}}(-)^{T_f-M_{T_f}}
\left(\matrix{T_f &1 &T_i \cr -M_{T_f} &\pm 1 &M_{T_i} \cr}\right)
<J_f;T_f\doubred i{\hat T}_{J;1}^{\rm el5}(q)\doubred J_i;T_i> \cr
          F_{MJ_5}^{\pm}(q)&\equiv{1\over{\sqrt{2J_i +1}}}(-)^{T_f-M_{T_f}}
\left(\matrix{T_f &1 &T_i \cr -M_{T_f} &\pm 1 &M_{T_i} \cr}\right)
<J_f;T_f\doubred {\hat T}_{J;1}^{\rm mag5}(q)\doubred J_i;T_i>\ \ .
\cr}\eqno\nexteq\nameeq\Exdax
$$
Again the factors $i$ have been
included so all form factors here are real where the same phase convention
mentioned above is adopted.
In the case of the axial--vector current, which is not conserved,
longitudinal projections
cannot be re--written in terms of other multipoles and consequently all
four types of form factor in Eq.~(\Exdax) are required in general.

The final set of form factors
required for a complete treatment of hadronic
matrix elements of the electroweak current are those associated with the weak
neutral current interaction in parallel with the above formalism.  For
extensive discussion of the inter--relationships that exist within
the full set of electroweak current matrix elements see Ref.~[Don79a]. Here
we only summarize the developments that lead to the remaining form factors
needed in the rest of the present work.  Let us use
Eqs.~(\EGtilde) to write the vector and axial--vector NC multipole
operators in the form
$$
\eqalignno{(\Ohat_J)^\sst{NC}\>&=\>\xivteo\Ohat_J^\sst{EM}(T=1)
+\sqrt{3}\xivtez\Ohat_J^\sst{EM}(T=0)+\xivz\Ohat_J(s)&\nexteqp
        \nameeq\EOjnc\cr
           (\Ohat_J^5)^\sst{NC}\>&=\>\xiateo\Ohat_J^5(3)
+\xiatez\Ohat_J^5(8) +\xiaz\Ohat_J^5(s)\ \ , &\sameeq\cr}
$$
where $\Ohat_J$ ($\Ohat_J^5$) is any one of the vector (axial--vector)
multipole operators introduced above.
Here the notation $(s)$ refers to the  multipole projections of
$V_\lambda^{(s)}$ and $A_\lambda^{(s)}$,
$(T=0,1)$ denotes projections of the isoscalar and
isovector EM currents, and $(3,8)$ correspond to projections of the
octet axial--vector currents. Operationally, one may take the isoscalar matrix
elements and form factors defined in Ref.~[Don79a] and replace them with the
corresponding quantities involving strange or octet neutral currents.  For
single--particle matrix elements (see Sect.~III.D.2) this procedure
entails making the following replacements:
$$
\eqalign{G_{E,M}^{T=0} &\longrightarrow G_{E,M}^{(s)} \cr
{1\over 2} G_A^{(0)} &\longrightarrow G_A^{(s)}\ {\rm or}\  G_A^{(8)}
\ \ , \cr}\eqno\nexteq
$$
where $G_A^{(0)}$ is the isoscalar axial--vector form factor used
in Ref.~[Don79a] and differs by the same factor of 1/2 as do the
vector form factors in Eqs.~(\Esachsa).

In analogy with Eqs.~(\Exdvec) we may then
define weak neutral current form factors as follows:
$$
\eqalignno{\tilde{F}_{XJ}&\equiv{1\over{2\sqrt{2J_i +1}}}
(-)^{T_f-M_T}\Bigl\{
\left(\matrix{T_f &0 &T_i \cr -M_T &0 &M_T \cr}\right) \cr
&\qquad\qquad \times \Bigl[\sqrt{3}\xivtez
<J_f;T_f\doubred {\hat O}_J(T=0)\doubred J_i;T_i>
+\xivz <J_f;T_f\doubred {\hat O}_J(s)\doubred J_i;T_i> \Bigr] \cr
&\qquad\qquad\qquad
+\left(\matrix{T_f &1 &T_i \cr -M_T &0 &M_T \cr}\right)
\xivteo <J_f;T_f\doubred {\hat O}_J(T=1)\doubred J_i;T_i> \Bigr\} \cr
&\equiv {1\over 2}\Bigl[ \sqrt{3}\xivtez F_{XJ}(T=0) + \xivz F_{XJ}(s)
+ \xivteo F_{XJ}(T=1) \Bigr]
&\nexteqp\nameeq\Encurff \cr
          \tilde{F}_{XJ_5}&\equiv -{1\over{2\sqrt{2J_i +1}}}
(-)^{T_f-M_T}\Bigl\{
\left(\matrix{T_f &0 &T_i \cr -M_T &0 &M_T \cr}\right) \cr
&\qquad\qquad \times \Bigl[
\xiatez <J_f;T_f\doubred {\hat O}_J^5(8)\doubred J_i;T_i>
+\xiaz <J_f;T_f\doubred {\hat O}_J^5(s)\doubred J_i;T_i> \Bigr] \cr
&\qquad\qquad\qquad
+\left(\matrix{T_f &1 &T_i \cr -M_T &0 &M_T \cr}\right)
\xiateo <J_f;T_f\doubred {\hat O}_J^5(T=1)\doubred J_i;T_i> \Bigr\} \cr
&\equiv -{1\over 2} \Bigl[ \xiatez F_{XJ_5}(8) + \xiaz F_{XJ_5}(s)
+ \xiateo F_{XJ_5}(T=1) \Bigr]
\ \ , &\sameeq\cr }
$$
where for clarity the $q$--dependences have been suppressed. Here
$X=C,\ L,\ E$ or $M$ with corresponding operators $\Ohat_J$
($\Mhat_J$, $\Lhat_J$, $\That^{\rm el}_J$, and $i\That^{\rm mag}_J$) for the
vector currents and $\Ohat_J^5$ ($i\Mhat_J^5$,
$i\Lhat_J^5$, $i\That^{\rm el5}_J$, and $\That^{\rm mag5}_J$) for the
axial--vector currents. The factors of $\pm$ 1/2 have been introduced to
keep the normalization and sign the same as that used previously in
Ref.~[Don79a]
(with strangeness matrix elements set to zero).  Employing Eqs.~(\Elwlim)
it is straightforward to verify that for the elastic C0 form factor one
has the following long wavelength limit:
$$
\eqalign{{\tilde F}_{C0} &\longrightarrow {1\over{\sqrt{4\pi}}}\times
{1\over 2} \Bigl[ \sqrt{3}\xi_V^{T=0}{A\over 2} +\xi_V^{T=1} M_T
\Bigr] \cr
&= {1\over{\sqrt{4\pi}}}\times {1\over 2} \Bigl[ Z \xi_V^p +N \xi_V^n
\Bigr] \ \ , \cr}\eqno\nexteq
$$
where we have defined the couplings
$$
\eqalign{\xi_V^p &={1\over 2}[\sqrt{3}\xi_V^{T=0}+\xi_V^{T=1}]\cr
\xi_V^n &={1\over 2}[\sqrt{3}\xi_V^{T=0}-\xi_V^{T=1}]\ \ . \cr}
\eqno\nexteq\nameeq\Ewncpn
$$
as in Eqs.~(\Exiprne).

We wish to emphasize that  our previous comments about the isoscalar
axial--vector form factor $\GATEZ$ also apply to the isoscalar axial--vector
multipole matrix elements (see Eq.~(\EGAtilde))
$<J_f;T_f\doubred {\hat O}_J^5(T=0)\doubred J_i;T_i> = $
$\xiatez$ $<J_f;T_f\doubred {\hat O}_J^5(8)\doubred J_i;T_i>
+ \xiaz <J_f;T_f\doubred {\hat O}_J^5(s)\doubred J_i;T_i> $.
In the case of $\nu$ scattering, $\xiatez$ is sufficiently small and
theoretically certain to render
$<J_f;T_f\doubred {\hat O}_J^5(T=0)\doubred J_i;T_i>$
effectively a meter of strangeness contributions. For PV electron scattering,
however, the two terms here have the same scale,
with theoretical uncertainty in the first term of comparable magnitude.
Moreover, many--body contributions associated with nuclear PV
introduce additional theoretical uncertainties beyond those associated with
the one--body (single--nucleon) NC, as discussed in Sect.~III.D.5. In fact,
a measurement of the matrix element
of $\Ohat_J^5(T=0)$ with PV electron scattering could in principle serve better
as a probe of nuclear PV rather than of the axial--vector strangeness
current (see Sects.~III.D.5 and IV.E).
In contrast, contributions from nuclear PV to $\nu$--scattering are negligible,
so that from a theoretical standpoint, neutrinos ought to provide a \lq\lq
cleaner" probe of the nuclear axial--vector strangeness current.

\medskip
\noindent\undertext{Selection Rules}
\medskip

All of the form factors defined above are matrix elements of either vector
or axial--vector multipole operators.  For states with good parity
quantum numbers $\pi_i$ and $\pi_f$, the operators must carry the total
parity $\pi=\pi_i \pi_f$ (see below for discussion of the extensions
which occur when parity--mixing in the nuclear states is considered).  The
form factors divide into two classes, {\it viz.,\/} natural parity
transitions, where $(-)^J=\pi$ and only the following occur:
$$
\eqalign{F_{CJ},\ F_{LJ},\ F_{EJ}\ \ &{\rm and}\ \ {\tilde F}_{CJ}\
{\tilde F}_{LJ}\ {\tilde F}_{EJ} \cr
F_{MJ_5}\ \ &{\rm and}\ \ {\tilde F}_{MJ_5}
\cr}\eqno\nexteqp
$$
and non--natural parity transitions,
where $(-)^{J+1}=\pi$ and only the following occur:
$$
\eqalign{F_{MJ},\ \ &{\rm and}\ \ {\tilde F}_{MJ} \cr
F_{CJ_5},\ F_{LJ_5},\ F_{EJ_5}\ \ &{\rm and}\ \ {\tilde F}_{CJ_5}\
{\tilde F}_{LJ_5}\ {\tilde F}_{EJ_5} \ \ .
\cr}\eqno\sameeq
$$
To illustrate, consider a transition from a nuclear ground state having
$J_i^{\pi_i}={3\over 2}^-$ to a final state having $J_f^{\pi_f}=
{5\over 2}^+$, implying that $1\leq J \leq 4$ and $\pi = -$.  The above
parity and angular momentum selection rules then allow only
$F_{C1}$, $F_{L1}$, $F_{E1}$, $F_{M1_5}$; $F_{M2}$, $F_{C2_5}$, $F_{L2_5}$,
$F_{E2_5}$; $F_{C3}$, $F_{L3}$, $F_{E3}$, $F_{M3_5}$; $F_{M4}$, $F_{C4_5}$,
$F_{L4_5}$, $F_{E4_5}$ and the analogous NC form factors (with tildes).

For elastic scattering one may make use of time--reversal invariance, in
addition to invoking parity and angular momentum conservation together
with hermiticity, to
limit the number of multipoles permitted to a smaller set [deF66, Don76]:
only the even--$J$ vector form factors $F_{CJ}$ and the odd--$J$ vector,
$F_{MJ}$, and axial--vector, $F_{LJ_5}$, $F_{EJ_5}$, form factors (and
their analogs with tildes) may be
nonzero.  Of course, from angular momentum conservation one has $0\leq
J \leq 2J_i$ for elastic scattering.  As discussed in Ref.~[Don76], if,
in contrast to our assumption here, the
second--class axial--vector tensor current were to be nonzero,
then in addition there could be odd--$J$ form factors
$F_{CJ_5}$ and ${\tilde F}_{CJ_5}$.  An example of using these general
symmetry properties, discussed in more detail in Sect.~IV.C,
is that of elastic scattering
from deuterium.  There the ground state has $J_i=1$ and consequently the
form factors permitted are $F_{C0}$, $F_{C2}$, $F_{M1}$, $F_{L1_5}$ and
$F_{E1_5}$, together with the analogous quantities with tildes.

\bigskip
\goodbreak
\noindent III.D.2.\quad NUCLEAR MATRIX ELEMENTS OF ONE--BODY OPERATORS

The electroweak currents in nuclei can usually be treated as one--body
operators with corrections from two--body meson--exchange currents (see
Sect.~III.D.3).  Any one--body operator can be written in the form
[Fet71]
$$
{\hat O}_{J M_J;T M_T}^{[1]}(q) = \sum_{\alpha \alpha'}<\alpha'|
O_{J M_J;T M_T}^{[1]}(q) |\alpha> c^{\dagger}_{\alpha'} c_{\alpha}\ \ ,
\eqno\nexteq\nameeq\Eonebda
$$
where we use $\alpha$ and $\alpha'$ to label any complete set of
single--particle wave functions.  Typically $\alpha= \{ a; m_{j_a}
m_{t_a} \}$ with $a= \{ n_a (l_a {1\over 2}) j_a; t_a={1\over 2} \}$. Also
$-\alpha\equiv \{ a; -m_{j_a} -m_{t_a} \}$.  Working with single--particle
states above and below the Fermi surface, $\epsilon_a > \epsilon_F$ and
$\epsilon_a \leq \epsilon_F$, respectively, we have
$$
c_{\alpha}= \theta (\epsilon_a -\epsilon_F) a_{\alpha} +
\theta (\epsilon_F -\epsilon_a) S_{\alpha} b^{\dagger}_{-\alpha} \ \ ,
\eqno\nexteq
$$
where $a_{\alpha}$ is a particle destruction operator and
$b^{\dagger}_{-\alpha}$ is a hole creation operator.  The phase factor
$S_{\alpha}\equiv (-)^{j_a -m_{j_a}}(-)^{{1\over 2} -m_{t_a}}$ is included to
maintain the irreducible tensor character of $c_{\alpha}$ and its
adjoint $c^{\dagger}_{\alpha}$.  Thus a one--body operator can be
expanded in a complete set of single--particle matrix elements
$<\alpha'| O_{J M_J;T M_T}^{[1]}(q) |\alpha>$ which multiply bilinear products
of creation and destruction operators --- any complete set will make
this expansion an identity.  The form factors defined in the previous
subsection all involve doubly--reduced matrix elements of multipole
projections of the currents.  For one--body operators these can be cast
in the following form:
$$
<J_f;T_f\doubred {\hat O}_{J;T}^{[1]}(q)\doubred J_i;T_i> =
\sum_{a,a'} \psi_{J;T}^{(fi)}(a'a) <a'\doubred O_{J;T}^{[1]}(q) \doubred a>
\ \ , \eqno\nexteq\nameeq\Exdden
$$
{\it i.e.,\/} as expansions over a complete set of doubly--reduced
single--particle matrix elements weighted by one--body density matrix
elements defined by
$$
\psi_{J;T}^{(fi)}(a'a) \equiv {1\over {\sqrt{2J+1}\sqrt{2T+1}}}
<J_f;T_f \doubred \bigl[ c^{\dagger}_{a'} \otimes {\tilde c}_a
\bigr]_{J;T} \doubred J_i;T_i > \ \ , \eqno\nexteq\nameeq\Exdpsi
$$
where ${\tilde c}_{\alpha}\equiv S_{\alpha} c_{-\alpha}$.  Treatments of
the single--particle matrix elements of the operators that occur in
semi--leptonic electroweak interaction studies can be found in
Refs.~[Don79b, Don80].

Equation (\Exdden) is basic to most treatments of nuclear matrix elements
of the electroweak current operators and their associated multipole
projections.  The many--body matrix elements (the left--hand side of the
equation) is expressed in terms of single--particle matrix elements and
specific numbers --- the one--body density matrix elements --- that
embody the nuclear structure physics content  (the right--hand side
of the equation).  The former are obtained as non--relativistic
reductions of the quantities discussed in Sect.~III.C with allowance
made for the fact that the single--particle wave functions used in
Eq.~(\Exdden) are usually to be viewed as solutions to the Hartree--Fock
approximation (see, for example, Ref.~[Fet71]), rather than as plane
waves as is appropriate for free single nucleons.  Details concerning
this connection may be found in Refs.~[deF66, Don79a].  Importantly in the
present context is the observation that in the one--body approximation,
strangeness
is contained in these single--nucleon matrix elements.  The nuclear
many--body dynamics are represented by the one--body density matrix elements
$\psi_{J;T}^{(fi)}(a'a)$.  Two approaches may be followed in obtaining
these quantities for specific nuclear transitions: one is to obtain
approximate solutions to the nuclear many--body problem (Hartree--Fock
approximation, Random Phase approximation (RPA), nuclear shell model
diagonalizations, {\it etc.,\/} as discussed for example in [Fet71]);
another is to use information from PC electron scattering and perhaps
$\beta$--decay to constrain the density matrix and thus permit one
to predict the form factors for (as yet) unmeasured electroweak
processes.  The latter approach is discussed in some detail in
Ref.~[Don79a and references therein].  Both approaches require truncating
the sums in Eq.~(\Exdden), an approximation that leads to some level of
uncertainty in the nuclear modeling and consequently impacts the degree
to which the single--nucleon content in the nuclear matrix elements can
be extracted.  In Sect.~IV we return to address this issue in specific
circumstances.

\bigskip
\goodbreak
\noindent III.D.3.\quad MESON--EXCHANGE CURRENTS

     Most modern attempts to treat nuclear structure microscopically
describe the nucleus in terms of the position and motion of the individual
nucleons.  But these treatments also include an interaction between the
nucleons mediated by the exchange of mesons.  Construction of the
electromagnetic and weak neutral currents of the nucleus must, therefore,
include both the nucleonic and the mesonic degrees of freedom.  However,
since the exchanged meson is emitted by one nucleon and eventually absorbed
on another, most treatments do not address the meson degrees of freedom
explicitly but instead introduce additional multi--nucleon current operators
to account for the contributions from the exchanged mesons.  A few examples
of diagrams contributing to these meson--exchange currents (MEC) are shown in
Fig.~3.8.

     Two--body operators are developed in a way that is completely
analogous to that in the previous section.  Any two--body operator can
be written in the form (cf. Eq.~(\Eonebda)):
$$
{\hat O}_{J M_J;T M_T}^{[2]}(q) = \sum_{\alpha \alpha'}
\sum_{\beta \beta'}<\alpha'\beta'|
O_{J M_J;T M_T}^{[2]}(q) |\alpha\beta> c^{\dagger}_{\alpha'}
c^{\dagger}_{\beta'} c_{\beta} c_{\alpha}\ \ .
\eqno\nexteq
$$
Thus any two--body operator can be
expanded in a complete set of two--particle matrix elements
$<\alpha'\beta'| O_{J M_J;T M_T}^{[2]}(q) |\alpha\beta>$ which multiply
quadrilinear products of creation and destruction operators.
The form factors defined in Sect.~III.C
all involve doubly--reduced matrix elements of multipole
projections of two--body current operators of this type, as well as of
the one--body operators discussed above. In analogy with Eq.~(\Exdden)
we have:
$$
\eqalign{
<J_f;T_f &\doubred {\hat O}_{J;T}^{[2]}(q)\doubred J_i;T_i> \cr
&= \sum_{(a',b')J_{a'b'}T_{a'b'}} \sum_{(a,b)J_{ab}T_{ab}}
\psi_{J;T}^{(fi)}((a',b')J_{a'b'}T_{a'b'};(a,b)J_{ab}T_{ab}) \cr
&\qquad\qquad\qquad\qquad \times
<(a',b')J_{a'b'}T_{a'b'}\doubred O_{J;T}^{[2]}(q) \doubred (a,b)J_{ab}T_{ab}>
\ \ , \cr}\eqno\nexteq\nameeq\Emecff
$$
where the two--body density matrix elements (Cf. Eq.~(\Exdpsi)) are
defined by
$$
\eqalign{
&\psi_{J;T}^{(fi)}((a',b')J_{a'b'}T_{a'b'};(a,b)J_{ab}T_{ab})
\equiv -{1\over {\sqrt{2J+1}\sqrt{2T+1}}} \cr
&\qquad \times <J_f;T_f \doubred \Bigl[ \bigl[ c^{\dagger}_{a'} \otimes
c^{\dagger}_{b'} \bigr]_{J_{a'b'};T_{a'b'}} \otimes \bigl[ {\tilde c}_a
\otimes {\tilde c}_b \bigr]_{J_{ab};T_{ab}} \Bigr]_{J;T}
\doubred J_i;T_i > \ \ . \cr}\eqno\nexteq\nameeq\Emecaa
$$
These density matrix elements depend on the detailed nature of the
nuclear states involved, $|i>$ and $|f>$.  Given, for example, a
shell--model description of the nuclear ground--state and some
specific excited state in the same nucleus, it is straightforward
in principle (although frequently rather more complicated in practice ---
see Ref.~[Dub75, Dub76]) to evaluate the matrix elements in Eq.~(\Emecaa).
To obtain the current matrix elements in Eq.~(\Emecff) it remains to
decide on specific forms for the two--particle matrix elements
$<(a',b')J_{a'b'}T_{a'b'}\doubred O_{J;T}^{[2]}(q)
\doubred (a,b)J_{ab}T_{ab}>$.

Following [Dub75, Dub76, Dub80] for completeness and to show the
characteristic structure of the two--particle current operators let us
record without proof the
EM pionic
MEC contributions corresponding to the diagrams in Fig.~3.8.  For the
pair current of Fi.g~3.8a one obtains
$$\eqalign{&J^{[2]}_{\mu}({\svec x}_1, {\svec x}_2 ;q)_{\rm pair} \cr
&\qquad = -e f^2 [{\svec \tau}(1)\times {\svec \tau}(2)]_3
\bigl( ({\svec \sigma}(1)\cdot {\svec u}_r)\sigma_k(2)e^{i{\svec q}\cdot
{\svec x}_2} +  \sigma_k(1)({\svec \sigma}(2)\cdot {\svec u}_r)
e^{i{\svec q}\cdot {\svec x}_1} \bigr) \cr
&\qquad\qquad\qquad   \times (1+x_{\pi}) {{e^{-x_{\pi}}}\over{x_{\pi}^2}}
\ \ \ \ \mu=k=1,2,3 \cr
&\qquad \approx 0\ \ \ \ \mu=0\ , \cr}\eqno\nexteq
$$
where $f$ is the pion--nucleon coupling,
$x_{\pi}\equiv m_{\pi}r$, and ${\svec u}_r$ is a unit vector in the
${\svec r}$--direction.  Corresponding to diagram (b) in Fig.~3.8 is the
pion--in--flight current:
$$\eqalign{
&J^{[2]}_{\mu}({\svec x}_1, {\svec x}_2 ;q)_{\rm pionic} \cr
&\qquad =
e(f/m_{\pi})^2 [{\svec \tau}(1)\times {\svec \tau}(2)]_3
({\svec \sigma}(1)\cdot {\svec \nabla}(1))
({\svec \sigma}(2)\cdot {\svec \nabla}(2)) \cr
&\qquad\qquad\qquad   \times \int_{-1/2}^{+1/2} dv\ [i{\svec q}rv +
{\svec x}]_k {{e^{-x}}\over{x}} \exp{i{\svec q}\cdot ({\svec R}-v{\svec r})}
\ \ \ \ \mu=k=1,2,3 \cr
&\qquad \approx 0\ \ \ \ \mu=0\ , \cr}\eqno\nexteq
$$
where ${\svec r}\equiv {\svec x}_1 -
{\svec x}_2$, ${\svec R}\equiv ({\svec x}_1 +{\svec x}_2)/2$, and
${\svec x}\equiv L_{\pi}{\svec r}$ with $L_{\pi}\equiv [m_{\pi}^2
-q^2(v^2-1/4)]^{1/2}$.
Finally, for the nucleon resonance term in
Fig.~3.8c one has
$$\eqalign{
&J^{[2]}_{\mu}({\svec x}_1, {\svec x}_2 ;q)_{\rm N^*} \cr
&\qquad = {{i m_{\pi}^3\mu^{T=1}}\over{6\mn x_{\pi}}} \Bigl\{
4h_2 \Bigl[ ({\svec q}\times {\svec u}_r)_k {\svec u}_r\cdot
(\tau_3(2){\svec \sigma}(2) e^{i{\svec q}\cdot{\svec x}_1} +
\tau_3(1){\svec \sigma}(1) e^{i{\svec q}\cdot{\svec x}_2} )
[x_{\pi}d/dx_{\pi} -1] \cr
&\qquad\qquad\qquad + \tau_3(2)({\svec q}\times {\svec \sigma}(2))_k
e^{i{\svec q}\cdot{\svec x}_1} + \tau_3(1)({\svec q}\times
{\svec \sigma}(1))_k e^{i{\svec q}\cdot{\svec x}_2} \Bigr] \cr
&\qquad -h_1 [{\svec \tau}(1)\times {\svec \tau}(2)]_3
\Bigl[ {\svec q}\times \Bigl( ({\svec \sigma}(1)\times{\svec \sigma}(2))
[e^{i{\svec q}\cdot{\svec x}_1}+e^{i{\svec q}\cdot{\svec x}_2}] \cr
&\qquad\qquad\qquad + [({\svec \sigma}(1)\times{\svec u}_r)({\svec
\sigma}(2)\cdot{\svec u}_r)
e^{i{\svec q}\cdot{\svec x}_1}-({\svec \sigma}(2)\times{\svec u}_r)
({\svec \sigma}(1)\cdot{\svec u}_r) e^{i{\svec q}\cdot{\svec x}_2}] \cr
&\qquad\qquad\qquad \times [x_{\pi}d/dx_{\pi}-1] \Bigr) \Bigr]_k \Bigr\}
{d\over{dx_{\pi}}}
\Bigl({{e^{-x_{\pi}}}\over{x_{\pi}}}\Bigr)
\ \ \ \ \mu=k=1,2,3 \cr
&\qquad \approx 0\ \ \ \ \mu=0\ , \cr}\eqno\nexteq
$$
where $h_1=$ 0.074 $m_{\pi}^{-3}$ and $h_2=$ 0.0658 $m_{\pi}^{-3}$ are
the couplings used [Che71b] to take into account both the $\Delta_{33}$
and Roper resonances.  Two--particle matrix elements may then be
obtained using appropriate single--particle wave functions, for instance,
using plane waves when treating quasielastic scattering or harmonic
oscillator wave functions when treating discrete states via the nuclear
shell model.

Two considerations are critical in understanding the role
of MEC in electron scattering.  The first is the range of the exchanged
meson.  Since MEC involve (at least) two nucleons and since the repulsive
hard core of the nucleon--nucleon interaction tends to keep nucleons apart,
one can on general grounds expect the longer--ranged meson exchanges to be
the most important.  In addition to this consideration,
however, one must also observe
the ``selection rules'' associated with each current.
{}From the consideration of range, for example, the above
single--pion exchange graphs are
expected to be the most important, yet they
contribute to lowest order only for the vector three--current and
the axial--vector charge and only to isovector transitions. The other terms
are higher order in ${\omega/m_N}$ and/or ${k/m_N}$.

     In has become fairly common to include the effects of the
one--meson--exchange currents (with $\pi$, $\rho$, and $\omega$ mesons) in
modern treatments of electron scattering from nuclei.  For transitions that
obey the MEC ``selection rules'' the MEC typically contribute as much as 10\%
in the region of maxima in transverse $(e,e^\prime)$ form factors.   In
the higher--$|Q^2|$ ``tail'' of
nuclear form factor, MEC usually
fall off more gradually than the
one--body contribution and hence
are relatively more important.  There are also special
cases where MEC are
expected to play an important role or indeed dominate due to some
``accidental'' cancellation in the nucleon contributions to the
currents ({\it e.g.,\/}
electrodisintegration of the deuteron near threshold).  In such cases
calculations of the
one--meson exchange currents generally provide a reliable description of the
data and give confidence in our ability to treat the
leading--order MEC accurately.

In the context of semileptonic NC interactions, the MEC corrections to the
isovector and non--strange isoscalar weak neutral vector currents are identical
to those entering the $T=0$ and $T=1$ EM currents. This feature follows
directly from Eq.~(\Ejncmu). There exist, however, additional MEC
considerations germane to NC matrix elements which do not enter the
analysis of the EM current. First, as noted above, the leading axial--vector
MEC's arise
in the time component of $\rbra{\>}J_{\mu 5}^\sst{NC}\rket{\>}$ rather
than the spatial components, as in the case of $\rbra{\>}J_\mu(T=1)
\rket{\>}$.  Moreover, the two--body axial--vector charge
operator, $\rohat_5^{[2]}$,
enters at the same order in $v/c$ ({\it i.e.}, $\kf/\mn$ or $\omega/\mn$)
as the one--body axial--vector charge operator. In contrast, the
leading MEC corrections
to the isovector vector currents enter at {\it relative} order $v/c$ with
regard to the one--body terms. This result carries a number of implications for
the analysis of semileptonic NC observables. First, the PV asymmetry
receives no axial--vector MEC corrections from pion exchange
to leading non--trivial order in
$v/c$. The reason is that there exist no EM Coulomb multipoles having the same
parity for a given $J$ as Coulomb and longitudinal projections of $\rohat_5$,
so that the latter do not enter the interference response, $W^{\rm (PV)}$
(see Eqs.~(\Fwltmul) below). Only the axial--vector three--current
enters this
response via products of the matrix elements
$\rbra{\>}\That^{{\rm el} 5}_J\rket{\>}
\rbra{\>}\That^{\rm mag}_J\rket{\>}^*$ or
$\rbra{\>}\That^{{\rm mag} 5}_J\rket{\>}
\rbra{\>}\That^{\rm el}_J\rket{\>}^*$. Consequently, axial--vector terms in
the PV asymmetry, which involves the ratio of $W^{\rm (PV)}$ and the EM
response,
receive MEC contributions primarily via the EM vector current matrix elements.
In contrast, the neutrino scattering cross section, which is second order in
the weak interaction, generally does contain products of axial--vector charge
matrix elements (see Eq.~(\Fneutff) below) and may, therefore, be
sensitive to axial--vector MEC's at leading non--trivial order.

        An additional set of MEC considerations involves the isospin and flavor
dependence of the vector current operators. In general, the two--body isovector
and isoscalar current operators possess different spin-- and
momentum--dependencies, so that two--body corrections to the one--body $T=1$
and
$T=0$ matrix elements will differ. Since the weak neutral and EM currents
contain different linear combinations of the isovector and isoscalar currents
(see Eq.~(\Ejncmu)), the corresponding matrix elements will also display
different relative sensitivities to MEC's. As a result, vector current terms
in $\alr$, which involves the ratio of weak neutral vector current and EM
current matrix elements, may retain a sensitivity to MEC corrections. In the
special case of pure isoscalar transitions, such as elastic scattering form
$^4$He, MEC contributions to the non--strange vector current terms in $\alr$
cancel, since the isoscalar NC and EM nuclear operators are identical apart
from the electroweak couplings, $\xi^{(T)}_\sst{V}$.

        The situation with the two--body corrections to the one--body strange
quark currents is more subtle. In this case, one must distinguish among the
so--called \lq\lq pair", \lq\lq mesonic", and \lq\lq isobar" MEC's illustrated
in Fig.~3.8. The pair currents (Fig.~3.8a) involve the
creation of a virtual $N\bar N$ pair by the exchanged $\gamma$ or $Z^0$.
Conventionally, one derives an effective two--body operator for this process
by retaining the negative energy part of the nucleon propagator and employing
the same nucleon form factor ({\it viz}, $G_\sst{E}$ or $G_\sst{M}$) as
appears in the one--body operator. Thus, for example, the leading $T=0$ and
$T=1$ Coulomb operators have the form
$$
\Mhat_J^{[1]}(T)+\Mhat_J^{[2]}(T)^{\rm pair} = G_\sst{E}^\sst{T}\Bigl[
\Ohat_J^{[1]}(T)+\Ohat_J^{[2]}(T)^{\rm pair}\Bigr]\ \ \ , \eqno\nexteq
$$
where the $\Ohat_J^{[1,2]}(T)$ are nuclear operators. In contrast, the
operators arising from the interaction of the $\gamma$ or $Z^0$ with the
exchanged meson (Fig.~3.8b) or excitation of a nucleon resonance (Fig.~3.8c)
involves different form factors. Again, in the case of the Coulomb operators,
one has
$$
\eqalign{\Mhat_J^{[2]}(T)^{\rm mesonic}&=F_\sst{M}\Ohat_J^{[2]}(T)^{\rm
mesonic}\cr
\Mhat_J^{[2]}(T)^{\rm isobar}&=G_\sst{E}^{(B,T)}\Ohat_J^{[2]}(T)^{\rm
isobar}\ \ \ ,\cr}\eqno\nexteq
$$
where $F_\sst{M}$ and $G_\sst{E}^{(B, T)}$ are form factors associated
with the charge of the exchanged meson and with the isobar transition and
$\Ohat_J^{[2]}(T)^{\rm mesonic,\ isobar}$ are the corresponding nuclear
operators. To illustrate further the impact of these corrections, we consider
PV elastic scattering from $^4$He, which is a pure isoscalar process. In this
case, one has for the weak NC Coulomb matrix element
$$
\eqalign{&\rbra{\hbox{g.s.}}\Mhat_0^{[1]}(T=0)\rket{\hbox{g.s.}}^\sst{NC}
=\sqrt{3}
\xivtez\Bigl\lbrace\GETEZ\rbra{\hbox{g.s.}}\Ohat_0^{[1]}(T=0)+
\Ohat_0^{[2]}(T=0)^{\rm pair}\rket{\hbox{g.s.}}\cr
&+ F_\sst{M}\rbra{\hbox{g.s.}}\Ohat_0^{[2]}(T=0)^{\rm mesonic}\rket{\hbox{
g.s.}} + G_\sst{E}^{(B,T=0)}\rbra{\hbox{g.s.}}\Ohat_0^{[2]}(T=0)^{\rm
isobar}\rket{\hbox{g.s.}}\Bigr\rbrace\cr
&+\xivz\Bigl\lbrace\GES\rbra{\hbox{g.s.}}\Ohat_0^{[1]}(T=0)+
\Ohat_0^{[2]}(T=0)^{\rm pair}\rket{\hbox{g.s.}}\cr
&+ F_\sst{M}^{(s)}\rbra{\hbox{g.s.}}\Ohat_0^{[2]}
(s)^{\rm mesonic}\rket{\hbox{
g.s.}} + G_\sst{E}^{(B,s)}\rbra{\hbox{g.s.}}\Ohat_0^{[2]}(s)^{\rm
isobar}\rket{\hbox{g.s.}}\Bigr\rbrace\ ,\cr}\eqno\nexteq
$$
where \lq\lq $s$" indicates a form factor or operator associated with
$\sbar\gamma_\mu s$ and where we have assumed the exchange of only one
type of meson and the excitation of only a single nucleon isobar for
simplicity. The EM Coulomb matrix element has a similar structure
$$
\eqalign{&\rbra{\hbox{g.s.}}\Mhat_0^{[1]}(T=0)\rket{\hbox{g.s.}}^\sst{EM}
=\Bigl\lbrace\GETEZ\rbra{\hbox{g.s.}}\Ohat_0^{[1]}(T=0)+
\Ohat_0^{[2]}(T=0)^{\rm pair}\rket{\hbox{g.s.}}\cr
&+ F_\sst{M}\rbra{\hbox{g.s.}}\Ohat_0^{[2]}(T=0)^{\rm mesonic}\rket{\hbox{
g.s.}} + G_\sst{E}^{(B,T=0)}\rbra{\hbox{g.s.}}\Ohat_0^{[2]}(T=0)^{\rm
isobar}\rket{\hbox{g.s.}}\Bigr\rbrace\  .\cr}\eqno\nexteq
$$
The one--body and pair operators multiplying $\GETEZ$ and $\GES$ are identical
since the corresponding diagrams have the same spin and isospin structure.
Those arising from mesonic and isobar graphs, however, do not give identical
operators for the isoscalar and strange--quark currents, since the
non--nucleonic
degrees of freedom in the nucleus display a different flavor structure than
the nucleons themselves. The authors of Ref.~[Sch90] have shown that for
low--to--moderate momentum transfer, the one--body and pair currents dominate
the EM matrix element:
$$
\eqalign{
\Bigl\vert F_\sst{M}\rbra{\hbox{g.s.}}\Ohat^{[2]}_0(T=0)^{\rm mesonic}
\rket{\hbox{g.s.}}&/\GETEZ\rbra{\hbox{g.s.}}\Ohat_0^{[1]}(T=0)+
\Ohat_0^{[2]}(T=0)^{\rm pair}\rket{\hbox{g.s.}}\Bigr\vert\cr
& \qquad <\!\!< 1\ \ \ .\cr}\eqno\nexteq
$$
Were a similar result to hold for the mesonic and isobar contributions to
the strange--quark matrix elements, then the MEC contributions to the
strange--quark term in $\alr(^4\hbox{He})$ would largely cancel, leaving
a dependence primarily on the ratio of single--nucleon form factors,
$\GES/\GETEZ$. A study of mesonic and isobar MEC contributions
to nuclear matrix elements of $\sbar\gamma_\mu s$ is in progress [Mus93b].

        A final class of MEC corrections to PV electron scattering arises from
nuclear PV. We postpone a discussion of these corrections until Sect.~III.D.5
below.

\bigskip
\goodbreak
\noindent III.D.4.\quad ISOSPIN--MIXING

Another issue is the assumption of an exact isospin symmetry at the
{\it nuclear}
level.  In the case of elastic PV electron scattering from $(J^\pi T) =
(0^+ 0)$ nuclei (Sect.~IV.B), the assumption of exact isospin symmetry
allows one to eliminate much of the dependence of the PV
asymmetry on the details of the nuclear structure of the ground
state (in the absence of $s$--quark contributions). The presence of
charge--symmetry--breaking (CSB) nuclear forces, such as
the Coulomb interaction between protons, reduces strong isospin to an {\it
approximate} symmetry and the effects of this approximation must be
understood and estimated.

     Isospin--mixing has been the subject of study throughout much of the
history of nuclear physics (see the discussion in [Don89]).
While an exact description of nuclear structure remains elusive,
enough phenomenological information is available to permit one to make
reasonable {\it estimates} of the effects of isospin--mixing
in PV $(\vec e,e^\prime)$ from a variety of nuclear targets.  Such estimates
are reported in Ref.~[Don89] for a number of relevant nuclei and
transitions. The authors conclude that the theoretical uncertainty associated
with isospin--mixing is generally large enough to introduce non--negligible
errors into Standard Model tests with PV electron scattering.
However, they also find that there are also more favorable cases
(including the critical elastic scattering from $(0^+0)$
states in very light nuclei discussed in Sects.~IV.B and V) where
isospin--mixing can be ignored at a level relevant to Standard Model tests.

     The approach taken in Ref.~[Don89] is to consider mixing of two states
$\ket{T_0}$ and $\ket{T_1}$ having exact isospin $T_0$ and $T_1$, respectively,
to produce states of nominal isospin
$\vert ``T_0"\rangle$  and $\vert ``T_1"\rangle$:
$$
\vert ``T_0"\rangle =\cos\chi_T \vert T_0\rangle + \sin\chi_T
\vert T_1\rangle\eqno\nexteqp\nameeq\Eisomix
$$
$$\vert ``T_1"\rangle = -\sin\chi_T \vert T_0\rangle + \cos\chi_T
\vert T_1\rangle\ \ . \eqno\sameeq
$$
Since matrix elements of the charge--symmetry--breaking piece of the nuclear
interaction in light nuclei are generally much smaller (up to a few hundred
keV) than the energy splitting between states of identical spins but
different isospins (typically a few MeV), the isospin--mixing can be treated
perturbatively.  In this case, $\chi_T$ is small and
$$
\cos\chi_T\approx 1\eqno\nexteqp
$$
$$
\sin\chi_T\approx \chi_T = \langle T_1\vert {\hat H}_{CSB}\vert
T_0\rangle/\Delta E\ \ , \eqno\sameeq
$$
where ${\hat H}_{CSB}$ is a charge--symmetry--breaking interaction
and
$\Delta E$ is the energy difference between the two states, giving
$$
\vert``T_0"\rangle \approx \vert T_0\rangle +\chi_T\vert
T_1\rangle\eqno\nexteqp
$$
and
$$
\vert``T_1"\rangle \approx -\chi_T\vert T_0\rangle +\vert
T_1\rangle\ \ .\eqno\nexteqp
$$
For elastic scattering, the ground state will have the minimum allowed
isospin $T_0 = \vert N-Z\vert/2$, so only higher isospin $T_1 > T_0$ need be
considered.  Furthermore, the transition operators we consider are either
isoscalar or isovector, so states with isospin $T_1 > T_0 +1$ will not be
connected to the dominant $\vert T_0\rangle$ terms in Eq.~(\Eisomix).
Thus we limit discussion to mixing with states of $T_1 = T_0+1$.  One can
also go beyond the simple two--state mixing of Eqs.~(\Eisomix) in first--order
perturbation theory simply by adding additional states of isospin $T_0 + 1$.

     In taking matrix elements between states of nominal isospin, the
isospin selection rules are only approximately valid and we obtain
additional terms of ${\hcal O}(\chi_T)$ that would be missing for matrix
elements between states of pure isospin.  To illustrate, consider
the case of electron scattering from a spin--0, (nominally) isospin--0
nucleus.  To form the asymmetry one needs only matrix elements of the $\hat
M_0$ projection of the electromagnetic and weak neutral currents.
For pure isospin states, only
the isoscalar operator contributes and we have (neglecting $s$--quark
contributions)
$$F_{CJ}(q) = \langle 0^+0\Vert \hat M_0(T=0) \Vert0^+0\rangle
\eqno\nexteqp\nameeq\Easyfei
$$
and
$$\tilde F_{CJ}(q) = \Bigl({1\over 2}\sqrt{3}\xi_V^{T=0}\Bigr)
\langle 0^+0\Vert \hat M_0(T=0) \Vert0^+0\rangle\ \ ,
\eqno\sameeq$$
so that in the PV asymmetry, which depends on the {\it ratio} of these
form factors (see Sect.~III.E), the single common matrix element
cancels and we obtain for this ratio
the (structure--independent) first factor in
Eq.~(\Easyfei b).  Including the isospin--mixed states with (small) $T=1$
components, however, allows for nonzero matrix elements of isovector
operators as well
$$\eqalign{F_{CJ}(q) =& \langle 0^+``0"\Vert \hat M_0(T=0)
+ \hat M_0(T=1) \Vert0^+``0"\rangle \cr
\approx&\langle 0^+0\Vert \hat M_0(T=0) \Vert0^+0\rangle
+2\chi_T \langle 0^+0\Vert \hat M_0(T=1) \Vert0^+1\rangle \cr}
\eqno\nexteqp$$
and
$$
\eqalign{\tilde F_{CJ}(q) =& {1\over 2}\langle 0^+``0"\Vert \sqrt{3}
\xi_V^{T=0}\hat M_0(T=0)
+ \xi_V^{T=1}\hat M_0(T=1) \Vert0^+``0"\rangle \cr
\approx& {1\over 2}\sqrt{3}\xi_V^{T=0}\langle 0^+0\Vert \hat M_0(T=0)
\Vert0^+0\rangle
+\chi_T \xi_V^{T=1}\langle 0^+0\Vert \hat M_0(T=1) \Vert0^+1\rangle \ \ .\cr}
\eqno\nexteqp$$
Two {\it different} structure--dependent matrix elements now enter the form
factors. Moreover,
the numerator and denominator in their ratio, which governs the
PV asymmetry,  contain {\it
different} linear combinations of these two matrix elements.  Keeping only
terms up to order $\chi_T$, one can write the ratio as
$$\alr\propto {1\over 2}\sqrt{3}\xi_V^{T=0} + \Gamma(q)\eqno\nexteq$$
with
$$\Gamma(q) = \chi_T (\xi_V^{T=1}-\sqrt{3}\xi_V^{T=0})
{\langle 0^+0\Vert \hat M_0(T=1) \Vert0^+1\rangle \over
\langle 0^+0\Vert \hat M_0(T=0) \Vert0^+0\rangle}\ \ .\eqno\nexteq
$$
For transitions involving a single multipole, analogous expressions obtain.
For other types of transitions where additional multipoles contribute, the
effects of isospin impurities can be treated in a similar (though
algebraically more complicated) manner, introducing small (of order
$\chi_T$) additional matrix elements for each multipole.

For elastic
scattering from light $(0^+0)$ nuclei, such as $^4$He, $^{12}$C, and
$^{16}$O, the authors of Ref.~[Don89] estimate $\Gamma(q) < 0.01$ over
the entire momentum transfer range
of practical interest for PV electron scattering experiments (see also
Sect.~IV.B). For heavier
$(0^+0)$ nuclei such as $^{28}$Si, the effect can be much larger with
$\Gamma$ increasing from 0 at $q=0$ (a consequence of the orthogonality of
the $T=0$ and $T=1$ states) to a few percent for 1 fm$^{-1}$ $< q < 3$
fm$^{-1}$.  This significant difference between the light and heavier
nuclei can be attributed to the relative inability of the former to support
an isovector (monopole) breathing mode. For the $C2$
(forward--angle) excitation
of the 4.44 MeV $(2^+0)$ first excited state in $^{12}$C, the
isospin--mixing correction $\Gamma(q)$ may
be as much as a few percent at intermediate $q$; this has important
implications for the energy resolution required for a elastic scattering
experiment.  Finally, for the $M1$ excitation of the 12.71 MeV $(1^+0)$
state (a transition frequently used to study the effects of isospin--mixing
with the relatively strong, relatively close in energy, isovector M1
excitation of the 15.11 MeV $(1^+1)$ state) the authors of Ref.~[Don89]
find effects as large as
$\Gamma \approx 2$ at low $q$ and $\Gamma \approx 0.3$ at intermediate $q$
and conclude that, were the figure--of--merit not so low, $(\vec
e,e^\prime)$ experiments would be excellent means for studying
isospin--mixing in such cases.

\bigskip
\goodbreak
\noindent III.D.5.\quad PARITY--MIXING AND THE ANAPOLE MOMENT

\def\HPV{{{\hat H}_\sst{PV}^{[2]}}}
\def\gnnmi{{g_\sst{NNM}^{(i)} }}
\def\hnnmi{{h_\sst{NNM}^{(i)} }}

        In a manner analogous to the mixing of states with good isospin
by charge--symmetry--breaking forces, the weak N--N interaction induces
small admixtures of opposite--parity states into states of a given, exact
parity. One consequence of such parity--mixing is the existence of
non--vanishing
nuclear observables normally forbidden by parity--invariance. In the case of
electron scattering, for example, a virtual photon emitted by the
electron could couple to a nuclear multipole having the \lq\lq wrong" parity
for a given transition. A similar process involving an atomic electron
interacting with the nucleus can give rise to a parity--violating atomic
Hamiltonian which, in turn, induces mixing between opposite--parity atomic
states. Such atomic parity--mixing would be manifested in the existence
of parity--forbidden atomic observables. These possibilities have motivated
suggestions for studying the weak nuclear force with charged lepton probes.
As noted in Sect.~II, one hopes that such studies might complement other
nuclear PV experiments and help to further test the conventional picture of
nuclear PV.

        In this conventional picture, contributions from nuclear PV to
semileptonic interactions occur via the diagrams in Fig.~3.9. In
Fig.~3.9a,b the exchange of the lightest pseudoscalar and vector mesons ---
with one vertex being parity--conserving and the other parity--violating ---
mixes nuclear states of good parity  which then couple to the virtual
photon. The weak quark--quark interaction mediated by short range ($\sim 0.002$
fm) $W^\pm$ and $Z^0$ exchanges occurs within the PV meson--nucleon vertex,
while the resultant PV N--N interaction occurs over a larger distance scale.
In the conventional framework, this N--N interaction takes the form
$$
\HPV = \sum_i \gnnmi\hnnmi\Ohat_{(i)}^{[2]}\ \ ,
\eqno\nexteq\nameeq\Ehpv
$$
where the $\gnnmi$ ($\sim 10$) are strong--interaction meson--nucleon
couplings, the $\hnnmi$ ($\sim 10^{-7}$) are the weak PV meson--nucleon
couplings, the $\Ohat_{(i)}^{[2]}$ are two--body nuclear operators, and the sum
over $i$ corresponds to the various exchanged mesons and isospin channels.
Typically, one includes only $\pi$--, $\rho$--, and $\omega$--exchange. Under
this
truncation, the different isospin components of the weak interaction between
light quarks ($T=0,1,2$) give rise to seven terms in the sum of Eq.~(\Ehpv)
corresponding to the parameters, $h_\pi$, $h_\omega^{0,1}$, $h_\rho^{0,1,2}$,
and $h_\rho^{1'}$. Since the two--body operators $\Ohat_{(i)}^{[2]}$ are
generally momentum--dependent, one must also include PV meson--exchange current
contributions (Fig.~3.9c) in matrix elements of the nuclear EM current in
order to satisfy the continuity equation: $Q^\lambda\bra{f}J_\lambda^\sst{EM}
\ket{i}=0$.

        In the simplest case of two--state parity--mixing, one has in analogy
with Eq.~(\Eisomix)
$$\eqalign{
\ket{``\pi^+"} &= \cos{\chi_P}\ket{\pi^+}+\sin{\chi_P}\ket{\pi^-} \cr
&\approx \ket{\pi^+}+\chi_P\ket{\pi^-}\ \ ,\cr}\eqno\nexteq\nameeq\Epmix
$$
where $\ket{\pi^+}$ and $\ket{\pi^-}$ are states of exact, but opposite,
parity and $\ket{``\pi^+"}$ is a state of nominal parity \lq\lq $\pi^+$".
The mixing angle is simply
$$
\chi_P=\bra{\pi^-}\HPV\ket{\pi^+}/\Delta E\ \ ,\eqno\nexteq
$$
with $\Delta E$ being the
difference in energy between $\ket{\pi^+}$ and $\ket{\pi^-}$. Generically, one
has $\chi_P\sim 10^{-7}$ (assuming typical nuclear level spacings) [Ade85], so
use of the perturbative expression (\Epmix) is valid. For nuclei having
nearly--degenerate opposite--parity states, the scale of $\chi_P$ may be
enhanced by an order of magnitude or more [Ade85]. In a realistic nucleus,
one must include mixing of the full spectrum of opposite--parity states into
$\ket{\pi^+}$, although for nuclei possessing a pair of nearly degenerate
opposite--parity levels, the use of two--state mixing may be a justifiable
approximation.

        To illustrate the impact of Eq.~(\Epmix) on semileptonic processes,
we consider an elastic process for which $\bra{\pi^+}\That^{\rm el}_{1\lambda}
\ket{\pi^+}_\sst{EM}=0=\bra{\pi^+}\That^{\rm el}_{1\lambda}
\ket{\pi^+}_\sst{EM}$, according to the selection rules discussed in
Sect.~III.D.1
(the subscript $EM$ denotes multipole projections of the EM current).
Matrix elements of $\That^{\rm el}_{1\lambda}$ in the parity--mixed
state $\ket{``\pi^+"}$, however, need not vanish. Rather, one has
$$
\bra{``\pi^+"}\That^{\rm el}_{1\lambda}\ket{``\pi^+"}_\sst{EM} =
2\hbox{Re}\Bigl\{\chi_P\bra{\pi^+}\That^{\rm el}_{1\lambda}\ket{\pi^-}_\sst{
EM}\Bigr\}\ \ .\eqno\nexteq\nameeq\Eelab
$$
The matrix element in Eq.~(\Eelab) corresponds to the coupling of a virtual
photon to the nucleus via the processes shown in Fig.~3.9a,b. The
meson--exchange process of Fig.~3.9c induce a two--body EM current operator
having the spacetime transformation properties of an axial--vector:
$$
J_{\lambda 5}^\sst{EM}[2]=\sum_i\gnnmi\hnnmi{\hat j}_{\lambda 5}^{[2]}
\ \ ,\eqno\nexteq\nameeq\Ejexch
$$
where the ${\hat j}_{\lambda 5}^{[2]}$ are two--body axial--vector current
operators. According to the selection rules discussed above, the
current of Eq.~(\Ejexch) will also contribute to elastic matrix elements
of $\That^{\rm el 5}_{1\lambda}$ as
$$
\bra{\lq\lq\pi^+"}\That^{{\rm el},\ [2]}_{1\lambda}\ket{\lq\lq\pi^+"}_\sst{EM}=
\bra{\pi^+}\That^{{\rm el5},\ [2]}_{1\lambda}\ket{\pi^+}+{\cal O}(\chi_P^2)
\ \ .\eqno\nexteq\nameeq\Eelc
$$
The full elastic transverse electric matrix element will be the sum of
Eqs.~(\Eelab) and (\Eelc). These two terms represent contributions of,
respectively, matrix elements of the normal (polar vector) EM current between
states of opposite parity and matrix elements of the two--body axial--vector
MEC between states of like parity.

        The analysis for nuclear transitions is similar to that for elastic
processes. In both cases, parity--mixing plus PV MEC's give rise to
non--vanishing EM multipole matrix elements normally forbidden in the absence
of nuclear PV. From the discussion of Sect.~III.D.1, we observe that
these parity--forbidden multipoles are precisely those which appear in matrix
elements of the nuclear axial--vector weak neutral three--current. In effect,
nuclear PV induces an axial--vector, conserved EM current which contributes in
tandem with the nuclear axial--vector NC to semileptonic processes involving
charged leptons:
$$
\bra{f}\Ohat_{J\lambda}^5\ket{i}_\sst{NC}\rightarrow
\bra{f}\Ohat_{J\lambda}^5\ket{i}_\sst{NC}+\beta
\bra{f}\Ohat_{J\lambda}^5\ket{i}_\sst{EM}\ \ ,\eqno\nexteq\nameeq\Emultmix
$$
where
$$
\beta = -{8\sqrt{2}\pi\alpha\over G_\mu Q^2}{Q_e\over\gve}\ \ .
\eqno\nexteq\nameeq\Ebeta
$$
Although the nuclear PV contribution is generically suppressed by
roughly $\alpha$ with respect to the tree--level, axial--vector NC interaction,
nuclear structure effects and/or a fortuitous suppression of the
axial--vector NC amplitude may enhance the relative importance of the nuclear
PV contribution.
\medskip
\goodbreak
\noindent\undertext{The Nuclear Anapole Moment}
\medskip

        A special case of relevance to atomic PV and low--energy PV electron
scattering is given by the low--$|Q^2|$ limit of the elastic
$\That^{\rm el}_{1\lambda}$ matrix element discussed above. The
requirements of current conservation, embodied in an \lq\lq extended"
version of Siegert's Theorem [Sie37, Fri84, Fri85], allows one to
decompose this operator as [Hax89]
$$
\That^{\rm el}_{1\lambda}=\Shat^{\rm el}_{1\lambda}+\Rhat^{\rm el}_{1\lambda}
\ \ ,\eqno\nexteq
$$
where $\Shat^{\rm el}_{1\lambda}$ is a piece constrained by current
conservation and $\Rhat^{\rm el}_{1\lambda}$ is an unconstrained remainder.
For elastic processes, matrix elements of $\Shat^{\rm el}_{1\lambda}$ vanish.
For low--$q^2$, the remainder has the form
$$
\Rhat^{\rm el}_{1\lambda}\rightarrow
-{i\over\sqrt{6\pi}}{q^2\over\mns}\ahat_\lambda +{\cal O}(q^4)\ \ ,
\eqno\nexteq\nameeq\Erelec
$$
where
$$
\ahat_\lambda\>=\>{\mns\over 9}\int d^3r r^2\{{\hat J}_\lambda^\sst{EM}(\rv)
+\sqrt{2\pi}[Y_2(\Omega_r)\otimes{\hat J}^\sst{EM}(\rv)]_{1\lambda}\}
\eqno\nexteq\nameeq\Eanap
$$
is the anapole moment operator. In Eq.~(\Eanap), ${\hat J}^\sst{EM}$ is the
full
EM three--current operator. Matrix elements of $\ahat_\lambda$ containing
the polar vector component of ${\hat J}^\sst{EM}$ are non--vanishing between
opposite--parity components of a state of mixed--parity
(as in Eq.~(\Eelab)), while those containing the
PV MEC operator are non--vanishing between same--parity components
(Eq.~(\Eelc)).
The sum of these matrix elements defines the nuclear anapole moment (AM).

        While a detailed discussion of the anapole moment can be found
elsewhere [Mus91, Hax89 and references therein], we make a few comments on
the form of Eqs.~(\Erelec) and (\Eanap). First, the matrix element of
Eq.~(\Erelec) vanishes for real photons ($q^2=0$ in the Breit frame) and
couples only to virtual photons entering semileptonic interactions with the
nucleus. Moreover, when multiplied by the $1/q^2$ appearing in the photon
propagator in the semileptonic amplitude, the matrix element
of Eq.~(\Erelec)
gives rise to a contact interaction between leptonic and hadronic currents
in co--ordinate space, like that
engendered by low--$q^2$ $Z^0$--exchange. The result is that transverse
electric projections of the axial--vector NC should be replaced by the
following combination
$$
\gve\bra{\>}\That_{1\lambda}^{\rm el5}\ket{\>}\rightarrow\gve\Bigl\{
\bra{\>}\That_{1\lambda}^{\rm el5}\ket{\>}+\xi_\sst{AM}
\bra{\>}\ahat_\lambda\ket{\>}\Bigr\}\ \ ,\eqno\nexteq\nameeq\Eaxnew
$$
where
$$
\xi_\sst{AM} = {Q_e\over\gve}\biggl[8\sqrt{\pi\over 3}\biggr]\biggl({\alpha
\over G_\mu\mns}\biggr)\approx 6.4\times 10^4\ \ .\eqno\nexteq\nameeq\Exian
$$

Third, the scale of $\rbra{\>}\ahat\rket{\>}$, typically $\sim 10^{-6}
\to 10^{-7}$, grows as $A^{2/3}$. This result is suggested by the
$r^2$ factor appearing in the integral (\Eanap) and has been shown rigorously
by the authors of Refs.~[Fla84, Hax89]. The scale of the anapole moment
can also be strongly enhanced in nuclei having a pair of nearly degenerate
opposite--parity states, where one of the states is the ground state. One
finds, then, that for heavy and/or nearly--degenerate nuclei the scale of
the axial--vector NC and nuclear anapole contributions to the low--$q^2$,
elastic electron--nucleus interaction can be commensurate [Fla84, Hax89].
In addition, the relative importance of the anapole contribution is
amplified in $T=0$ nuclei, whose axial--vector NC matrix elements are
suppressed. Consequently, were it possible to separate contributions
from the axial--vector
NC and anapole moment to semileptonic processes involving such nuclei,
one could in principle impose new constraints on nuclear PV via a determination
of $\rbra{\>}\ahat\rket{\>}$. This possibility is discussed more fully in
Sect.~IV.E.

        Finally, from a formal standpoint, it is not possible to define the
AM of an elementary particle (lepton or quark) or nucleon unambiguously,
since it depends on the choice of electroweak gauge parameter [Mus91].
The contributions made by the electron, quark, and nucleon AM's to
scattering amplitudes is more appropriately included in the full set of
axial--vector radiative corrections, $R_\sst{A}$. The many--body nuclear AM is,
in contrast, gauge--parameter independent and, in principle, distinguishable
from the $R_\sst{A}$ due to its $A^{2/3}$ scaling behavior [Mus91].

\vfil\eject

%

\def\qvecsq{\vec q^{\mkern2mu\raise1pt\hbox{$\scriptstyle2$}}}

\def\xiva{{\xi_\sst{V}^{(a)}}}

\def\evec{{\vec e}}
\def\sigpin{{\Sigma_{\pi\sst{N}}}}

\def\alrzer{{A^0_\sst{LR}}}  
\def\gtilan{{\tilde G_\sst{A}^\sst{N}}}
\def\gtilen{{\tilde G_\sst{E}^\sst{N}}}
\def\gtilmn{{\tilde G_\sst{M}^\sst{N}}}

\def\gtilep{{\tilde G_\sst{E}^p}}

\def\gtilenn{{\tilde G_\sst{E}^n}}

\def\xivp{{\xi_\sst{V}^p}}
\def\xivn{{\xi_\sst{V}^n}}

\def\lams{{\lambda_\sst{E}^\sst{(s)}}}

\def\msubt{{M_\sst{T}}}
\def\tsubr{{T_\sst{R}}}
\def\thetr{{\theta_\sst{R}}}

\def\gaeight{{G_\sst{A}^{(8)}}}
\def\gathree{{G_\sst{A}^{(3)}}}


\noindent{\bf III.E.\quad Lepton Scattering from Nucleons and Nuclei}
\medskip

In this section we discuss the basic formalism required in studying
inclusive lepton scattering from nucleons and nuclei.  In particular,
we summarize in Sect.~III.E.1 the expressions that will prove to be useful
in the rest of the article for parity--conserving and --violating electron
scattering, followed by discussions of the figure--of--merit
for PV electron scattering in Sect.~III.E.2. The formalism presented
here is meant to provide a general framework in which the detailed treatment
of specific cases om Sect.~IV can be carried out.
Nevertheless, to help in introducing the essential features of PC/PV
electron scattering we begin the discussion of several important examples
in this section.
Specifically, we introduce the formalism for elastic scattering from
spin--0, spin--1/2, and spin--1 nuclei as well as the essential features of
quasielastic scattering from nuclei.  More detailed discussions of these
cases are contained in Sects.~IV.A--C and IV.F. We conclude this section
by summarizing the formalism for treating inclusive neutrino scattering
from nucleons and nuclei (Sect.~III.E.3), again postponing more detailed
discussions to a later section (Sect.~IV.J).

For all cases of inclusive, semi-leptonic scattering from nucleons and
nuclei, the cross section is proportional to the contraction of a leptonic
tensor, $\eta_{\mu\nu}$, and hadronic tensor, $W^{\mu\nu}$. General covariance
and symmetry considerations require this contraction to have the following
form [Don85, Wal75]:
$$
\eta_{\mu\nu}W^{\mu\nu} \sim V_{CC} R^{CC} +2 V_{CL} R^{CL}
+V_{LL} R^{LL} +V_T R^T -\lambda V_{T'} R^{T'}\ \ , \eqno\nexteq\nameeq\Eresp
$$
where $\{C,L,T\}$ stand for $\{$charge, longitudinal, transverse$\}$ and
involve the $\mu,\nu =$ 0, 3, and 1 or 2 components of the electroweak
currents, respectively (as usual, we choose the $z$--axis to be along
$\vec q$).  The $V$'s are lepton kinematical factors and the $W$'s are
hadronic response functions; $\lambda=\pm$ 1 is the incident lepton's
helicity.  In the extreme relativistic limit (ERL) where the lepton
masses may be neglected with respect to their energies the lepton
kinematical factors become [Don85, Wal75]:
$$
\eqalign{V_{CC} &\longrightarrow 1 \times \cos^2 {\theta\over 2} \cr
V_{CL} &\longrightarrow -(\omega/q)\times \cos^2 {\theta\over 2} \cr
V_{LL} &\longrightarrow (\omega/q)^2\times \cos^2 {\theta\over 2} \cr
V_{T} &\longrightarrow v_T \times \cos^2 {\theta\over 2} \cr
V_{T'} &\longrightarrow v_{T'} \times \cos^2 {\theta\over 2}\ \ , \cr}
\eqno\nexteq\nameeq\Eerll
$$
where $q=|\vec q|$ and $\omega$ are the magnitude of three--momentum and the
energy transferred to the hadronic system. The four--momentum transfer is
then $Q^\mu = (\omega,\vec q\/)$. Carrying the overall factor
$\cos^2 {\theta\over 2}$, where $\theta$ is the lepton scattering angle, the
ERL transverse lepton kinematical factors are given by
$$
\eqalignno{
v_T &= {1\over 2} \left| {Q^2\over q^2}\right| + \tan^2 {\theta\over 2}
&\nexteqp\nameeq\Elept \cr
v_{T'} &= \sqrt{ \left| {Q^2\over q^2}\right| + \tan^2 {\theta\over 2}}
\ \tan {\theta\over 2}\ \ . &\sameeq \cr}
$$
The response functions in Eq.~(\Eresp) in the general case contain
products (with the appropriate sum over
spin and isospin quantum numbers) of matrix elements of the nuclear
currents.
$R^{CC}$, $R^{CL}$, $R^{LL}$ involve products of matrix elements of
nuclear vector currents with nuclear vector currents and of nuclear
axial--vector currents with nuclear axial-vector currents
while $R^{T}$ contains a product of nuclear vector with nuclear
axial--vector
current matrix elements.
For conserved vector currents (CVC) the
purely vector $CC$, $CL$ and $LL$ responses are proportional to each other;
then, defining another ERL lepton kinematical factor
$$
\eqalignno{
v_L &= \left| {Q^2\over q^2}\right|^2\ \ , &\sameeq \cr}
$$
the first three terms in Eq.~(\Eresp) can be combined into one:
$$
\Bigl[ V_{CC} R^{CC} +2 V_{CL} R^{CL} +V_{LL} R^{LL} \Bigr]_{VV}
\longrightarrow v_L R^L \times \cos^2 {\theta\over 2} \ \ ,
\eqno\nexteq\nameeq\Ecvc
$$
where $R^L\equiv R^{CC}=(q/\omega) R^{CL}=(q/\omega)^2 R^{LL}$ for
purely vector--vector (VV) contributions.

\goodbreak
\bigskip
\noindent III.E.1.\quad BASIC ELECTRON SCATTERING FORMALISM

     In electron scattering, both graphs of Fig.~3.1
enter and, in principle,
$(e,e^\prime)$ experiments probe the structure of both the electromagnetic
and weak neutral currents.  However, for $|Q^2| <\!\!< \mzs$,
the photon--exchange amplitude of Fig.~3.1a is generally several
orders of magnitude larger than the $Z^0$--exchange amplitude in Fig.~3.1b and
the latter can be safely neglected.  In that
case, the doubly--differential $(e,e^\prime)$ ERL cross section
may be written [deF66]
$$
{d^2\sigma\over d\Omega\,d\epsilon'}
= \sigma_M
\left\{ v_L R^L (q,\omega) + v_T R^T (q,\omega)\right\} \equiv \sigma_M
{W}^{\rm (EM)}(q,\omega,\theta)\ \ . \eqno\nexteq\nameeq\Edtsige
$$
The incident electron is assumed to have
four--momentum $K^{\mu}=(\epsilon,{\vec k})$ with three--momentum $k\equiv
|\vec k|$ and energy $\epsilon=\sqrt{k^2+m_e^2}\ $ (where $\epsilon=k$
in the ERL); similarly, the scattered
electron has four--momentum $K^{\prime\mu}=(\epsilon',{\vec k}')$. The
four--momentum transfer is then given by
$Q^\mu \equiv (K-K^\prime)^\mu = (\omega,\vec q\/)$, so that
$\omega = \epsilon - \epsilon^\prime$ and $q = \vert\vec q\/\vert =
\vert \vec k - \vec k^\prime\vert$. The kinematics of electron scattering
require that $Q^2 = Q_{\mu}Q^{\mu} = \omega^2 - q^2
\leq 0$. In Eq.~(\Edtsige)
the Mott cross section for electron scattering from a point
unit charge at angle $\theta$ is given by
$$
\sigma_M = \left[ {\alpha\cos{\theta\over 2} \over 2\epsilon \sin^2
{\theta\over 2}}\right]^2 \eqno\nexteq\nameeq\Esigmm
$$
and the lepton kinematical factors are given in Eqs.~(\Elept).  As discussed
above, since the electromagnetic current is purely vector in nature only
the longitudinal (really a combination of $\mu=$ 0 and 3 components, see
above) and transverse response functions, $R^L$ and $R^T$, respectively,
enter. These embody the hadronic matrix elements of the
electromagnetic current operators discussed in Sects.~III.C and III.D and
depend only on $q$ and $\omega$ (or, equivalently, only on $Q^2$ and
$\nu\equiv\omega$ in the language more commonly used in high--energy physics).
In writing the above expressions we have assumed the ERL
for the electron ($\epsilon, \epsilon^\prime \gg m_e$), which,
while unnecessary, does simplify the formalism somewhat (see Refs.~[deF66,
Don85, Don86b] for discussions of the leptonic tensor when the ERL is not
invoked).  We have also restricted the treatment to
the first--order plane--wave Born approximation; this is
usually adequate for treating scattering from the low--$Z$ targets that
we will mostly be considering here. However, where necessary, distortion of the
incoming and outgoing electron waves can and should be included.

     For elastic and inelastic scattering to discrete nuclear states,
$\omega$ is fixed by the excitation energy and momentum transfer $q$
and it is conventional to use the singly--differential
cross section [deF66]
$$
{d\sigma\over d\Omega} = 4\pi\sigma_M f_{\rm rec}^{-1} F^2(q,\theta)
\ \ ,\eqno\nexteq\nameeq\Edisc
$$
where the factor involving $f_{\rm rec}=1+(2\epsilon/
M_T)\sin^2{\theta\over2}$
accounts for recoil of the target nucleus of mass $M_T$.  The total form
factor $F^2$ is the sum of longitudinal and transverse contributions
$$
F^2 (q,\theta) = v_L F^2_L (q) + v_T F^2_T (q)\ \ , \eqno\nexteq\nameeq\Effdis
$$
where $F^2_L$ and $F^2_T$
are given by
$$
\eqalign{F^2_L (q)&= \sum_{J\geq0} F^2_{CJ} (q) \cr
F^2_T (q)&= \sum_{J\geq1} \lbrack F^2_{EJ} (q) + F^2_{MJ} (q)
\rbrack \ \ , \cr}\eqno\nexteq\nameeq\Efltmul
$$
using the form factors introduced in Sect.~III.D.1.
The number of terms entering each of these sums depends on how many multipole
projections are allowed by the angular momentum and parity quantum numbers of
the initial and final nuclear states (see Sect.~III.D.1).

     In order to detect the presence of the very small weak neutral current
contributions to electron scattering, one must search for a characteristic
signature that only occurs when such contributions are present.
Since electromagnetism obeys an exact parity symmetry
while the neutral current amplitude contains a parity--violating
piece, one such
signature would be a difference between cross sections for scattering of
electrons longitudinally polarized parallel ($+$ or $R=$ right--handed) and
anti--parallel ($-$ or $L=$ left--handed) to
their momenta, as this difference is parity--violating.  Based on the same
arguments about the relative size of the
photon--exchange and $Z^0$--exchange amplitudes used above to neglect the
latter in unpolarized scattering, we expect this parity--violating asymmetry
to be dominated by the interference of the two amplitudes.  That is, our
focus will be on the PV asymmetry
$$
\eqalign{\alr &=
\left\{ {d^2\sigma^+\over d\Omega\,d\epsilon'} - {d^2\sigma^-\over d\Omega
\,d\epsilon'} \right\}
\bigg/
\left\{ {d^2\sigma^+\over d\Omega\,d\epsilon'} + {d^2\sigma^-\over d\Omega\,
d\epsilon'} \right\} \cr
&\equiv\alrzer \times
{{W^{\rm (PV)} (q,\omega,\theta)}\over {W^{\rm (EM)} (q,\omega,\theta)}}\ \ ,
\cr}\eqno\nexteq\nameeq\Ealrtwo
$$
where [Don88, Don89, Mus92a]
$$W^{\rm(PV)} (q,\omega,\theta)= v_L R^L_{AV} (q,\omega) + v_T R^T_{AV}
(q,\omega) + v_{T'} R^{T'}_{VA} (q,\omega) \eqno\nexteq\nameeq\Ewpvtwo
$$
and
$$
\alrzer\equiv {G_\mu\left| Q^2\right|\over 2\pi\alpha\sqrt{2}} \ \ .
\eqno\nexteq
$$
The parity--violating helicity--difference cross section in Eq.~(\Ealrtwo)
involves a product of the electromagnetic amplitude of Eq.~(\Emem) and the
parity--violating NC amplitude of Eq.~(\Empv).  The latter arises from product
terms of leptonic vector currents $\times$ hadronic
axial--vector currents and {\it vice versa}.  The subscripts on the response
functions in Eq.~(\Ewpvtwo) reflect this fact and
identify which currents are involved: the subscript
$AV$ denotes axial--vector leptonic and vector hadronic currents, whereas
$V\!A$ indicates the vector leptonic and axial--vector hadronic currents.
Note the structure of the PV response: since one has a product of
electromagnetic (purely vector) current hadronic matrix elements and weak
neutral current hadronic matrix elements (with both vector and axial--vector
contributions), the only terms in the general electroweak response
(Eq.~(\Eresp)) which can occur are those labeled ``$L$'', ``$T$'' and
``$T'\ $''.  In particular, since no hadronic axial--vector $\otimes$
hadronic axial--vector
contributions are retained (they occur only in the square of the
$Z^0$--exchange amplitude in Fig.~3.1b, which is ignored here), the CVC
result in Eq.~(\Ecvc) can be invoked.  An additional consequence of this
is that the only place where hadronic
axial--vector currents do occur is in the
``$T'\ $'' term which involves only transverse projections. In Sect.~III.E.3
we will draw comparisons of the different structure that occurs in
discussing neutrino scattering.

For scattering to discrete states, the asymmetry is defined similarly:
$$
\eqalign{\alr &=
\left\{ {d\sigma^+\over d\Omega} - {d\sigma^-\over d\Omega
} \right\}
\bigg/
\left\{ {d\sigma^+\over d\Omega} + {d\sigma^-\over d\Omega
} \right\} \cr
&\equiv \alrzer \times
{{W^{\rm (PV)} (q,\theta)}\over {F^2 (q,\theta)}}\ \ , \cr}
\eqno\nexteq\nameeq\Edisca
$$
where
$$
W^{\rm(PV)} (q,\theta)= v_L W^L_{AV} (q) + v_T W^T_{AV} (q) +
v_{T'} W^{T'}_{VA} (q)\ .\eqno\nexteq
$$
Here, in a form similar to Eqs.~(\Efltmul), we have [Don89]
$$
\eqalign{W^L_{AV} (q) &= -g_A^e\sum_{J\geq0} F_{CJ}(q)\tilde F_{CJ}(q) \cr
W^T_{AV} (q) &= -g_A^e\sum_{J\geq1} \lbrack F_{EJ}(q) \tilde F_{EJ}(q) +
F_{MJ}(q) \tilde F_{MJ}(q) \rbrack \cr
W^{T^\prime}_{VA} (q) &= -g_V^e
\sum_{J\geq1} \lbrack F_{EJ}(q) \tilde F_{MJ_5}(q) +
F_{MJ}(q) \tilde F_{EJ_5}(q) \rbrack \ \ . \cr}\eqno\nexteq\nameeq\Ewltmul
$$
Again, the form factors which occur here have already been introduced in
Sect.~III.D.1: the tilde indicates which quantities arise from matrix
elements of the weak neutral current, while the ``5'' indicates which are
axial--vector form factors. The factors $g_A^e$ and $g_V^e$ in Eqs.~(\Ewltmul)
come from the coupling of the leptonic current to
the exchanged $Z^0$ [see Eqs.~(\Emnc) and (\Empv)].
Recall also that, as argued above,
only transverse projections of the axial--vector current enter in PV electron
scattering (to order $G_{\mu}$).

Let us rewrite these results in terms of the quantity
$$
{\cal E} \equiv \left[ 1 + 2|q^2/Q^2|\tan^2 {\theta\over 2}
\right]^{-1} \eqno\nexteq\nameeq\Elongpl
$$
which characterizes the angle--dependences in the cross section and
asymmetry: ${\cal E}\rightarrow 1$ for $\theta\rightarrow 0^\circ$ and
${\cal E}\rightarrow 0$ for $\theta\rightarrow 180^{\circ}$.
Then $v_L/v_T=2|Q^2/q^2|{\cal E}$ and $v_{T'}/v_T =
\sqrt{1-{\cal E}^2}$. The (PC) electron scattering cross section
in Eq.~(\Edisc) may then be written
$$
{{d\sigma}\over{d\Omega}}={{4\pi\alpha^2}\over{|Q^2|}}
\Bigl({{\epsilon^{\prime}}\over{\epsilon}}\Bigr)f^{-1}_{\rm rec}
{1\over{1-{\cal E}}}\Bigl[2|Q^2/q^2|{\cal E}F^2_L(q)+F^2_T(q)\Bigr]
\eqno\nexteq\nameeq\Edsinel
$$
and the PV asymmetry in Eq.~(\Edisca) may be re--expressed as
$$
\alr=\alrzer \times
\Bigl[{{2|Q^2/q^2|{\cal E}W^L_{AV}(q)+W^T_{AV}(q)+\sqrt{1-{\cal E}^2}
W^{T^{\prime}}_{VA}(q)}\over{2|Q^2/q^2|{\cal E}F^2_L(q)+F^2_T(q)}}\Bigr]\ \ .
\eqno\nexteq\nameeq\Especa
$$

\bigskip
\goodbreak
\noindent \undertext{Elastic scattering from $(J^\pi T) = (0^+0)$ nuclei}

The simplest example of the formalism introduced above is that of
elastic scattering from a spin--0, isospin--0 target, since then only
monopole form factors can occur.  From Eqs.~(\Efltmul) and (\Ewltmul)
one has [Don89]
$$
\eqalign{F^2_L(q) &= F^2_{C0} \cr
W^L_{AV} (q) &= -g_A^e F_{C0}(q) {\tilde F}_{C0}(q) \cr
F^2_T (q) &= W^T_{AV}(q) = W^{T'}_{VA}(q) = 0 \ \ , \cr}\eqno\nexteq
$$
and hence the hadronic ratio in Eq.~(\Edisca) simply involves the ratio
of the two monopole form factors, $F_{C0}$ (electromagnetic) and
${\tilde F}_{C0}$ (weak neutral current):
$$
{\alr \over \alrzer} = -g_A^e \times
{{{\tilde F}_{C0}(q)}\over{F_{C0}(q)}} \ \ , \eqno\nexteq\nameeq\Espinz
$$
which depends on $q$, but is independent of $\theta$.
Furthermore, in the limit that the ground state is an exact $T=0$ state,
only the isoscalar pieces of these currents can contribute,
since the isovector piece cannot connect $T=0$ initial and final states.
In the absence of strange--quark contributions, as can be seen from
Eqs.~(\Exdvec) and (\Encurff), the only surviving nuclear
matrix elements are proportional to each other
with the coupling $\sqrt{3}\xi_V^{T=0}/2$ as the constant of
proportionality.  In this case the nuclear matrix elements then {\it
cancel} in the ratio in Eq.~(\Espinz), leaving a $q$--independent
number; accordingly, the predicted asymmetry is {\it
independent of the details of nuclear structure} [Fei75, Wal77].

     The presence of strangeness and the consideration of the higher--order
electroweak processes will introduce $q$--dependent, structure--dependent
terms and yield the full form of the hadronic ratio [Mus92a]:
$$
{\alr \over \alrzer} = - {1 \over 2} \left\{ \sqrt{3}\xi_V^{T=0}
 \left[ 1 +\Gamma(q)\right] +
\xi_V^{(0)} {{F_{C0}(s)}\over{F_{C0}(T=0)}}
\right\}\ \ .\eqno\nexteq\nameeq\Ehefour
$$
Here $\Gamma(q)$ is introduced to account for the effects of the breaking of
isospin symmetry (see Sect.~III.D.4). The final term in Eq.~(\Ehefour)
isolates the strange--quark contribution to the (monopole) charge density.
The leading--order approximation using one--body operators (see
Sect.~III.D.2) and working to the appropriate order in $Q^2$ (as discussed
in more detail in Sect.~IV.B) yields for the ratio of form factors in
Eq.~(\Ehefour) simply the ratio of the corresponding single--nucleon
form factors:
$$
{{F_{C0}(s)}\over{F_{C0}(T=0)}}={G_E^{(s)}\over G_E^{T=0}} \ \ .
\eqno\nexteq\nameeq\Estrzer
$$
The role of PV elastic electron scattering from a spin--0, isospin--0 nuclear
target in a program of hadronic neutral current studies is discussed in
detail in Sect.~IV.B.

\bigskip
\goodbreak
\noindent \undertext{Elastic scattering from the proton}

For elastic electron scattering from a spin--1/2 target the problem is
somewhat more complicated than that described above.  Using the results
of Sect.~III.D.1 for elastic scattering, we know in this case that
the form factors to be considered are $F_{C0}$, $F_{M1}$,
${\tilde F}_{C0}$, ${\tilde F}_{M1}$, ${\tilde F}_{E1_5}$
and ${\tilde F}_{L1_5}$ and from the general discussions above we need not
consider the last, leaving five basic form factors that enter into the
cross sections and asymmetry. From Eqs.~(\Efltmul) and (\Ewltmul) we
have
$$
\eqalign{F^2_L(q) &= F^2_{C0} \cr
F^2_T(q) &= F^2_{M1} \cr
W^L_{AV} (q) &= -g_A^e F_{C0}(q) {\tilde F}_{C0}(q) \cr
W^T_{AV} (q) &= -g_A^e F_{M1}(q) {\tilde F}_{M1}(q) \cr
W^{T'}_{VA} (q) &= -g_V^e F_{M1}(q) {\tilde F}_{E1_5}(q)
\ \ . \cr}\eqno\nexteq
$$
When dealing with elastic scattering from
the nucleon as a special case of a spin--1/2 target it is conventional
to use the fully relativistic analogs of the formalism summarized
above and so to express
$F^2$ directly in terms of the electric and magnetic Sachs form factors:
$$
F_p^2 (\tau,{\cal E})\equiv 4\pi F^2 (\tau,{\cal E}) = {1\over(1+\tau){\cal E}}
\Bigl({\cal E} [\GEp(\tau)]^2 +\tau[\GMp(\tau)]^2\Bigr)\ \ ,
\eqno\nexteq\nameeq\Eproff
$$
where, instead of writing the total form factor as a function of $q$ and
$\theta$ as in Eq.~(\Effdis), we write it equivalently as a function
of $\tau=|Q^2|/4\mns$ and $\cal E$ (see Eq.~(\Elongpl)).
In a similar way, the PV response takes the relatively simple form [Don88,
Mus92a]
$$
\eqalign{4 \pi W^{\rm (PV)}(\tau,{\cal E})  = -
{1\over 2(1+\tau){\cal E}}\bigg(&g_A^e \Bigl[{\cal E}\GEp(\tau)\geptil(\tau)
+\tau\GMp(\tau)\gmptil(\tau) \Bigr] \cr
&+g_V^e\sqrt{1-{\cal E}^2}\sqrt{\tau(1+\tau)}\GMp(\tau)
{\tilde\GAp}(\tau)\bigg)\ .\cr}\eqno\nexteq\nameeq\Epropv
$$
The PV asymmetry is then proportional to the ratio of responses in the
last two equations.
Inserting the Standard Model results for the leptonic couplings (see
Table~3.1), using Eqs.~(\EGtilde) for the proton ($\tau_3= 1$)
and inserting these results into Eqs.~(\Eproff) and (\Epropv) allows us to
write the PV asymmetry as
$$
\eqalign{\alr ({\vec e}p) &= -{1\over 2}\alrzer \cr
&\times \Bigg\lbrace  \xi_V^p\
+ \Big[{\cal E}\GEp\{\xi_V^n \GEn +\xi_V^{(0)}\GES\}
+\tau\GMp\{\xi_V^n \GMn+\xi_V^{(0)} \GMS\}\cr
&\qquad -(1-4\sstw)\sqrt{1-{\cal E}^2}
        \sqrt{\tau(1+\tau)}\GMp{\tilde\GAp}\Big]
/[{\cal E}(\GEp)^2+\tau(\GMp)^2]\Bigg\rbrace\ \ ,\cr}
\eqno\nexteq\nameeq\alrprotpn
$$
where we have suppressed the $\tau$--dependences for clarity and used
the weak NC couplings labeled ``p'' or ``n'' defined in Eqs.~(\Ewncpn).
In obtaining this result, we have assumed the nucleon to be
a state of pure isospin in order to rewrite the isoscalar and isovector
neutral current form factors in terms of
the proton and neutron form factors
[Eqs.~(\EGtilde)].  We discuss several potentially interesting
experiments for PV electron scattering from the proton
and explore the various kinematic dependences and limits
of Eq.~(\alrprotpn) in detail in
Sect.~IV.A.

\bigskip
\goodbreak
\noindent \undertext{Elastic scattering from the deuteron}

Next let us consider elastic scattering from spin-1 nuclei such as the
deuteron.  Using the results of Sect.~III.D.1, we have as form factors
in this case the following: $F_{C0}$, ${\tilde F}_{C0}$, $F_{C2}$,
${\tilde F}_{C2}$, $F_{M1}$, ${\tilde F}_{M1}$ and ${\tilde F}_{E1_5}$
for a total of seven form factors (again, as above, ${\tilde F}_{L1_5}$
does not enter in PV electron scattering).  Equations~(\Efltmul) and
(\Ewltmul) yield
$$
\eqalign{F^2_L(q) &= F^2_{C0}(q) + F^2_{C2}(q) \cr
F^2_T(q) &= F^2_{M1}(q) \cr
W^L_{AV}(q) &= -g_A^e \Bigl[F_{C0}(q){\tilde F}_{C0}
+F_{C2}(q){\tilde F}_{C2}\Bigr]\cr
W^T_{AV}(q) &= -g_A^e F_{M1}(q){\tilde F}_{M1}(q) \cr
W^{T'}_{VA}(q) &= -g_V^e F_{M1}(q) {\tilde F}_{E1_5}(q)\ \ . \cr}
\eqno\nexteq
$$
Using the Standard Model values for the leptonic couplings (Table~3.1)
we then obtain for the PV asymmetry
$$
\alr=-\alrzer\times \Bigl\{ {{ v_L \bigl[ F_{C0}{\tilde F}_{C0} +
F_{C2}{\tilde F}_{C2} \bigr] + F_{M1} \bigl[ v_T {\tilde F}_{M1} +
(1-4\sstw)v_{T'} {\tilde F}_{E1_5} \bigr] } \over {
v_L \bigl[ F_{C0}^2 + F_{C2}^2 \bigr] + v_T F_{M1}^2 }}
\Bigr\} \ \ , \eqno\nexteq\nameeq\Easydeut
$$
where again we have suppressed the $q$--dependences for clarity. At the
tree--level in the absence of strangeness and assuming good isospin
symmetry we again have that the weak NC vector form factors are all
proportional to their EM counterparts and that the isoscalar
axial--vector form factor is zero, which yields the simple result
$$
-2{\alr \over \alrzer} \rightarrow \Delta_{(1)}\equiv \sqrt{3}\xi_V^{T=0}
\ \ . \eqno\nexteq\nameeq\Easylow
$$
In Sect.~IV.C we return to treat the case of elastic scattering from the
deuteron in more detail, including there the complete expressions which
incorporate radiative corrections and isospin--mixing together with
discussions of how studies of this case this might serve
to help in shedding light on the strangeness content of the nucleon.

Following the review presented here of the basic formalism for
elastic scattering from spin--0, spin--1/2 and spin--1 systems the next
logical step is perhaps to consider inelastic excitations of discrete
nuclear states.  Since in general this subject is considerably more
complicated than the formalism presented above (often having many more
multipole
form factors entering in the PC and PV cross sections), we choose not to
continue along this path, but to postpone discussion of other discrete
nuclear transitions until Sect.~IV.D.  Instead, we now consider excitations
into the nuclear continuum, {\it viz.,\/} quasielastic electron scattering.

\bigskip
\goodbreak
\noindent \undertext{Quasielastic scattering}

     While quasielastic scattering from a nuclear target is
complicated by the details of the nuclear structure involved, the
role of final--state interactions, {\it etc.,\/} it is instructive to take as a
starting point the ``static'' approximation in which one invokes the
plane--wave impulse approximation [deF83].   In this case the QE cross section
is given as an incoherent sum over the nucleons in the nucleus and involves
integrals of the following general form
$$
d\sigma ({\rm QE}) =\sum_i{\int dE_i\, dk_i\, S_i (E_i, {\svec k}_i)
d\sigma_i (E_i, {\svec k}_i)\delta(\omega-E_{\rm final}+E_{\rm initial})}
\ \ ,\eqno\nexteq\nameeq\Eqecr
$$
where $S_i (E_i,k_i)$ is the spectral function, {\it i.e.,\/} the probability
of finding the i$^{\rm th}$ nucleon (proton or neutron) moving with momentum
${\svec k}_i$ and energy $E_i$. The quantity $d\sigma_i (E_i, {\svec k}_i)$
is the half--off--shell electroweak
cross section for the i$^{\rm th}$ nucleon and is also a function of q,
$\omega$ and the electron helicity. The $\delta$--function
enforces overall energy conservation and contains the difference in nuclear
energies, $E_{\rm final}-E_{\rm initial}$, which in turn depends on q, $E_i$
and ${\svec k}_i$.  Next, we assume that for the kinematics of interest in QE
scattering, the $\sigma_i$ are each strongly peaked
about some common ${\svec k}_i \equiv {\svec p}$ and $E = \sqrt{p^2 +
m_N^2}$, corresponding to having the struck nucleon on--shell and moving with
momentum $\svec p$. In this case, one has
$$
\eqalign{
\int dE_i\, d{\svec k}_i\,& S_i (E_i, {\svec k}_i) d\sigma_i (E_i, {\svec k}_i)
\delta(\omega-E_{\rm final}+E_{\rm initial}) \cr
&\approx d\sigma_i (E,{\svec p}) \int dE_i\, d{\svec k}_i\, S_i (E_i
,{\svec k}_i )\delta(\omega-E_{\rm final}+E_{\rm initial})
\ \  . \cr}\eqno\nexteq\nameeq\Eqeint
$$
In forming the parity--violating asymmetry we require the ratio of
helicity--difference and helicity--sum cross sections and so the integral in
Eq.~(\Eqeint) will cancel, leaving only the single--nucleon cross sections,
$d\sigma_i (E,{\svec p})$ in the ratio.  Finally, to obtain the crudest
approximation (and so get some feeling for the form of the asymmetry)
we take ${\svec p}=0$ and thus evaluate the single--nucleon cross sections in
the rest frame of the struck nucleon.

In the static approximation we then find that $W^{\rm (EM)}$ is simply
proportional to the sum of the response of $Z$ (on--shell) protons and $N$
(on--shell) neutrons [Don92, Mus92a]:
$$
W^{\rm (EM)} \propto
{\cal E}\left[ Z (G^p_E)^2 + N (G^n_E)^2\right] + \tau \left[
Z(G^p_M)^2 + N (G^n_M)^2\right]\ \ ,\eqno\nexteq\nameeq\Eqeem
$$
while the parity--violating response is
$$
\eqalign{-2W^{\rm (PV)} &\propto
{\cal E}\left[ Z G^p_E \tilde G^p_E + N G^n_E \tilde G^n_E\right]
+ \tau \left[ Z G^p_M \tilde G^p_M + N G^n_M \tilde G^n_M\right]\cr
&- (1-4\sstw) \sqrt{1 - {\cal E}^2} \sqrt{
\tau(1+\tau)}\, \left[ Z G^p_M \tilde G^p_A + NG^n_M \tilde
G^n_A\right]\ \ , \cr }\eqno\nexteq\nameeq\Eqepv
$$
where we have suppressed the $\tau$--dependences in the single--nucleon
form factors and have used the Standard Model leptonic couplings (Table~3.1).
The static model PV QE asymmetry is then given by
$$
\eqalign{\alr(QE)_{\rm static}  &= -{1\over 2}\alrzer\times
\biggl\{ {\cal E} \left[ Z G^p_E \tilde G^p_E + N G^n_E \tilde G^n_E\right]
+ \tau \left[ Z G^p_M \tilde G^p_M + N G^n_M \tilde G^n_M\right]\cr
&- (1-\sstw)\sqrt{1 - {\cal E}^2} \sqrt{
\tau(1+\tau)}\, \left[ Z G^p_M \tilde G^p_A + NG^n_M \tilde
G^n_A\right]\biggr\} \cr
&\times \left\{ {\cal E}\left[ Z (G^p_E)^2 + N (G^n_E)^2\right] + \tau \left[
Z(G^p_M)^2 + N (G^n_M)^2\right]\right\}^{-1}\ \ . \cr}
\eqno\nexteq\nameeq\Eqeasy
$$
This result should be compared with the expression for the asymmetry for
elastic scattering from the proton, Eq.~(\alrprotpn), since the static
model naturally reverts to the proton elastic asymmetry when $Z=1$ and
$N=0$ (or to the elastic neutron asymmetry when $N=1$ and $Z=0$).
The consequences of extending the treatment of PV quasielastic scattering
beyond
this simple static approximation are discussed in more detail in
Sect.~IV.F.

Similar considerations are applicable for excitation energies beyond the
QE peak.  In particular, the region dominated by quasi--free excitation
of the $\Delta$(1232) has received some attention; our discussion of this
problem is postponed until Sect.~IV.G.

\bigskip
\goodbreak
\noindent III.E.2.\quad FIGURE--OF--MERIT FOR PV ELECTRON SCATTERING

     While $A_{LR}$ depends directly on the matrix elements of the
hadronic neutral current and presents a clear signature of the presence of
these currents, it is a small quantity and is challenging to measure
experimentally.
The precision with which it may be determined depends on details of the
experimental configuration (luminosity,
detector resolution, beam polarization, {\it
etc.\/}) as well as on the kinematic conditions under which a measurement is
carried out.  The latter also influence the degree to which various quantities
of interest, such as $\xi_\sst{V}^p$ or hadronic form factors, enter
theoretical predictions of $\alr$.  Consequently, a non--trivial correlation
exists between kinematical constraints imposed by considerations of
experimental (statistical precision)
issues and by considerations of interpretability.
In many cases, these different sets of considerations conspire to
constrain the optimal kinematical regime for performing a measurement.  It is
important, then, to understand the kinematical conditions imposed by the
requirement that the statistical uncertainty in $A^{\rm exp}_{LR}$ be
sufficiently small to make a given measurement theoretically meaningful.

        To this end, we review the standard
figure--of--merit (FOM) for PV electron scattering
which, when maximized, corresponds to a minimal ($\delta \alr/ \alr
)_{\rm statistical}$. To derive the FOM, we write the asymmetry as
$$
\alr \equiv \Delta /N_0\ \ ,\eqno\nexteq
$$
where $\Delta = N_+ - N_-$, $N_0=N_++N_-$, and $N_{+(-)}$ is the number of
events with electrons polarized parallel (anti--parallel) to their incident
momenta.  The statistical error in $\alr$ is, then
$$
\delta \alr = \left| \left( \delta A^+_{LR} \right)^2 + \left( \delta
A^-_{LR}\right)^2 \right|^{1/2}\ , \eqno\nexteq
$$
where
$$
\delta A^{+(-)}_{LR} \equiv {\partial \alr \over \partial N_{+(-)}} \delta
N_{+(-)}\ \ . \eqno\nexteq
$$
Using $\delta N_{+(-)} \equiv \sqrt{N_{+(-)}}$ it is straightforward to
obtain
$$
\delta \alr = {1\over \sqrt{N_0}} \left[ 1 - A^2_{LR}\right]^{1/2}
\ \ . \eqno\nexteq
$$
For the range of electron energy and momentum transfer
of interest here, $|\alr| <\!\!<1$ so that to an
excellent approximation $\delta \alr\cong 1/ \sqrt{N_0}$.
The relative error in $\alr$, then, is\footnote{*}{Henceforth, we take
$\delta\alr/\alr$ to be non--negative, that is, to denote the absolute value
of the fractional error.}
$$
{\delta \alr\over \alr} = \left[ N_0A^2_{LR}\right]^{-1/2}\
\ .\eqno\nexteq
$$
Writing $N_0 = \left( d\sigma/d\Omega\right)\ \Delta\Omega\ {\cal L}\ T_0$,
where $d\sigma/d\Omega$ is the
helicity--independent differential cross
section,
$\Delta\Omega$ is the detector solid angle, ${\cal
L}$ is the luminosity, and
$T_0$ is the running time, we have
$$
\eqalign{{\delta \alr\over \alr} &= \left[ {\cal F}
X_0\right]^{-1/2} \cr
{\cal F} &= \left( {d\sigma\over d\Omega}\right) A^2_{LR} \cr
X_0 &= {\cal L}\ \Delta\Omega \ T_0\ \ . \cr} \eqno\nexteq\nameeq\Edoabl
$$
The quantity ${\cal F}$ is the figure--of--merit.  It depends only on
intrinsic properties of the target and lepton probe and on the relevant
kinematic variables ($\{\epsilon$, $\epsilon'$ and $\theta\}$ or equivalently
$\{q$, $\omega$ and $\theta\}$).  In particular, the results
given above for discrete states in Eqs.~(\Edsinel) and (\Especa) allow one
to write (recall: $\omega$ is fixed by the excitation energy and $q$
in this case)
$$
{\cal F}(q,\theta) ={{\cal F}_0(q,\theta)\over{1-{\cal E}}}
{{\Bigl[2|Q^2/q^2|{\cal E}W^L_{AV}(q)+W^T_{AV}(q)+\sqrt{1-{\cal E}^2}
W^{T^{\prime}}_{VA}(q)\Bigr]^2}
\over{\Bigl[ 2|Q^2/q^2|{\cal E}F^2_L(q)+F^2_T(q)\Bigr]}}\
\ ,\eqno\nexteq\nameeq\Efinel
$$
where for convenience the overall scale
$$
{\cal F}_0(q,\theta)\equiv {{G^2}\over{2\pi}}|Q^2|\Bigl({{\epsilon^{\prime}}
\over{\epsilon}}\Bigr)f^{-1}_{\rm rec}\ \ \eqno\nexteq\nameeq\Efscale
$$
has been introduced (see the discussions to follow in Sect.~IV).
Extrinsic experimental conditions
are contained in the quantity $X_0$.  For a given value of the latter,
$\delta \alr/\alr$ is a decreasing function of ${\cal F}$.  The values of
${\cal L}$ and $\Delta \Omega$ attainable also depend on properties of the
target and kinematical conditions, as discussed in more detail in Sect.~V.
In analyzing the do--ability/interpretability correlation in Sect.~IV,
we take reasonable values of $X_0$
and work primarily with ${\cal F}$.

Before proceeding to specific examples we make several observations
concerning the quantities given above for the typical
conditions that apply when studying discrete nuclear states. First,
since $F_{L,T}^2$, $W_{AV}^{L,T}$ and
$W_{VA}^{T^{\prime}}$ all contain the square of a characteristic
nuclear form factor ($\equiv F_{\rm nuc}$), the
figure--of--merit $\cal F$, given in Eq.~(\Efinel),
is proportional to $F_{\rm nuc}^2$. To a fair approximation, the $q$-dependence
of this quantity can be parameterized as
$F_{\rm nuc}^2\approx \exp [-(q/q_0)^2]$, where
$q_0\approx$ 250 MeV/c $\times A^{-1/6}$ gives the value of $q$ for which
$F_{\rm nuc}^2$ falls off by a factor of $e$ from its $q=0$ value.
For instance, taking $A=12$
yields $q_0 =$ 165 MeV/c as the characteristic 1/e scale at which $\cal F$
falls off with $q$. Of course, there is usually other
non--trivial momentum transfer dependence beyond this (see below);
however, this parameterization
sets the rough scale for the $q$-dependence, and one should expect
that for momentum transfers significantly larger than $q_0$
the figure--of--merit will not be sufficient to render
measurements practical.  Secondly, when
one applies these general expressions to situations involving
discrete transitions the energy transfer $\omega$ is fixed once the
three--momentum transfer $q$ and excitation energy $E_x$ are specified:
$\omega = \sqrt{q^2+(M_T+E_x)^2}-M_T \cong E_x+q^2/2M_T$.
Since the three--momentum transfer must usually be kept rather low (say below
$\approx$ 200--300 MeV/c except for the lightest nuclei, using the above
estimates), the recoil energy $q^2/2M_T$ is only a few MeV or less.
The excitation energies of discrete
nuclear states characteristically fall in the range [0---10] MeV and hence
this range is also appropriate for $\omega$.  As a consequence typically
$\omega<\!\!< q$ and the combination $|Q^2/q^2|=1-(\omega/q)^2$ in the above
equations may be taken to be approximately unity except when the
three--momentum transfer is very small.  Furthermore,
$\epsilon^{\prime}/\epsilon$ and $f_{\rm rec}$ satisfy the following
inequalities
$$
\eqalign{{{1-\omega/q}\over{1+\omega/q}}\leq {\epsilon^{\prime}\over
\epsilon}&\leq 1 \cr
1 \leq f_{\rm rec}&\leq 1 + {{q+\omega}\over M_T}\cr}
\eqno\nexteq\nameeq\Efscalex
$$
and thus it can be argued that under typical conditions they
are both also nearly unity. These arguments imply that in studying
discrete--state transitions via PV electron scattering under typical
conditions the scale factor ${\cal F}_0$ in Eq.~(\Efscale)
is nearly constant for
fixed momentum transfer but varying scattering angle.  The angle--dependence in
the figure--of--merit is then isolated essentially in the factors
$\cal E$ in Eq.~(\Efinel).  Importantly this expression contains the
overall factor $(1-{\cal E})^{-1}$ which varies quite rapidly for small
scattering angles, where ${\cal E}$ is only slightly less than unity, and leads
us to expect that the figure--of--merit will usually be largest for small
$\theta$.

In Figs.~3.10--3.12 the figure--of--merit calculated using Eq.~(\Efinel)
is shown
as a function of incident electron energy at fixed scattering angle
for several selected nuclear transitions.  In particular, results are given
for elastic scattering from $^{12}$C and $^4$He, for inelastic scattering
to the $J^{\pi}T=2^+0$ (4.44 MeV), $1^+0$ (12.71 MeV) and $1^+1$ (15.11 MeV)
levels in $^{12}$C and, to place the results in context, for elastic
scattering from the proton.  In each
case the electromagnetic form factors have been determined directly from
experimental cross sections [Car80, Fla78, Fla79, Fro67, Jan72, McC77,
Reu82, Sic70]
(simple parameterizations have been
used and the parameters so introduced adjusted to produce fits to the data).
We have set all strangeness form factors to zero in producing these results,
although, when specific cases involving potential nonzero strangeness
content are discussed below, the figures--of--merit are recomputed
incorporating these additional contributions before arriving at
numerical estimates of the fractional uncertainties in the quantities of
interest.

Three angles have been selected: a very forward angle (10$^{\circ}$) which
is characteristic of the limits that can be reached by the spectrometers
being built at CEBAF, a typical forward angle (30$^{\circ}$) and a typical
backward angle (150$^{\circ}$).  In the figures $\cal F$ is given for the
range in $\epsilon$ corresponding roughly to $0\leq q \leq 500$ MeV/c.
To interpret the results given here it is helpful to recall Eq.~(\Edoabl)
where the fractional statistical
precision obtainable for the asymmetry is given by $[{\cal F}X_0]^{-1/2}$,
with $X_0={\cal L}\ \Delta\Omega\ T_0$.  As in our previous discussions,
let we assume that the electron beam is 100\% polarized,
$\Delta\Omega=10$ msr, $T_0=1000$ hr $=3.6\times 10^6$ s and the luminosity
is given by ${\cal L}[^{12}\hbox{C}]=1.25\times10^{38}$ cm$^{-2}$ s$^{-1}$ or
${\cal L}[^{4}\hbox{He},\ ^{1}\hbox{H}]=
5\times10^{38}$ cm$^{-2}$ s$^{-1}$.  These assumptions yield:
$X_0[^{12}$C$]=4.5\times10^{42}$ cm$^{-2}$ and
$X_0[^4$He, $^1$H$]=1.8\times10^{43}$ cm$^{-2}$.  These values set
the scale for the fractional precision in $A_{LR}$, namely
$$
\eqalign{
{\cal F}= &10^{-39}\ {\rm cm}^2 {\rm sr}^{-1}\quad\longleftrightarrow
\quad|\delta A_{LR}/A_{LR}| \cong \ 1.5\%\ (^{12}{\rm C}),\ \ \ 0.75\%\
(^4{\rm He},\ ^1{\rm H})\cr
&10^{-41}\ {\rm cm}^2 {\rm sr}^{-1}\quad\longleftrightarrow
\qquad\qquad\qquad\qquad 15\%, \qquad\qquad\  7.5\% \cr
&10^{-43}\ {\rm cm}^2 {\rm sr}^{-1}\quad\longleftrightarrow
\qquad\qquad\qquad\qquad 150\%, \qquad\qquad 75\%
\ \ .\cr}\eqno\nexteq
$$
It is then straightforward to determine the fractional precision with
which specific transitions could be studied given the above experimental
conditions.  For elastic scattering from $^{12}$C at the peak of the
figure--of--merit in Figs.~3.10--3.12 we find $|\delta A_{LR}/A_{LR}|=$
0.7\% (10$^{\circ}$), 2.2\% (30$^{\circ}$) and 31\% (150$^{\circ}$).
Essentially the same values are obtained for elastic scattering from
$^4$He, since the smaller $Z$ is just compensated
for by the slower fall--off
of the nuclear form factor when comparing $^4$He with $^{12}$C.  For the
excitation of the $2^+0$ state in $^{12}$C, under conditions where its
figure--of--merit peaks, we find $|\delta A_{LR}/A_{LR}|=$
2\% (10$^{\circ}$), 6\% (30$^{\circ}$) and 86\% (150$^{\circ}$).  Likewise
for the excitation of the $1^+$ states in $^{12}$C where their respective
figures--of--merit peak we have the following: for the $1^+1$ state we find
$|\delta A_{LR}/A_{LR}|=$
26\% (10$^{\circ}$), 67\% (30$^{\circ}$) and 163\% (150$^{\circ}$), whereas
for the $1^+0$ state we find $|\delta A_{LR}/A_{LR}|=$
142\% (10$^{\circ}$), 413\% (30$^{\circ}$) and 1197\% (150$^{\circ}$).  Of
course, if a larger solid angle detector having sufficient resolution to
permit the separation of the transition of interest were to be built, then the
values of fractional precision in the PV asymmetry given here would be lower.
For example, given a detector with $\Delta\Omega=$ 0.16 sr,
the above numbers would all be decreased by a
factor of four.  However, even were such a detector to be realized, some of
the results presented above must still be regarded as rather uninteresting
because of the large statistical uncertainties in the asymmetry that would be
incurred. In context the results at forward angles and high energies (for the
range of momentum transfers
displayed in the figures) illustrate an observation that will be
elaborated in Sect.~IV.A: the proton asymmetry and figure--of--merit
are atypically small for such kinematics.  As noted in those discussions,
this situation
is due to the smallness of all of the terms which contribute.  The
asymmetry in Eq.~(\alrprotpn) involves $1-4\sstw$, $G_E^n$,
$G_E^{(s)}$, $\tau$ or
$0.092\sqrt{\tau(1+\tau)(1-{\cal E}^2)}$, all of which are suppressed for such
kinematics.  In contrast, for example, elastic scattering from $J=T=0$
nuclei instead involves $-4\sstw$ and therefore from this factor
alone one expects the figure--of--merit in this case to exceed
that of the proton
by about two orders of magnitude. In addition, the many--body nuclear form
factor diminishes the figure of merit for a nuclear target as $q$ increases.
In the case of elastic scattering, however, the coherence factor $Z^2$
offsets this suppression for momentum transfers below the characteristic
value $q_0$ introduced above.

\bigskip
\goodbreak
\noindent III.E.3.\quad BASIC NEUTRINO SCATTERING FORMALISM

     Development of the expressions for neutrino and
antineutrino NC scattering, $(\nu_l,\nu'_l)$ and $({\bar\nu}_l,{\bar\nu'}_l)$,
respectively, and for charge--changing neutrino and antineutrino reactions,
$(\nu_l,\ell^-)$ and $({\bar\nu}_l,\ell^+)$, respectively, where $\ell$
labels the flavor of lepton, closely follows the derivation of the electron
scattering cross sections above.  In this case, however, the
photon--exchange amplitude cannot contribute to leading order in electroweak
couplings (cf. Fig.~3.1); only the diagrams in
Fig.~3.13 involving $Z^0$ exchange (neutral current) or $W^{\pm}$ exchange
(charge--changing current) between leptons and hadrons enter.
The leading--order contributions to the cross section are then
${\cal O}(G_{\mu}^2)$ and involve the
weak interaction current matrix elements bilinearly. Consequently,
the contributions to the
general response functions in Eq.~(\Eresp) are now somewhat more involved than
in the case of PV response for electron scattering, which contain only a
linear dependence on weak NC matrix elements. Moreover, the neutrino scattering
$CC$, $CL$, $LL$ and $T$ responses all involve both products of nuclear
vector currents and products of nuclear axial--vector currents
whereas the $T'$
response, as before, contains the interferences between the nuclear vector
currents and the nuclear axial--vector currents.
The development of the various cross sections for
charge--changing and neutral current weak interaction processes involving
neutrinos and antineutrinos has been discussed at length in previous
review articles (see, {\it e.g.,\/} Ref.~[Don79a] where many of the
conventions used are similar to those employed in the present work).  Since
the main focus of this work is PV electron scattering, we shall
not discuss these other processes in as much detail.  Rather our intent
in this section is only to bring out the strong parallelism that exists in
approaching
the wider class of semi--leptonic electroweak interaction processes
and to provide a basis for the discussions in Sect.~IV.J where
in context some of the main implications of neutrino scattering are
summarized.

For the present discussions let us again assume the ERL, in which case
Eqs.~(\Eerll) apply in general and Eq.~(\Ecvc) may be used for the VV
responses. The analog of Eq.~(\Edtsige) for neutrino and antineutrino
scattering with exchange of a $Z^0$ may be written [Don79a]
$$
{d^2\sigma\over d\Omega\,d\epsilon'}
= \sigma_0
\left\{
 S^L (q,\omega)_0 + v_T R^T (q,\omega)_0 \pm
v_{T'} R^{T'} (q,\omega)_0
\right\}\ \ , \eqno\nexteq\nameeq{\Enuddsig}
$$
where $\sigma_0$ is the elementary cross section (the neutrino scattering
analog of the Mott cross section in Eq.~(\Esigmm))
$$
\sigma_0 = \left[ {G_\mu\epsilon^\prime\cos{\theta\over 2} \over \pi\sqrt{2}
}\right]^2\ \ \eqno\nexteq
$$
and the $+/-$ sign on the third term corresponds to neutrino/antineutrino
scattering (see also Ref.~[Alb93b]).  The leptons are labeled as in
Sect.~III.E.1.  The responses
in Eq.~(\Enuddsig) are distinguished from their analogous electron
scattering counterparts by the subscript ``0'' (denoting neutral current
weak interaction processes, see below).  From our general discussions at the
beginning of this section we can write
$$
\eqalign{ S^L (q,\omega)_0 &= v_L \Bigl[R^L (q,\omega)_0\Bigr]_{VV} \cr
&\qquad + \Bigl[R^{CC} (q,\omega)_0 - \Bigl({\omega\over q}\Bigr)
R^{CL} (q,\omega)_0 + \Bigl({\omega\over q}\Bigr)^2
R^{LL} (q,\omega)_0 \Bigr]_{AA} \cr
R^T (q,\omega)_0 &= \Bigl[R^T (q,\omega)_0\Bigr]_{VV} +
\Bigl[R^T (q,\omega)_0\Bigr]_{AA} \cr
R^{T'} (q,\omega)_0 &= \Bigl[R^{T'} (q,\omega)_0 \Bigr]_{VA}\ \ . \cr}
\eqno\nexteq\nameeq\Eneutz
$$
It is straightforward to generalize these expressions to incorporate ERL
charge--changing neutrino reactions.  The structure is basically the
same with the replacements
$$
\eqalign{\sigma_0 &\longrightarrow \sigma_{\pm}=2\sigma_0 \cr
({\rm Response})_0 &\longrightarrow ({\rm Response})_{\pm}\ \ , \cr}
\eqno\nexteq\nameeq\Eccneut
$$
where the subscripts ``$\pm$'' correspond to $(\nu_l,\ell^-)$ and
$({\bar\nu}_l,\ell^+)$ reactions, respectively, and where the factor 2
is conventional [Wal75]. Furthermore, $\beta$--decay
and charged--lepton capture can be added to the set of processes that can be
inter--related, as can charge--changing neutrino reactions in circumstances
where the ERL cannot be invoked (see Ref.~[Don79a] for details).

     For elastic and inelastic scattering to discrete states one has, in
analogy with Eq.~(\Edisc):
$$
{d\sigma\over d\Omega} = 4\pi\sigma_0 f_{\rm rec}^{-1} {\tilde F}^2 (q,\theta)
\ \ , \eqno\nexteq\nameeq\Eneutcr
$$
where $f_{\rm rec}$ has been defined above and where,
in analogy with the electron scattering formalism discussed above,
we define a
neutrino scattering form factor
$$
{\tilde F}^2 (q,\theta) = {\tilde F}^2_L (q) + v_T {\tilde F}^2_T (q)
\pm v_{T'} {\tilde H}_{T'} (q)
\ \ . \eqno\nexteq\nameeq\Eneusign
$$
In this case the
$T'$ contributions have been denoted ${\tilde H}_{T'}$ rather than,
say, ${\tilde F}^2_{T'}$ since they can be positive or negative.
As in the latter case, it is possible to perform a generalized Rosenbluth
decomposition of the form factor into a
sum of terms using the form factors defined in Sect.~III.D:
$$
\eqalign{ {\tilde F}^2_L (q) &= v_L \sum_{J\geq0} {\tilde F}_{CJ}^2 (q)
+ \sum_{J\geq0} (\tilde F_{CJ_5} (q)
+ {\omega\over q}{\tilde F}_{LJ_5} (q))^2 \cr
{\tilde F}^2_T (q) &=  \sum_{J\geq1} \lbrack {\tilde F}^2_{EJ} (q) +
{\tilde F}^2_{MJ} (q) \rbrack + \sum_{J\geq1} \lbrack {\tilde F}^2_{EJ_5} (q) +
{\tilde F}^2_{MJ_5} (q)\rbrack \cr
{\tilde H}_{T'} (q) &= -2 \sum_{J\geq1} \lbrack {\tilde F}_{EJ} (q)
{\tilde F}_{MJ_5} (q) + {\tilde F}_{MJ} (q) {\tilde F}_{EJ_5} (q)
\rbrack \ \ , \cr}\eqno\nexteq\nameeq\Eneutff
$$
where, as before, the nuclear states are assumed to have good angular
momentum and parity quantum numbers.  In Eqs.~(\Eneutff) we have
substituted the results for the neutrino couplings from Table~3.1:
$g_V^{\nu}=-g_A^{\nu}=1$.  Note the contrast with PV electron scattering
(Eq.~(\Ewltmul)): in the case of PV electron scattering the axial--vector
contributions are suppressed by the ratio $|g_V^e/g_A^e|= 1 - 4\sstw \cong$
0.092, whereas for neutrino scattering there is no suppression and the
corresponding ratio is unity.  Analogous expressions may be written
for charge--changing neutrino reactions [Don79a] by using the
$F_{XJ}^{\pm}$ and $F_{XJ_5}^{\pm}$ form factors introduced in Eqs.~(\Exdax)
and multiplying by 2, as in Eq.~(\Eccneut).

We conclude this section by considering two special cases of neutrino
scattering: elastic
scattering from $(J^\pi T) = (0^+0)$ nuclei and elastic scattering from
the nucleon.  Treatment of all other cases is postponed until
Sect.~IV.J.
Concerning the type of measurement which might be attempted, a few
words are appropriate at this point.  Since in the scattering of neutrinos
or antineutrinos
the outgoing lepton cannot be detected, some alternative signature must
be found.  For inelastic scattering, in certain cases the de--excitation of
the final--state nuclear level by, say, emission of $\gamma$--rays can be used
to study the neutral current neutrino excitation (see Ref.~[Don79a]
and Sect.~IV.J.5).  However, for elastic
scattering all that happens is recoil of the target when it is struck by
the neutrino or antineutrino.  In specific circumstances (see Ref.~[Don83]
and the further discussion in Sects.~IV.J.3 and IV.J.4) detection of the
recoiling target may prove feasible for
elastic scattering from nuclei; the present situation for elastic scattering
from the proton is briefly reviewed in Sect.~IV.J.2.

For elastic scattering one has the following:
$$
\eqalign{ \tsubr &= 2 \msubt\tau = {|Q^2|\over 2 \msubt} = \omega \cr
        q &= 2 \msubt \sqrt{\tau(1+\tau)} \cr
         \epsilon &= \msubt \Bigl[\cos\thetr\sqrt{1+2 \msubt/\tsubr}
          -1\Bigr]^{-1}\cr
        &= \msubt\Bigl[\cos\thetr\sqrt{1+1/\tau}-1\Bigr]^{-1}\cr
        \sin^2 \theta/2 &= { {\cos^2 \thetr} \over {\cos^2 \thetr
          + (1+\epsilon/\msubt)^2 \sin^2 \thetr } } \ \ , \cr}
\eqno\nexteq\nameeq\Ekinconvx
$$
where $\epsilon$ is the energy of the incident neutrino (or antineutrino),
$\theta$ is the neutrino scattering angle and $\tsubr$ and $\thetr$ are the
angle and kinetic energy of the final--state recoil, respectively.  As usual,
$\msubt$ is the target mass. The
lepton scattering angle $\theta$ can take on any value from $0^\circ$ to
$180^\circ$ and accordingly the recoil angle goes between $90^\circ$
and $0^\circ$.

\bigskip
\goodbreak
\noindent \undertext{Elastic scattering from $(J^\pi T) = (0^+0)$ nuclei}

As in the discussions above of PV elastic electron scattering, the analysis
of neutrino scattering from
spin--0, isospin--0 nuclei is simplified considerably since only
isoscalar monopole form factors can occur.  One finds that
$$
\eqalign{{\tilde F}^2 (q,\theta) &= {\tilde F}^2_L (q)
= v_L {\tilde F}_{C0}^2 (q) \cr
{\tilde F}^2_T (q) &= {\tilde H}_{T'}(q) = 0 \ \ , \cr}\eqno\nexteq
$$
where the neutral current form factor can be written in terms of the
electromagnetic form factor (compare Eq.~(\Ehefour)):
$$
{\tilde F}_{C0} (q) = {1\over 2} \left\{ \sqrt{3}\xi_V^{T=0}
\left[ 1 +\Gamma(q)\right]F_{C0} (q) +
\xi_V^{(0)} {F_{C0}(s)}
\right\}  \ \ .\eqno\nexteq\nameeq\Ehefourneu
$$
The Rosenbluth factor $v_L$ is given by
$$
v_L=(1+\tau)^{-2}={ {[1 + 2{\epsilon / \msubt}
+({\epsilon / \msubt})^2 \sin^2 \thetr ]^2} \over { [ 1+\epsilon/\msubt ]^4} }
\ \ . \eqno\nexteq
$$
Equation~(\Eneutcr) may be used together with the form factor above
to obtain the cross section for elastic scattering from
$0^+0$ targets; for fixed $\epsilon$ this
may be expressed in the following form:
$$
{d\sigma(\epsilon)\over dQ^2} = 2G_\mu^2 \Bigl({\msubt\over \epsilon}\Bigr)^2
\Bigl[ (\epsilon/\msubt -\tau)^2 - \tau(1+\tau) \Bigr] {\tilde F}^2 (q) \ \ ,
\eqno\nexteq\nameeq\Eelassneu
$$
differential in the 4--momentum transfer, or
$$
{d\sigma(\epsilon)\over d\Omega_\sst{R}} = {8\over \pi}G_\mu^2
\epsilon^2 {(1+\epsilon / \msubt})^4
{{\sin^2 \thetr \cos \thetr}\over{[1 + 2{\epsilon / \msubt}
+({\epsilon / \msubt})^2 \sin^2 \thetr]^3 }}   {\tilde F}^2 (q)\ \ ,
\eqno\nexteq\nameeq\Eelassneua
$$
differential in the recoil solid angle.

One issue to which we return in Sect.~IV.J involves the hadronic
couplings occurring in Eqs.~(\Ehefour) and (\Ehefourneu).  These are the same
at tree level, but have different radiative corrections, as discussed in
Sect.~III.A.  In fact, if sufficiently high precision PV electron
scattering and neutrino scattering experiments could be undertaken, then
interesting information on the radiative corrections might emerge.
The practical difficulties of achieving sufficient precision, however,
make this goal a very difficult one to attain.

\bigskip
\goodbreak
\noindent \undertext{Elastic scattering from the nucleon}

     For elastic scattering from the nucleon we have
expressions analogous to those in Eqs.~(\Eproff) and (\Epropv) where the
results are given as functions of $\tau$ and $\cal E$:
$$
\eqalign{{\tilde F}_N^2 (\tau,{\cal E}) \equiv
16\pi &{\tilde F}^2 (\tau,{\cal E})= {1\over(1+\tau){\cal E}}\Bigl({\cal E}
[{\tilde G}_E^N (\tau)]^2 +\tau[{\tilde G}_M^N(\tau)]^2 \cr
&+(1 + \tau) [{\tilde G}_A^N(\tau)]^2
\mp 2\sqrt{1-{\cal E}^2}\sqrt{\tau(1+\tau)}{\tilde G}_M^N(\tau)
{\tilde G}_A^N(\tau) \Bigr) \cr}\eqno\nexteq\nameeq\Eneutnuc
$$
and where $N=p$ or $n$ with, as usual, the upper (lower) sign corresponding to
neutrino (antineutrino) scattering. As before (compare Eq.~(\alrprotpn))
the neutral current form factors may be re--expressed in terms of the
electromagnetic and strangeness form factors using Eqs.~(\EGtilde).

For the charge--changing reactions $p({\bar\nu}_l,\ell^+)n$ and
$n(\nu_l,\ell^-)p$ we have similar expressions for the cross sections,
although of course only isovector form factors enter:
$$
\eqalign{F_{N\pm}^2 (\tau,{\cal E})\equiv
2\pi &F^2_{\pm}(\tau,{\cal E}) = {1\over(1+\tau){\cal E}} \Bigl({\cal E}
[G_E^{T=1}(\tau)]^2 + \tau[G_M^{T=1}(\tau)]^2 \cr
&+(1 + \tau) [G_A^{T=1}(\tau)]^2
\mp 2\sqrt{1-{\cal E}^2}\sqrt{\tau(1+\tau)}G_M^{T=1}(\tau)
G_A^{T=1} (\tau)\Bigr)\ \ , \cr}\eqno\nexteq
$$
where the upper (lower) sign corresponds to the reaction
$\nu_l + n \rightarrow p + \ell^-$
(${\bar\nu}_l + p \rightarrow n + \ell^+$).

Continuing with elastic scattering, the cross section may be obtained
using the form factor given in Eq.~(\Eneutnuc) with the general formula
in Eq.~(\Eelassneu), where $\msubt\rightarrow \mn$,
$\tsubr\rightarrow T_\sst{N}$ (the recoiling nucleon's kinetic energy) and
$\thetr\rightarrow \theta_\sst{N}$ (the recoiling nucleon's angle with respect
to the neutrino beam direction). In addition, it is straightforward to
re--express the answer in terms of the familiar Mandelstam variables
$s=(P+K)^2$, $t=(K-K')^2=Q^2$ and $u=(P-K')^2$, where $P^{\mu}=
(\mn,0)$ is the struck nucleon's 4--momentum in the laboratory system (the
other momenta have been defined above). It is straightforward to show that
$$
\Bigl({s-u\over\mns}\Bigr) =4\Bigl({\epsilon\over\mn}-\tau\Bigr)
\eqno\nexteq
$$
and hence that the quantity $\cal E$ in Eq.~(\Eneutnuc) (see Eq.~(\Elongpl))
may be written
$$
{\cal E}= \Bigl[{ ({{s-u}\over\mns})^2 -16\tau(1+\tau) }\Bigr]\Bigl/
          \Bigl[ { ({{s-u}\over\mns})^2 +16\tau(1+\tau) }\Bigr]
\ \ . \eqno\nexteq
$$
The cross section may then be cast in the form frequently encountered in
the literature:
$$
{d\sigma(\epsilon)\over dQ^2}\>=\>{G_\mu^2\mns\over 8\pi\epsilon^2}\Bigl[A\pm
\Bigl({s-u\over\mns}\Bigr)B+\Bigl({s-u\over\mns}\Bigr)^2C\Bigr]\ \ ,
\eqno\nexteq\nameeq\Edsignup
$$
where the upper (lower) sign corresponds to the neutrino (anti--neutrino)
cross section.  Using the above results the following identifications may
be made:
$$
\eqalign{A(\tau)&=\tau\bigl[(1+\tau)(\gtilan)^2
-(\gtilen)^2+\tau(\gtilmn)^2\bigr] \cr
         B(\tau)&=-\tau\gtilan\gtilmn \cr
         C(\tau)&={1\over 16}\Bigl({1\over 1+\tau}\Bigr)
\bigl[(1+\tau)(\gtilan)^2+(\gtilen)^2+\tau(\gtilmn)^2\bigr]
        \ \ . \cr}\eqno\nexteq
$$

In Sect.~IV.J.2 we discuss the case of the nucleon in somewhat
more detail.  One situation discussed there involves elastic scattering
from hydrogen. Another involves quasielastic scattering
in nuclei where clearly the same approach followed in Sect.~III.E.1 and
developed in Sect.~IV.F for QE PV electron scattering could also be
pursued for neutrino scattering and neutrino reactions.  In fact, the
analog of Eq.~(\Eqecr) and the developments that follow can be used
directly for the charge--changing neutrino reactions (merely replacing the PC
or PV $p(e,e')p$  and $n(e,e')n$ cross sections with the
$p({\bar\nu}_l,\ell^+)n$ and $n(\nu_l,\ell^-)p$ cross sections).  However,
for the neutral current neutrino processes again some signature must be
identified. For scattering in the quasielastic region one expects that
a proton or neutron will be found in the final--state with energy--momentum
corresponding roughly to elastic scattering from one of the nucleons in the
nuclear ground state.  Thus, one needs a different kind of inclusive
cross section from those discussed heretofore, namely one where the
final--state lepton kinematics are integrated over while the final--state
ejected nucleon kinematics are not.  Such $(\nu_l,N)$ and $({\bar\nu}_L,N)$
cross sections are not
``total cross sections'' in the hadronic variables, in that such details as
the propagation of the outgoing nucleon must be accounted for, placing
more stringent demands on the nuclear model than is required for the rest of
the inclusive cross sections discussed in this work (which are all ``total
cross sections'' in the hadronic variables).  As a consequence of this
difference we have chosen to stop at this point and postpone further treatment
of quasielastic neutrino scattering to Sect.~IV.J.2.  Instead, we now return
to the main theme of this work and continue with more detailed discussions
of PV electron scattering in the next section.

\vfil\eject

%

\secnum=4
\neweq

%

\def\qvecsq{\vec q^{\mkern2mu\raise1pt\hbox{$\scriptstyle2$}}}

\def\xiva{{\xi_\sst{V}^{(a)}}}

\def\evec{{\vec e}}
\def\sigpin{{\Sigma_{\pi\sst{N}}}}

\def\xivp{{\xi_\sst{V}^p}}
\def\xivn{{\xi_\sst{V}^n}}
\def\lambestr{ {\lambda_\sst{E}^{(s)} } }

\noindent{\bf IV.\quad SPECIFIC CASES}
\bigskip

	In Sect.~IV we discuss a variety of specific cases where, using
semileptonic electroweak interaction processes, the hadronic neutral current
matrix elements may be explored, on the one hand, or
where Standard Model tests may be performed, on the other. We begin in
Sects.~IV.A and IV.B with the important studies of PV elastic electron
scattering from the proton and from $(J^\pi T)=(0^+ 0)$ nuclei,
respectively, and then continue in Sect.~IV.C with a discussion of the
role that PV elastic scattering from $^2$H could play. Other discrete
nuclear transitions initiated by inelastic PV electron scattering are
treated in Sect.~IV.D, while in Sect.~IV.E we briefly discuss the axial--vector
response and the anapole moment, also involving discrete nuclear states.
Quasielastic PV electron scattering is treated in Sect.~IV.F and PV
electroexcitation of the $\Delta(1232)$ is discussed in Sect.~IV.G. In the
final three subsections of this major section we provide connections to
scattering processes in other kinematic regimes (PV deep--inelastic
electron scattering in Sect.~IV.H) and to other electroweak processes
(atomic PV in Sect.~IV.I and $\nu$--scattering in Sect.~IV.J) as these
exhibit different sensitivities to hadronic content and to specifics of
the underlying electroweak theory than does PV electron scattering and
hence yield complementary information.

\bigskip
\noindent{\bf IV.A.\quad Elastic Scattering from the Proton}
\medskip

        At first glance, one would expect a measurement of the PV asymmetry
for elastic scattering of polarized electrons from nucleons to provide the
most direct and conceptually straightforward probe of the nucleon's weak
neutral current. Several factors, however, conspire to render the
interpretation of PV electron--nucleon scattering considerably more complicated
than it might first appear. The lack of a free neutron target, for example,
implies
either restricting oneself to PV elastic $\evec p$ scattering or turning
to $A>1$ targets in order to access the NC of the neutron. The latter option,
of course, introduces nuclear physics considerations into the interpretation
of the PV asymmetry. Even in the case of $\alr(\evec p)$, however, one must
account for the interplay between various hadronic form factors, hadronic
uncertainties in radiative corrections, and the physics of the underlying
electroweak gauge theory, as well as practical questions relating experimental
doability to theoretical interpretability. The bottom line is that PV elastic
$\evec p$ is a fundamental component of any attempt to probe the nucleon's
NC, but that by itself, it is not sufficient for providing all the information
one might want to acquire.

        To illustrate the rationale for this conclusion, we return to
the expression for the PV asymmetry given in Eq.~(\alrprotpn)

$$
\alr(\evec p) = a_0 \tau {W^{\rm (PV)}\over F^2}\ \ ,
$$
where $a_0\approx 3.1\times 10^{-4}$ and where the hadronic ratio is given by
$$
\eqalignno{{W^{\rm (PV)}\over F^2} =
& \xivp + \Big[{\cal E}\GEp\{\xivn\GEn+\xivz\GES\}&\cr
        &\>+\tau\GMp\{\xivn\GMn+\xivz\GMS\}&\nexteq\nameeq\Eeptot\cr
&\>-\sqrt{1-{\cal E}^2}
        \sqrt{\tau(1+\tau)}(1-4\sstw)\GMp{\tilde\GAp}\Big]
\Big/[{\cal E}(\GEp)^2+\tau(\GMp)^2]\ \ \ .&\cr}
$$
The kinematic dependence of the various terms in Eq.~(\Eeptot) suggests a
variety of possible $\alr(\evec p)$ measurements. A very low--$|Q^2|$
experiment ({\it e.g.,\/} $ q\lapp\ \hbox{few}\ 100 \ \hbox{MeV}/c$),
in which the contribution of the hadronic form factors is
minimized, might allow one to extract $\xivp$ and test the Standard Model.
The motivation for a moderate momentum transfer measurement ({\it e.g.\/},
$\hbox{few}\ 100\ \hbox{MeV}/c\to \hbox{few GeV}/c$) would be to determine
the nucleon form factors, particularly those associated with the
strange--quark currents.
The $\theta$--dependence of the PV asymmetry suggests that one might consider
a forward--angle measurement, which would constrain the electric form
factors, $\GES$ and $\GEn$ and a combination of intermediate-- and
backward--angle measurements at the same $Q^2$ with the intent of separating
the axial--vector and weak magnetic responses, ideally
allowing a determination of $\GMS$. The experiments discussed in
Refs.~[McK89, Bec91, Bei91b, Fin91] include
two of these prospective measurements,
and they would provide the first experimental bounds on the low--$|Q^2|$
behavior of $\GMS$ and $\GES$.

     As we illustrate below, however, completion
of such a set of $\evec p$ measurements would
not be sufficient to constrain the
strangeness form factors to arbitrary precision, although it would provide
valuable experimental limits.
Specifically, the contribution from
the axial--vector response persists to sufficiently forward angles that a
Rosenbluth--type separation of the weak magnetic and axial--vector form
factors does not
appear possible. Consequently, large theoretical uncertainties in
the radiative corrections for the axial--vector current term
impose an intrinsic theoretical bound
on the precision to which $\GMS$ may be
determined from $\alr({\svec e}p)$.  A similar limit on the achievable
precision in $\GMS$ is imposed by
the experimental uncertainty in $\GMn$, which
also contributes to $\alr(\evec p)$ at backward angles.
Moreover, this ``intrinsic'' uncertainty in
$\GMS$ enters the extraction of $\GES$ from forward--angle
$\alr({\svec e}p)$ measurements. Assuming realistic experimental
conditions, one finds that the resulting uncertainty in $\GES$ is somewhat
larger than the theoretical error quoted for the model calculation of
Ref.~[Jaf89]. This conclusion also carries implications for prospective
$\alr(\evec p)$ electroweak tests.  Tighter limits on $\GES$ would be
required before one could hope to extract $\xivp$ from a low--$|Q^2|$
$\alr(\evec p)$ measurement to 10\% accuracy (as assumed in Fig.~2.4).
Thus, a program of NC studies with polarized protons would require scattering
from $A>1$ targets to complement elastic PV $\evec p$ experiments.

        We now proceed to discuss these results in more detail, drawing
essentially from the treatment in Ref.~[Mus92a].

\bigskip
\noindent IV.A.1.\quad BACKWARD--ANGLE SCATTERING
\medskip

In the $\theta\to 180^o$ limit, ${\cal E}\to 0$ and Eq.~(\Eeptot)
simplifies to
$$
{W^{\rm (PV)}\over F^2}\>\longrightarrow\>\xivp +
        \xivn{\GMn\over\GMp}+\xivz{\GMS\over\GMp}
- (1-4\sstw)\sqrt{{1\over\tau}+1}{\tilde \GAp
        \over\GMp}\ \ .\eqno\nexteq\nameeq\Eapbalim
$$
Conservation of angular momentum and
electron helicity (in the $\epsilon >\!\!>\me$
regime) require the proton to flip its spin, so that only the magnetic and
axial--vector terms contribute.
The first term in Eq.~(\Eapbalim) comes from the piece of
$\gmptil$ proportional to $\GMp$; hence the form factor dependence cancels
with the $\GMp^2$ of the denominator.  The second and third terms
arise from the $\GMn$ and $\GMS$ terms in $\gmptil$.
The SAMPLE experiment, presently underway at
MIT/Bates [McK89], will measure the backward--angle $\evec p$ asymmetry,
thereby constraining the strangeness magnetic moment,
$\mustr\equiv\GMS(0)$. The anticipated experimental uncertainty corresponds
a limit on $\mustr$ of $|\delta\mustr| < 0.22$.
A second--generation SAMPLE experiment having smaller experimental error
might hope to tighten this limit.
There are, however, other factors that must be considered.  Up to radiative
corrections, $\GMn$ and $\GMS$ enter the asymmetry with equal weight
($\xivn=\xivz=-1$ at tree level; see Table~3.2 and Eq.~(\Exiprne b)), so that
an
accurate extraction of $\GMS$ requires very accurate knowledge of $\GMn$.
Of even greater concern are the complications introduced by the
theoretical uncertainties in the axial--vector form factor appearing in
Eq.~(\Eapbalim).
Uncertainties in $\tilde\GAp$ arise from both $\eta_s$ and $\GdipA$ (see
Sect.~III.C)
as well as from the radiative corrections, $\RAp$. At the SAMPLE kinematics,
uncertainties in the axial--vector dipole mass parameter
induce roughly a one percent error
in $\tilde\GAp$; the corresponding induced error in $\GMS$ is negligible for
the present purposes. The impact of uncertainties in $\eta_s$ (see
Table~2.2) is included in the projected SAMPLE limits on $\GMS$. In
principle, the LSND neutrino oscillation experiment underway at
LAMPF [Lou89] could reduce the uncertainty in $\eta_s$, although with the
expected 20\% statistical uncertainty in the elastic $\nu$--$p$ cross section
little improvement, if any, would be made over the uncertainty quoted in
Table~2.2.\footnote{*}{The BNL error quoted in Table~2.2 does {\it not}
include the impact of the $\eta_s$--$M_\sst{A}$ correlation. The LSND
experiment
would eliminate the latter source of uncertainty in $\eta_s$.}
The error in the radiative correction, $\RAp$, appears to be more problematic.
As discussed in Sect.~III, this quantity contains significant theoretical
uncertainty in the case of PV $\evec p$ scattering. Since
this uncertainty is associated with low--energy hadronic contributions, one
has little hope of calculating its magnitude with significantly better
precision than given by the estimate of Ref.~[Mus90].
For the present purposes, then, we consider this error to be intrinsic.
Together with uncertainty in $\mustr$, it induces a fractional change in
the backward--angle asymmetry of
$$
{\delta\alr\over\alr}\>\approx\>{\delta\RAp\over 5}-
{\delta\mustr\over 3}\eqno\nexteq\nameeq\Epprotd
$$
at the SAMPLE kinematics. Thus, even if a \lq\lq perfect" SAMPLE experiment
(zero percent experimental uncertainty) were possible, one would not be able
to constrain $\mustr$ to better than
$$
\delta\mustr
\>\approx\>0.6\ \delta\RAp\>\approx\>\pm0.12\eqno\nexteq\nameeq\Edmfromdr
$$
according to the estimates of Ref.~[Mus90]. The relation between
$\delta\RAp$ and $\delta\mustr$ for nonzero experimental uncertainty
is indicated by the correlation plot of Fig.~4.1. As we illustrate
below, this
theoretical uncertainty may propagate through other experiments,
introducing limitations on $\GES$ determinations as well as precision
Standard Model tests.

     One might hope to reduce the uncertainties  due to lack of knowledge
of $\tilde \GAp$ by using the different $\theta$--dependences in
Eq.~(\Eeptot) to
separate the axial--vector term.
Since ${\cal E} \to 1$ for $\theta \to 0^\circ$, the
importance of the axial--vector term should be reduced for sufficiently small
angles.  As Fig.~4.2 illustrates, however, ${\cal F}_A$, the
fraction of $\alr$ due to
the axial--vector term, does not begin to
decrease significantly until $\theta\lapp 40^\circ$ (depending on the
energy).
At this point the fraction ${\cal F}_L$ contributed by the longitudinal term
increases and one can no longer reliably apply the backward--angle limit to
isolate $\GMS$.

A second possibility
for separating the magnetic and axial--vector terms,
a backward--angle
measurement at higher energy, also depends on the different
kinematical factors
appearing with the second and third terms in Eq.~(\Eeptot).  Since $\tau =
(\epsilon\epsilon'/\mns)
\sin^2{\theta\over 2}$ for elastic scattering with extremely relativistic
electrons, the axial--vector term decreases in importance relative to the
magnetic term for increasing energy at backward angles.  Thus, one might
hope to decrease
the sensitivity of $\alr(\evec p)^{\theta\to 180^\circ}$
to the more uncertain axial--vector term by performing an
experiment at higher energy than envisioned for SAMPLE.  This gain, however, is
soon offset by a decrease in the FOM for sufficiently high energies, resulting
from a falloff in the proton magnetic form factor with increasing
$\vert Q^2\vert $. In
Fig.~4.3 we plot $(\delta \alr/\alr)_{\rm stat}$
as a function of $\epsilon$ at $\theta =
175^\circ$ for fixed experimental conditions
(a 1000--hour experiment, a luminosity
of $5\times 10^{38}$~cm$^{-2}s^{-1}$, a beam polarization of 100\%, and a 1 sr
solid angle).  Also shown is ${\cal F}_A$
as a function of $\epsilon$ at $\theta =
175^\circ$.    From these two curves we observe that although ${\cal F}_A$
decreases by a factor of two in going from $\epsilon =250$ to 1000~MeV,
$(\delta \alr/\alr)_\sst{\rm stat}$ increases by nearly the same amount.

     We conclude, then, that neither of these approaches
will significantly improve the limits on $\mu_s$ attainable from
a backward--angle $\alr({\svec e}p)$ measurement.  A potentially more promising
alternative is to measure either $R_A^p$ or $\mu_s$ with some other target
for which they are not as strongly correlated
as in the case of
intermediate-- or backward--angle $\alr({\svec e}p)$. Elastic scattering from
the deuteron and quasielastic scattering constitute two such possibilities,
as discussed in Sects.~IV.C and IV.F below.

\bigskip
\goodbreak
\noindent IV.A.2.\quad FORWARD--ANGLE SCATTERING
\medskip

For $\theta\to 0^o$, one has ${\cal E}\approx 1$ and the
axial--vector term in Eq.~(\Eeptot) becomes negligible, leading to
$$
\eqalign{{W^{\rm (PV)}\over F^2} \longrightarrow
& \xivp +
\Big[\GEp\{\xivn\GEn+\xivz\GES\}\cr
        &\>+\tau\GMp\{\xivn\GMn+\xivz\GMS\}
\Big]
\Big/[(\GEp)^2+\tau(\GMp)^2]\ \ . \cr} \eqno\nexteq\nameeq\Eepfor
$$
At moderately low momentum transfers
($\vert Q^2 \vert <\!\!< 4 \ (\hbox{GeV}/c)^2$, so that $\tau <\!\!< 1$ )
we can simplify the discussion by keeping only the terms through O($\tau$):
$$
{W^{\rm (PV)}\over F^2}\>\longrightarrow\>\xivp
        -\>\tau\Bigl[\mu_n
        \xivn-\rhostr\xivz
        \>-\>\mu_p\{\mu_n\xivn+\mustr\xivz\}\Bigr]
        \>+\>\hcal{O}(\tau^2)\
        \ .\eqno\nexteq\nameeq\Eepforlow
$$
An extraction of the term of {\cal O}$(\tau)$ in Eq.~(\Eepforlow) would
constrain a linear combination of $\rhostr$ and $\mustr$.
A series of $\alr(\evec p)$ measurements, carried out at different
values of $\tau$ and designed to extract this term of $\hcal{O}(\tau)$, has
been discussed as a possibility for CEBAF [Bec91, Bei91b, Fin91]. As
Eq.~(\Eepforlow) demonstrates, however, the placement of constraints
on $\GES$ from such an extraction would require one to account for
correlations between uncertainties in all of the contributions of
{\cal O}$(\tau)$.

To evaluate the impact of
these various correlations on a $\GES$--determination,
the authors of Ref.~[Mus92a] have performed a sensitivity
analysis for a prospective measurement carried out at the kinematics of
Ref.~[Nap91]:  $\epsilon =2.6\ \hbox{GeV}$ and $0.1 \leq
\vert Q^2\vert\leq 0.3\ $ \gevocsq.
In this regime, the strongest correlation occurs between $\rhostr$ and
$\mustr$. Assuming
 zero percent experimental uncertainty, for example, the
uncertainties in these two parameters are related by
$$
\delta\rhostr=-\mu_p(1+\lambestr\tau)\delta\mustr\ \ .\eqno\nexteq
$$
For either $\tau=0$ or $\lambestr=0$, the uncertainty in $\alr$ due to $\delta
\mustr$ is weighted by a factor of $\mu_p\approx 2.79$ relative to the
error induced by $\delta\rhostr$. For nonzero $\tau$ or $\lambestr$, the
impact of $\delta\mustr$ on $\delta\rhostr$ may be enhanced. For example,
taking
$\lambestr=\lamn=5.6$ and $\vert Q^2\vert=0.2$ \gevocsq, corresponding to
$\tau=0.06$, gives
$$
\delta\rhostr\approx -3.7\delta\mustr\ \ . \eqno\nexteq
$$

     The $\rhostr$--$\mustr$ correlation taken from Ref.~[Mus92a] for a
more realistic experiment
($\delta\alr/\alr)_{\rm experiment}\not=0$) is displayed in Fig.~4.4.
The error bands
correspond to the combined statistical and systematic errors projected for
the measurement discussed in Ref.~[Nap91]. From this analysis a limit of
$\delta\mustr \approx 0.22$ from a backward--angle experiment will allow
extraction of $\GES$ at a level of $\delta \GES \approx \GEn$,
while the limit of Eq.~(\Edmfromdr),
$\delta\mustr \approx 0.12$ would give
$\delta\GES \approx 0.7\GEn$. The latter uncertainty is somewhat larger
than the theoretical uncertainty in the prediction of Ref.~[Jaf89] and
somewhat smaller than the total prediction of the broken SU(3) Skyrme
model (see Table~2.3).

     For comparison, the authors of Ref.~[Mus92a] also examined
the $\GES$--$\GEn$, $\GES$--$\GMn$
and $\GES$--$\sstw$
correlations. Since $\GES$ and $\GEn$ enter the terms of
$\hcal{O}(\tau)$ with a relative weighting of 1:1 (up to radiative
corrections), any uncertainty in
$\GEn$ induces an identical uncertainty in $\GES$. The former
is presently about $\pm 0.5\GEn$ and might be somewhat reduced by the
completion of the current
MIT/Bates experiments [Mil88].
One might ultimately hope for another
factor of two or so improvement in $\delta\GEn$ within
the next decade [Mad85, Are88, Mad89, Jon91, Chu88, Mil89, Ala91].
Lack of knowledge in $\GMn$ may eventually be more problematic
than uncertainty in $\GEn$, due to the pre--multiplying factor of $\mu_p$ in
Eq.~(\Eepforlow).  While the
error in $\GMn$ may be reduced to about 5\% by
the end of the decade [Jou88], for example,
it is roughly three times more important than the
uncertainty in $\GEn$ as far as a $\GES$--determination is
concerned.  Thus, a 5\% uncertainty in $\GMn$ will introduce more uncertainty
in $\GES$ than would a 10\% uncertainty in $\GEn$.  Moreover, the
error in $\GMn$ would be roughly comparable with that corresponding to the
$\vert\delta\mustr\vert < 0.12$ limit.
Finally, for $\vert Q^2\vert=0.2$ (GeV$/c)^2$ and
$\delta\alr/\alr=0$, a one percent error in $\sstw$ generates
an uncertainty in $\GES$ of $\delta\GES\approx\pm 0.15\ \GEn$.
Thus, both the $\GEn$ and $\sstw$
uncertainties in $\GES$ are smaller than the error induced
by $\mustr$ when the latter is limited by the radiative correction uncertainty
or by the $\GMn$ uncertainty.

     There is little doubt that a set of measurements of the asymmetry for PV
$(\evec,e^\prime)$ from the proton will be a keystone in the determination of
the contribution of $s$ quarks to the vector current of
the nucleon. Given the large theoretical uncertainties associated with
the axial--vector radiative corrections, $\alr(\evec p)$ is less suitable
as a probe of strangeness axial--vector current of the nucleon. Indeed, the
former uncertainty limits the extent to which a series of $\alr(\evec p)$
measurements could constraint $\GES$ and $\GMS$. Additional, complementary
measurements will be needed to fully elucidate the role of the strange
quarks in the nucleon.

\goodbreak
\bigskip
\noindent IV.A.3.\quad STANDARD MODEL TESTS
\medskip

In the limit that $|Q^2|\to 0$, the hadronic ratio $W^{\rm (PV)}/F^2$
for forward--angle scattering is
simply proportional to the proton neutral current coupling, $\xi_V^p =
(1-4\sstw) (1+\RVp)$. As indicated in Sect.~II, a 10\% determination
of this quantity could nicely complement atomic PV or PV elastic
$^{12}$C$(\evec, e)$ scattering as a low--energy probe of physics beyond
the Standard Model. At $Q^2=0$, this 10\% figure translates directly into
a 10\% determination of $\alr(\evec p)$.
Any actual experiment must be carried out at $Q^2\not=0$, since $\alr$
vanishes with $Q^2$. In this case, one has, to lowest non--trivial order in
$\tau$ one has
$$
{\delta\xivp\over \xivp}=\Bigl[1+\tau\Bigl({B\over\xivp}\Bigr)\Bigr]\Bigl(
        {\delta\alr\over\alr}\Bigr)-\tau\Bigl({\delta B\over\xivp}\Bigr)\ \ ,
\eqno\nexteq\nameeq\Eepstanda
$$
where
$$
B\>\equiv
        \>-\mu_n\xivn+\rhostr\xivz
        \>+\mu_p\Bigl[\mu_n\xivn+\mustr\xivz\Bigr]\ . \eqno\nexteq\nameeq
\Eepstandb
$$
The quantity $\delta B$ represents the error in $B$ from all sources,
including $\GES$, $\GMS$, $\GEn$, $\xivn$, and  $\xivz$. We now
consider these sources of error.

First, we note that the errors induced in $\xivp$ by
$\delta B$ and $\delta\alr/\alr$ depend on the value of $\tau$. The impact of
these errors is minimized as one decreases the value of $\tau$, as
indicated by Eq.~(\Eepstanda).\footnote{*}{Note that both $\xivp$ and
$B$ are positive quantities.}
However, the FOM also decreases with $\tau$, so that
the achievable statistical precision in $\alr$ improves for {\it increasing}
$\tau$. These two competing features of $\alr(\evec p)$ conspire to
restrict severely the kinematical region in which one should attempt to
carry out a 10\% determination of $\xivp$ . In order to quantify these
statements, we plot in Fig.~4.5 curves of $(\delta\alr/\alr)_\sst{\rm stat}$
versus $\tau$ for
constant values of the scattering angle. We have assumed a
luminosity of $5\times 10^{38}$ cm$^{-2}$ s$^{-1}$, a 100\% beam
polarization, a solid angle of
0.01 sr and a running time of 1000 hours. These conditions are within
reasonable expectations for what might be achievable at  CEBAF within the next
decade (see Sect.~V.C).
For each curve, increasing $\tau$ corresponds to increasing incident
electron energy. On the
same graph we plot the minimum $(\delta\alr/\alr)_\sst{\rm stat}$
needed in order to keep
the error in $\xivp$ below ten percent (dashed line). The latter is
derived from Eqs.~(\Eepstanda) and (\Eepstandb) assuming
tree--level Standard Model couplings and setting
$\delta B=\GMS=\GES=0$.

        From Fig.~4.4 one can see that an experiment should have
$\epsilon \rapp $ 1100 MeV and $\theta\lapp 15^o$ to
extract $\xivp$ at the 10\% level.  However, one cannot go to arbitrarily
large values of $\epsilon$ without introducing problematic errors from
$\delta B$.
For example, in an experiment carried out at $\theta=10^o$, one needs
$\tau\geq 0.011$. Under these conditions, the
uncertainty in $\GES$ must be $\delta\GES\leq 0.47\GEn$
in order to keep the error induced in $\xivp$ by $\delta B$ to less than ten
percent. From our previous discussion, it does not appear
possible to determine $\GES$ to this precision with a series of $\evec p$
experiments alone.  Further analysis of Ref.~[Mus92a]
suggests that, anticipating
a 10\% uncertainty in $\GEn$ and a 5\% uncertainty in $\GMn$, it is the
uncertainty from the strange--quark form factors that would most severely
limit a determination of $\xivp$.
Thus, to extract $\xivp$ to ten percent, one would
need to improve the precision in
$\GES$.

        To realize this objective, one has a number of options: (i) a
direct measurement of $\GES$ with PV electron scattering from another
target (see, {\it e.g.}, Sect.~IV.B); (ii) a more precise direct
measurement of $\GMS$ with PV electron scattering from another target,
thereby reducing the $\GMS$--induced error in an $\alr(\evec p)$
determination of $\GES$ (see, {\it e.g.}, Sect.~IV.C); or (iii) a
measurement of $\tilde\GAp$ using PV QE electron scattering (see
Sect.~IV.F), thereby reducing the $\GMS$ and $\GES$ uncertainties
associated with a series of $\alr(\evec p)$ measurements. Alternatively,
one could live with the larger $\GES$ uncertainty and perform an
$\alr(\evec p)$ electroweak test at more forward angles than assumed
in the foregoing analysis. For $\delta\GES \approx0.7\GEn$, for
example, a measurement at $\theta\lapp 8^\circ$ would be needed. Such
a measurement lies beyond the capabilities of presently envisioned
facilities at CEBAF, and construction of a new detector would be
required.

We conclude this section by pointing to one issue which arises in the
extraction of constraints on new physics from a determination of $\xivp$,
namely, the appearance of theoretical hadronic uncertainties in the
radiative correction, $\RVp$. In order to arrive at the limits on
$S$ and $T$ displayed in Fig.~2.4, one would need to keep the error
in this correction to $|\delta\RVp(\hbox{had})|\lapp 0.1$. This radiative
correction
arises primarily from hadronic loops in the $Z^0$--$\gamma$ mixing tensor and
from hadronic intermediate states in the $Z^0$--$\gamma$ \lq\lq box" diagrams.
The authors of Ref.~[Mar84] have estimated the former using a dispersion
analysis
and find a corresponding uncertainty in the radiative correction of
$\delta\RVp(Z-\gamma \hbox{ mixing})\approx 0.02 $.
Hadronic contributions to the
box diagram correction have also been estimated in Ref.~[Mar84] for the case
of atomic PV. These authors
considered proton intermediate states,
whose dominant contribution to $\RVp$ is suppressed by a factor of
$(1-4\sstw)$. As a rough and, perhaps, liberal estimate, one might equate
the hadronic error in $\delta\RVp$ with their estimate of this ``Born"
contribution to $\RVp$, resulting in an estimate of $\delta\RVp=\pm 0.01$,
well below the problematic limit. Contributions from higher--lying intermediate
states, however, need not carry the $1-4\sstw$ factor and could introduce
significantly larger uncertainty. Further study of these contributions is
warranted for both PV electron scattering and atomic PV.

\vfil\eject

%

\def\qvecsq{\vec q^{\mkern2mu\raise1pt\hbox{$\scriptstyle2$}}}

\def\xiva{{\xi_\sst{V}^{(a)}}}

\def\evec{{\vec e}}
\def\sigpin{{\Sigma_{\pi\sst{N}}}}

\def\lamEs{\lambda_\sst{E}^{(s)}}
\def\alrzer{{A_\sst{LR}^0}}

\noindent{\bf IV.B.\quad Elastic Scattering from Spin--0 Nuclei}

     In turning from the nucleon to nuclear targets one might imagine at least
two sources of additional complications.  The first would be the extra
layer of strong--interaction physics manifested in details of
nuclear structure;  the second would be the variety of and interplay
between the numerous multipole matrix
elements that can contribute for arbitrary
initial-- and/or final--state nuclear angular momenta.  Indeed,
as discussed later in Sect.~IV.D, such
complications coupled with practical considerations such as energy
resolution probably render most PV $({\vec e},e^\prime)$
experiments involving transitions between discrete nuclear states dubious
at best for
exploring the nucleonic current including its potential $s{\bar s}$ content,
although such experiments
might be of great value for studies of nuclear structure.  However, there
are also circumstances where specific features of nuclear
structure can be used advantageously to provide simplification and/or to
isolate or at least emphasize physics of interest.  The best example of
this is elastic PV electron scattering from a spin--0, isospin--0 nucleus.
In general, elastic scattering has
an advantage over inelastic scattering in that the FOM, being proportional to
the EM cross section, is enhanced in the forward direction and low--$|Q^2|$
by the $Z^2$ coherence factor in the monopole form factor. As Figs.~3.10--3.12
illustrate, this results in $\cal F$ being very large and
consequently, the
attainable statistical precision is improved by a factor of $1/Z$ with
respect to $\evec p$ or inelastic scattering to a discrete state
for a given luminosity and
running time. Second, in the absence of strangeness and in the
limit of exact isospin symmetry $\left( [{\svec T}, H_{\rm nuc}]=0\right)$,
elastic matrix elements of $J^\sst{NC}_\mu$ and $J^\sst{EM}_\mu$ are
proportional for these targets, since only the $T=0$ components of the currents
contribute. Accordingly, as discussed in Sect.~III.E.1, in this limit the
matrix elements cancel from the hadronic ratio $W^{\rm (PV)}/F^2$, leaving
only the particle physics coupling, $\sqrt{3}\,\xi^{T=0}_V$.  Moreover,
$\alr(0^+0)$ is also free from magnetic and axial--vector contributions,
since this target has $J=0$.

To begin our discussion of what happens when one proceeds beyond this
simple limit, let us substitute the Standard Model couplings from Tables~3.1
and 3.2 into Eq.~(\Ehefour) to obtain
$$
-2{\alr\over\alrzer} = 4 \sin^2\theta_W
 \left[ 1 + R^{T=0}_V +\Gamma(q)\right] +
[1+\rvz] {{F_{C0}(s)}\over{F_{C0}(T=0)}} \ \ .
\eqno\nexteq\nameeq\Ezerostd
$$
The $Q^2$--dependence here arises from three terms:
$\Gamma$, the correction that represents the breaking of isospin symmetry
discussed in Sect.~III.D.4, the ratio
${{F_{C0}(s)}/{F_{C0}(T=0)}}$
that comes from the presence of nonzero strangeness components in the
hadronic neutral current
and the radiative corrections $R_V^{T=0}$ and $R_V^{(0)}$ which were
introduced in Sect.~III.B.
In the absence of the corrections, which will be discussed in more detail
in the following subsections, the right--hand side of Eq.~(\Ezerostd)
yields the especially simple answer $4\sstw$ and accordingly elastic
scattering from spin--0 nuclei was
suggested as a place to test the Standard Model [Fei75, Wal77]. One
experiment of this type has been attempted, the $^{12}$C(${\vec e},
e$) measurement at MIT/Bates, whose detailed results have been presented in
Ref.~[Sou90a] (see also the summary of these results in Sect.~V.B.3).  In this
pioneering effort
$\alr(0^+0)\approx 1.7$ ppm was measured with 23\% statistical and
3\% systematic errors.  Future, higher--precision experiments of this
type appear to afford the possibility of exploring in more detail
all of the various facets of Eq.~(\Ezerostd) and, accordingly, we shall
follow the discussions in Refs.~[Mus92a, Don89] and
focus on these future issues. We begin by exploring the interpretation and
implications of
this basic equation for the extraction of the strange--quark contribution
to the hadronic neutral current (Sect.~IV.B.1) and then return to discuss
such PV elastic scattering measurements as Standard Model tests
(Sect.~IV.B.2).  Finally, in Sect.~IV.B.3 we consider
elastic scattering from spin--0 nuclei with $N\neq Z$ ({\it i.e.,\/} $T>0$
nuclei) as a new means to study their ground--state neutron distributions.
Discussion of inelastic scattering to discrete nuclear states, including
$J>0$ nuclei, will be deferred to Sect.~IV.D.

Before entering into these more detailed discussions let us summarize
some of the basic conclusions.
As discussed in Ref.~[Mus92a], high--$|Q^2|$ forward--angle measurements are
potentially sensitive enough to $G^{(s)}_E$ to permit a determination of this
form factor. For ${}^4$He, experiments carried out at moderately--forward
angles
($\theta\sim 30^\circ$) for energies in the regime $0.2 \lapp
\epsilon\lapp 1.1$~GeV (below the first diffraction minimum) and
$1.3\lapp \epsilon\lapp 2.0$~GeV (above the first diffraction minimum)
could allow a $G^{(s)}_E$--determination with an uncertainty of roughly half
what is attainable with $\alr({\svec e}p)$ measurements, up to presently
unquantified theoretical uncertainties.  Results for $^{12}$C are similar,
although the useful
energy ranges are somewhat narrower. Experiments at more forward
angles
and higher energies would also achieve the same end.
Viewed as a Standard Model test, the low--$|Q^2|$
forward--angle projections given in Ref.~[Mus92a] show that $\alr(0^+0)$
is predominantly sensitive to $\sin^2\theta_W$ and
possible contributions from ``non--standard,'' degenerate, heavy--fermion
doublets [Mar90, Pes90], although potentially large and theoretically
uncertain dispersion
corrections could seriously cloud the interpretability of a $\lapp 1$\%
measurement, as discussed in Sect.~II.A.  Furthermore, that lack of knowledge
in $G^{(s)}_E$ places
tight requirements on the kinematic regime for which a 1\%
measurement of $\alr(0^+0)$ would be interpretable as a Standard Model test.
For both ${}^4$He
and ${}^{12}$C, a low--to--intermediate energy measurement would be required,
whereas an experiment performed at CEBAF energies (2--4 GeV) would introduce
$G^{(s)}_E$--uncertainties at greater than the 1\%
level.  Experiments at very--forward and moderately--forward angles
both appear to be possible.  Finally, it appears to be quite feasible to
extract the rms radius of the ground--state neutron distribution for a wide
range of nuclei to a precision of about 1\% using PV elastic electron
scattering.

\bigskip
\goodbreak
\noindent IV.B.1\quad SENSITIVITY TO THE STRANGENESS FORM FACTOR $G_E^{(s)}$

PV elastic electron scattering from 0$^+$0 nuclei might ultimately
allow a determination of $G^{(s)}_E$ to higher precision than is possible with
${\svec e}p$ scattering.  By carrying out an $\alr(0^+0)$ experiment at
higher values of $|Q^2|$, one enhances the importance of the
strangeness--dependent term of Eq.~(\Ezerostd) relative to the leading
term [Bec89].  By going to sufficiently high--$|Q^2|$,
one might also explore the non--leading $Q^2$--dependence of $\GES$.
As discussed in Sect.~III.D.4,
for favorable $N=Z$ nuclei such as $^4$He and $^{12}$C, the first of the
corrections in Eq.~(\Ezerostd), that relating to isospin--mixing ($\Gamma$),
has been conservatively estimated
in Ref.~[Don89] to remain less than 0.01 over the entire
range of $Q^2$ of interest for extraction of strange--quark contributions
(limited on the high--$|Q^2|$ side by the rapid falloff of the nuclear form
factor and, correspondingly, the FOM --- see Figs.~3.10--3.12).  Moreover, the
$Q^2$--dependence of $R_V^{T=0}$ and $R_V^{(0)}$ should be relatively weak.
Thus, depending on the size of the strangeness form factor
$F_{C0}(s)$, measurement of the PV
asymmetry for elastic scattering from such light 0$^+$0
nuclei may be expected to provide a relatively simple method for isolating
the strangeness contribution to the nuclear vector current.

To understand how well elastic scattering from $0^+$ nuclei can be used
to learn about $G_E^{(s)}$, it is necessary to probe more deeply into
the uncertainties that compete with the strangeness dependence.  Let us
begin with the nuclear many--body problem.  If one treats the EM and NC
form factors as arising from matrix elements of one--body operators
(see Sect.~III.D.2), then from Eq.~(\Exdden) the relevant matrix elements
can be written
$$
<0^+;T_0\doubred {\hat O}_{0;T}^{[1]}(q)\doubred 0^+;T_0> =
\sum_{a} \psi_{0;T}(a^2) <a\doubred O_{0;T}^{[1]}(q) \doubred a>
\ \ , \eqno\nexteq\nameeq\Elzera
$$
where we assume in general that the ground state has isospin $T_0$. The
one--body density matrix elements can be shown to be [Don84]
$$
\eqalign{ \psi_{0;0}(a^2) &= { \sqrt{2T_0+1}\over{\sqrt{2(2j_a+1)}} }
\bigl[ N_p(a) + N_n(a) \Bigr] \cr
\psi_{0;1}(a^2) &= -{\sqrt{(T_0+1)(2T_0+1)}\over{\sqrt{6T_0(2j_a+1)}} }
\bigl[ N_p(a) - N_n(a) \Bigr]\ \ , \cr}\eqno\nexteq\nameeq\Elzerb
$$
where $N_{p,n}$ are the occupation numbers of the single--particle level
labeled $a$ (see Sect.~III.D.2). The single--particle matrix elements in
Eq.~(\Elzera) contain all of the momentum transfer dependence in the
nuclear form factors.  In particular, the coherence in the many--body
matrix elements arises from the fact that the $T=0$ density matrix
elements above involve a sum over protons and neutrons.  Thus, at low
momentum transfer the $T=0$ nuclear matrix elements are proportional to
the sum over all of the nucleons in the nucleus, {\it i.e.,\/} to $A$.
In contrast, the $T=1$ density matrix elements involve proton--neutron
differences and usually lead to smaller isovector effects.  These
arguments are independent of the nature of the electroweak probe ---
that aspect of the problem is contained entirely in the single--particle
matrix elements.  In particular, since in comparing the EM and NC
form factors the latter differ only by the hadronic weighting factors in
Table~3.2 and by the inclusion of the appropriate single--nucleon form
factors, $G_E^T$ with $T=0,1$ and $G_E^{(s)}$, one has the following
proportionality relationship involving the nuclear many--body monopole
form factors and the single--nucleon form factors:
$$
\eqalign{ { {F_{C0}(s)}\over{F_{C0}(T=0)} } &= { {G_E^{(s)}}\over{G_E^{T=0}} }
\cr
&= 2\rho_s\tau\xi_E^{(s)} \cr}\eqno\nexteq\nameeq\Emonobas
$$
using the parameterizations in Sect.~III.C.  Thus the ratio of {\it nuclear}
form factors is simply a ratio of {\it single--nucleon} form factors in
this one--body--operator approximation (see Sect.~III.D.3 for a brief
discussion of two--body meson exchange current corrections).

Before proceeding further, let us make the following comment about the
nuclear monopole form factors: in computing matrix elements of the one--body
charge operator, $\rohat^{(1)}$, one faces an ambiguity as to whether to use
Dirac and Pauli ($F_1, F_2$) or Sachs ($G_\sst{E}, G_\sst{M}$) form factors
(as above). Since the operator $\rohat^{(1)}$ arises as a non--relativistic
reduction of the covariant nucleon currents, the nature of this ambiguity
is most clearly characterized by expanding the operator in powers of $v/c$.
We carry this expansion out to second order in $v/c$, since $F_1$ and
$G_\sst{E}$ differ by terms of ${\cal O}(\tau)$. For a spin--0 nucleus, the
spin--dependent part of $\rohat^{(1)}$ may be neglected, so that the time
component of the single--nucleon current involves
$$
F_1-\tau F_2 -{1\over 2}\tau F_1
	\>=\> G_\sst{E}(1-{1\over 2}\tau)+{\cal O}(\tau^2)\eqno\nexteq
$$
up to terms of ${\cal O}(v/c)^2$ associated with spinor normalization.
The factor $1-\coeff{1}{2}\tau$ multiplying $G_\sst{E}$
cancels from the hadronic ratio and yields the result in Eq.~(\Emonobas) to
${\cal O}(\tau^2)$.
In short, the use of $G_\sst{E}$ form factors effectively accounts for all
relativistic corrections through order $\tau$,
with truncation errors entering only at order $\tau^2$. Since $\GES$ is
proportional
to $\tau$, the error involved in making this approximation is only of ${\cal
O}(\tau)$ relative to the leading term. Additional corrections associated with
relativistic many--body dynamics should enter at the same  order in $\tau$
or $v/c$. For further discussion, see Ref.~[Mus92a].

Within the context of the approximation in Eq.~(\Emonobas), one can analyze
the errors in a potential determination of
$G^{(s)}_E$ in terms of the ``extended'' Galster parameterization of
$G^{(s)}_E$ (see Eqs.~(\Egalstr) and (\Egalstrxi)) to obtain
$$
\eqalign{{\delta G^{(s)}_E\over G^n_E} = -\left( {G_D^V\over G^n_E}\right)
\Bigl[ &\left\{ 2\sin^2\theta_W \left[ 1 + R^{T=0}_V + \Gamma\right] +
\left[ 1 + R^{(0)}_V \right] \rho_s \tau \xi_\sst{E}^{(s)}\right\} \left(
{\delta \alr\over \alr}\right) \cr
&- 2 \sin^2\theta_W \left\{ \left(
{\delta\sin^2\theta_W\over \sin^2\theta_W} \right) \left[ 1 + R^{T=0}_V +
\Gamma\right] + \delta R^{T=0}_V + \delta\Gamma\right\}\Bigr]\ \ , \cr}
\eqno\nexteq\nameeq\Ezersen
$$
where we have used $\GEn$ to set the size scale for $\GES$.
{}From this result we observe that the sensitivities of
the form factor $G^{(s)}_E$ to
$\delta R^{T=0}_V$, $\delta\Gamma$, and $\left(
\delta\sin^2\theta_W/\sin^2\theta_W\right)$ are roughly the same.
For $|\rho_s \tau\xi_\sst{E}^{(s)}|
<\!\!< 1$,
the sensitivity to $\left( \delta \alr/\alr\right)$
is also similar to that of the other uncertainties.

     Neglecting for the moment the uncertainties in $\sstw$, $\Gamma$, and
$R_V^{T=0}$, let us consider only the correlation between $\delta\GES$ and
$\delta\alr$.  Fig.~4.6 shows the statistical uncertainties for elastic
scattering from $^4$He as a function of $\tau$ for different scattering
angles under the experimental conditions of ${\cal L}=5\times 10^{38}$
cm$^{-2}$s$^{-1}$, $P_e = 100\%$, $T=1000$ hours, and $\Delta\Omega = 0.01$
sr and $\Delta\Omega = 0.16$ sr.  The curves in the two cases
are similar, although the use of the larger solid angle for $\theta\approx
30^{\circ}$ would permit a slightly higher level of precision than that
attainable in the small solid angle regime.
The break at $\tau\approx 0.11$
corresponds to the diffraction minimum in the elastic form factor of
$^4$He.  Also shown are the uncertainties in $\alr$ required to reach
limits on $\GES$ of $|{\delta\GES/\GEn}| = 1.4$ (dashed line --- corresponding
to that expected from a forward--angle $\vec ep$ experiment with
$\delta\mu_s\approx 0.22$) and $|{\delta\GES/\GEn}| = 0.7$
(dotted line --- roughly the best one could expect
from $\vec ep$ under the circumstances discussed in Sect.~IV.A).  In both
cases $\lamsE$ has been chosen to be
zero (see Sect.~III.C). Clearly,
there are both low--$|Q^2|$ ($\tau\approx 0.04$) and somewhat higher--$|Q^2|$
($\tau\approx 0.2$) regions where PV elastic scattering from $^4$He holds
the promise of a much better determination of $\GES$ than even the best
conditions for $\vec ep$ (see below). Similar plots for elastic scattering from
$^{12}$C are also given in Ref.~[Mus92a] and similar conclusions can be drawn.
In general, however, the kinematic ranges over which elastic scattering is
preferable to $\vec ep$ are much greater for $^4$He than for $^{12}$C and
therefore, other conditions being equal, $^4$He provides the greater
flexibility.  This results from the fact that, although the cross section
appearing in the FOM
is larger at low--$|Q^2|$ for ${}^{12}$C than for ${}^4$He due to the
$Z^2$ coherence factor in the charge form factor, the $^{4}$He
cross section falls off less rapidly with $|Q^2|$ (see Figs.~3.10--3.12).
Moreover, a larger
achievable luminosity has been assumed for a helium target.

     As can be seen from Fig.~4.6, the statistical error in
$\alr(0^+0)$ that
must be achieved to match the best $\vec ep$ determination of $\GES$ are
approximately 5\% at $\tau = 0.04$ and 25\% at $\tau = 0.2$.  These are to
be compared with the other sources of possible uncertainty in $\GES$ from
Eq.~(\Ezersen).
The isospin symmetry breaking correction $\Gamma$ is very model--dependent
but, for the favorable cases such as $^4$He and $^{12}$C, even if
$\delta\Gamma \approx \Gamma$, this uncertainty is still unimportant, since
$\Gamma\rightarrow 0$ rapidly as $|Q^2|\rightarrow 0$ and should be no larger
than 0.01 at the higher values of $\tau$ considered.
The error in $\sstw$ is currently (see Sect.~II.C.1) ${\delta\sstw/\sstw}
\approx$ 1--2\% and can be expected to be better than 1\% in the next few
years.

Note that the maximum allowable statistical uncertainty grows with
energy for constant $\delta\GES/\GEn$.
Since the weighting of other sources of
uncertainty is roughly the same as that of $\delta \alr/\alr$ in their
impact on $G^{(s)}_E$ in Eq.~(\Ezersen), the maximum allowable
$(\delta\alr/\alr)_{\rm stat}$ also sets the scale for the maximum allowable
uncertainty from other sources.  For example, consider an attempt to measure
$\rho_s$ to roughly $\pm 0.7$ precision.  For an experiment
performed at $30^\circ$ and $\tau=0.04$ (near the first minimum in
Fig.~4.6b), a 1\%
error in $\sin^2\theta_W$ would be larger than $(\delta\alr/\alr)^{\rm stat}$
and would, therefore, rule out a determination of $\rhostr$ to this
precision.  However, at $\tau = 0.2$ (the second minimum in Fig.~4.6b),
the same error in $\sin^2\theta_W$ falls well below the maximum
allowable statistical error and should not, therefore, be problematic in this
case.

        A second way to analyze the prospective value of such
determinations of $\GES$ is
to consider the constraints imposed on the $\rhostr$ and $\lamEs$ parameters
in the ``Galster'' parameterization of $\GES$ (see Sect.~III.C).
In Ref.~[Mus92a] prospective $^4$He measurements were analyzed for kinematics
corresponding to $\tau = 0.04$ and $0.2$ at $\theta=30^\circ$.
Since $\GES$ is as yet not known, the authors of Ref.~[Mus92a]
considered different parameterizations for $\GES$; here we employ two of those,
{\it viz.,\/}
(A) $(\lamEs, \rhostr) = $ ($\lamn$,\ 0) and (B) $(\lamEs,
\rhostr) = $($\lamn$,
-2), where the value of $\rhostr=-2$ corresponds roughly
to the average prediction of Ref.~[Jaf89].
The results of this analysis are shown in Fig.~4.7.  For comparison,
the constraints from the $\evec p$ determination of
$\GES$ discussed previously are also included. The
bands in the figure correspond to the
uncertainty in $\rhostr$ associated with the uncertainty in $\alr$ for
given values of $\lamEs$.

        As Fig.~4.7 illustrates, it does not appear to be possible to constrain
the value of $\lamEs$ in the event that
$\rhostr$ is zero (parameterization A above).
Moreover, in this case, the lower--energy measurement gives
the tightest constraints in the $(\lamEs, \rhostr)$ plane. No new information
is added by a higher--energy experiment. In contrast, if $\rhostr\not=0$, it
is in principle possible to restrict the uncertainty in both $\rhostr$ and
$\lamEs$ with a combination of ${}^4$He measurements. In none of these cases,
however, would the $\evec p$ results add significant information to
the series of helium measurements. We consider in Sect.~IV.F.3 the
possibility that a high--$|Q^2|$ QE measurement would contribute additional
constraints to those of Fig.~4.7.

Several other features of these results are worth noting. First, the
statistical precision in $\alr$ associated with the lower--energy constraints
is roughly one percent (see Ref.~[Mus92a]). Since other sources of uncertainty
enter into a determination of $\delta\GES$ at roughly the same level as
$\delta\alr/\alr$, as mentioned above in discussion Eq.~(\Ezersen), one
might worry about the impact of these as yet unquantified uncertainties ---
particularly those associated with dispersion corrections ---
on the low--energy bands in Fig.~4.7. In contrast, the higher--energy
measurement
corresponds to $\delta\alr/\alr$ on the order of 10\%, so that theoretical
uncertainties would only become problematic when they reach this scale.

\bigskip
\goodbreak
\noindent IV.B.2\quad STANDARD MODEL TESTS

     In the absence of $s$--quark contributions, radiative corrections, and
isospin--mixing, the PV asymmetry of Eq.~(\Ezerostd) is simply proportional
to $\sstw$ and historically PV electron scattering from $0^+0$ nuclei was
viewed as a test of the Weinberg--Salam model and a measure of $\sstw$ (see,
{\it e.g.,\/} Refs.~[Fei75, Wal77]).  Use
of such an experiment for a contemporary test of the Standard Model would
require a measurement accurate enough to allow determination of $R_V^{T=0}$
to within $\pm 0.01$.  Such a requirement demands a statistical uncertainty
in $\alr$ of $\delta\alr/\alr \approx 1\%$, as well as knowledge of the
other terms in Eq.~(\Ezerostd), $\Gamma (q)$ and $\GES/G_E^{T=0}$, to the same
level of accuracy --- we now proceed to consider the latter.

As discussed in Sect.~III.D.4, drawing upon the work of Ref.~[Don89] we
expect $|\Gamma (q)| < 0.01$ at least for the lightest $0^+0$ nuclei such as
$^4$He and $^{12}$C.  The correction $\Gamma(q)\rightarrow 0$ as
$q\rightarrow 0$, since the isovector charge operator
${\hat\rho}^{T=1}\rightarrow {\hat T_3}^{\rm tot}$
in this limit and since the states are eigenstates of this
operator. At moderate values of momentum transfer, however, $\Gamma(q)$ need
not be small.
In particular, for nuclei heavier than $^{16}$O the correction can be several
percent; however, for nuclei in the $1s$-- and $1p$--shells the correction
is somewhat
smaller because of the difficulty of sustaining an isovector breathing
mode in the relevant nuclear shell model space (see Ref.~[Don89]).
The most favorable cases for PV electron scattering studies appear to be
$^4$He and $^{12}$C. Of the two targets, ${}^4$He is  advantageous from the
standpoint of interpretability. The first excited state in ${}^4$He lies at
20.1~MeV as compared to 4.44~MeV in ${}^{12}$C, thereby introducing less
likelihood
of contamination from inelastic events.  Furthermore, luminosity loss over the
length of the target is less problematic for ${}^4$He than for ${}^{12}$C
(see Sect.~V.C). On the other hand,
the ${}^{12}$C FOM is larger than that of ${}^4$He at very low--$|Q^2|$ due
to the larger value of $Z$, thereby increasing the range of energies over
which a 1\% Standard Model test could be performed.

As shown in Fig.~4.5, uncertainty in the
strange--quark contribution, even when limited by the ``best'' determination
of $\GES$ possible from a $\vec ep$ experiment as discussion in Sect.~IV.A,
renders a determination of $R_V^{T=0}$ at the level of $\delta R_V^{T=0}$
$\approx \pm 0.01$ impossible.  Only in a very
limited kinematic range for small $\tau$ can one expect to reduce the
uncertainties to the necessary level.  As $\tau\rightarrow 0$,
$\GES\rightarrow 0$ and the significance of the
strange--quark term, even if $\delta\GES \approx \GES$, is reduced.  On the
other hand, the asymmetry $\alr$ also $\rightarrow 0$ as $\vert
Q^2\vert\rightarrow 0$ and the statistical uncertainty attainable for given
experimental conditions begins to increase at very small $\tau$.  For fixed
$\lamEs$, it is possible with either the high-- or low--$|Q^2|$ ${}^4$He
$\GES$ determination to reduce the uncertainty in $\rhostr$ to well below
what is needed to make a 1\% $\sstw$ determination for the
kinematic ranges discussed above.  Once $\lamEs$ is allowed to vary, however,
this statement no longer holds, as discussed in Ref.~[Mus92a].
However, from that work it was concluded that a
low--$|Q^2|$ $\GES$ determination using $^4$He could be sufficient to keep the
$\GES$--induced error below a problematic level for a ${}^{12}$C Standard
Model test.

To evaluate the prospect of running a low--$|Q^2|$ experiment, the
$\left(\delta A/A\right)_{\rm stat}-\tau$ relationship for potential
${}^{12}\hbox{C}({\svec e},e)$ and ${}^4\hbox{He}(\evec, e)$
experiments under conditions possible at MIT/Bates, Mainz, or CEBAF was
analyzed in Ref.~[Mus92a].  The resultant curves are shown in Fig.~4.8.
In arriving at these plots, luminosities of
${\cal L}[{}^{12}\hbox{C}]= 1.25\times 10^{38}\,\hbox{cm}^{-2} s^{-1}$ and
${\cal L}[{}^{4}\hbox{He}]= 5\times 10^{38}\,\hbox{cm}^{-2} s^{-1}$, 100\% beam
polarization, and a running time of 1000 hours were assumed, together with
two choices of solid angle, $\Delta \Omega = $ 0.01 and 0.16 sr.
The solid curves give the statistical uncertainty $\delta
\alr/\alr$ as a function of $\tau$, for different values of $\theta$.
The two
sets of dashed curves give the $G^{(s)}_E$--induced error in $\alr$ for two
different values of $\rho_s$,  1.4 and 0.7 (where the former
corresponds to the ``ideal'' ${\svec e}p$ limits discussed in Sect.~IV.A).
In order to perform a
precision $(\lapp 1\%$) Standard Model test, both the statistical and
$G^{(s)}_E$--induced uncertainties must fall below 1\%.  These requirements
thus determine the appropriate kinematics.

{}From these results, we observe first that although the low--$|Q^2|$ FOM is
larger
for ${}^{12}$C than for ${}^4$He, owing to the difference in $Z$, the larger
luminosity achievable for ${}^4$He compensates for most of this difference,
leading to roughly comparable
$(\delta \alr/\alr)_{\rm stat}$ in most cases.  In
the very--forward--angle regime, experiments on these targets would
need to be carried out at $\theta\lapp 15^\circ$ and at energies in the
400 to 800~MeV range.  The lower bound is set by the requirement that $(\delta
\alr/ \alr)_{\rm stat}
 \le 1\%$, while the upper limit is determined by the uncertainty
in $\rho_s$.  The range of allowable energies, for a given angle, is slightly
larger for ${}^{12}$C than for ${}^4$He.
The general characteristics for the intermediate--angle case are similar.
Here, the decrease in FOM, relative to very forward angles, is compensated by
the larger solid angle.  However, lower energies are needed in order to
keep the statistical and $\delta \alr/\alr$ error below 1\%.

We emphasize that the limits corresponding to $\delta\rho_s = \pm 1.4$ apply
only in the event that a sequence of high precision, forward-- and
backward--angle
$\alr({\svec e}p)$ measurements are performed.  In particular, the
$\evec p$ experiments proposed thus far would not be sufficient to reach
this level.

     Even if an experimental determination of
$R_V^{T=0}$ can be accomplished at the required level, one must be able to
interpret the experimental value.  As discussed in Sect.~III.B, $R_V^{T=0}$
receives contributions from at least three possible sources:
one--quark radiative
corrections in the Standard Model, non--Standard--Model physics, and hadronic
processes beyond the one--quark terms.  The first of these is in principle
calculable to arbitrary accuracy given values for the $t$--quark and Higgs
masses with a much stronger dependence on the former than the latter.
Physics beyond the minimal Standard Model may be expressed in terms of the
$S$ and $T$ parameters described in Sects.~II.C and
III.B.  The relevant contribution
here enters roughly as 0.016 $S$ so that a determination of
$R_V^{T=0}$ to $\delta R_V^{T=0} \approx \pm 0.01$ would limit
the uncertainty in $S$ to
$\delta S < 0.6$ assuming the other contributions to $R_V^{T=0}$ were
precisely known.  Atomic PV experiments using, {\it e.g.,\/} $^{133}$Cs, show a
similar sensitivity to $S$; current results from atomic PV give $S =
-2.7\pm2.3$.

     Unfortunately, it is not presently clear that one can calculate the
hadronic contribution to $R_V^{T=0}$ with sufficient accuracy to permit a
competitive determination of $S$.
The largest such contribution probably arises from the
dispersion corrections discussed in Sect.~III.B.  In particular, since
$\alr$ is a {\it ratio} of the $Z^0$--exchange to the photon--exchange
term in the one--boson exchange approximation, it is the {\it difference}
between two--photon dispersion correction and the photon--$Z^0$ dispersion
correction that enters $R_V^{T=0}$.
As discussed in Sect.~III.B these two
corrections have significantly different analytic structures and
$Q^2$--dependences.  Furthermore, analysis of
a recent $(e,e)$ experiment
on $^{12}$C and a comparison of $e^+$ and $e^-$ scattering from both
$^{12}$C and $^{208}$Pb suggest that the two--photon dispersion corrections
may enter at the level of a few percent, a significantly larger
contribution than previously estimated theoretically.  Therefore one clearly
needs to develop a greater confidence in the reliability of theoretical
calculations of such dispersion corrections before one can expect to study
physics beyond the minimal Standard Model.

\bigskip
\goodbreak
\noindent IV.B.3\quad DETERMINING THE GROUND--STATE NEUTRON DISTRIBUTION

     In the previous discussion we have relied upon the isospin symmetry of
$T=0$ nuclei to select only the isoscalar piece of the hadronic current.
In the more general case of $N\neq Z$ (but still spin--0) nuclei,
$T\neq 0$ and consequently both the isoscalar and isovector currents enter.
In this case using Eq.~(\Encurff a) one has
$$
-{\alr\over\alrzer} = { {{\tilde F}_{C0}}\over {F_{C0}} } =
{ {\sqrt{3}\xi_V^{T=0}F_{C0}(T=0)
+\xi_V^{T=1}F_{C0}(T=1)
+\xi_V^{(0)}F_{C0}(s) }
\over
{2[F_{C0}(T=0) + F_{C0}(T=1)]} }
\ \ .\eqno\nexteq\nameeq\Ejzntz
$$
For convenience let us define the following sum and difference of the
$T=0,1$ monopole form factors:
$$
F_{C0}(\pm)\equiv F_{C0}(T=0) \pm F_{C0}(T=1) \eqno\nexteq
$$
and then Eq.~(\Ejzntz) may be re--written using Eqs.~(\Ewncpn)
$$
\eqalign{-2{\alr\over\alrzer} &= \xi_V^p + \Bigl( { {\xi_V^n F_{C0}(-)
+ \xi_V^{(0)} F_{C0}(s)} \over {F_{C0}(+)} } \Bigr) \cr
&\equiv (\xi_V^p + {N\over Z} \xi_V^n) \Bigl\{1 + {\tilde\Gamma}(q)
\Bigr\} \ \ , \cr}\eqno\nexteq\nameeq\Enzasym
$$
where following Ref.~[Don89] we have introduced the $q$--dependent quantity
$$
{\tilde\Gamma}(q) = \xi'_V \Bigl\{ \Bigl[ 1 - {{F_{C0}(-)/N}\over{F_{C0}(+)/Z}}
\Bigr] - \Bigl[ {{\xi_V^{(0)}}\over{\xi_V^n}} {{Z F_{C0}(s)}\over{N F_{C0}(+)}}
\Bigr] \Bigr\}\ \ . \eqno\nexteq\nameeq\Egtilneu
$$
Here
$$
\xi'_V \equiv - \Bigl[ 1 + {{Z \xi_V^p}\over{N \xi_V^n}} \Bigr]^{-1}\ ,
\eqno\nexteq
$$
which only varies slightly with the choice of target ($\xi'_V \approx -1.1$
at tree--level for all $0^+$ nuclei).  Note also that
$\xi_V^{(0)}/\xi_V^n=1$ at tree--level.

If initially we ignore the strangeness form factor and use the definitions
of the form factors written in terms of Fourier transforms of ground--state
matrix elements of the appropriate density operators (see, for example,
Refs.~[deF66, Don75, Don79a]), we have (for one--body operators)
$$
\eqalign{F_{C0}(+) &= {1\over\sqrt{4\pi}}\int
d{\vec x}\ j_0(qx)\ \rho_p ({\vec x}) \cr
F_{C0}(-) &= {1\over\sqrt{4\pi}}\int
d{\vec x}\ j_0(qx)\ \rho_n ({\vec x})\ \ , \cr}
\eqno\nexteq
$$
where $\rho_p(\rho_n)$ is the proton (neutron) density normalized to
$Z(N)$.  Then ${\tilde\Gamma}(q)$ in Eq.~(\Egtilneu) becomes
$$
{\tilde\Gamma}(q) = \xi'_V \Bigl\{ 1 -
{ {\int d{\vec x}\ j_0(qx)\ \rho_n({\vec x})/N} \over
{ \int d{\vec x}\ j_0(qx)\ \rho_p({\vec x})/Z} }
\Bigr\} \eqno\nexteq\nameeq\Egtilnos
$$
or, equivalently, the PV asymmetry involves
$$
-2{\alr\over\alrzer} = \xi_V^p + \xi_V^n
{ {\int d{\vec x}\ j_0(qx)\ \rho_n({\vec x})} \over
{ \int d{\vec x}\ j_0(qx)\ \rho_p({\vec x})} }\ \ .
\eqno\nexteq\nameeq\Ernoverrp
$$
At tree--level, the first term in Eq.~(\Ernoverrp) becomes
$\xi_V^p  = (1-4\sstw) \approx
0.092$, while the coefficient of the second term
$\xi_V^n = -1$ and accordingly the second term is dominant.  The Fourier
transform of the proton density occurring in the denominator in the second
term of Eq.~(\Ernoverrp) is determined by unpolarized electron
scattering.  Thus, {\it measurement of the PV asymmetry in this case is
nearly a direct measurement of the (Fourier transform of the) neutron
density.}

     The ratio of the Fourier transforms of the neutron and proton
densities occuring in the dominant term in Eq.~(\Ernoverrp) has some
interesting properties.  For pure $T=0$ nuclei, $\rho_n=\rho_p$, this ratio
is 1 and Eq.~(\Ernoverrp) yields the simple result $\sqrt{3}\xi_V^{T=0}$
discussed above (as expected).
For $T=0$ nuclei with small isospin impurities introduced {\it
e.g.,} by Coulomb interactions between protons, $\rho_n\approx\rho_p$ and
deviations of this ratio from unity are related to a nonzero $\Gamma$ term in
Eq.~(\Ezerostd).  For $T>0$ (usually $N>Z$) nuclei, the ratio deviates from
1 even at $q=0$.  In addition, if $\rho_n$ and $\rho_p$ have different
radial dependences, the Fourier transforms of these densities will have
different diffraction structure with diffraction minima occurring at
different $q$.  In this case the PV asymmetry will show rather dramatic
deviations from the simple monotonic increase with momentum transfer that
obtains for $\rho_n/N=\rho_p/Z$.  This ``beat'' phenomenon thus leads to a very
sensitive dependence of $\alr$ to {\it differences} in the proton and
neutron densities.

Including the strangeness contribution in Eq.~(\Egtilneu)
does not significantly alter the conclusions reached here.  Instead of
$\rho_n$ in Eqs.~(\Egtilnos) and (\Ernoverrp) it is only necessary to
make the replacement
$$
\rho_n\longrightarrow [\rho_n]_{\rm eff}\equiv \rho_n +
{1\over 2} \{ \rho_p + \rho_n \} \Bigl( {{G_E^{(s)}}\over{G_E^{T=0}}}
\Bigr) \ \ , \eqno\nexteq\nameeq\Erhowstr
$$
again treating the density operators as one--body operators.  For the
very small values of momentum transfer where one might attempt to
extract the neutron distribution (see below) the second term in Eq.~(\Erhowstr)
is only about a percent or so of the first term, the one that involves the
neutron distribution.  Of course, were extremely high--precision called for,
then the second term could be treated as a correction with $G_E^{(s)}$ as
a quantity that would have been determined from the other measurements
described elsewhere in the present work.

Examples of how sensitive a probe of the ground--state neutron
distribution PV elastic electron scattering can be are shown in Fig.~4.9
which is taken from Ref.~[Don89].
Despite many years of effort using a variety of experimental probes,
detailed knowledge of neutron densities in nuclei remains extremely
limited.  The authors of Ref.~[Don89] have concluded that a measurement of
the PV asymmetry for $^{208}$Pb at $q\approx 0.5$ fm$^{-1}$ could provide a
determination of the radius of the neutron density to an accuracy of 1\%;
they also conclude that such an experiment is eminently feasible given
reasonable beam and detector parameters.
Because such experiments are both difficult and
costly it is unlikely that $(\vec e,e)$ will be used in an extensive
program to map out neutron densities in a variety of nuclei.  However, it
should be possible to use it to provide critical ``benchmark'' measurements
against which to compare both theoretical predictions and other
experimental probes of neutron densities.

\vfil\eject

%

\def\qvecsq{\vec q^{\mkern2mu\raise1pt\hbox{$\scriptstyle2$}}}

\def\xiva{{\xi_\sst{V}^{(a)}}}

\def\evec{{\vec e}}
\def\sigpin{{\Sigma_{\pi\sst{N}}}}

\def\alrzer{{A_\sst{LR}^0}}
\def\mevoc{{MeV/c}}

\catcode`@=11 
\def\lsim{\mathrel{\mathpalette\@versim<}}
\def\gsim{\mathrel{\mathpalette\@versim>}}
\def\@versim#1#2{\lower0.2ex\vbox{\baselineskip\z@skip\lineskip\z@skip
  \lineskiplimit\z@\ialign{$\m@th#1\hfil##\hfil$\crcr#2\crcr\sim\crcr}}}
\def\taud{{\tau_{\rm d}}}
\catcode`@=12 

\noindent{\bf IV.C.\quad Elastic Scattering from the Deuteron }
\medskip

As with elastic scattering from $(0^+0)$ targets, elastic $^2\hbox{H}(\evec,
e)$ scattering serves as an isospin \lq\lq filter", since the ground state
is nominally $T=0$. Thus, as with the $(0^+0)$ cases discussed in the
previous section, one might hope to eliminate some of the uncertainties
associated with the multitude of form factors entering $\evec p$ elastic
scattering through the use of a deuterium target. On the other hand, since
this nucleus has spin--1, the elastic asymmetry depends on isoscalar magnetic
dipole, axial--vector dipole and Coulomb quadrupole form factors as well as the
Coulomb monopole form factors entering
$\alr(0^+0)$. For this reason, one or more measurements of $\alr(^2\hbox{H})$
could nicely complement the experiments discussed in the foregoing sections as
a means for further constraining the strangeness form factors. In fact, the
isoscalar character of this nucleus enhances the sensitivity of
$\alr(^2\hbox{H})$ to $\GMS$ at backward angles by roughly
$\mu_p/\mu^\sst{T=0}$ over the corresponding sensitivity of $\alr(^1\hbox{H})$,
while reducing the impact of theoretical uncertainties in the axial--vector
contribution. Moreover, uncertainties associated with the nuclear wave function
appear to be tolerably small at low--$|Q^2|$. Consequently, one might hope to
improve the constraints on $\mustr$ by a factor of two over the \lq\lq ideal"
constraints with $\evec p$ elastic (see Eq.~(\Edmfromdr)). Elastic
scattering from deuterium does present a challenge to experimental
resolution, given the
small (2.22 MeV) binding energy of the ground state. It appears, however, that
an experiment which admits a non--negligible amount of inelastic contribution
would not seriously impair the extraction of interesting constraints on
$\GMS$. Furthermore, while no analogous enhancement factor arises in the
longitudinal
contribution to  $\alr(^2\hbox{H})$, it appears that a moderate--energy,
moderately--forward--angle measurement of this asymmetry could permit
tighter constraints on $\GES$ than are possible with $\evec p$ elastic
scattering alone.

	To illustrate the rationale for these conclusions, we return to the
expressions for the deuteron asymmetry given in Sect.~III.E (see
Eq.~(\Easydeut)).  Let us extend the leading--order result in Eq.~(\Easylow)
to the following:
$$
-2{\alr\over \alrzer}\equiv \delone+\deltwo+\delthree \ \ ,
\eqno\nexteq
$$
where $\delone$ gives the tree--level Standard Model contribution in the
absence of strangeness, $\deltwo$ contains the strangeness contribution to
deuteron vector current form factors, and $\delthree$ contains the
axial--vector terms.
The latter two terms vanish in the absence of strangeness and electroweak
corrections.
In terms of the quantities introduced in Sect.~III, we have
$$
\eqalign{\delone&=
                \sqrt3 \xi_V^{T=0} \gae
                 = -4\sstw[1+R_V^{T=0}+\Gamma]\cr
         \deltwo&=
                \xi_V^{(0)}\gae\left[
                          v_L\left(F_{C0}(T=0) F_{C0} (s)
                                + F_{C2}(T=0) F_{C2} (s) \right)
                        + v_T F_{M1}(T=0) F_{M1} (s)
                                        \right] \cr
	&\qquad\qquad\qquad \ \times\left[v_L(F_{C0}^2(T=0)
	+F_{C2}^2(T=0))+v_TF_{M1}^2(T=0)
                                \right]^{-1}\cr
         \delthree&=-v_{T'}\gve F_{M1}(T=0) \left[
                          \xi_A^{T=0}F_{E1}^5(8)
                        + \xi_A^{(0)}F_{E1}^5(s)
                        + \beta F_{E1}^5(AM)
                                        \right] \cr
	&\qquad\qquad\qquad \ \times\left[v_L(F_{C0}^2(T=0)
	+F_{C2}^2(T=0))+v_TF_{M1}^2(T=0)
                                \right]^{-1}\ .\cr}
\eqno\nexteq\nameeq\EDeld
$$
We have neglected in $\delthree$ a term that is second--order in the nominally
small quantities $\xi_A^{T=0}$ and $F_{M1}(s)$.
The term containing $\beta F_{E1_5}(AM)$ is
generated by the many--body nuclear anapole moment (see Sect.~III.D.5), where
$F_{E1_5}(AM) = \rbra{\rm g.s.}\That^{\rm el 5}_1\rket{\rm
g.s}^\sst{EM}$ and $\beta$ is defined in Eq.~(\Ebeta).
The AM is one of a number of many--body effects which do not enter in
scattering
from single nucleons. Nor does it arise in scattering from $(0^+0)$ targets,
since it requires a target having nonzero spin.

	The term $\delone$ is analogous to the leading term in the $(0^+0)$
asymmetry. It arises from the piece of the isoscalar neutral vector
current proportional to $J_\mu^\sst{EM}(T=0)$ and is, therefore, nominally
independent of nuclear physics. Nuclear corrections enter via both
$\rvtez$ ({\it e.g.}, dispersion corrections) and $\Gamma$, which represents
the mixing of $T\not= 0$ continuum states into the deuteron ground state.
For purposes of this discussion, we will take the nucleus--independent parts
of $\delone$ to be sufficiently well determined from experiments in other
sectors and focus on $\deltwo$ and $\delthree$. In what follows, we let
$\deltwoe$ and $\deltwom$ denote the contributions to $\deltwo$ from
$\GES$ and $\GMS$, respectively.

      After casting the results of Ref.~[Pol90] in the formalism outlined
in Sect.~III, we obtain the following impulse approximation (IA) expressions
for the deuteron form factors:
$$
\eqalign{\sqrt{4\pi}\fcz(T=0)&=\sqrt{3(1+\taud)}\Bigl\{\GETEZ\Bigl[
	 (1-\coeff{2}{3}\taud)\dc+\coeff{2}{3}\taud\dme
	   +\coeff{4}{9}\taud^2\dq\Bigr] +\coeff{2}{3}\taud
	   \GMTEZ\dmm\Bigr\}\cr
         \sqrt{4\pi}\fct(T=0)&=\sqrt{2(1+\taud)}\taud\Bigl\{\GETEZ
	 \Bigl[(1+\coeff{2}{3}\taud)\dq-\dc
	 +\dme\Bigr] +\GMTEZ\dmm\Bigr\}\cr
         \sqrt{4\pi}\fmo(T=0)&=-
	 {q\over\md}\sqrt{1+\taud}(\geiso\dme+\gmiso\dmm)\cr
         \sqrt{4\pi}F_{E1_5}&=-\GATEZ\dmm \ \ , \cr
}
\eqno\nexteq\nameeq\Efmuld
$$
where $\taud\equiv |Q^2|/4\md^2$ and
where $\dc , \dq , \dmm , \dme $ and $ D_A$ are integrals dependent on
the S-- and D--state components of the deuteron wave function defined in
Ref.~[Pol90]. The vector NC form factors $\tilde F_\sst{C0}$,
$\tilde F_\sst{C2}$, and $\tilde F_\sst{M1}$ have identical forms to
the EM form factors in Eq.~(\Efmuld) with
the replacements $\GETEZ\to\geteztil$
and $\GMTEZ\to\gmteztil$. Making
use of these formulae, we now consider $\Delta_{(2,3)}$ at forward and
backward angles. Results for other kinematic regimes may be found in
Ref.~[Pol90].

\bigskip
\goodbreak
\noindent IV.C.1\quad BACKWARD--ANGLE SCATTERING

Let us begin by considering backward--angle PV electron scattering. Using
Eqs.~(\EDeld) and (\Efmuld) and the fact that $v_\sst{L}/v_\sst{T} <\!\!< 1 $
as $\theta\to 180^o$ we have
$$
\eqalign{\deltwo&\rightarrow -\left\{
                {{G}_M^{(s)}\over G_M^{T=0} }
                 + \lambda{{G}_E^{(s)}\over G_M^{T=0} }
                                \right\}
                                \left[1-\delnuc+R_V^{(0)}
                                \right]\ \ ,\cr
}
\eqno\nexteq\nameeq\EDlimd
$$
where $\lambda \equiv D_M^E/D_M^M$, and
$$
\eqalign{
\delnuc\ \equiv\ &{{\lambda(G_E^{T=0}/G_M^{T=0})}\over{
                        1+\lambda(G_E^{T=0}/G_M^{T=0})}
                        }\ \approx \ \lambda(G_E^{T=0}/G_M^{T=0})\ \ . \cr
}
\eqno\nexteq\nameeq\Ednucd
$$
The terms in Eqs.~(\EDlimd) and (\Ednucd)
arise purely from the ratio of magnetic deuteron form factors,
$F_{M1}(s)/F_{M1}(T=0)$.
Contributions from the Coulomb form factors
are highly suppressed by $v_\sst{L}/v_\sst{T}$ at backward angles. The
presence of the nucleon form factors $\GETEZ$ and $\GES$
in the deuteron magnetic
form factors arises from magnetic projections of the convection part of the
one--body  vector current:
$$
{\vec J}(q)_{\rm conv}=G_\sst{E}(q^2){(\pv+\ppv)\over 2\mn}
\eqno\nexteq\nameeq\Ejconvd
$$
and similarly for the strange--quark vector current.
Since the contribution of Eq.~(\Ejconvd)
to the magnetization, ${\vec m}(\rv)=\coeff{1}{2}(
\rv\times{\vec J})$, is proportional to the orbital angular momentum operator,
contributions containing $\geiso$ and $\GES$ to $F_{M1}(T=0)$ and
$F_{M1}(s)$
vanish for a pure S--wave. Hence, the
contributions involving $G_E$
are weighted by the integral $\dme$, which depends only on
the D--state component of the deuteron wave function. The correction factors
$\delnuc$ and $\lambda$ thus {\it vanish} in the
limit that D--wave components of the deuteron are neglected.

        Simple estimates for the scale of terms in Eqs.~(\EDlimd) and
(\Ednucd) may
be obtained by considering sufficiently small values of $q$ such that
terms of order $(qr/2)^2$ and higher in the Bessel functions appearing
in the $D$--integrals are negligible.  Since the percent of D--wave in the
deuteron is generally considered to be $\leq 10\%$ of the total, we
have that the ratio $\dme/\dmm$ is approximately
$$
\lambda = {\dme\over\dmm}\approx
\Bigl({3\over 4}\Bigr){\%\hbox{D-wave}\over\%\hbox{S-wave}}\lsim 0.08\ \ .
\eqno\nexteq
$$
Moreover, at $\vert Q^2\vert=0$, one has
$\geiso(0)=\coeff{1}{2}$ and $\gmiso(0)=$
$\coeff{1}{2}(\mu_p+\mu_n)\approx0.44$. Hence, at $\vert Q^2\vert=0$,
$$
\delnuc\approx\
\biggl({\geiso\over\gmiso}\biggr)\biggl({\dme\over\dmm}\biggr)\lsim 0.1\ \ .
\eqno\nexteq\nameeq\Ednucappd
$$
The scale of $\deltwom(\theta\to 180^o)_\sst{IA}$ is thus given by
$\mustr/\muiso$ with a small ($\leq 10\%$) nuclear physics correction.
The model--dependence of this correction has been analyzed in Ref.~[Pol90],
using nonrelativistic deuteron wave functions.
For example, the quantity $\delnuc$  does depend on momentum
transfer, running from $\approx 0.05$ at $\vert Q^2\vert
=0$ to $\approx 0.2$ at
$|Q^2| =$ 0.5 \gevocsq. The latter number is quite dependent on
the details of the deuteron wave function. The Bonn
potential yields a value of $0.13$, while the Paris potential gives
$0.23$, and Reid soft core $0.29$. (None of these nonrelativistic
models should be considered to be totally reliable at momentum transfers
much above
this scale, in any case.)  At low--$|Q^2|$, however, $\delnuc$ is
constrained to the value in Eq.~(\Ednucappd) by static
properties of the deuteron,
primarily the ratio of D-- to S--wave content.

The result of Eq.~(\EDlimd) has two notable features. First, the presence
of $\muiso$ rather than $\mu_p$ in the denominator of $\deltwom$
enhances the backward--angle $\mustr$ signal by roughly a factor of six
over the corresponding signal in $\alr(\evec p)$. Second, the small
magnitude of the nuclear physics correction to $\alr(\theta\to
180^o)_\sst{IA}$ implies that theoretical uncertainties in the deuteron
wave function should not introduce serious theoretical ambiguities
into the interpretation of a backward--angle measurement.  As
discussed above, for $|Q^2| < 0.1$ \gevocsq, $\delnuc \lapp 0.1$
in all deuteron models considered, with an uncertainty of less than
$\pm 0.03$.

{}From Eqs.~(\EDlimd) and (\Ednucd) we also observe that
the backward--angle strangeness
radius signal is suppressed with respect to the $\mustr$ term
by the same factor which minimizes the nuclear
physics correction in $\deltwom$. Specifically,
$$
{\deltwoe\over\deltwom}\Biggr\vert^\sst{IA}_{\theta\to 180^o}\approx
\> {\tau\rho_s\over\mustr} \lambda \approx 0.56\tau
\eqno\nexteq
$$
for small $|Q^2|$ and using the values for $\mustr$ and $\rhostr$ from
Ref.~[Jaf89].  If
these estimates are realistic,  one would need to go to extremely large
values of $|Q^2|$ in order for $\deltwoe$ to compete with $\deltwom$ at
backward angles. Consequently, unless the ratio $|\rho_s/\mustr|$ is much
larger than the prediction of
Ref.~[Jaf89], a nonzero value of the strangeness radius should pose no
serious difficulty for the extraction of a value of $\mustr$ from low--$|Q^2|$,
$\theta\to 180^o$ measurements.

Following a similar line of reasoning as above, we arrive at the
following low--$|Q^2|$, backward--angle signal from the axial--vector term of
$\delthree$:
$$
\eqalign{\delthree(\theta\to 180^o)_\sst{IA}&\approx
        \gve\left({v_{T'}\over v_T}\right) {M_D\over q}\sqrt{1+\taud}\cr
        &\quad\times\left\{
        {\tilde G_A^{T=0}\over G_M^{T=0}} - {Q_e\over\gve}
                \left[ 16\pi\alpha \over \sqrt3 G_\mu m_N^2\right]
                {{\rbra{D}\hat a \rket{D}} \over {G_M^{T=0}D_M^M}}
               \right\}[1-\delnuc]\ \ , \cr}
\eqno\nexteq
$$
where $\rbra{D}\hat a \rket{D} $ is the elastic matrix element of
the anapole operator defined in Eq.~(\Eanap)

It is interesting to compare the relative
sensitivities of $\alr(^1\hbox{H})$ and $\alr(^2\hbox{H})$ to the
dominant sources of uncertainty. In the latter case, one has
$$
{\delta \alr\over\alr}\approx 2\delta\mustr +{1\over 5}{\md\over q}
\delta\GATEZ\ \ , \eqno\nexteq
$$
which, for an experiment performed at the SAMPLE kinematics ($q\approx
300$ MeV/c) leads to
$$
{\delta \alr\over\alr}\approx 2\delta\mustr +{6\over 5}\delta\GATEZ\ \ .
\eqno\nexteq
$$
Comparing this result with Eq.~(\Epprotd),
one observes that the elastic
deuteron asymmetry is much more sensitive to $\mustr$ than is
$\alr(^1\hbox{H})$. Moreover, although $\mustr$ and
uncertainties in the axial--vector form factors are correlated at the
same level in both cases, one has that $\delta\GATEZ <\!\!< \delta\GATEO$
from the effects of radiative corrections. Thus, in contrast to the
situation with $\evec p$ scattering,  the impact of axial--vector radiative
correction uncertainties on a determination of $\mustr$ with $\evec D$
scattering appears to be less problematic.

To ascertain the potential constraints on $\mustr$ one might achieve with
a measurement of $\alr(^2\hbox{H})$, we show in Fig.~4.10 the fractional
statistical uncertainty in  the asymmetry as a function of energy for
fixed scattering angle, assuming reasonable experimental conditions (see
figure caption). On the same plot we give the induced uncertainty from
$\mustr$ associated with the \lq\lq ideal" SAMPLE determination of
this parameter. From these curves it is apparent that one might hope to
improve upon the best case SAMPLE results by a factor of two by
performing a large--solid--angle elastic $\evec D$ experiment at $\epsilon
\approx $ 200 MeV.

Such an experiment, of course, would have to confront the issue of resolution.
For a detector like SAMPLE, it is not possible to
resolve the ground state and consequently the asymmetry would
receive contributions from excitation of continuum states (see
Ref.~[Hwa86] for a discussion of $\alr(^2\hbox{H})$ near threshold). In
Ref.~[Mus92a] the impact of contributions from excitation of the first
continuum state ($^1$S$_0$) on a determination of $\mustr$ was analyzed.
A simple estimate for $q=300$ MeV/c leads to
$$
\delta\mustr\approx {3\over 20}{f^{\rm inel}\over f^{\rm el}}\Bigl[\delta\rateo
+\delta\Delta_\sst{MEC}\Bigr]\ \ , \eqno\nexteq
$$
where $f^{\rm el}$ ($f^{\rm inel}$) is the fraction of the total cross section
arising from elastic (inelastic) scattering, $\delta\rateo$ is the theoretical
uncertainty in the isovector axial--vector radiative corrections, and
$\delta\Delta_\sst{MEC}$ is a correction due to meson--exchange currents.
Using the estimate of $\delta\rateo = \pm 0.2$ from Ref.~[Mus90]
one has $\delta\mustr\approx 0.03\ (f^{\rm inel}/f^{\rm el})$ from radiative
correction uncertainties. One expects uncertainties
in the meson--exchange correction to contribute at a similar, if not smaller,
level. To leading order in $v/c$, the latter contribute only to
the magnetic transition matrix element appearing in the denominator of
the asymmetry. This contribution, dominated by $\pi$--exchange at low
momentum transfer, is well constrained by experimental knowledge of
threshold deuteron electrodisintegration [Are82, Auf85, Ber71, Gan72,
Hoc73, van91].  Hence,
to the extent that one can experimentally restrict inelastic
contributions to those arising from the isovector transition to the first
continuum state, the inclusion of inelasticities should not seriously
weaken the $\mustr$ constraints attainable from purely elastic $\evec D$
scattering.

\bigskip
\goodbreak
\noindent IV.C.2\quad FORWARD--ANGLE SCATTERING

        In the extreme forward direction, the axial--vector term
$\delthree$ vanishes
since $v_{T'}\propto\tan\theta/2$. Thus, we need only consider
$\deltwom$ and $\deltwoe$ in this kinematical regime. In contrast to the
backward--angle limit, where $v_\sst{T}>\!\!>v_\sst{L}$,
the $\theta\to 0^o$ vector current terms
receive contributions from both the charge and magnetic form
factors. In this limit, using similar arguments as employed previously to
estimate the scale of the
wave function integrals, one has from Eqs.~(\EDeld)
and (\Efmuld) the following IA expressions for $\deltwom$ and $\deltwoe$:
$$
\eqalign{\deltwom(\theta\to 0^o)^\sst{IA}&\approx
                -2{\GMS G_M^{T=0}\over (G_E^{T=0})^2} \taud
\cr
         \deltwoe(\theta\to 0^o)^\sst{IA}&\approx
                -{\GES\over G_E^{T=0}}
                 \ \ . \cr}
\eqno\nexteq\nameeq\EDelfd
$$
In arriving at Eqs.~(\EDelfd),
we have omitted terms of ${\cal O}(\taud)$ as well as those
involving the $\dq$ integral.
Since this integral, which depends only on the D--state component of the
wave function, vanishes as $Q^2$, it should be negligible in comparison with
$\dc$, $\dmm$, and $\dme$ at small values of momentum transfer. We have
also neglected  small additional nuclear physics corrections $\sim \delnuc$.

Comparing Eqs.~(\EDelfd) and (\EDlimd), we observe that
$\deltwom(\theta \to 0^o)^\sst{IA}$ is suppressed with respect to the
corresponding
backward--angle limit by roughly $ 0.4\ (q/\md)^2$. On the other hand,
$\deltwoe$, which is suppressed with respect to the ratio
$\GES/\geiso$ by $\lambda\approx 0.1$ at backward angles, now differs
from $\GES/\geiso$ by only a small, $Q^2$--dependent nuclear physics
correction. Thus, one expects the $\rhostr$ signal as $\theta\to 0^o$ to be
relatively much stronger in comparison with the $\mustr$ signal than at
backward angles. Moreover, since $\GES$ vanishes as $Q^2$ for small
momentum transfer, the ratio $\deltwoe/\deltwom$
should be largely independent of momentum transfer in the low--$|Q^2|$ limit.
In particular, neglecting the small nuclear physics corrections and taking
$\md\approx 2\mn$, we have
$$
{\deltwoe(\theta\to 0^o)^\sst{IA}\over
         \deltwom(\theta\to 0^o)^\sst{IA}}\approx
                {2}{\rho_s\over \mustr}
         {\GETEZ\over\GMTEZ} \approx 15
\eqno\nexteq
$$
using the estimates of Ref.~[{Jaf89}].  Thus, the strangeness radius signal at
forward angles could be relatively significant.

	To estimate the constraints that a forward--angle $\alr(^2\hbox{H})$
measurement could place on $\rhostr$, we have considered a 1000 hour
measurement at $\theta=30^\circ$, assuming a luminosity of
${\cal L}=5\times 10^{38} {\rm cm}^{-2}{\rm s}^{-1}$, solid angle
$\Delta\Omega=0.16$ sr, and beam polarization $P_e=100\%$. Under
these conditions, the most stringent constraints on $\rhostr$ would be
obtained at energy $\epsilon\approx $ 800 MeV, for which the statistical
uncertainty would be roughly 2\% . The resulting uncertainty on
$\rhostr$ would be about a factor of five smaller than is achievable
with an \lq\lq ideal" series of $\alr(\evec p)$ measurements and comparable
to the constraints attainable from a measurement of $\alr(^4\hbox{He})$
at low--$|Q^2|$. The technical feasibility and impact of inelastic
contributions
remain to be analyzed.

In summary, the use of $\vec e D$ scattering as an additional
constraint on the nucleon strangeness form factors has several
advantageous features: enhanced sensitivity due to
the nucleon's small isoscalar magnetic moment and its small axial--vector
isoscalar form factor, weaker sensitivity to axial--vector radiative
corrections than in the free proton case, and the possibility of a
kinematic separation of $\mu_s$ effects from $\rho_s$ effects. Moreover,
the impact of nuclear wave function uncertainties appears to be
negligible at low--$|Q^2|$. Finally, in the case of backward--angle
scattering, it appears that an experiment could integrate into some of
the inelastic region without seriously affecting a determination of
$\mustr$. These features suggest that further theoretical study of
contributions from the anapole moment, dispersion corrections, and
relativistic effects as well as experimental analysis of technical
feasibility appear warranted.

\vfil\eject

%

\def\qvecsq{\vec q^{\mkern2mu\raise1pt\hbox{$\scriptstyle2$}}}

\def\xiva{{\xi_\sst{V}^{(a)}}}

\def\evec{{\vec e}}
\def\sigpin{{\Sigma_{\pi\sst{N}}}}

\def\alrzer{{A_\sst{LR}^0}}

\noindent {\bf IV.D.\quad Other Discrete Nuclear Transitions}
\medskip
\nobreak
In this section we consider a few examples of PV electron scattering
involving discrete nuclear transitions other than those discussed
above.  The treatment presented here is not meant to be comprehensive:
while the procedures followed are straightforward extensions of the
formalism applied above to elastic scattering from spin--0 systems
(see Sect.~IV.B), it is generally necessary to deal with several
multipole matrix elements when considering transitions between nuclear
states which have non--trivial angular momenta and this significantly
complicates the analysis.  On the one hand, the ability to select specific
transitions with good quantum numbers for the initial and final nuclear
states can, in principle, be an advantage for then the focus can be
placed on particular pieces of the electroweak current.
On the other hand, given the proliferation of different electroweak matrix
elements that can occur in general, the
description of the asymmetry now usually becomes more dependent on details
of the nuclear modeling than is the case for the very special monopole
situation that occurs for elastic scattering from spin--0 nuclei.  Both of
these aspects of the problem will be illustrated in the discussions to
follow using specific nuclear transitions.

It is clear that under certain circumstances the PV asymmetry
may be quite sensitive to the single--nucleon
content and therefore it would appear that choosing the transition and
kinematics judiciously might provide a means to explore, for instance, the
nucleon's strangeness form factors (given that the nuclear many--body
uncertainties are not too large).  However, as discussed in detail below,
this is usually not the case.  The reason is not one of principle, but of
practice: for most discrete nuclear transitions of interest other than
elastic scattering, which is coherent and so has a large cross section, the
figure--of--merit introduced in Sect.~III.E is very small. As a consequence,
unfortunately it is rather
unlikely that high--precision determinations of the PV asymmetry can be
contemplated for most discrete nuclear transitions in the foreseeable future.

\bigskip
\goodbreak
\noindent IV.D.1\quad ILLUSTRATIVE EXAMPLES OF TRANSITIONS IN $N=Z$ NUCLEI

To make the treatment reasonably tractable in this subsection let us restrict
our attention to $N=Z$ nuclei with ground states having
angular momentum and isospin 0 with positive parity.  The discrete excited
states will be assumed to have good quantum numbers $J^{\pi}T$ and
accordingly specific transitions may be classified in the following
ways: first, the transitions are either isoscalar ($T=0$) or isovector
($T=1$).  Secondly, they fall into two classes, natural parity
transitions where $\pi=(-)^J$:
$$
\eqalign{
F^2_L &= F_{CJ}^2 \cr
F^2_T &= F_{EJ}^2 \cr
W^L_{AV} &= -F_{CJ}{\tilde F}_{CJ} \cr
W^T_{AV} &= -F_{EJ}{\tilde F}_{EJ} \cr
W^{T^{\prime}}_{VA} &= (1-4\sin^2 \theta_W) F_{EJ}{\tilde F}_{MJ_5}
\cr}\eqno\nexteqp\nameeq\Einelpi
$$
and non--natural parity transitions where $\pi=(-)^{J+1}$ and one has
$$
\eqalign{
F^2_L &= 0 \cr
F^2_T &= F_{MJ}^2 \cr
W^L_{AV} &= 0 \cr
W^T_{AV} &= -F_{MJ}{\tilde F}_{MJ} \cr
W^{T^{\prime}}_{VA} &= (1-4\sin^2 \theta_W) F_{MJ}{\tilde F}_{EJ_5}
\ \ ,\cr}\eqno\sameeq
$$
using Eqs.~(\Efltmul) and (\Ewltmul) in Sect.~III.E.1.
Clearly the special case of elastic scattering from spin--0 systems discussed
in Sect.~IV.B is contained in Eq.~(\Einelpi a), {\it i.e.,\/} when only Coulomb
monopole form factors are retained: $F^2_L = F_{C0}^2$,
$W^L_{AV} = -F_{C0}{\tilde F}_{C0}$ and $F^2_T = W^T_{AV} =
W^{T^{\prime}}_{VA} =0$.  In the discussions to follow we shall compare
the figures--of--merit for a few selected transitions with that for elastic
scattering from spin--0 nuclei and hence it is appropriate to re--state
the result for the latter (see also Sect.~IV.B):
$$
{\cal F}={\cal F}_0{{\cal E}\over{1-{\cal E}}}
2|Q^2/q^2|{\tilde F}_{C0}^2 \ \ .\eqno\nexteq\nameeq\Efinelz
$$

The general behavior of the FOM has been discussed in
Sect.~III.E.2.  There elastic scattering from the proton, $^4$He and
$^{12}$C was compared with three particular representative inelastic
excitations in $^{12}$C.  Let us now examine the asymmetries and
figures--of--merit for these illustrative examples in somewhat more detail to
bring out the reasons why the specific results shown in Figs.~3.10--3.12 are
obtained.

\bigskip
\goodbreak
\noindent\undertext{Non--Natural Parity Transitions}

Let us start with non--natural parity transitions, since this case is somewhat
simpler than the natural parity case as no
Coulomb matrix elements occur.  The cross section is proportional to
the square of the magnetic form factor ($\propto F_{MJ}^2$), whereas
from Eq.~(\Especa) the asymmetry is
$$
{\alr\over\alrzer} = -[{\tilde F}_{MJ}-
\sqrt{1-{\cal E}^2}(1-4\sin^2 \theta_W){\tilde F}_{EJ_5}]/F_{MJ}
\ \ .\eqno\nexteq
$$
Since in general $F_{MJ}$ and ${\tilde F}_{MJ}$ have different
$q$--dependences and hence different locations of their diffraction
minima, it is clear that the asymmetry can become quite large near
a minimum in $F_{MJ}$.  While that may seem at first sight to be
advantageous for studying PV electron scattering, it should be
remembered that the FOM characterizes the precision
with which the asymmetry can be determined.  Equation~(\Efinel) yields
$$
{\cal F}={{\cal F}_0\over{1-{\cal E}}}\Bigl[{\tilde F}_{MJ}-
\sqrt{1-{\cal E}^2}(1-4\sin^2 \theta_W){\tilde F}_{EJ_5}\Bigr]^2
\ \ ,\eqno\nexteq\nameeq\Efnnp
$$
which does nothing special as one passes through the diffraction
minima of $F_{MJ}$.  Said in other words: it does no good to have
a large asymmetry but a small cross section if they conspire as they
do here to yield only a typical small figure--of--merit (see below), for then
using Eq.~(\Edoabl) the
fractional uncertainty in the asymmetry is large. As we
shall see below, under typical conditions the asymmetry itself is not
sufficiently sensitive to, for example, details of the strangeness or
axial--vector single--nucleon form factors to overcome this large fractional
uncertainty.

As particular examples, let us consider further the $J=1$, $T=0,1$ cases
shown in
Figs.~3.10--3.12 and use the following to characterize the form factors that
enter: $F_{M1}(T)\equiv -(q/M)G_M^T K_{M1}$,
${\tilde F}_{M1}(T)\equiv -(q/M){\tilde G}_M^T {\tilde K}_{M1}$ and
${\tilde F}_{E1_5}(T)\equiv -2{\tilde G}_A^T {\tilde K}_{E1_5}$, where
we have isolated the overall dependences on $q$
which must occur, as well as specific choices for the dependences on the
single--nucleon form factors (see below).
Isospin--mixing is ignored in this section.
$K_{M1}$, ${\tilde K}_{M1}$ and ${\tilde K}_{E1_5}$ contain all of the
remaining $q$--dependences; all may be nonzero at low momentum transfer.
The normalizations are chosen here in order that for specific
spin--flip--dominated cases, where the convection current contributions
are unimportant,  one has $K_{M1}={\tilde K}_{M1}={\tilde K}_{E1_5}$.
Specifically, for the excitation of the $1^+1$ state at tree level
we have from Eq.~(\Efnnp)
$$
{\cal F}={{\cal F}_0\over{1-{\cal E}}}\Bigl[0.546\ F_{M1}\Bigr]^2
\times \Bigl(1+\sqrt{1-{\cal E}^2}\Bigl\{ {{85\ {\rm MeV/c}}\over{q}} \Bigr\}
\Bigl[ {{G_D^A}\over{G_D^V}} \Bigr]\Bigl[ {{{\tilde K}_{E1_5}}\over{K_{M1}}}
\Bigr] \Bigr)^2 \ \ .\eqno\nexteq
$$
Taking ${\tilde K}_{E1_5}=K_{M1}$ for this spin--flip--dominated transition
(which is borne out by more detailed analysis; see Ref.~[Don79a]) together with
the minimum fractional uncertainties deduced from Figs.~3.10--3.12 we obtain
the following for the
fractional uncertainties of the axial--vector contributions to the
asymmetry: 300\% (10$^{\circ}$), 320\% (30$^{\circ}$) and
465\% (150$^{\circ}$). Clearly some dramatic (unanticipated) improvement
in experimental do--ability would have to occur before interesting
information concerning the isovector axial--vector form factor could be
extracted.

Similar conclusions are reached for the excitation of the $1^+0$ state in
$^{12}$C. Treating this as an eigenstate of isospin (which is known not to be
the case: see Refs.~[Don79a, Fla79] where isospin--mixing is discussed),
we have from Eq.~(\Efnnp)
$$
{\cal F}={{\cal F}_0\over{1-{\cal E}}}\Bigl[F_{M1}\Bigr]^2
\times {1\over 2}\Bigl( \Bigl[ {{{\tilde G}_M^{T=0}}\over{G_M^{T=0}}}\Bigr]
\Bigl[ {{{\tilde K}_{M1}}\over{K_{M1}}} \Bigr]
-\sqrt{1-{\cal E}^2}\Bigl\{ {{173\ {\rm MeV/c}}\over{q}} \Bigr\}
\Bigl[ {{{\tilde G}_A^{T=0}}\over{G_M^{T=0)}}} \Bigr]
\Bigl[ {{{\tilde K}_{E1_5}}\over{K_{M1}}} \Bigr]
\Bigr)^2 \ \ .\eqno\nexteq
$$
For simplicity let us assume that this transition is also
spin--flip--dominated (this is not as good an approximation as it is for
the excitation of the 15.11 MeV level in $^{12}$C, as discussed in
Ref.~[Don79a]; however it is adequate for the present purposes where we
merely wish to set the general scale of the problem), in which case the
ratios of $K$'s are both unity.  Furthermore, if we for example set the vector
strangeness form factor of the nucleon to zero and take the EMC value for the
axial--vector strangeness form factor (see Eq.~(\Egalstra)), we also have
${\tilde G}_M^{T=0}/G_M^{T=0}=-0.918$ and
${\tilde G}_A^{T=0}(0)/G_M^{T=0}(0) = -0.441\pm 0.070\pm 0.105$.
We then obtain the following
fractional uncertainties in the latter when the $1^+0$ figures--of--merit at
their first (second) maxima are employed: 3070\% (3245\%) for 10$^{\circ}$,
3240\% (3365\%) for 30$^{\circ}$ and 4705\% (4734\%) for 150$^{\circ}$.
If instead we ignore the axial--vector contribution (which can be made
small by working in the forward direction) and allow for a nonzero magnetic
strangeness contribution, then we have
${\tilde G}_M^{T=0}/G_M^{T=0}=-0.918\ (1 + 2.5\mu_s)$ using our
parameterization of $G_M^{(s)}$.  The fractional uncertainty in $\mu_s$ is
$|\delta A_{LR}/A_{LR}|/|\mu_s|$, implying from the numbers given above
that the magnetic strangeness could only be defined to several $\times$
100\% using the $1^+0$ state in $^{12}$C which is not competitive with
other cases discussed in Sects.~IV.A--C.
Clearly no useful information is likely to be forthcoming from studies of
this transition.

\bigskip
\goodbreak
\noindent\undertext{Natural Parity Transitions}

The natural parity class of transitions is somewhat more complicated. In
this case the PC cross section (Eq.~(\Edsinel)) involves both Coulomb and
electric form factors ($\propto 2|Q^2/q^2|{\cal E}F_{CJ}^2+F_{EJ}^2$) and the
hadronic ratio is given by
$$
W^{({\rm PV})}/F^2 =-\Bigl[{{2|Q^2/q^2|{\cal E}F_{CJ}{\tilde F}_{CJ}
+F_{EJ}{\tilde F}_{EJ}-\sqrt{1-{\cal E}^2}(1-4\sin^2 \theta_W)
F_{EJ}{\tilde F}_{MJ_5}}\over{2|Q^2/q^2|{\cal E}F_{CJ}^2+F_{EJ}^2}}
\Bigr]\ \ .\eqno\nexteq
$$
Using Eq.~(\Efinel) the FOM can be written in the following way:
$$
{\cal F}={{\cal F}_0\over{1-{\cal E}}}{{\Bigl[2|Q^2/q^2|{\cal E}
{\tilde F}_{CJ}+\gamma_J[{\tilde F}_{EJ}-
\sqrt{1-{\cal E}^2}(1-4\sin^2 \theta_W){\tilde F}_{MJ_5}]\Bigr]^2}\over
{2|Q^2/q^2|{\cal E}+\gamma_J^2}}
\ \ ,\eqno\nexteq\nameeq\Einfom
$$
where for convenience we have defined $\gamma_J\equiv F_{EJ}/F_{CJ}$.
In the long wavelength limit ($q\rightarrow 0$) the Coulomb and electric
multipole matrix elements are related by the continuity equation [deF66],
which implies that $\gamma_J\rightarrow -\sqrt{(J+1)/J}(\omega/q)$ in this
limit.  For example, in Figs.~3.10--3.12 the figure--of--merit is given for
the $2^+0$ state at 4.44 MeV in $^{12}$C.  The peak value of $\cal F$
occurs at $q\approx$ 265 MeV/c which yields $|\gamma_2|\approx$
$2.0\times 10^{-2}$ using the above approximation.  Experiment [Fla78]
gives $3.9\times 10^{-2}$, implying that the long wavelength limit, while
not precisely correct at such values of momentum transfer, is in fact a
reasonable measure of the relative importance of Coulomb and transverse
multipole contributions.  In either case, at least for this specific
transition, $|\gamma_2|$ is rather small and Eq.~(\Einfom) can roughly be
approximated by
$$
{\cal F}={{\cal F}_0\over{1-{\cal E}}}2|Q^2/q^2|{\cal E}
{\tilde F}_{CJ}^2 \eqno\nexteq
$$
except when $\cal E$ becomes very small ({\it i.e.,\/} at rather
backward angles) and consequently when the transverse multipoles
must be retained.  As seen in Figs.~3.10--3.12 at large $\theta$ the
FOM is diminished and consequently we shall focus on
forward--angle scattering.  There the FOM has the same
form as that found for elastic scattering (Eq.~(\Efinelz)) up to corrections of
order $\gamma_J$ for natural--parity transitions that behave as the
4.44 MeV transition in $^{12}$C does.  The requirement that must be met
is that the Coulomb/convection current currents must dominate.  Examples
with this character include the rather collective low--lying $2^+$ states
such as the 4.44 MeV level in $^{12}$C and the usual (convective) giant
dipole resonance. In contrast, certain
cases exist which are spin--flip dominated, such as the spin--flip
giant dipole resonance [Don68, Don70]; for these more detailed
analyses would need to be performed.

The general observation to be drawn from studies of natural--parity
transitions is
that they apparently offer no special advantage over elastic scattering
and, in fact, have the detriment that the FOM in the
latter case is generally more favorable due to the coherence of elastic
scattering.  A possible exception to this statement could come from
using high--energy forward--angle scattering to probe the weak neutral
current in nuclei at somewhat higher values of $q$ than may be permitted
with elastic scattering.  To be specific, in $^{12}$C
the elastic form factor falls off with increasing momentum transfer
sufficiently rapidly that the $2^+0$ competes successfully with it at
$q$ greater than about 265 MeV/c and then, for instance, using the results
for 10$^{\circ}$ (Fig.~3.10) for the $2^+0$ transition we find
$|\delta A_{LR}/A_{LR}|=$ 5.1\% at $\epsilon =$ 2.5 GeV ($q=$ 435 MeV/c).
The forward--angle asymmetry involves the ratio
${\tilde G}_E^{T=0}/G_E^{T=0}=-0.918\ (1+2.2[G_E^{(s)}/G_E^{T=0}])$
and, in contrast to the magnetization--dominated $1^+0$ case discussed
above, has only very small strangeness sensitivity because of the
suppression of $G_E^{(s)}$ at low momentum transfer.  Using the above
conditions this would result in a 45\% determination of $G_E^{(s)}$
(taking model (B) discussed in Sect.~IV.B.1 for the electric
strangeness form factor of the nucleon).  While this might constitute a
good alternative to elastic scattering, it should be remembered that
the transverse multipole contributes at some level (quantified above for
the $2^+0$ transition in $^{12}$C, which is one of the few that have
been adequately studied using PC electron scattering) and consequently
that additional nuclear model dependence enters the problem for any
case other than $0^{\pm}\rightarrow 0^{\pm}$ transitions where only the
monopole can contribute.

\bigskip
\goodbreak
\noindent IV.D.2\quad NUCLEAR TRANSITIONS: DISCUSSION AND SUMMARY

A few additional comments concerning discrete nuclear transitions are in
order.  One pertains to the question of experimental resolution in
studies of PV and PC electron scattering. As discussed in Sect.~V, when
extreme values of luminosity are demanded (implying thick targets) and
large solid angles are required to make high--precision measurements of
$A_{LR}$ it may not be possible to resolve individual discrete states.
Instead, under some conditions, it may be necessary to sum over a few
transitions to obtain a total asymmetry
$$
A_{LR}^{\rm tot}=\sum_i p_i A_{LR,i}\ \ ,\eqno\nexteq
$$
where $A_{LR,i}$ is the asymmetry for the i$^{\rm th}$ transition and
$p_i\equiv (d\sigma/d\Omega)_i/[\sum_j (d\sigma/d\Omega)_j]$ with
$(d\sigma/d\Omega)_i$ the PC cross section for the i$^{\rm th}$ transition.
An example where this might occur is when the $2^+0$ transition in $^{12}$C
is summed together with elastic scattering.  Again using the results
for scattering at forward angles shown in Fig.~3.10, we see that only about
6\% of the total asymmetry comes from the inelastic transition.  Of
course, for very high precision studies of the PV asymmetry this may be
significant.  For instance, as discussed in Sect.~IV.B, one issue that
arises is that of isospin--mixing, which, being theoretically uncertain
(see the treatment in Ref.~[Don89]), could lead to an intolerable uncertainty
in the total asymmetry.  Clearly the problem is only worse with poorer
resolution where more transitions must be summed and accordingly where
a larger fraction of the total asymmetry comes from inelastic excitations.

A related comment also involves summing over several transitions.  If
the range of excitation involves only $T=0$ states in $N=Z$ nuclei
({\it e.g.,\/} takes into account only states below 15.11 MeV in $^{12}$C,
ignoring isospin--mixing problems), then the tree--level asymmetry is
proportional to
$$
\Bigl(W^{({\rm PV})}/F^2\Bigr)^{\rm tot}=0.454 + \sum_i p_i
\Bigl(W^{({\rm PV})}/F^2\Bigr)_i^{(s)}\ \ ,\eqno\nexteq\nameeq\Einwtot
$$
where ``(s)'' indicates contributions coming only from terms with
nonzero strangeness.  Insofar as complications from isospin--mixing can
be ignored (or modeled successfully) the total asymmetry could be used
to obtain some indication about the presence or absence of strangeness: if
experiment yielded the result 0.454 for Eq.~(\Einwtot), then it would at
least be suggestive that the strangeness contributions were small.  Of
course, accidentally they could cancel in the sum on the right--hand side
of the equation and no conclusion could be drawn.  Such a measurement
might serve to get a first glimpse at the importance of strangeness in
electron scattering in that, being summed over several levels, it would have
the maximal FOM attainable for a given nucleus.

In this section we have focused on discrete transitions in $N=Z$ nuclei, since
the analysis is reasonably tractable and serves to set the scale for nuclei
in general.  Odd--$A$ nuclei can, of course, also be investigated, although
the proliferation of multipoles that occurs for non--trivial values of
$J_i^{\pi_i}$ and $J_f^{\pi_f}$ takes us beyond the intent of the present
work.  Only one case will be mentioned --- that of elastic scattering
from spin--1/2 targets (see also Sect.~IV.E).  Focusing first on
forward--angle scattering where the
FOM is expected to be large, we have for this case that
the PV asymmetry is proportional to
$$
W^{({\rm PV})}/F^2={{\rho_C + \delta^2 \rho_M}\over
{1 + \delta^2}}\ \ ,\eqno\nexteq
$$
where $\rho_C\equiv -{\tilde F}_{C0}/F_{C0}$, $\rho_M\equiv
-{\tilde F}_{M1}/F_{M1}$ and $\delta\equiv -|q^2/Q^2| F_{M1}/\sqrt{2}F_{C0}$.
In the long wavelength limit we have the following:
$\rho_C \rightarrow 0.5\ (N/Z)-0.046$ neglecting strangeness, $\rho_M$ is
of order unity and
$\delta\rightarrow (q/2M)\times(\mu/Z)$, where $\mu$ is the magnetic
moment of the nuclear ground state in nuclear magnetons. (Recall also
from the general arguments above that $|Q^2/q^2|\approx 1$.) Consider a
case such as $^{207}$Pb where $\mu=$ 0.584.  Using a momentum transfer of
about 100 MeV/c ({\it i.e.,\/} employing the estimate $q_0$ given above)
yields $\delta\cong 3.8\times 10^{-4}$ and therefore for such high--$Z$
cases the transition is completely dominated by the C0 multipole.  Even
for a case as light as $^{13}$C (with $\mu=$ 0.7022, $q_0\cong$ 163 MeV/c)
one finds $\delta\cong 1.0\times 10^{-2}$ and the magnetic effects
are typically only a percent or so at low momentum transfer. For
backward--angle scattering the axial--vector dipole form factor also
plays a role.  Defining $\rho_E\equiv {\tilde F}_{E1_5}/F_{M1}$
the complete hadronic ratio may be written
$$
W^{({\rm PV})}/F^2={{\rho_C + \left[ \rho_M +(1-4\sstw) \sqrt{1-{\cal E}^2}
\rho_E \right] \delta^2/{\cal E} }\over
{1 + \delta^2/{\cal E} }}\ \ .\eqno\nexteq
$$
Backward--angle scattering corresponds to ${\cal E} <\!\!< 1$ and thus
the factor $\sqrt{1-{\cal E}^2}$ may safely be taken to be unity.  For
the remaining $\cal E$--dependence two regimes exist: one is at extreme
angles where ${\cal E} <\!\!< \delta^2 $, in which case the hadronic
ratio goes to $\rho_M +(1-4\sstw) \rho_E$, yielding a relatively large
sensitivity to the axial--vector form factor.  However, this regime is
very hard to study.  For example, even for a nucleus as light as $^{13}$C
one has $\delta\sim 10^{-2}$ and therefore the angle at which
${\cal E}=\delta^2$ is so far backward ($178^{\circ}$) as to prove
impractical.  The other regime is where $\delta^2<\!\!< {\cal E}<\!\!<1$.
This corresponds to angles between about 90$^{\circ}$ and 170$^{\circ}$,
using the fact that ${\cal E}\approx 1/3$ for $\theta = 90^{\circ}$.
In this regime the axial--vector term is not suppressed appreciably
because of the factor $\sqrt{1-{\cal E}^2}$; however, the factor
$\delta^2/{\cal E}$ is unfortunately rather small ({\it e.g.,\/}
about 1\% at $\theta\approx 165^{\circ}$ and even smaller at more forward
scattering angles).  Consequently, it would appear that it will be rather
difficult to learn anything about the axial--vector contributions to
elastic scattering if they enter with typical single--particle
strengths.  On the other hand, if these contributions are enhanced by
collective effects (see the discussions of the anapole moment in
Sect.~IV.E), then it may be possible to study them in this way.

The arguments
here can be generalized to include elastic scattering from nuclei with
spins greater than 1/2.  The basic conclusion will be unchanged: at low--$q$
and forward angles where the FOM is large, the elastic
scattering PV asymmetry will come almost entirely from the C0 multipole in
all but the lightest nuclei.  This implies that elastic scattering can be
used to determine the radius of the ground--state neutron distribution
even in odd--$A$ nuclei (see the discussions in Sect.~IV.B).  Of special
interest in this regard is the case of $^{133}$Cs because of its importance
in studies of atomic PV (see Sect.~IV.I).

Our conclusions for the prospects of studying $A_{LR}$ for discrete
nuclear transitions are mixed: on the one hand, the case of elastic
scattering (from spin--0 nuclei especially, but also as mentioned above from
nuclei with spins greater than zero) is likely to be quite important as
it provides a way to isolate the effects of $G_E^{(s)}$ from the other
nucleon form factors and consequently will complement studies of the
proton (see Sects.~IV.A and B).  A few transitions such as the excitation
of the $2^+0$ state at 4.44 MeV in $^{12}$C might add to the elastic
scattering studies and perhaps early, poor resolution experiments could
yield some information about the rough level at which strangeness
enters; however, in most cases we have found that the figure--of--merit
for excitation of discrete states is too low for there to be much hope
that useful information on the electroweak currents will be forthcoming
from PV electron scattering in the foreseeable future.

\vfil\eject

\def\rvec{{\vec r}}
\def\kappatil{{\tilde\kappa}}
\def\Ftil{{\tilde F}}

\noindent{\bf IV.E \quad Axial Hadronic Response}
\medskip

Thus far, we have focused primarily on ways in which nucleon structure
physics and the underlying lepton--quark electroweak interactions manifest
themselves in the hadronic response functions. As alluded to earlier, however,
these response functions are also sensitive to the many--body physics
of the target nucleus. In this respect, one case of particular interest
involves the piece of the PV hadronic response which involves nuclear matrix
elements of the axial current [Don86b]:
$$
W_{VA}^{T'}\propto\hbox{Re}\Bigl\lbrace\bra{f}J_{+1}^\sst{EM}\ket{i}
\bra{f}J_{+1,\ 5}^\sst{NC}\ket{i}^* -\bra{f}J_{-1}^\sst{EM}\ket{i}
\bra{f}J_{-1,\ 5}^\sst{NC}\ket{i}^* \Bigr\rbrace
$$
where \lq\lq$\pm 1$" refer to the transverse components of the currents and
the subscript \lq\lq 5" denotes the hadronic weak NC. As discussed in
Section III.D.5, the presence of PV $NN$ interactions in the nucleus requires
one to replace multipole matrix elements of $J_{\lambda 5}^\sst{NC}$ with
the sum $\bra{f}\Ohat^5_J\ket{i}^\sst{NC}+\beta\bra{f}\Ohat_J^5
\ket{i}^\sst{EM}$, where $\beta$ is given in Eq.~(\Ebeta). The many--body
physics of interest --- in this case the PV $NN$ force --- is contained in
$\bra{f}\Ohat_J^5\ket{i}^\sst{EM}$. Although we postpone a more detailed
discussion of atomic PV until Section IV.I, we note that a special
case of $\bra{f}\Ohat_J^5\ket{i}$ --- the nuclear anapole moment ($f=i$
and $\Ohat_J^5\to\That_1^{{\rm el} 5}$) --- is also accessible with atomic PV
observables. In tandem with the hadronic axial NC, the anapole moment (AM)
induces a nuclear spin--dependent (NSD) PV atomic Hamiltonian
$$
{\cal H}^\sst{PV}_{\rm atom}(\hbox{NSD}) = {G_\mu\over 2\sqrt{2}}{\tilde
\kappa}{\vec I}\cdot
\int\ d{\vec x}
\ \hat\psi_e^\dagger(\vec x) {\vec\alpha}\psi_e(\vec x)
\delta(\rvec)\eqno\nexteq\nameeq\Eatomanap
$$
where $\vec\alpha$ is the vector of Dirac matrices acting in the space of
lepton spinors, $\vec I$ is the nuclear spin, $\rvec$ is the electron
coordinate, and the strength $\tilde\kappa=\kappatil_\sst{NC}+
\kappatil_\sst{AM}$ is determined by nuclear PV ($\kappatil_\sst{AM}$) and
the weak NC interaction ($\kappatil_\sst{NC}$) [Fla80]. The interaction
in Eq.~(\Eatomanap) mixes atomic states of good parity, thereby giving rise
to PV atomic observables, such as optical rotation or circular polarization
of incident radiation [Nov75]. Since $\kappatil_\sst{NC}\sim
\kappatil_\sst{AM}$ for heavy and/or nearly degenerate nuclei, a measurement
of an appropriate atomic PV observable can provide access to nuclear PV.

        In seeking to study nuclear PV with charged leptons, one faces two
challenges: (a) finding a nucleus in which the contribution from nuclear
PV ({\it e.g.}, the AM or $\bra{f}\Ohat_J^5\ket{i}^\sst{EM}$) is at least
commensurate with that of the axial NC, and (b) to find an observable
in which the total axial contribution is large enough to be observed
experimentally. For both atomic PV and PV electron scattering, meeting
the latter requirement is a non--trivial exercise. Fortunately, in order to
place interesting new constraints on nuclear PV, one need not measure the
axial response with the same degree of high--precision as required by
electroweak tests or probes of nucleon strangeness. With these comments
in mind, we discuss three cases which have received some attention as possible
means for studying nuclear PV.

\medskip
\noindent\undertext{Atomic PV in heavy atoms}
\medskip

        The experimental difficulty in achieving sufficient precision to
observe atomic PV can be mitigated by performing experiments with heavy
atoms. To illustrate the reasons why, consider the mixing matrix element
between $S_{1/2}$ and $P_{1/2}$ states in an atom such as Cs. In the limit
that only single--electron states enter this mixing, one has [Nov75]
$$
\eqalignno{\bra{S_{1/2}}&{\cal H}^\sst{PV}_{\rm atom}(\hbox{NSID})+
{\cal H}^\sst{PV}_{\rm atom}(\hbox{NSD})\ket{P_{1/2}}\cr
&\cr
&={iG_\mu\over 2\sqrt{2}}Z^2 {\cal N}\Bigl\lbrace Q_\sst{W}-
\kappatil\bigl[{2\gamma +1\over 3}\bigr]g_i[F(F+1)-I(I+1)-\coeff{3}{4}\bigr]
\Bigr\rbrace\ \ \ ,&\nexteq\nameeq\Espmix\cr}
$$
where NSID and NSD refer to the nuclear spin--independent (Eq.~(\FHpva))
and spin--dependent (Eq.~(\Eatomanap)) components, respectively,
of the atomic Hamiltonian; $Q_\sst{W}$ is the weak charge defined in
Eq.~(\Fapvme); {\cal N} is a structure--dependent overall normalization
factor; $\gamma=\sqrt{1-\alpha^2 Z^2}$; $g_i\bra{\hbox{g.s.}}\sigv_n
\ket{\hbox{g.s.}}=\bra{\hbox{g.s.}}\vec I\ket{\hbox{g.s.}}$,
with $\bra{\hbox{g.s.}}\sigv_n\ket{\hbox{g.s.}}$
being the spin of an unpaired nucleon and $\bra{\hbox{g.s.}}
\vec I\ket{\hbox{g.s.}}$ being the total nuclear spin in the ground state;
and $\vec F=\vec I+\vec J$ is the total atomic angular momentum. The NSD
part of the electron--nucleus interaction is extracted
from PV observables which
depend on $F$, such as transitions between different hyperfine levels. The
overall $Z^2$ factor appearing in Eq.~(\Espmix), in tandem with the
coherence enhancement of $Q_\sst{W}$ ($\approx N$) raises the scale of
the NSID contribution to the mixing matrix element to an observable level
for heavy atoms. In contrast, the NSD contribution receives no such
coherence enhancement and is thus suppressed by roughly a factor of $N$ with
respect to the NSID part. Moreover, in the absence of nuclear PV, one
has for atomic Cs $\kappatil=\kappatil_\sst{NC}\approx\coeff{2}{9}(1-4\sstw)
\approx 0.022$. Thus, the NSD part can be down by almost two orders of
magnitude relative to the NSD contribution. Given this difference in scale
between the two terms, it is significant that a value of $\kappatil=0.42
(23)$ has been extracted from the recent $^{133}$Cs atomic PV experiment
carried out by the Boulder group [Noe88, Blu90]. In that experiment, the
NSD contribution was determined from measurements of the rates for transitions
between different hyperfine levels involving parity--mixed $6S_{1/2}$ and
$7S_{1/2}$ atomic levels. In the same experiment, a value for $Q_\sst{W}$
with $\approx 2\%$ experimental error was obtained. The experimenters also
report that their data implies a 97\% probability that $\kappatil$ is larger
than zero. The experimental result has the same sign and order of magnitude
as the prediction of Ref.~[Fla84], in which a single--particle effective
PV nuclear Hamiltonian and Woods--Saxon potential were used. The large
scale predicted for $\kappatil_\sst{AM}$, relative to that of $\kappatil_\sst{
NC}$, results from two factors: (i) the nuclear PV contribution contains
no $\gve=-1+4\sstw$ suppression factor, and (ii) an nuclear enhancement
resulting from the $A^{2/3}$ scaling behavior of the nuclear AM. A new
calculation of $\kappatil_\sst{AM}$ using the full two--body PV nuclear
Hamiltonian and including PV MEC's (see Section III.D.5), is in progress
[Hax93].

        On the experimental side, the prospects for better limits on
$\kappatil_\sst{AM}$ appear encouraging. From a new atomic Cs
measurement in progress
at Boulder, one expects to reduce the experimental uncertainty by a factor
of four or five over the previous Cs determination [Wie93]. It is hoped that
future Cs experiments will bring about another factor of five improvement
in the anapole limit. In principle, one might hope to extract a limit
on $\kappatil_\sst{AM}$ from the atomic Pb PV experiment nearing completion
in Seattle. However, the natural Pb used in that experiment mostly consists
of $^{208}$Pb, whose ground state is spin--0 and cannot support an anapole
moment; the natural abundance of $^{207}$Pb (spin--$\coeff{1}{2}$) is only
about 20\%. Consequently, the expected scale of the contribution from
the $^{207}$Pb AM to the PV observable in this experiment is an order of
magnitude smaller than the level of statistical precision. A future atomic
PV measurement by the same group using thallium, however, is likely to
be sensitive to the Th AM at the level expected from theoretical predictions
[Lam93].

\medskip
\noindent\undertext{Elastic PV electron scattering}
\medskip

The possibility of studying nuclear PV with elastic PV electron scattering has
been explored in some detail by Serot for the case of $^{13}$C
[Ser79].  Since this nucleus has spin $J=\coeff{1}{2}$, the elastic asymmetry
receives contributions from Coulomb, magnetic dipole, and the $J=1$ axial
transverse electric multipole (see Eq.~(\Exdvec) of Section III.E).
Contributions from nuclear PV arise only in the last of these multipoles,
which can be decomposed according to Eq.~(\Emultmix) as
$$
\Ftil_{E1_5}\longrightarrow \Ftil_{E1_5}^\sst{NC}+\beta F_{E1_5}^\sst{EM}
\ \ \ .\eqno\nexteq\nameeq\Efetil
$$
For elastic scattering, conservation of the EM current implies that the EM
contribution to $F_{E1_5}$ is finite as $Q^2\to 0$. In this limit,
the EM term in Eq.~(\Efetil) behaves as
$$
F_{E1_5}^\sst{EM}\propto \rbra{\hbox{g.s.}}i\Rhat^{\rm el}_1\rket{
\hbox{g.s.}}\rightarrow {1\over\sqrt{6\pi}}{q^2\over\mns}\rbra{\hbox{g.s.}}
\ahat\rket{\hbox{g.s.}}\ \ \ .\eqno\nexteq\nameeq\Efetilem
$$
The leading $q^2$--dependence on the right side of Eq.~(\Efetilem)
cancels the $1/q^2$ (in the Breit frame) contained in $\beta$, rendering a
finite contribution to Eq.~(\Efetil) at zero $q^2$. One consequence of this
result is that even in the presence of nuclear PV, the $L$, $T$, and $T'$
contributions to the $^{13}$C PV response vary with $\theta$ for fixed
$\epsilon$ in a manner entirely analogous to the case of a proton target,
for which the axial contribution is largest at backward angles and lower
energies.

        Nuclear PV contributions to $\alr(^{13}\hbox{C})$ at back angles
were calculated in Ref.~[Ser79] using a one--body average of
${\hat H}^{[2]}_\sst{PV}$ and the [Des80]
value for $h_{\sst{NN}\pi}
\sim 5\times 10^{-7}$.
Only the $\pi$--exchange part of the PV Hamiltonian was included, and
contributions to $\Ftil_{E1_5}^\sst{EM}$ were not explicitly accounted for.
Nevertheless, the estimates of Ref.~[Ser79] should provide a rough guide
as to the scale and kinematic dependences of various contributions to
$\alr(^{13}\hbox{C})$. The results of this work indicate that the nuclear
PV and total NC (vector and axial vector) contributions to
$\alr(^{13}\hbox{C})$ at backward angles are comparable for low incident
electron energies ($\epsilon\lapp 100$ MeV), while the total NC contribution
dominates by one or two orders of magnitude for larger energies
($\epsilon\rapp 200$ MeV). Thus, the low--energy regime appears to be the
most appropriate for elastic scattering studies of nuclear PV. To estimate
the possible precision with which such a measurement could be made, we
consider a 1000 hour experiment at $\epsilon=50$ MeV and backward angles,
a luminosity ${\cal L}[^{13}\hbox{C}]=1.25\times 10^{38}{\rm cm}^{-2}{\rm
s}^{-1}$, solid angle $\Delta\Omega = 1$ sr, and 100\% beam polarization as
an illustrative case. Under these conditions, one expects a statistical
uncertainty in the asymmetry of 40\% . Were all other contributions to the
asymmetry known to infinite precision, and were $\pi$--exchange the dominant
contribution, one could then determine $h_{\sst{NN}\pi}$ with an uncertainty
of $\pm 2\times 10^{-7}$. Although the latter figure is somewhat larger
than the range on $h_{\sst{NN}\pi}$ obtained from the $^{18}$F experiment,
it is sufficiently small to provide one with a test of the $^{18}$F result.

        A more detailed analysis of this case would require a calculation
using the full, two--body PV Hamiltonian and including PV MEC's. In addition,
one would require an analysis of the $\theta$--dependence of the different
contributions to the asymmetry, given the large solid angle required to
obtain the above--mentioned statistical precision. We also note that if
the low--energy nuclear PV contribution to elastic scattering is dominated
by the leading $q^2$--dependence of matrix elements of
$\Rhat^{\rm el}_{1\lambda}$, then the $A^{2/3}$ scaling behavior of this
operator suggests that a heavier target could be more favorable than
$^{13}$C.

\medskip
\noindent\undertext{Inelastic PV electron scattering}
\medskip

        Two cases of inelastic PV electron scattering have been analyzed
as possible means of studying nuclear PV: the excitation of the 15.11
MeV $(1^+,1)$ state in $^{12}$C [Ser79] and electrodisintegration of the
deuteron [Hwa81]. In the former case, which is a pure isovector transition,
parity and angular momentum selection rules restrict the set of allowed
multipole form factors to the following: (a) $F_{M1}$, $\Ftil_{M1}$,
$\Ftil_{E1_5}$, $\Ftil_{C1_5}$, and $\Ftil_{L1_5}$ for transition matrix
elements between components of the ground state and 15.11 MeV state having
positive parity; (b) $F_{C1}$ ($F_{L1}$) and  $F_{E1}$ for EM transitions
involving a state of negative parity mixed into either the ground state or
excited state; (c) $F_{C1_5}$ ($F_{L1_5}$) and $F_{E1_5}$ arising from
PV EM MEC's; and (d) $\Ftil_{C1}$ ($\Ftil_{L1}$), $\Ftil_{E1}$, and
$\Ftil_{M1_5}$ for NC transition matrix elements
involving a negative--parity component of either
the ground state or excited state. The latter
set of form factors (d), as well as all other Coulomb and longitudinal
multipoles in (b) and (c), appear in $\alr$ only at second order in the weak
interaction, so one may neglect them for present purposes. Hence, only the
form factors $F_{M1}$, $\Ftil_{M1}$, $\Ftil_{E1_5}$, $F_{E1}$, and
$F_{E1_5}$ appear at leading non--trivial order in the weak interaction.
Of these form factors, only the last two contain information on nuclear
PV. In the analysis of Ref.~[Ser79], which did not include PV MEC
contributions, no contribution from $F_{E1_5}$ appears.

        The remaining form factor sensitive to nuclear PV, $F_{E1}$, which
arises from parity--mixing in the ground state and 15.11 MeV state,
receives contributions from matrix elements of both $\Shat^{\rm el}_{1\lambda}$
and $\Rhat^{\rm el}_{1\lambda}$. This situation contrasts with the elastic
case, for which matrix elements of $\Shat^{\rm el}_{1\lambda}$ vanish.
Consequently, $F_{E1}$ need not vanish as $Q^2$ for small $Q^2$, as it must
for elastic scattering. An important consequence of this fact is that the
parity--mixing contribution to $\alr$ does not vanish as $\theta\to 0^\circ$,
whereas the weak NC contributions do vanish in this limit. In fact, the
results of Ref.~[Ser79] indicate that the parity--mixing term dominates the
asymmetry at forward angles. For sufficiently low incident energies
($\epsilon\lapp 50$ MeV), this dominance of the nuclear PV component persists
to backward angles. To illustrate, consider scattering at $\epsilon=30$ MeV
and $\theta = 30^\circ$. At these kinematics, the parity--mixing part of the
asymmetry is roughly an order of magnitude larger than the weak NC terms.
Moreover, under the same experimental conditions as assumed above for
$^{13}$C, but with a solid angle of 0.16 sr, one has $(\delta\alr/\alr)_{\rm
stat}\approx 0.13$. Hence, assuming all other contributions are known with
sufficient precision, a measurement of $\alr$ under these conditions could
allow for a significant improvement in the present constraints on $h_{\sst{NN}
\pi}$ and possibly help in resolving the current discrepancy.

        In their analysis of PV electrodisintegration of the deuteron, the
authors of Ref.~[Hwa81] find that the weak NC and nuclear PV contributions
to the asymmetry have comparable magnitude for low incident electron
energies ($\epsilon\lapp 50$) MeV and low relative final state $np$
energies ($E_{\rm rel}\lapp 1$ MeV), assuming the [Des80] values for the
$\hnnm$. In this analysis, a Reid soft--core potential [Rei68] was used to
describe the strong $NN$ interaction. Gauge invariance was maintained through
the use of Siegert's Theorem [Sie37] and explicit inclusion of MEC's. The
authors also suggest that by measuring the asymmetry at a variety of kinematic
conditions, one could in principle constrain different linear combinations
of the $\hnnm$. The experimental doability ({\it i.e.},
$(\delta\alr/\alr)_{\rm
stat}$) of such measurements, however, was not analyzed, and it is not
presently know what statistical precision would be achievable.

\medskip
\noindent\undertext{Summary}
\medskip

        From the foregoing discussion, it appears that a combination of atomic
PV and $\alr$ measurements may produce useful new constraints on the
conventional model of the PV weak $NN$ force. From both a theoretical and
experimental perspective, progress in this direction is more advanced for
atomic PV than for PV electron scattering. In the former case, new experiments
which should improve on the present atomic cesium anapole limits are planned
by the Boulder and Seattle groups [Wie93, Lam93]. In the latter case,
low--energy, moderate--to--forward angle measurements of $\alr$ for nuclear
transitions appear most promising, based on initial studies. Further analysis,
including a search for the most favorable cases, more sophisticated nuclear
calculations, and a detailed study of experimental doability, appears to
be warranted.

\vfil\eject

%

\def\qvecsq{\vec q^{\mkern2mu\raise1pt\hbox{$\scriptstyle2$}}}

\def\xiva{{\xi_\sst{V}^{(a)}}}

\def\evec{{\vec e}}
\def\sigpin{{\Sigma_{\pi\sst{N}}}}

\def\alrzer{{A^0_\sst{LR}}}
\def\alrone{{a_\sst{LR}}}
\def\mevoc{{MeV/c}}

\noindent{\bf IV.F.\quad Quasielastic Scattering}
\medskip
\nobreak
While PV elastic scattering from $(0^+ 0)$ targets and from the deuteron hold
out the
possibility of determining $G^{(s)}_E$ and $G^{(s)}_M$ more precisely than is
possible with ${\svec e}p$ scattering alone, as discussed in Sects.~IV.B and
IV.C, a precision  $\alr({\rm QE})$
measurement might allow one to measure $R^{T=1}_A$, which is
responsible for a large portion of the theoretical hadronic uncertainty in
the SAMPLE determination of $\mustr$ (see Sect.~IV.A).  More generally,
QE PV scattering has the
attraction that if the kinematics are chosen carefully, the resultant
figures--of--merit can be rather large. Ideally, one
would carry out an experiment at precisely quasi--free kinematics, in which
case cross sections and helicity--differences are predominantly given by the
corresponding quantities for individual nucleons.  To the extent that the
process is sufficiently ``quasi--free'' and so has controllably small nuclear
model uncertainties from final--state interactions, meson--exchange currents,
{\it etc.,\/} then QE scattering may provide valuable
information about the single--nucleon form factors themselves.

A few general comments about PV QE scattering are in order. Let us
re--write the PV asymmetry in Eq.~(\Ealrtwo) in the following form involving
ratios of the relevant response functions:
$$
\alr({\rm QE})=\alrzer\times \Bigl\{ {{ 2|Q^2/q^2|{\cal E} \Bigl[{R^L\over
{R^T}}\Bigr] \Bigl[{R^L_{AV}\over{R^L}}\Bigr] +
\Bigl[{R^T_{AV}\over{R^T}}\Bigr] + \sqrt{1-{\cal E}^2}
\Bigl[{R^{T'}_{AV}\over{R^T}}\Bigr] }\over{ 1 + 2|Q^2/q^2|{\cal E}
\Bigl[{R^L\over{R^T}}\Bigr] }} \Bigr\} \ \ , \eqno\nexteq\nameeq\Easyqegen
$$
where as usual ${\cal E}$ is given by Eq.~(\Elongpl). Each of the responses
here may be decomposed into $T=0$ and $T=1$ contributions (see below).
First, we note
from the results presented in Ref.~[Don92] (see also Ref.~[Bei91a]) that
$\alr({\rm QE})$ is dominated by the isovector transverse response and is,
therefore, rather insensitive to $G^{(s)}_E$, $G^{(s)}_M$ and $G^{(s)}_A$,
which contribute to the isoscalar longitudinal and transverse responses,
respectively.  The isovector part of the $R^{T'}_{VA}$ response,
however, depends on $\tilde G^{T=1}_A$, so that a measurement of
backward--angle PV QE scattering should serve in constraining the problematic
$R^{T=1}_A$. Second, as discussed in more detail below, the longitudinal
responses are expected to be suppressed with respect to the dominant
isovector transverse responses and to be highly sensitive to
isospin--dependent nuclear correlations at moderate momentum transfers
[Don92]. Consequently, forward--angle
PV QE scattering may offer a new window on nuclear dynamics. Third, in the
event that the electric strangeness form factor is unsuppressed at
$|Q^2|\sim$ few \gevocsq, it may overwhelm the correlation effects in the
PV longitudinal response and hence could be determined by measuring
$\alr({\rm QE})$ at high momentum transfer and forward angles.

Recent work [Don92] within the context of the relativistic Fermi
gas model bears out these expectations and suggests that experiments may
be feasible in which the QE PV asymmetry could be determined to a
fractional precision of about 1--2\%.  To reach such high precision will
likely require integrating over some region around the QE peak.
While this integration may
increase the do--ability of the experiment, it may also introduce
contaminations from physics beyond the quasi--free approximation, including
significant final--state interaction effects at low excitation energies,
differences in reaction mechanism as in the case of pion production
and the effects of two--body meson exchange currents.  One must ask,
then, whether theoretical uncertainties associated with these contaminations
would cloud the interpretation of
$\alr({\rm QE})$ measurements at a problematic level. Although extensions
for complex nuclei
beyond the initial relativistic Fermi gas modeling undertaken thus far
will have to be pursued\footnote{*}{Initial steps in this direction have
been taken in Refs.~[Hor93a, Hor93b] and [Alb93a].} before
definitive answers can be obtained, it is
nevertheless encouraging that these initial studies indicate that the
asymmetry can be relatively insensitive to the above uncertainties, at
least for specific, well--chosen kinematics.  In particular, the fact that
the asymmetry involves ratios of responses as in Eq.~(\Easyqegen) and is
therefore not as critically dependent
on details of the nuclear model as are the individual responses suggests
that the QE region might be a relatively good one for high--precision PV
studies. Also supporting these expectations is
the case of QE PV scattering from $^2$H which has recently been studied in
detail [Had92]. It provides a special situation in
which the nuclear modeling can be undertaken at a more sophisticated level
than is generally the case for many--body systems.  In Ref.~[Had92] realistic
$NN$ potentials were employed in obtaining bound $^2$H and continuum $np$
wave functions for use in calculating electroweak current matrix elements.
By comparing with models that contain only plane--wave final states (the
plane--wave impulse approximation, PWIA, and the plane--wave Born
approximation, PWBA), it was found in that work that $\alr({\rm {^2}H,\ QE})$
is rather insensitive to final--state interaction effects for
intermediate--to--high values of momentum transfer (say, above $\sim$ 400
\mevoc). We shall return to discuss some of these results below.

To obtain some feeling for the issues involved in studies of PV QE scattering,
let us continue with the static approximation introduced in Sect.~III.E.
Writing each form factor in Eqs.~(\Eqeem)--(\Eqeasy) in terms of
the decomposition given in
Eqs.~(\EGtilde) we have
$$
\eqalign{Z G^p_E \tilde G^p_E &+ NG^n_E \tilde G^n_E \cr
&=  (Z+N) \Bigl\{ G^{T=0}_E \left[ \sqrt{3} \xi_V^{T=0} G^{T=0}_E
+ \xi_V^{(0)} G^{(s)}_E \right]
+ G^{T=1}_E \left[ \xi_V^{T=1} G^{T=1}_E \right] \Bigr\} \cr
&+ (Z-N) \Bigl\{
G^{T=1}_E \left[ \sqrt{3} \xi_V^{T=0} G^{T=0}_E + \xi_V^{(0)}
G^{(s)}_E \right]
+ G^{T=0}_E \left[ \xi_V^{T=1} G^{T=1}_E
\right] \Bigr\} \cr}\eqno\nexteqp\nameeq\Eqezngt
$$
(and likewise with $E\rightarrow M$) for the combinations involving only
vector form factors, together with
$$
\eqalign{Z G^p_M \tilde G^p_A &+ NG^n_M \tilde G^n_A \cr
&=  (Z+N) \Bigl\{ G^{T=0}_M \left[ \xi_A^{T=0} G^{(8)}_A +
\xi_A^{(0)} G^{(s)}_A \right]
+ G^{T=1}_M \left[ \xi_A^{T=1} G^{(3)}_A \right] \Bigr\} \cr
&+ (Z-N) \Bigl\{
G^{T=1}_M \left[ \xi_A^{T=0} G^{(8)}_A + \xi_A^{(0)}
G^{(s)}_A \right]
+ G^{T=0}_M \left[ \xi_A^{T=1} G^{(3)}_A
\right] \Bigr\} \cr}\eqno\sameeq
$$
for the combinations involving both vector and axial--vector form factors.
The denominator in the asymmetry involves the EM form factors:
$$
Z(G_E^p)^2  + N(G_E^n)^2 = (Z+N)\Bigl\{(G_E^{T=0})^2 + (G_E^{T=1})^2\Bigr\}
+(Z-N)\Bigl\{2 G_E^{T=0} G_E^{T=1} \Bigr\} \eqno\nexteq\nameeq\Eisoem
$$
and likewise with $E\rightarrow M$.
First, we note that the mixed isovector--isoscalar
components in Eqs.~(\Eqezngt) and (\Eisoem) are suppressed with respect to
the pure isovector or isoscalar pieces by $(Z-N)/(Z+N)$. Hence, in making
the following arguments (but not in the results taken from Ref.~[Don92]) we
neglect these terms.  Second, we note that in the transverse responses in
Eqs.~(\Eqezngt)
involving $G^{T}_M$ and $\tilde G^{T}_M$  or $\tilde G^{T}_A$, the
isovector component is enhanced with respect to the
isoscalar by a factor of $G^{T=1}_M/G^{T=0}_M\approx 5$ at low--$|Q^2|$.
Hence,
the effect of $\mustr$, appearing in $G^{T=0}_M \tilde G^{T=0}_M$ will be
less important, relative to the leading magnetic term, than it is in the
elastic ${\svec e}p$ PV response (see Eq.~(\Epropv)).  A similar remark
applies to $G^{(s)}_A$, which appears in the combination
$G^{T=0}_M \tilde G^{(s)}_A$ in Eq.~(\Eqezngt) and is suppressed with respect
to
the leading $G^{T=1}_M \tilde G^{T=1}_A$ axial--vector response.
Finally, we note that in the simple static approximation
where incoherent sums over protons and neutrons occur,
the PV longitudinal response is highly suppressed with respect to
the transverse responses (however, see the further discussion in
Sect.~IV.F.2).   This feature is most easily seen by writing the PV
longitudinal response in the form
$$
\eqalign{ {1\over 2} \Bigl(
ZG_E^p\tilde G_E^p + N G^n_E \tilde G^n_E \Bigr) &= ZG^p_E \left[ \left( 1-
4\sin^2\theta_W\right] G^p_E - G^n_E - G^{(s)}_E\right] \cr
&+ N G^n_E \left[ \left(1
- 4 \sin^2\theta_W \right) G^n_E - G^p_E - G^{(s)}_E\right]\ \
,\cr}\eqno\nexteq\nameeq\smallqe
$$
where radiative corrections have been omitted for simplicity.
What would be the largest component --- the term containing
 $ \left( G^p_E\right)^2$
--- is suppressed by the $(1-4\sin^2\theta_W)\approx 0.092$ premultiplying
factor.  The next largest component, given by $(Z+N) G^p_E G^n_E$, is small at
low--$|Q^2|$ due to the smallness of $G^n_E$.  In the case of the magnetic
response  ($E\to M$ in Eq.~(\smallqe)),
on the other hand, no such form factor suppression arises. Hence,
one expects the transverse isovector response to dominate $\alr({\rm QE})$,
especially given the large scale of $G^{T=1}_M$ appearing in this term.

\bigskip
\goodbreak
\noindent IV.F.1.\quad BACKWARD--ANGLE QE SCATTERING:
${\tilde G}_A^{T=1}$--SENSITIVITY

Let us begin by discussing the possibility
that a measurement of $\alr({\rm QE})$ might
eliminate one of the sources of uncertainty in the interpretation of
$\alr({\svec e}p)$, namely, the radiative correction
$R^{T=1}_A$ contained in
$R^p_A$.  According to the estimates of Ref.~[Mus90], one has
$R^{T=1}_A/ R^{T=0}_A \cong -3$, with uncertainties $\delta
R^{T=0}_A\cong {{+0.04}\atop{-0.03}}$ and $\delta R^{T=1}_A =
{{+0.18}\atop{-0.20}}$.  Hence, a
sufficiently precise measurement of $\tilde G^{T=1}_A
= -g_\sst{A}^{(1)}G_\sst{D}^\sst{A}(1+{\rateo})$ could
eliminate most of the theoretical uncertainties associated with the
axial--vector
contribution to a SAMPLE--type experiment [Bei91a].  For scattering at backward
angles $\alr({\rm QE})$ depends primarily on the transverse PV response
functions, $R_{AV}^T$, $R_{VA}^{T^{\prime}}$ (see Eq.~(\Ewpvtwo)) and the
transverse PC electromagnetic response function, $R^T$ (see Eq.~(\Edtsige)),
with only small effects from longitudinal contributions.
As discussed above, all of these transverse contributions are dominated by
isovector spin--flip currents.  As $\theta \rightarrow 180^0$ the leptonic
kinematic factors attain the limit $v_{T^{\prime}}/v_T \rightarrow 1$
(equivalently, ${\cal E} \rightarrow 0$ in Eq.~(\Elongpl)) and consequently
one obtains from Eq.~(\Easyqegen) only two terms,
$${\alr (QE)\over \alrzer} \rightarrow
{{R_{AV}^T} \over {R^T}} + {R_{VA}^{T^{\prime}} \over {R^T}}\ \ .
\eqno\nexteq\nameeq\Erqelim
$$
The first(second) involves the $\tilde G_M$($\tilde G_A$) dependence in
Eq.~(\Eqepv).  Typically the second term is about one quarter of the first,
and thus the backward--angle asymmetry derives about 20\% of its strength
from the response which contains the axial--vector currents.  Given that the
PV QE scattering asymmetry can be measured to $\sim$1--2\% (see
Ref.~[Don92]), this implies
that the axial--vector contributions would be determined to $\sim$5--10\% if
the first, purely vector pieces were perfectly known.  Of course, these
vector contributions are uncertain at some level, since they contain form
factors which are only known with finite precision.  Additionally, the
nuclear modeling itself entails some uncertainty (see also Sect.~IV.F.2 and
Ref.~[Don92] for some
discussion of the level of confidence that might be expected when specific
nuclear models are employed).  To the extent that one does not incur too
much uncertainty from the nuclear modeling, the backward--angle asymmetry can
then be used to shed light on the interplay of the $G_M^{p,n}$,
$G_M^{(s)}$ and $G_A$ form factors.  As noted above, the primary reason for
using PV QE scattering ({\it i.e.,\/} from a nucleus) together with elastic
scattering from the proton is that the interplay of the form factors is
different in the two cases and may permit separations of the effects from
the various form factors to be disentangled.  Clearly, using only elastic
${\svec e} p$ scattering alone involves too many form factors and too few
observables to permit such separations to be made.

Let us expand a little further on these ideas.  Writing the transverse
ratios in Eq.~(\Erqelim) in terms of the single--nucleon form factors
({\it i.e.,\/}
taking ${\cal E} \rightarrow 0$ in Eq.~(\Eqeasy) and substituting for
the $\tilde G$ form factors --- again we invoke the static
approximation in making these arguments), we note the following:
(1) the denominators in Eq.~(\Erqelim) are
proportional to $Z (G_M^p)^2 + N (G_M^n)^2$; (2) the non--strange part of
the first (vector) ratio involves the combination $A G_M^p G_M^n$ in its
numerator, whereas the magnetic strangeness content occurs there in the
form $(Z G_M^p + N G_M^n) G_M^{(s)}$; and (3) the axial--vector content in
the second term appears in the combinations
$(Z G_M^p - N G_M^n) {\tilde G_A^{T=1}}$ and
$(Z G_M^p + N G_M^n) {\tilde G_A^{T=0}}$, where any strangeness
axial--vector form factor occurs in the latter.  We may then make at least two
important observations.  First, for elastic scattering from the proton,
obtained from these expressions by setting $Z=1$ and $N=0$ (compare
Eqs.~(\Eproff) and (\Epropv)), the only
dependence on $G_M^n$ occurs in the numerator of the hadronic ratio.  In
contrast, for PV QE scattering where typically $N\sim Z$, such dependence
occurs both in the numerator and the denominator of the hadronic ratio.
As a consequence, the effect of having limited precision from PC electron
scattering on $G_M^n$ is diminished somewhat in the nuclear case.  This
fact is illustrated in panel (a) of Fig.~4.11, which is taken from
Ref.~[Don92] (note that these results were obtained using the full
relativistic Fermi gas model, not just the static approximation).
The correlation of $\tilde G_A^{T=1}(0)$ with
$\rho_{Mn}$ is shown for $^1$H, $^{12}$C and $^{184}$W (see also below).
Clearly the correlation is weaker for the nuclear cases than for the proton.
Second, any occurrence of $G_M^{(s)}$ or $\tilde G_A^{T=0}$ enters
multiplied by the combination $Z G_M^p + N G_M^n$, whereas the
isovector axial--vector form factor, $G_A^{T=1}$, enters multiplied by
$Z G_M^p - N G_M^n$.  Thus, the relative importance of the two
classes of contributions is governed by
$$
{{Z G_M^p + N G_M^n}\over{Z G_M^p - N G_M^n}}\sim
{{Z \mu_p + N \mu_n}\over{Z \mu_p - N \mu_n}}\ \ .
\eqno\nexteq\nameeq\Eqerlowq
$$
This yields 1 for elastic scattering from the proton and, for example, in
$N=Z$ nuclei such as $^2$H or $^{12}$C a much smaller value, 0.187.  Thus,
effects from $G_M^{(s)}$ and $\tilde G_A^{T=0}$ are suppressed in PV QE
scattering from nuclei, possibly permitting the focus to be placed on
the unsuppressed form factor, $\tilde G_\sst{A}^{T=1}$.  One may take these
observations
to their natural extreme and choose a nucleus which has $N : Z$ very
close to $\mu_p : \mu_n$ and consequently yields nearly zero for the ratio in
Eq.~(\Eqerlowq).  For example, in Ref.~[Don92] the target $^{184}$W was
chosen for
discussion; then the ratio becomes -0.009 and the suppression is virtually
complete.  The correlations of $\tilde G_A^{T=1}(0)$
with $\mu_s$ and $g_A^{(s)}
\equiv G_A^{(s)}(0)$ are
shown in panels (b) and (c) of Fig.~4.11 for the three nuclear targets
discussed.  Clearly these are rather strong correlations for the
case of the proton, as noted previously in Sect.~IV.A where we have
discussed the problem
of making an unambiguous determination of $G_M^{(s)}$ in a SAMPLE--type
experiment,
{\it i.e.,\/} without incurring some uncertainty from the axial--vector
form factors.  For nuclei, where only very weak correlations are seen to
occur, it should be possible to focus on the isovector
axial--vector form factor and accordingly remove at least this source of
uncertainty.  Of course, $\tilde G_\sst{A}^{T=1}$ is interesting in its own
right and
PV QE scattering from appropriately chosen nuclei may help to shed light
on it.

The special case of QE scattering from deuterium may be used to
illustrate these ideas in a different way.  Let us concentrate on the form
factors $G_A^{T=1}$, $G_M^{(s)}$ and
$G_E^{(s)}$.  Using the parameterizations discussed in Sect.~III.C
(specifically, taking $\lambda_E^{(s)}=0$) we may write
the PV asymmetry in the following form:
$$
\alr \equiv \alrone \left[ 1 - b_A
\tilde G_A^{T=1}(\vert Q^2\vert=0) + b_M \mu_s
+ b_E \rho_s \right] \ \ ,\eqno\nexteq\nameeq\Qeparam
$$
where the numbers $(\alrone; b_A, b_M, b_E)$ reflect the way the
asymmetry depends on these three particular form factors.
Using the results from
Ref.~[Had92] for $q=$ 500 \mevoc\ and $\theta=150^{\circ}$ (corresponding
to an incident electron energy of 321 \mevoc\ at the QE peak) the form
factor dependences are the following (see Eq.~(\Qeparam)):
$$
\eqalign{(\alrone; b_A, b_M, b_E)
&= (-1.54\times 10^{-5}; 0.197, -0.461, -0.005) \ \hbox{proton elastic} \cr
&= (-2.07\times 10^{-5}; 0.166, -0.070, -0.003) \ \hbox{deuteron QE --
FSI(SdT)}
\cr
&= (-2.07\times 10^{-5}; 0.166, -0.068, -0.003) \ \hbox{deuteron QE -- FSI(Y)}
\cr
&= (-2.06\times 10^{-5}; 0.168, -0.073, -0.002) \ \hbox{deuteron QE -- PWIA\ .}
\cr}\eqno\nexteq
$$
As expected, the effects due to $G_E^{(s)}$ are very small at backward angles.
All of the models for deuterium represented here (see the
discussions above) give answers which
are very similar and again show the rather weak model dependence for
these (favorable) kinematics.  Our expectations concerning
the differences between the proton and deuteron cases are borne out in
detail.  Importantly, the relative dependence on the magnetic strangeness
form factor (embodied in the parameter $\mu_s$) is
more than six times stronger in the proton case than for the deuteron.
Clearly the latter case is relatively more sensitive to the isovector
axial--vector form factor (represented by the parameter
$\tilde G_A^{T=1}(0)$). A
high enough precision measurement of the QE PV asymmetry in deuterium
should help in defining
${\tilde G}_A^{T=1}$ and so, used in concert with elastic PV scattering from
the
proton, permit $G_M^{(s)}$ to be determined
with better precision than is possible with the proton alone.

\bigskip
\goodbreak
\noindent IV.F.2.\quad ISOSPIN--DEPENDENT CORRELATIONS

One  source of nuclear physics uncertainties
is the impact of nuclear correlations (see, {\it e.g.,\/} Ref.~[Alb90]).  Such
correlations may be responsible to some degree for the well--known failure
of the Coulomb sum rule for PC QE scattering.  Since the various nuclear
response functions display
different sensitivities to inclusion of final--state
interaction and meson--exchange current effects, and since these have different
isospin--dependences, we expect the individual nuclear ingredients in the PV
asymmetry to be isospin--dependent as well.  In particular, the transverse
responses (the PC T--response and PV T-- and T$^{\prime}$-- responses in
Eqs.~(\Erqelim)) are
dominated by nuclear matrix elements of isovector spin--flip operators,
$\sim \sigma\tau$ (see, {\it e.g.,\/} Ref.~[Don92]).  This can be made
clear by writing the ratios of response functions in Eq.~(\Easyqegen) as
follows (here we take $N=Z$ to simplify the expressions):
$$
{R^T_{AV}\over{R^T}} =-{1\over 2}\xi_V^{T=1} \Bigl\{ 1 + {{R^T_{AV}(T=0)}
\over{R^T_{AV}(T=1)}} \Bigr\} \Bigl\{ 1 + {{R^T(T=0)}\over{R^T(T=1)}}
\Bigr\}^{-1}\ \ , \eqno\nexteq
$$
where the PV ratio in the static model is given by
$$
{ {R^T_{AV}(T=0)} \over {R^T_{AV}(T=1)} } = \Bigl( { {G_M^{T=0}} \over
{G_M^{T=1}} }
\Bigr)^2 {1\over {\xi_V^{T=1}} }\Bigl[ \sqrt{3}\xi_V^{T=0} + \xi_V^{(0)}
{ {G_E^{(s)}} \over {G_E^{T=0}} } \Bigr] \eqno\nexteqp\nameeq\Eqetrans
$$
and where the corresponding EM ratio is
$$
{ {R^T(T=0)} \over {R^T(T=1)} } = \Bigl( { {G_M^{T=0}} \over {G_M^{T=1}} }
\Bigr)^2 \cong 0.035\ \ . \eqno\sameeq
$$
Thus, the isoscalar effects can be expected to contribute to these ratios
only at about the 3\% level.
Isospin correlations will modify the relative amounts of the responses that
are isoscalar and isovector and hence change the ratios
$R^T_{AV}(T=0)/R^T_{AV}(T=1)$ and $R^T(T=0)/R^T(T=1)$
in Eqs.~(\Eqetrans) from their static model values.  However, even a reasonably
large change of these ratios is still a minor effect on $R^T_{AV}/R^T$
because of the isovector dominance. Consequently,
isospin correlations
are not expected to have much effect on the
asymmetry as long as the transverse responses are dominant, {\it viz.\/} at
backward scattering angles.  Similar expressions may be written for
$R^{T'}_{VA}/R^T$ which is likewise isovector dominated.

The situation for the longitudinal PC and PV nuclear responses is
quite different --- there the balance of isoscalar--to--isovector content is
such that changes from the static model predictions for the relative amounts
of each are expected to be quite large [Don92].  For instance, in the PC QE
longitudinal
response the balance is approximately 1:1 and so if isoscalar correlations
effectively provide a reduction from the na{\"\i}ve answer (say by moving
strength to low excitation energies), then roughly half of this response
will be affected proportionately, whereas the transverse responses will be
almost unchanged.
The PV QE longitudinal response is especially sensitive to isospin
correlations and consequently might provide
an interesting new window on such many--body effects. To see how this
arises, let us again assume that $N=Z$ for simplicity and
write the isoscalar/isovector parts of $R^L_{AV}$ in the following way:
$$
R^L_{VA} (T) = {\cal N}(1+\Delta_T) G^{(T)}_E \tilde G^{(T)}_E\ \ ,
\eqno\nexteq
$$
where ${\cal N}$ is an overall normalization and $\Delta_T$ with $T=0,1$
represent the amounts that the correlations cause the actual responses to
deviate from their static model values. One then has
$$
R^L_{AV}(T=0) + R^L_{AV}(T=1)
= {\cal N} \left\{ (1 + \Delta_1) G^{T=1}_E \tilde G^{T=1}_E +
(1+\Delta_0) G^{T=0}_E \tilde G^{T=0}_E\right\}\ .\eqno\nexteq
$$
Writing the $G^{(T)}_E$ and $\tilde G^{(T)}_E$ in terms of $G^{n,p}_E$
and $\tilde G^{n,p}_E$ using the inverse of Eqs.~(\smallqe) gives
$$
\eqalign{R^L_{AV} &= {{\cal N}\over 2} \biggl\lbrace \left[ 1 + {1\over 2}
(\Delta_0 +
\Delta_1)\right] \left[ G^p_E \tilde G^p_E + G^n_E \tilde G^n_E\right]\cr
       &\qquad\qquad + {1\over 2}
(\Delta_0 - \Delta_1) \left[ G^p_E \tilde G^n_E + G^n_E\tilde
G^p_E \right] \biggr\rbrace\cr}\eqno\nexteq\nameeq\Eqecorr
$$
and inserting the tree--level Standard model results (Eq.~(\EGtilde)) we have
that the first term in Eq.~(\Eqecorr) is proportional to
$$
\left( - 1 +4\sin^2\theta_W \right)
\left[ (G^p_E)^2 + (G^n_E)^2\right] +
2G^p_E G^n_E + \left( G^p_E + G^n_E\right) G^{(s)}_E\ \ ,
\eqno\nexteq\nameeq\Eqecorrx
$$
which is suppressed with respect to order unity
by factors of $(-1+4\sin^2\theta_W)$, $G^n_E$, or
$G^s_E$ (compare Eq.~(\smallqe)).  The second term, which is sensitive to the
difference in
isospin--dependent correlation effects, is proportional to
$$
\left( G^p_E\right)^2 + \left( G^n_E\right)^2 +
2 ( - 1 + 4\sin^2\theta_W) G^p_E G^n_E + \left(
G^p_E + G^n_E\right) G^{(s)}_E\ \ .\eqno\nexteq\nameeq\Eqecorry
$$
Due to the presence of the first term, Eq.~(\Eqecorry) can be a factor of
10 or more larger than Eq.~(\Eqecorrx) for
low--$|Q^2|$.  Hence, one sees that the effect of differences
in isospin--dependent correlations can be significant in $R^L_{AV}$. A
measurement of $\alr({\rm QE})$ sufficiently sensitive to $R^L_{AV}$ might,
therefore, provide an effective probe of these correlations.  Recent work
[Alb93a] has been aimed in part at exploiting this
sensitivity to nuclear dynamics.

\bigskip
\goodbreak
\noindent IV.F.3.\quad FORWARD--ANGLE QE SCATTERING: $G_E^{(s)}$--SENSITIVITY

An exception to the strong nuclear model dependence found in forward--angle
scattering can occur.  As noted in Ref.~[Don92], it may turn out that the
strangeness electric form factor is unsuppressed at high momentum transfer
($\lambda_E^{(s)} \approx 0$)
and consequently dominates in the longitudinal response (see Eq.~(\smallqe)).
A special (calculable) case where such high--$|Q^2|$, forward--angle studies
have recently been explored [Had92] is that of QE scattering from deuterium.
At $q= 1$ \gevoc\
and $\theta=12.5^{\circ}$ the following results were obtained in that work:
$$
\eqalign{(\alrone; b_A, b_M, b_E)
&= (-3.43\times 10^{-5}; 0.035, -0.495, -0.172) \ \hbox{proton elastic} \cr
&= (-5.25\times 10^{-5}; 0.028, -0.080, -0.085) \ \hbox{deuteron QE --
FSI(SdT)}
\cr
&= (-5.40\times 10^{-5}; 0.027, -0.073, -0.078) \ \hbox{deuteron QE -- FSI(Y)}
\cr
&= (-4.87\times 10^{-5}; 0.032, -0.082, -0.108) \ \hbox{deuteron QE -- PWIA\ .}
\cr}\eqno\nexteq
$$
The proton results are for elastic scattering (see Sect.~IV.A), whereas
the deuteron results
are for kinematics corresponding to the QE peak; in both cases the incident
electron energy is 4.36 GeV.  The scattering angle was taken to be the
minimum possible with the spectrometers that are being built in Hall A at
CEBAF.  For the deuteron, three sets of results are given, {\it viz.,\/}
two with different $NN$ potentials labelled FSI(SdT) and FSI(Y) and the
last for the PWIA.  Details of the two potentials can be found in
[deT73, deT75, Cot76]
(SdT) and [Bre67, Sea68] (Y). Here the two FSI models differ in $\alrone$
by less than 3\% and yet the relative
effect of $G_E^{(s)}$ could be as large as 16--17\% in the FSI models (or
22\% in the PWIA) if $\rho_s= -2$ as in the model of Ref.~[Jaf89].
The differences between the FSI and PWIA results are due to the
non--relativistic expansion procedure
used in the former; for the purpose of predicting the asymmetry at high
momentum transfer the relativistic PWIA should be more reliable.  Comparing
the proton and deuteron results, we see rather striking differences.  In
elastic scattering from the proton at forward scattering angles the
effects from $G_M^{(s)}$ are quite important and any significant uncertainty
in this form factor from backward--angle determinations will propagate
into connected uncertainty in $G_E^{(s)}$, no matter how precisely the
small angle proton asymmetry is measured.  However, taken together
with the QE PV asymmetry on the deuteron where the magnetic strangeness
dependence is much weaker, new information about the form factor
dependences could be extracted.  As the results of Ref.~[Had92]
suggest, it should be possible to undertake such comparisons with rather
high confidence in the deuterium QE predictions.

\bigskip
\goodbreak
\noindent IV.F.4.\quad SUMMARY

Thus far only one quasielastic PV electron scattering experiment has
been performed, the pioneering measurement on $^9$Be at Mainz [Hei89].
For future studies at extreme luminosities there appear to be several
attractive features of high--precision PV QE scattering to explore.
Foremost is likely the possibility of extracting information on $R_A^{T=1}$
from backward--angle scattering at modest momentum transfers and
consequently relatively low electron energies ($\sim$ 300 -- 500 \mevoc ).
As a choice of target, the deuteron has the merit of being ``calculable''
to the degree that at moderate energies, where relativistic effects are
believed to be relatively unimportant, the nuclear physics uncertainties
in the modeling of the asymmetry are likely to be under control.  Even for
heavier nuclei, the expectation is that the nuclear model uncertainties are
rather small for the purely transverse responses and thus for the
asymmetry at backward scattering angles.  More theoretical work will be
required before one can be certain about the scale at which nuclear
dynamics (from final--state interactions, meson--exchange currents and
relativistic corrections) effect the asymmetry; more importantly, it will
be necessary to quantify the level of uncertainty in modeling the nuclear
dynamics if or when high--precision measurements become feasible.

Forward--angle PV QE electron scattering may also prove to be interesting.
There the possibility of learning about isospin correlations exists, since,
as discussed in the previous subsections, the various contributions to
the PV longitudinal response at the
level of the static approximation are all very small, whereas the terms
which arise when isospin correlations are present are about one order of
magnitude larger.  Consequently, even relatively small amounts of
isospin correlation will be significantly magnified and appear as large
changes in the forward--angle PV asymmetry.  Aside from studies of such
nuclear many--body effects, another potential circumstance where
forward--angle scattering might prove interesting is that of adding
information on $G_E^{(s)}$ at high momentum transfer (see also Sects.~IV.A
and IV.B).  In the
event that the electric strangeness form factor is unsuppressed at
large--$|Q^2|$ it could become the dominant contribution to the
longitudinal response (since all other terms are relatively small, as
mentioned above), having only the isospin correlation effects with which
to compete.

Ultimately, it may be necessary to study a given nucleus over a range
of kinematics and/or to study several different nuclei to disentangle
the nuclear many--body effects from those that relate directly to the
properties of the nucleon. By comparing the PV asymmetry measured at
low-- and high--$|Q^2|$, for forward-- and backward--angle scattering,
in each case as a function of $\omega$ should help in quantifying the
level at which nuclear modeling uncertainties enter.  For example, from recent
work [Alb93a] it appears that the nuclear modeling
could be tested by choosing the kinematics wisely (in particular, to
emphasize the
aspects such as isospin correlations which need to be understood better)
and then used with increased confidence for other kinematics where
the nuclear many--body effects are less important and where the
single--nucleon form factor dependences are best revealed.

In Sect.~V we return to discuss the future experimental program for PV QE
electron scattering as it is presently perceived.

\vfil\eject

%

\def\qvecsq{\vec q^{\mkern2mu\raise1pt\hbox{$\scriptstyle2$}}}

\def\xiva{{\xi_\sst{V}^{(a)}}}

\def\evec{{\vec e}}
\def\sigpin{{\Sigma_{\pi\sst{N}}}}

\def\alrzer{{A_\sst{LR}^0}}
\def\Ttil{{\tilde T}}

\noindent {\bf IV.G.\quad The Nucleon--to--Delta Transition}
\medskip

In this section, we consider the excitation of the
$\Delta(1232)$ resonance, $\vec e + N \rightarrow e + \Delta$. Since
this is an isovector transition, measurements of the PV
asymmetry could allow direct extraction of both the vector and
axial--vector isovector couplings. As discussed in Sect.~III.B, the radiative
corrections due to heavy--quark contributions are quite different for an
isovector transition than for both isoscalar and elastic nucleon
scattering, involving the $S$ and $T$ parameters with almost equal weight.
In analogy to the elastic cases considered above, such PV electroproduction
measurements might provide a possible test of the Standard Model that is
essentially independent of details of the underlying hadronic
physics. This is only true to the extent that (i) kinematics (or neutrino
data) can be used to eliminate axial--vector terms; (ii) the corresponding
electromagnetic transition matrix elements which contribute to the
non--resonant background are well measured; and (iii) hadronic
(target--dependent)
contributions to electroweak radiative corrections are understood.
One advantage afforded by $\alr(N\to\Delta)$ is that uncertainties from
isoscalar contributions
({\it e.g.,\/} nucleonic $s$--quark content) are suppressed, in contrast to
the elastic cases considered above. Alternatively, if electroweak
couplings are taken as given input,
PV $\Delta$--production may be useful as a means to {\it measure}
the weak transition matrix elements, giving information on
difficult--to--measure isovector amplitudes and on the currently uncertain
axial--vector transition strength.

The $N\to\Delta$ asymmetry has previously been calculated assuming
elastic $\Delta$--production in the high--energy [Cah78] and
intermediate--energy [Jon80, Nat82] regimes. ``Elastic'' here
means treating the $\Delta$ as a stable (spin--$3\over 2$) particle.
Existing data on photo-- and electro--production of pions indicate,
however, that there is a non--negligible background to the
$\Delta$--production (see Ref.~[Moo78]), and since this
contains both isovector and isoscalar
pieces, one needs an estimate of the size and uncertainty of these
contributions. Such estimates have been performed in the context of specific
nuclear models [Li82, Rei87]. Here, we focus on what hadronic
model--independent statements can be made. To set the scale, we note the
limits this asymmetry could place on the $S$ and $T$ parameters
mentioned above. From Fig. 2.4 , $\alr(N\to\Delta)$ would need to be measured
at the several percent level or better, in order to be effective in
complementing atomic PV and the other PV electron scattering experiments
discussed earlier.

Since existing data on the $N\to\Delta$ transition have been obtained from
$N(e,e'\pi)N$ and $N(e,e'N)\pi$ experiments,
we consider pion electroproduction below the 2--pion threshold,
ignoring electroweak radiative corrections.  One can write a
general expression for the weak and electromagnetic cross sections in a
multipole expansion [Adl68, Pol87, Ras89].
Near the $\Delta(1232)$--resonance, the vector  magnetic dipole
multipole ($\equiv M_{1+}$) dominates, leading to a very
simple expression when neglecting backgrounds
(cf., Eqs.~(\FApi) and (\FFsDthr) below).
In addition, all multipoles can be expanded in terms of their
isospin structure: namely, we can decompose a generic electromagnetic
multipole,
$T$, (suppressing all spin/parity labels) as [Adl68]
$$
\eqalign{
T_{n\pi^+}\ \equiv\ &{\textstyle{1\over\sqrt2}}T^{3\over2} +
                        T^{1\over2} -\sqrt2 T^0 \cr
T_{p\pi^0}\ \equiv\ & T^{3\over2} -{\textstyle{1\over\sqrt2}} T^{1\over2}
                        + T^0\cr
T_{n\pi^0}\ \equiv\ & T^{3\over2} -{\textstyle{1\over\sqrt2}} T^{1\over2}
                        - T^0\cr
T_{p\pi^-}\ \equiv\ & {\textstyle{1\over\sqrt2}}T^{3\over2} +
                        T^{1\over2} +\sqrt2 T^0\ \ ,\cr
}\eqno\nexteq\nameeq\ETnpi
$$
where $T^0$ is isoscalar, and $T^{1\over2}$ and $T^{3\over2}$ are
linearly independent isovector multipoles, going to final states with
isospin ${1\over2}$ and $3\over2$ respectively.  For pure
$\Delta$--production, of course, $T^0 = T^{1\over2} = 0$.
Relating the vector NC multipoles to the EM multipoles via Eqs.~(\Ejncmu),
one has
$$
\eqalign{
\Ttil^0\ =\  &\sqrt3 \xi_V^{T=0}\  T^0 \cr
\Ttil^{{1\over2},{3\over2}} \ = \ &\xi_V^{T=1}\
                                T^{{1\over2},{3\over2}}\ \ .\cr
}\eqno\nexteq\nameeq\ETweak
$$

If one performed a coincidence pion--electroproduction experiment and
measured final charge states, the total cross section for a
proton target would be  obtained by adding $n \pi^+$ and $p \pi^0$ cross
sections incoherently. Such an asymmetry measurement would be misleading
for purposes of studying PV effects, however,
since PC helicity--differences can also occur (specifically, the
so--called 5$^{\rm th}$ response function will be nonzero and generally
much larger than the PV observables; see Ref.~[Ras89]).
For inclusive electron scattering the helicity--differences are
only PV and hence unambiguous. The latter, however, are less selective
in that all final states must be considered; this lack of specificity
means that background effects are less easily controlled (see below).
Since the theoretical analysis of the coincidence measurement
is somewhat simpler and will permit us to bring out the main features of
potential PV studies in the $\Delta$ region, we focus on
this case in what follows, knowing full well that extensive analyses of
the inclusive reactions will be required as well. The asymmetry involves the
interference of
weak and electromagnetic amplitudes, and each vector term in the
multipole expansion will then contribute as
$$
[\Ttil^*T]_{p\pi^0} +  [\Ttil^*T]_{n\pi^+} =
\xi_V^{T=1}\left(|T_{p\pi^0}|^2 + |T_{n\pi^+}|^2 \right)
+ 2\xi_V^n (T_{p\pi^0}-\sqrt2T_{n\pi^+})^*T^0\ \ ,
\eqno\nexteq\nameeq\ETweakII
$$
using Eq.~(\Exiprne a). Here Eq.~(\ETweakII) is derived using Eqs.~(\ETnpi)
and (\ETweak) for the NC
multipoles, and then adding and subtracting the isoscalar multipole,
$T^0$, to the NC isovector terms, in order to pull out the overall
isovector factor of $\xi_V^{T=1}$ in the first term.
This term
is then {\it exactly} proportional to the (unpolarized) electromagnetic
cross section.  The remainder should be quite small. Since
from Eq.~(\ETnpi) one has $T_{p\pi^0}-\sqrt{2}T_{n\pi^+} \propto T^0 -
T^{1\over2}/\sqrt{2}$, the final term in Eq.~(\ETweakII) is a product
of two background amplitudes with final--state isospin ${1\over2}$.

The resulting electron asymmetry is given by a sum of three terms (we use
a superscript $\pi$ to indicate single--pion production),
$$
\alr(N\to\Delta) = -{1\over 2}\alrzer \times
        \Bigl\{\Delta^\pi_{(1)}+\Delta^\pi_{(2)} +
\Delta^\pi_{(3)}\Bigr\}\ \ ,
\eqno\nexteq
$$
where $\Delta^\pi_{(1)}$ gives the Lorentz vector, isovector
contributions (both resonant and non--resonant), corresponding to the
first term of Eq.~(\ETweakII) above;  $\Delta^\pi_{(2)}$ gives the remaining
isospin--${\textstyle {1\over 2}}$ channel, Lorentz vector, non--resonant
background piece;  $\Delta^\pi_{(3)}$ gives the axial--vector contributions,
both resonant and non--resonant; and $\alrzer$ has been introduced
in Sect.~III.E. The terms are given explicitly
by
$$
\Delta^\pi_{(1)}\ =\ \  g_A^e \xi_V^{T=1} \eqno\nexteqp\nameeq\EDpi
$$
$$
\eqalign{
F^2\Delta^\pi_{(2)} &= -2 g_A^e \xi_V^n \sum_l\,{\rm Re} \cr
&\times \Biggl\{ v_T \biggl[l(l+1)^2 \left(
{\textstyle{3\over\sqrt2}}
M_{l+}^{0*} M_{l+}^{1\over2} -3|M_{l+}^{0}|^2 \right)
+ l^2(l+1)\left( {\textstyle{3\over\sqrt2}}
        M_{l-}^{0*}M_{l-}^{1\over2}-3|M_{l-}^{0}|^2 \right)\cr
& +(l+2)(l+1)^2\left( {\textstyle{3\over\sqrt2}}
        E_{l+}^{0*} E_{l+}^{1\over2}-3|E_{l+}^{0}|^2 \right)
+l^2(l-1)\left( {\textstyle{3\over\sqrt2}}
                E_{l-}^{0*} E_{l-}^{1\over2} -3|E_{l-}^{0}|^2\right)
                         \biggr]        \cr
& + v_L \biggl[(l+1)^3 \left( {\textstyle{3\over\sqrt2}}
	S_{l+}^{0*}S_{l+}^{1\over2}-3|S_{l+}^{0}|^2 \right)
+ l^3 \left( {\textstyle{3\over\sqrt2}}
	S_{l-}^{0*}S_{l-}^{1\over2}-3|S_{l-}^{0}|^2 \right)
                        \biggr] \Biggr\} \cr }\eqno\sameeq
$$
$$
\eqalign{ F^2\Delta^\pi_{(3)}\ =\ \ &
        2 g_V^e\ v_{T'}\, \sum_l\,{\rm Re}
                \biggl[l(l+1)^2{\tilde E}_{l+}^{5*}M_{l+}
                        - (l+1)^2(l+2){\tilde M}_{l+}^{5*}E_{l+}\cr
&\qquad\qquad\qquad\ \ - l^2(l+1){\tilde E}_{l-}^{5*}M_{l-}
                      +l^2(l-1){\tilde M}_{l-}^{5*}E_{l-}
                 \biggr]\ \ ,\cr }\eqno\sameeq
$$
where the superscripts indicate the isospin
decomposition of Eq.~(\ETnpi),
the subscripts indicate  the angular momentum and parity of the
multipoles, and we have used Eq.~(\ETweakII)
to separate out the first and second
terms.  The $E$'s, $M$'s, and $S$'s are transverse electric, transverse
magnetic, and longitudinal multipoles, respectively [Adl68, Pol87, Ras89].
The isospin structure of the axial--vector term $\Delta_{(3)}^\pi$ is
not explicitly decomposed in Eq.~(\EDpi), as there is no electromagnetic
analog from which to extract information.

At the $\Delta(1232)$--resonance, and in the high--energy limit,
$\Delta_{(1)}^\pi$ dominates  the non--resonant backgrounds.
In this case, the asymmetry takes on the
the particularly simple form
$$
\alr(N\to\Delta) = -{1\over 2}\alrzer\times
\xivteo\ \ , \eqno\nexteq\nameeq\EApi
$$
which is the isovector analog of the form for purely isoscalar
transitions, Eq.~(\Easylow). At lower energies,
there does exist one significant difference,
however.
At forward angles, the axial--vector contribution to $\alr(N\to\Delta)$ does
not vanish, as is the case for elastic scattering. This arises because of
the different kinematics: since $2M_p \omega = M_{\Delta}^2 - M_p^2
+ |Q^2|$, in the limit $\theta\rightarrow 0$,
$Q^2\rightarrow 0$, we have
$$
\eqalign{
q \rightarrow\ \omega \rightarrow &\  {{M_{\Delta}^2-M_p^2}\over 2M_p} \ne 0\cr
v_{T'}/v_T\ \rightarrow &\
{{\epsilon^2-\epsilon'^2}\over {\epsilon^2+\epsilon'^2}}
\ne 0\ \ .\cr}\eqno\nexteq
$$
In this limit, at large $\epsilon$, the relative contribution of
the axial--vector term is indeed suppressed by $1/\epsilon$, but at threshold
($\epsilon'\rightarrow 0$) the ratio $v_{T'}/v_T$ is unity. Consequently, the
axial--vector transition matrix element contributes  at forward
angles, suppressed at low energies only by the coefficient $g_V^e$.

Ignoring the isoscalar background term $\Delta^\pi_{(2)}$, but
retaining the resonant axial--vector term results in the formula of
Ref.~[Jon80].  Because the axial--vector terms are all suppressed by $\gve$,
one can reasonably hope to estimate their
contribution  by including just the  $1^+$ resonant amplitude. Doing so
gives
$$
F^2\Delta_{(3)}^\pi =
        8 g_V^ev_T'\,{\rm Re}{\tilde E}_{1+}^{5*}M_{1+}\ \ ,
\eqno\nexteq\nameeq\EFsDthr
$$
which requires only the one axial--vector multipole, ${\tilde E}_{1+}^5$.
The latter
has been studied using charge--changing neutrino cross sections, but is
still not very well known (see Refs.~[Adl68, Zuc71, Lle72, Sch73, Rei87,
Kit90]).
Using several different possible parameterizations of this axial--vector term
gives {\it variations} in the small--angle asymmetry of 1.5--2.5\% at 4
GeV incident electron energy.  The corresponding uncertainty in $\xivteo$
is also 1.5--2.5\%. While the relative
contribution from the axial--vector term
does not vanish at small angles, higher beam energies decrease the
strength of this term. Dropping $\epsilon$ from 4 to 1 GeV  increases the
above mentioned  uncertainties to closer to 10\%. However, raising
$\epsilon$ to 20 GeV (while keeping $Q^2$ fixed) reduces the sensitivity to
the axial--vector term by roughly a factor of five, bringing the
uncertainties due to this term below the 1\% level.

One  practical limitation to  a model--independent approach, where the
multipoles appearing in Eq.~(\EDpi) are to be taken from electromagnetic
and charge--changing neutrino scattering, is
that the inelastic electromagnetic transition multipoles appearing, for
example, in $\Delta_{(2)}^\pi$, must be broken down into isoscalar and
isovector pieces separately. Such a separation is difficult for
the $\Delta$--resonance, requiring detailed neutron data to accomplish a full
separation of the three independent isospin pieces. Because of the complete
cancellation of nucleon structure from the $\Delta_{(1)}^\pi$ term,
however, this isospin decomposition is only relevant for the
non--resonant background terms.

Part of the background piece $\Delta_{(2)}^\pi$ can be re--written
in terms of electromagnetic cross sections; namely, that part which
contains terms of the form $T^0T^{1\over2}$. Multiplying by the
appropriate kinematic
coefficients yields exactly ${\textstyle {1\over
2}}(\sigma^\gamma_n-\sigma^\gamma_p)$, with $\sigma^\gamma$ the full
electromagnetic differential cross section. The remaining piece of this
background is purely isoscalar, going as $|T^0|^2$. For a PV coincidence
measurement, then, one in principle requires only
electromagnetic cross section data, rather than
the detailed multipole decomposition, plus information on the
explicitly isoscalar background amplitudes, to evaluate
$\Delta_{(2)}^\pi$.\footnote{*}{This separation is completely arbitrary,
however,
as there is no {\it a priori} reason to assume the $T^0$ background
multipoles should
be significantly smaller than the $T^{1\over2}$ multipoles.}

In order to set the scales involved one can use photoproduction data
which have been broken down into isospin components to get an estimate of the
effect of non--resonant background on the asymmetry at $Q^2=0$.  The
background is approximately 3--4\% of the total photoproduction cross
section, implying a similar amount for the asymmetry. This result, however,
depends sensitively on the poorly known neutron
cross section: a 10\% shift in $\sigma_n^{\gamma}$ would raise the
background contribution to almost 7\% of the asymmetry.
Thus, the uncertainty from background contributions appears to be
non--negligible
even at  $Q^2=0$. At large--$|Q^2|$, the uncertainties become larger. This
feature is due in part to
the difficulty in extracting detailed information
from $\nu$--scattering, which means that the $Q^2$--dependence of the
axial--vector form factors in the term $\Delta_{(3)}^\pi$ is uncertain.  Also,
the $\Delta$--resonance  drops in strength  relative to the backgrounds
with growing momentum transfer, making the separation of non--resonant
contributions,
$\Delta_{(2)}^\pi$, more problematic.  It seems likely that one would
want to use theoretical models to predict and correct for this effect.

This discussion has focused on the particular case of incoherent
summation of final charge states to illustrate the general methods of
calculation.  To reiterate,
false asymmetries in coincidence measurements mean that
one should probably consider {\it inclusive} ${\vec e}$--scattering, in
which case the background piece analogous to $\Delta_{(2)}^\pi$ will
also include interference terms between the background and resonance
multipoles, and thus may not be as small as the pure background
considered here.  An analysis of this case will appear in forthcoming
work [Pol92b].

On the experimental side, a proposal was recently made to measure
$\alr(N\to\Delta)$ at SLAC [Lou92a]. The asymmetry in that case would be
determined at kinematics corresponding to $\epsilon = 12.9$ GeV,
$|Q^2|= 2$ \gevocsq, and $\theta = 6.5^\circ$ with a
precision of 15\% in the asymmetry. From Fig.~2.4, we observe
that roughly an order--of--magnitude improvement in precision would
be needed to make the use of such a measurement as a probe of new
physics competitive with
others. The scale of the axial--vector contribution
at these kinematics is well below the experimental uncertainty.
A less definitive statement can be made about the non--resonant
background contributions, although at lower energies their
contribution is less than 10\%. While the existing proposal appears
to reach the limits of precision attainable at SLAC, the higher
beam current and larger solid angle attainable in CEBAF Hall A
could make a high--precision $\alr(N\to\Delta)$ measurement
feasible there. For example, a 1000 hour Hall A experiment
for $\epsilon\sim 4$ GeV ($|Q^2|\rapp 0.6$ $(\hbox{GeV}/c)^2$)
could yield a 1\% asymmetry measurement if an 80\% beam polarization
were achievable [Lou92b]. A combination of forward--angle measurements
might then afford a separation of the axial--vector, background, and/or
new physics contributions. The prospects for such a scenario at
backward angles appears somewhat less promising. The maximum achievable
precision for the asymmetry in this regime is roughly 5\% .

We conclude that a high--accuracy PV measurement in the
$\Delta$--regime, as a means to measure $\xivteo$ in a model--independent
way, is currently limited by uncertainties in $\nu$-- and
electro--production data.  A more thorough multipole and isospin
decomposition is needed, as well as better data on neutrino
pion--production.  Higher beam energies provide an advantage in
eliminating the uncertain axial--vector contributions, provided $|Q^2|$ is
not so large that the resonance peak gets washed out.  With current
experimental electroproduction uncertainties, one will have
to rely on model--dependent analyses to approximate the background.
Since the total background and axial--vector contributions appear to be small
at moderate momentum transfers,  the resulting model dependence of the
extraction should be fairly weak. Turning the argument around, however,
the axial--vector transition strength is itself of theoretical interest,
and lower energy and/or large angle $\Delta$--production may provide
useful information on this resonant multipole.


\vfil\eject



\def\atil{{\tilde a}}

\noindent{\bf IV.H. \quad Parity--Violating Deep Inelastic Scattering}

Thus far, we have concentrated on semileptonic NC studies at low-- and
intermediate--energies. A considerable degree of insight into the structure
of the lepton--quark electroweak interaction has also been derived from
experiments carried out in the deep inelastic regime. Although a detailed
discussion of these experiments (see {\it e.g.}, Ref.~[Ama87])
lies beyond the scope of the present
article, we briefly treat
one case of historical interest. Indeed, the first PV electron scattering
experiment, performed with a deuterium target at SLAC [Pre78, Pre79],
illustrates the different set of physics issues encountered in higher--energy
semileptonic NC scattering as compared with the NC processes treated
elsewhere in this article. In what
follows, we make no pretense of providing a complete discussion; our intent,
rather, is to illustrate the aforementioned comparison with lower--energy
experiments.

        At the energy and momentum transfers associated with the
SLAC experiment ({\it viz.,\/} $\langle\omega\rangle\approx 5$ GeV, $
|\langle Q^2\rangle |\approx 1.3$ \gevocsq\ [Pre79]),
the nuclear target is unlikely to retain its identity in the final state.
Moreover, the wavelength of the virtual vector--boson probe
($\gamma^*$, $Z^{0*}$)
is much smaller than the size of the nucleon, and the timescale for
its interaction with a given quark is much shorter than the timescale
associated with the strong quark--quark interaction [Bjo69]. Thus,
to a good approximation, the scattering
may be treated as an incoherent process involving the nucleon's constituents.
In the parton picture, these constituents are point--like valence and sea
quark--partons, characterized by momentum distribution functions $f_q(x)$,
where
$$
x \equiv {-Q^2\over 2\mn\omega}\eqno\nexteq\nameeq\bjx
$$
is the Bjorken scaling variable (note that the quantity $\omega$ used in
this work is usually called $\nu$ in high--energy physics). The SLAC data
were centered about a value of $x=0.165$.
In the infinite momentum frame, $x$ is
the fraction of longitudinal momentum carried by a given parton. The
distribution functions satisfy the sum rule
$$
\int_0^1 dx \sum_q x f_q(x) = 1\ \ \ .\eqno\nexteq
$$
The differential cross section for scattering from a nuclear target is then
given by
$$
{d\sigma(eN\to eX)\over dx dy}=\sum_q x f_q(x) \Bigl({d\sigma\over dx
dy}\Bigr)_{e-q}\ \ \ ,\eqno\nexteq\nameeq\discross
$$
where
$$
y\equiv {p\cdot Q\over p\cdot K}\>\mathrel{\mathop{\longrightarrow}^{\rm
lab}}\>
1-{\epsilon'\over \epsilon} \eqno\nexteq
$$
and where the differential cross section appearing on the right--hand side of
Eq.~(\discross) is for scattering of the electron from a given quark--parton
$q$. The helicity independent and dependent $e-q$ cross sections are
proportional to the contraction of the lepton tensors (in the ERL)
$$
\eqalign{L_{\mu\nu}^\sst{\rm (EM)}&=2Q_e^2(K_\mu K_\nu'+K_\nu K_\mu'
-g_{\mu\nu}K\cdot K')\cr
L_{\mu\nu}^\sst{\rm (EM-NC)}(h)&=2hQ_e\gae (K_\mu K_\nu'+K_\nu K_\mu'
-g_{\mu\nu}K\cdot K')\cr
&\qquad\qquad - 2ihQ_e\gve\epsilon_{\mu\nu\alpha\beta}K^\alpha K^{\beta\prime}
\ \ , \cr}\eqno\nexteq
$$
where $h$ is the electron helicity, with the corresponding helicity independent
tensors $W_{\mu\nu}^\sst{\rm (EM)}$ and $W_{\mu\nu}^\sst{\rm (EM-NC)}$
associated with
the quark--parton. The resulting PC and PV cross sections depend on the
variable
$y$. It is straightforward to work out the asymmetry as a function of $x$ and
$y$
$$
\alr(x,y) = A_\sst{LR}^0 \times {W^\sst{\rm (PV)}(x,y)\over
W^\sst{\rm (EM)}(x)}\ \ , \eqno\nexteq
$$
where the deep inelastic hadronic ratio is given by [Cah78]
$$
{W^\sst{\rm (PV)}\over W^\sst{\rm (EM)}}= {\sum_q f_q(x)Q_q\Bigl\{
\gae\gvq+\gaq\gve\Bigl[{1-(1-y)^2\over 1+(1-y)^2}\Bigr]\Bigr\}\over
2\sum_q f_q(x) Q_q^2}\ \ \ . \eqno\nexteq\nameeq\disasym
$$
Note how, in contrast to the situation that occurs in low-- and
intermediate--energy scattering to discrete states,
where the hadron structure physics enters via one-- and many--body form
factors (see Sects.~III.D and III.E), the deep inelastic hadronic ratio
depends on target structure via
the distribution functions $f_q(x)$. The reason for this difference is
essentially contained in the incoherent approximation embodied in
Eq.~(\discross) (summing probabilities rather than amplitudes --- in this
regard, it is more akin to the quasielastic scattering discussed in
Sects.~III.E and IV.F). Note also that,
in general, the structure of the $e-q$ electroweak interaction enters through
the $\gvq$, $\gaq$ {\it etc.}, rather than the through the isospin couplings
$\xiva$.

        As with lower--energy scattering, it is possible to eliminate much
of the hadronic physics dependence of Eq.~(\disasym)
through a judicious choice of
target. In the case of the deuteron, which is symmetric in valence
$u$ and $d$ quarks, one has $f_u^\sst{V}(x)=f_d^\sst{V}(x)$. In this case,
the hadronic ratio may be written in the simple form
$$
{W^\sst{\rm (PV)}\over W^\sst{\rm (EM)}}= {9\over 10}\biggl\{
\atil_1+\atil_2\Bigl[{1-(1-y)^2\over 1+(1-y)^2}\Bigr]\biggr\}\ \ \ ,
\eqno\nexteq\nameeq\disasymb
$$
where
$$
\eqalign{\atil_1&=\coeff{1}{3}\gae(2\gvu-\gvd)=-\coeff{1}{3}(3{\tilde\alpha}
        +{\tilde\gamma})\cr
         \atil_2&=\coeff{1}{3}\gve(2\gau-\gad)=-\coeff{1}{3}(3{\tilde\beta}
        +{\tilde\delta})\cr}\eqno\nexteq\nameeq\ata
$$
at tree--level and neglecting sea--quark contributions. One may account for the
latter as well as for the presence of \lq\lq non--standard" physics by through
correction factors as
$$
\eqalign{\atil_1&=(1-\coeff{20}{9}\sstw)[1+R_1(\hbox{st'd})+
        R_1(\hbox{new})+R_1(\hbox{had})]\cr
         \atil_2&=(1-4\sstw)[1+R_2(\hbox{st'd})+
        R_2(\hbox{new})+R_2(\hbox{had})]\ \ \ , \cr}\eqno\nexteq\nameeq\atb
$$
where the $R_{(i)}(\hbox{st'd})$ are radiative corrections in the Standard
Model
and the $R_{(i)}(\hbox{new})$ represent contributions from physics beyond
the Standard Model.  In the $S$ and $T$ framework one has approximately
$$
\eqalign{R_1(\hbox{new})&=0.02\ T-0.017\ S\cr
         R_2(\hbox{new})&=0.155\ T-0.206\ S\ \ \
.\cr}\eqno\nexteq\nameeq\disnew
$$
The hadronic structure corrections, obtained by keeping the sea--quark
contributions in Eq.~(\disasym), have the form
$$
\eqalign{R_1(\hbox{had})&\approx{\sum_q Q_q\gvq F_q^\sst{S}(x)\over f_q^\sst{V}
(x)(Q_u\gvu+Q_d\gvd)}-{\sum_q Q_q^2 F_q^\sst{S}(x)\over f_q^\sst{V}(x)(Q_u^2
+Q_d^2)}\cr
 R_2(\hbox{had})&\approx{\sum_q Q_q\gaq \delta f_q^\sst{S}(x)\over f_q^\sst{V}
(x)(Q_u\gau+Q_d\gad)}-{\sum_q Q_q^2 F_q^\sst{S}(x)\over f_q^\sst{V}(x)(Q_u^2
+Q_d^2)}\ \ \ ,\cr} \eqno\nexteq\nameeq\dishad
$$
where $F_q^\sst{S}\equiv f_q^\sst{S}+f_{\bar q}^\sst{S}$ and
$\delta f_q^\sst{S}\equiv
f_q^\sst{S}-f_{\bar q}^\sst{S}$ are the sum and difference, respectively, of
the sea quark ($q$) and anti--quark ($\bar q$) distributions. The reason that
the difference $\delta f_q^\sst{S}$ appears in $R_2(\hbox{had})$ rather
than the sum as in $R_1(\hbox{had})$ is that both $Q_q$ and $\gvq$ change
sign on going from quark to anti--quark, whereas the sign of $\gaq$ remains
the same [Cah78]. In the case of $u$ and $d$ quarks, the valence distribution
is {\it defined} as $f_q^\sst{V}(x)=f_q(x)-f_{\bar q}(x)$, so that the
$\delta f_q^\sst{S}(x)\equiv 0$ for $q=u,\ d$.

        The foregoing results merit several observations. First, from
Eq.~(\disasymb) one observes that a measurement of $\alr(x,y)$ as a function of
$y$ allows a separate determination of $\atil_1$ and $\atil_2$. At the
simplest level, these two quantities are hadronic physics independent
(neglecting the $R_{(i)}(\hbox{had})$). Thus, a separation of the $\atil_{(i)}$
allows one either to test various electroweak models at tree--level
(by specifying the $\gvq$, $\gaq$, {\it etc.}) or to place constraints
on the two linear combinations of the \lq\lq model--independent"
$e-q$ couplings appearing in Eq.~(\ata). Such a separation was performed
using the SLAC data, leading to the constraints (see, e.g., Refs.~[Pre78,
Pre79, Com83])
$$
\eqalign{{\tilde\alpha}+\coeff{1}{3}{\tilde\gamma}&=-0.60\pm 0.16\cr
         {\tilde\beta}+\coeff{1}{3}{\tilde\delta}&=0.31\pm 0.51\ \ . \cr}\eqno
\nexteq\nameeq\atc
$$
Alternately, one may assume the Standard Model forms for the $\atil_{(i)}$
and extract a value of $\sstw$. The SLAC results were found to be consistent
with the Standard Model with $\sstw=0.224\pm 0.020$.

        While the 9\% determination of $\sstw$ from the SLAC experiment
represented a triumph for this first PV electron scattering experiment,
present and future electroweak tests in other sectors are approaching
precision of 1\% or better. A deep inelastic $\evec D$ measurement at this
level of precision could be sensitive to physics beyond the Standard Model,
such as would be characterized, for example, by $S$ and $T$ appearing in
the $R_{(i)}(\hbox{new})$. To extract meaningful constraints on these
parameters, however, one must also make a reliable determination of the
structure corrections, $R_{(i)}(\hbox{had})$.

        An indication of the scale of these corrections may be obtained by
employing a parameterization of the distribution functions fit to EMC data
[Slo88, Sch88]. In this parameterization, one neglects contributions from
heavy quarks ($c$, $b$, $t$) in the sea and assumes equality of sea quark and
anti--quark distributions for the three lightest quarks:
$$
f_u^\sst{S}=f_\ubar^\sst{S}=f_d^\sst{S}=f_\dbar^\sst{S}=f_s^\sst{S}=
f_\sbar^\sst{S}\equiv f^\sst{S}\ \ \ .\eqno\nexteq
$$
Hence, one has $\delta f_q^\sst{S}=0$ and $F_q^\sst{S}=2f^\sst{S}$ in
Eq.~(\dishad).  In order to obtain the valence distributions for the
deuteron, we assume good isospin symmetry for the nucleon, so that
$f_u^\sst{V}(\hbox{proton})=f_d^\sst{V}(\hbox{neutron})$ and $f_d^\sst{V}(
\hbox{proton})=f_u^\sst{V}(\hbox{neutron})$. Thus, one has $f_q^\sst{V}(
\hbox{deuteron})=\coeff{1}{2}[f_u^\sst{V}(\hbox{proton})+f_d^\sst{V}(
\hbox{proton})]$. From the fits of Ref.~[Slo88] we then obtain the
$R_{(i)}(\hbox{had})$ as functions of $x$, as shown in Fig. 4.12.

        As expected, the $R_{(i)}(\hbox{had})$ become negligible for $x\rapp
0.2$. Even though $xf_q^\sst{V}(x)$ and $xf_q^\sst{S}(x)$ both vanish as
$x\to 1$, the sea distribution falls off more rapidly with $x$, becoming
negligible for $x\rapp 0.2$, while the valence distributions persist at an
appreciable level to somewhat larger values of $x$. Under the assumptions
employed in parameterizing the $f_q^\sst{V,S}$, it is the ratio $f^\sst{S}/
f_q^\sst{V}$ which governs the $x$--dependence of the $R_{(i)}(\hbox{had})$,
so that the latter become vanishingly small for $x\rapp 0.2$. Ideally, then,
one would perform a future measurement of the deep--inelastic
$\alr(^2\hbox{H})$
in a kinematic region for which $x\rapp 0.2$ if one were interested in
constraining $S$ and $T$ or other possible extensions of the Standard Model.

        The mean value of $x$ for SLAC $\alr(^2\hbox{H})$ measurements was
somewhat below $0.2$. To illustrate the impact of hadronic uncertainties
at the SLAC kinematics, we consider a measurement at $x=0.15$.
At this kinematic point, the values of the
$R_{(i)}(\hbox{new})$ would be the same as the $R_{(i)}(\hbox{had})$ if
one had $S=-1.4$ and/or $T=1.18$ ($i=1$) and $S=0.51$ and/or $T=-0.68$
($i=2$). Of course, it is the uncertainty in the $R_{(i)}(\hbox{had})$ rather
than their overall scale that is potentially problematic for the extraction
of limits on $S$ and $T$. At $x=0.15$, for example, a non--negligible
uncertainty
is introduced by the sea--quark distribution functions, parameterized
in Ref.~[Slo88] by the form $xf^\sst{S}(x)=C_3(1+\gamma)(1-x)^\gamma$. We take
the quoted uncertainty in $\gamma$ as a rough indication of the overall
uncertainty in the $f^\sst{S}(x)$. The corresponding uncertainties induced
in $S$ and $T$ via the $\delta R_{(i)}(\hbox{had})$ are $(\delta S, \delta
T)\approx (0.46, 0.39)$ $(i=1)$ and $(\delta S, \delta T)\approx (0.17, 0.23)$
($i=2$), where the values for $\delta S$ are obtained assuming $\delta T=0$
and vice--versa.These values do not change appreciably at smaller $x$ and fall
off by a factor of ten for $x\rapp 0.4$. Moreover, they are significantly
smaller than the prospective low--energy constraints of Fig. 2.4. A much
larger degree of uncertainty in the constraints on $S$ and $T$ is
introduced by the experimental error. For the SLAC data, the experimental
uncertainty in the asymmetry ($\sim$ 10\%) is generally more than an
order of magnitude larger than the uncertainty introduced by the quark
distribution functions. Consequently, the constraints on $S$ and $T$ from
the SLAC experiment are somewhat loose.

        A future, more precise measurement of the deep--inelastic deuterium
asymmetry could tighten these constraints without requiring substantial
improvements in one's knowledge of the quark distribution functions. Such
a measurement has recently been proposed for SLAC [Bos92]. The proposed
experiment would measure the deep inelastic deuterium asymmetry at several
kinematic points having $x\rapp 0.2$ with overall uncertainties in the
asymmetry ranging from one to two percent. In this kinematic range, the impact
of $f_q(x)$ uncertainties is negligible, but uncertainties associated with
higher--twist contributions could be worrisome [Bos92]. The proposal also
called for measurements of the asymmetry at several points with $x\lapp 0.2$,
where the sea--quark contributions become important. The goal of such
measurements would be to obtain two new observables which, when combined with
other DIS data, could permit a separation of light--quark
distribution functions.
Such a separation might shed light on the violation of the Gottfried Sum
Rule [Got67] reported by NMC [Ama91]. Although the proposal [Bos92] was not
approved by the 1992 SLAC PAC, a revised version will be submitted to the
upcoming PAC.

\vfil\eject

%

\def\qvecsq{\vec q^{\mkern2mu\raise1pt\hbox{$\scriptstyle2$}}}

\def\xiva{{\xi_\sst{V}^{(a)}}}

\def\evec{{\vec e}}
\def\sigpin{{\Sigma_{\pi\sst{N}}}}

\noindent {\bf IV.I.\quad Atomic Parity Violation}
\medskip

Atomic PV provides a class of experiments which is
sensitive to a rather different set of correction factors and
uncertainties than the PV electron scattering observable, $A_{LR}$,
considered so far [Bou74, For80, For84, Bou86]. Because of
the very small momentum transfers
involved in atomic transition matrix elements,
atomic PV is generally {\it less} sensitive to
unknown (or poorly known) hadronic form factors and certain types of
radiative correction uncertainties than is PV electron scattering.
On the other hand, new difficulties do arise,
including atomic wave function uncertainties [Blu90] and nuclear structure
uncertainties. In general, atomic PV appears to be nicely complementary
to PV electron scattering, and may continue to prove extremely useful
in yielding high--precision Standard Model tests.

To illustrate, consider the PV atomic Hamiltonian which mixes
opposite--parity atomic states, and leads to the presence of atomic PV
observables:
$$
\eqalign{
{\hat {\cal H}}_{\rm atom}^{PV}\ =&\ {G_\mu\over 2\sqrt2} \int\ d{\vec x}
\ \hat\psi_e^\dagger(\vec x) \gamma_5\hat \psi_e(\vec x)\rho^{NC}(\vec x) +
\cdots\ ,\cr}\eqno\nexteq\nameeq\EHpva
$$
where $\hat\psi_e(\vec x)$ is the electron field and $\rho^{NC}(\vec
x)$ is the Fourier transform of the matrix element of the charge (0)
component of the weak neutral current operator
(see Eqs.~(\Erjmult a), (\EOjnc a), and (\Encurff)).
We omit the
contribution of the three--vector part of the nuclear current in
Eq.~(\EHpva), since it is highly suppressed due to the  small
effective momentum transfer involved in atomic transitions. The leading
term given in Eq.~(\EHpva) is also enhanced relative to omitted nuclear
axial--vector
terms, at least in heavy atoms, due to the coherent behavior of the
nuclear charge operator.

Following Refs.~[For90] and [Mus92a], we write the matrix element of
$\hat{\cal H}_{\rm atom}^{PV}$ between atomic $S_{1\over2}$ and
$P_{1\over2}$ states in the form $\bra{P} \hat\psi_e^\dagger(\vec x)
\gamma_5\hat \psi_e(\vec x) \ket{S} = {\cal N}C_{sp}(Z)f(x)$, where
$\cal N$ is a calculable overall normalization factor which depends slightly
on the nuclear charge radius, $C_{sp}(Z) $ is an atomic
structure--dependent function, and $f(x) = 1 - {1\over2}(x/x_0)^2 +
\cdots$ gives the spatial dependence of the electron axial--vector charge
density.  In a simplified model where a charge--$Z$ nucleus is taken as a
sphere of constant electric charge density out to radius $R$, one has
$x_0 = R/Z\alpha$, neglecting small corrections involving the electron
mass. In this case, the atomic matrix elements of Eq.~(\EHpva) become
$$
\eqalign{
\bra{P}{\hat {\cal H}}_{\rm atom}^{PV} \ket{S}\ =&\ {G_\mu\over 2\sqrt2}
{\cal N}C_{sp}(Z) \left[
        Q_W^{(0)} + \Delta Q_W^{(n,p)} + \Delta Q_W^{(s)} + \Delta Q_W^{(T)}
                        \right]
+\cdots,\cr} \eqno\nexteq\nameeq\Eapvme
$$
where
$$
\eqalignno{
Q_W^{(0)}\ =&\ {1\over 2}
              \left(Z-N \right)\xi_V^{T=1} + {1\over 2}
        \sqrt3\left(Z+N \right) \xi_V^{T=0}, &\nexteqp\nameeq\EDQw\cr
\Delta Q_W^{(n,p)}\ =&\
        {\textstyle {1\over2}}[\sqrt3\xi_V^{T=0}+ \xi_V^{T=1}]
        \bra{T_0}| \sum_{k=1}^A {\textstyle{1\over2}}[1+\tau_3(k)]h(x_k)
                        |\ket{T_0}, &\cr
        &\qquad + {\textstyle {1\over2}}[\sqrt3\xi_V^{T=0}-\xi_V^{T=1}]
        \bra{T_0}| \sum_{k=1}^A {\textstyle{1\over2}}[1-\tau_3(k)]h(x_k)
                        |\ket{T_0}, &\sameeq\nameeq\Eqwb\cr
\Delta Q_W^{(s)}\ =&\
        -\xi_V^{(0)}\left(\rho_s\over 4m_N^2\right)
        \bra{T_0}| \sum_{k=1}^A \nabla_k^2 h(x_k)
                        |\ket{T_0},  &\sameeq\nameeq\Eqwc\cr
\Delta Q_W^{(T)}\ =&\
        \lambda\xi_V^{T=1}\left[
        \bra{T_0}|\sum_{k=1}^A  h(x_k)\tau_3(k)
                        |\ket{T_1}  + (T_1 \leftrightarrow T_0)\right]
        +\cdots  &\sameeq\nameeq\Eqwd\cr
}
$$
are contributions to the so--called \lq\lq weak charge",
with $h(x) \equiv f(x)-1$, and where $\bra{T_0}|\hat{\cal
O}|\ket{T_0}$ denotes reduced matrix elements of a nuclear operator
$\hat{\cal O}$ in a nuclear ground state having nominal isospin $T_0$.

The term in Eq.~(\EDQw a) is the leading contribution to the weak
charge usually discussed in treatments of atomic PV.
Including Standard Model radiative corrections in the $\overline{\hbox{MS}}$
renormalization scheme, as well as some possible (new) heavy--quark physics
corrections, leads to
$$
\eqalign{Q_W^{(0)}&=Z(1-4{\bar x})[1+\RVp(\hbox{st'd})+\RVp(\hbox{new})]\cr
	 &\quad-N[1+\RVn(\hbox{st'd})+\RVn(\hbox{new})]
	 +\Delta Q^{\rm new}_{\rm tree}(N,Z)\ \ . \cr}
\eqno\nexteq\nameeq\EQWz
$$
{}From Ref.~[Mar90] one has that
$$
\eqalign{
\bar x \ \equiv&\  \sin^2\hat\theta_W(M_Z) =
                0.2323 \pm 0.0007
}\eqno\nexteq\nameeq\Exbarms
$$
gives the weak mixing angle and
$$
\eqalign{\RVp(\hbox{st'd})&\approx -0.054\pm 0.033\cr
	 \RVn(\hbox{st'd})&\approx -0.0143\pm 0.0004\cr}
\eqno\nexteq\nameeq\ERatom
$$
are the Standard Model one--loop corrections in the $\overline{\hbox{MS}}$
scheme. The errors shown in Eqs.~(\Exbarms) and (\ERatom) result
from uncertainties in experimental input parameters, and also in
evaluations of one--loop diagrams. We note that since the effective momentum
transfer for atomic transitions is so small, uncertainties associated with
two--photon dispersion corrections should be negligible. This situation
contrasts with that of PV electron scattering from nuclei, where these
corrections introduce a non--negligible source of theoretical ambiguity.
Contributions from $S$ and $T$, which signal the presence of
non--Standard Model physics entering loops,
are contained in the $R_\sst{V}^{p,n}(
\hbox{new})$ as indicated in Eq.~(\ERsubv). In contrast,
the term $\Delta Q_{tree}^{new}$ accounts for new physics contributions
arising at tree--level, {\it e.g.} coming from extra $Z^0$ bosons. For
example, in SO(10) models with the exchange of an additional \lq\lq
$Z_\chi$" boson with no $Z-Z_\chi$ mixing, one has
$$
\Delta Q_{\rm tree}^{Z_\chi} \approx 0.4(2N+Z)\mws/M^2_{Z_\chi}
\eqno\nexteq\nameeq\EQzchi
$$
as given in Ref.~[Mar90].\footnote{*}{In the $Z_\chi$ model, and
indeed in all such models with an extra U(1) symmetry arising from
$E_6$, the vector coupling of the extra neutral boson to $u$ quarks is zero.
Consequently, the ratio of neutron to proton couplings in Eq.~(\EQzchi)
is exactly 2:1, since only $d$ quarks contribute.} The value of
$\Delta Q_{\rm tree}^{Z_\chi}$ one would extract from existing Cs atomic
PV measurements, {\it if all other radiative and heavy physics were
ignored}, is $2.2\pm 1.6\pm 0.9$. This value corresponds in the SO(10)
model to $M_{Z_\chi}\approx 500$ GeV.

It has recently been suggested [Dzu86, Mon90, Mar90]
that one perform measurements of the
weak charge for atoms along an isotopic chain rather than for a single
isotope. In the ratio $Q_W(Z,N')/Q_W(Z,N)$, the coefficient $C_{sp}(Z)$
exactly cancels, thereby eliminating much of the atomic physics uncertainties.
Moreover, this ratio will carry a different sensitivity to new physics
than does $Q_W$ for a single isotope.
Due to the largeness of $N/Z$ in heavy atoms, the weak charge of
Eqs.~(\EDQw) and (\EQWz) is almost completely independent of $T$. The
isotope ratio, however, carries the same {\it relative} dependence on
$S$ and $T$ as do various Z$^0$--pole observables. Isotope ratios also provide
a different linear combination of $S$, $T$, and $\Delta Q_{\rm tree}^{\rm
new}$ than does $Q_W$ for a single isotope. Thus, with a combination of
isotope ratio measurements and measurements of other NC observables ({\it
e.g.},
$\alr$), one might hope to disentangle contributions from various types of
physics beyond the Standard Model.

In order for such a scenario to be realized, theoretical uncertainties
associated with additional contributions to the weak charge must be
resolved. In particular,
the term $\Delta Q_W^{(n,p)}$ in Eq.~(\Eqwb) carries a dependence on the
ground--state neutron radius, $R_n$.  The impact of uncertainties in
$R_n$, and indeed in the full spatial neutron distribution, on the use of
atomic PV for high--precision electroweak tests has been discussed in
some detail in Refs.~[For90, Pol92a]. To illustrate, we use
here the simplifying assumptions of Eq.~(\Eqwb) for the case of atomic
cesium, on which the most recent high--precision experimental
measurements have been made, and for which the atomic physics
calculations can be done to quite high accuracy. One then  needs to
know $R_n$ to roughly 10\% to reduce the uncertainty induced in $S$
to $\approx \pm 0.6$ (the equivalent to a 1\% $\alr(0^+0)$ measurement).
For $^{208}$Pb, another promising future
experimental possibility,  one needs to know $R_n$ to roughly 4\%.
This requirement is fairly stringent, and the reliability of
existing nuclear model predictions for $R_n$ may begin to be questioned
at such a level.  In Ref. [Don89], the idea of using PV elastic
electron scattering to determine $R_n$ was explored (see also
Sect.~IV.B). For the case of
lead, the conclusion was that a 1\% determination of $R_n$ is possible,
requiring under 1000 hours of beam time with experimental conditions
which could in principle be provided at CEBAF. It seems likely that a
similar determination of $R_n$ could be made for cesium.
In the case of isotope ratio measurements, one requires
knowledge of the {\it change} in neutron radii between isotopes, which
presents additional challenges to theory and/or electron scattering
experiment.

Eq.~(\Eqwc) gives the leading contribution to $Q_W$ from $G_E^{(s)}$.
For the case of $^{133}$Cs, one has under the assumptions leading
to Eqs.~(\EDQw d) [Mus92a]
$$
\Delta Q_W^{(s)} = -\Bigl({3\over 4}\Bigr)Z^2 A^{1/3}\alpha^2\Bigl({1\over
\mn r_0}\Bigr)^2\delta\rhostr\ \ , \eqno\nexteq\nameeq\Eqwstrange
$$
where $r_0\approx 1.1$ fm and where only the leading term in $h(x)$ has
been retained. For a value of $\rhostr$ on the order of the prediction of
Ref.~[Jaf89], Eq.~(\Eqwstrange) would lead to an error in $\sstw$ from
$Q_W^{(0)}$ of roughly 0.1\%, about an order--of--magnitude smaller than
the dominant theoretical errors associated with atomic and nuclear structure.
Given the  constraints on $\GES$ likely to be achieved with $\alr$
measurements, nucleon strangeness uncertainties should remain well below
a problematic level for electroweak tests with atomic PV. We note further
that given the simple additive nature of Eq.~(\Eqwc), $\GES$ contributions
to isotope ratios are further suppressed from their already small
impact on $Q_W$. We also point out that
additional contributions to $Q_W$ arising from the
single--nucleon EM charge radii are discussed
elsewhere [Pol92a], but as an example, the net effect of the
nucleon's known internal electromagnetic structure results in a
correction of $Q_W$ for $^{133}$Cs of only approximately 0.1\%.

Eq.~(\EDQw d) gives the leading contributions to $Q_W$ from  isospin
impurities in the nuclear ground state.   We have shown explicitly only
the contribution to $\Delta Q_W^{(T)}$ arising from the mixing of a
single state of isospin $T_1$ into the ground state of nominal isospin
$T_0$ with strength $\lambda$. Further discussion can be found in
Ref.~[Mus92a], but this term
appears likely to be very small in most cases of interest.

There exists another class of possible atomic PV experiments,
namely those with muonic atoms. Such experiments may be achievable at
the 1--10\% level in the future at PSI [Lan91]. A recent discussion of
the effects
of radiative corrections and non--Standard Model physics on light muonic atoms
may also be found in Ref.~[Lan91].  The possibility
of performing PV experiments with
heavy muonic atoms has interest from a somewhat different perspective.
Because the ratio of Bohr radii
for muonic and electron atoms goes as $a_0^e/a_0^\mu= m_\mu/m_e\approx 207$,
the muon is much more tightly bound than the corresponding electron for
a given set of orbital quantum numbers. Consequently, one might expect
an enhanced sensitivity of $Q_W(\mu)$ to short--range effects, such as
the ground--state neutron distribution and the nucleon strangeness radius.
In the case of the latter, one may estimate the scale of this effect
by solving the Dirac equation for a single charged lepton moving in the
field of a nuclear sphere of uniform charge and retaining the lepton
mass [Mus92b].
The result is to make the replacement $x_0=R/Z\alpha\to [3R/4m_\mu
Z\alpha]^{1/2}$ in $h(x)$, which yields an enhancement of $Q_W^{(s)}(\mu)$
over $Q_W^{(s)}(e)$ by roughly $4m_\mu R/3Z\alpha$. For atomic cesium,
this enhancement factor is $\approx 8$, making $Q_W(\mu Cs)$ nearly as
sensitive to $\rhostr$ as is $\alr(\evec p)$. For light muonic atoms,
on the other hand, the sensitivity to $\rhostr$ is too weak to be observable.
In the case of heavy muonic atoms, where nucleon strangeness may be
observable, nuclear structure uncertainties also become correspondingly
more important. A combination of atomic PV and electron
scattering experiments (PV and PC) would then appear necessary in order
to separate nuclear structure and
strangeness contributions. A more thorough investigation of this possibility
is in progress [Mus93d].
{}From a theoretical standpoint, muonic atoms
have the additional advantage of being essentially a one--lepton problem,
making a determination of the muon wave function much more straightforward
than for electron atoms.

In summary, apart from questions of atomic theory and important
uncertainties in nuclear structure, the unique dependencies on new
physics means that atomic PV experiments can yield additional
high--precision Standard Model tests, unattainable from direct high--energy
experiments at the Z$^0$--pole. In addition, the relative insensitivity
to certain classes of radiative corrections, and hadronic form
factors,  makes atomic PV nicely complementary to PV electron
scattering.

\vfil\eject

%

\def\qvecsq{\vec q^{\mkern2mu\raise1pt\hbox{$\scriptstyle2$}}}

\def\xiva{{\xi_\sst{V}^{(a)}}}

\def\evec{{\vec e}}
\def\sigpin{{\Sigma_{\pi\sst{N}}}}

\def\alrzer{{A^0_\sst{LR}}}  
\def\gtilan{{\tilde G_\sst{A}^\sst{N}}}
\def\gtilen{{\tilde G_\sst{E}^\sst{N}}}
\def\gtilmn{{\tilde G_\sst{M}^\sst{N}}}

\def\gtilep{{\tilde G_\sst{E}^p}}

\def\gtilenn{{\tilde G_\sst{E}^n}}

\def\xivp{{\xi_\sst{V}^p}}
\def\xivn{{\xi_\sst{V}^n}}

\def\lams{{\lambda_\sst{E}^\sst{(s)}}}

\def\msubt{{M_\sst{T}}}
\def\tsubr{{T_\sst{R}}}
\def\thetr{{\theta_\sst{R}}}

\def\gaeight{{G_\sst{A}^{(8)}}}
\def\gathree{{G_\sst{A}^{(3)}}}

\def\gtilan{{\tilde G_\sst{A}^\sst{N}}}
\def\gtilen{{\tilde G_\sst{E}^\sst{N}}}
\def\gtilmn{{\tilde G_\sst{M}^\sst{N}}}

\def\gtilep{{\tilde G_\sst{E}^p}}

\def\gtilenn{{\tilde G_\sst{E}^n}}

\def\xivp{{\xi_\sst{V}^p}}
\def\xivn{{\xi_\sst{V}^n}}

\def\gaeight{{G_\sst{A}^{(8)}}}
\def\gathree{{G_\sst{A}^{(3)}}}

\noindent {\bf IV.J.\quad Neutrinos}
\medskip

        The subject of neutrino scattering from nuclei is sufficiently
broad and detailed that one could devote an entire review article to it
alone. Indeed, when considering the full range of neutrino beam energies,
one encounters a variety of physics issues which may be addressed in
this way, including charge--changing deep inelastic scattering measurements
aimed at testing the Standard Model or probing quark distributions
at high--energies and low--to--intermediate energy
inelastic excitations of nuclear levels as a means of studying
the nuclear weak current. Given the existence in the literature of other
reviews of semileptonic neutrino reactions (see, for example,
[Don79a]), we limit the
present discussion to those aspects of low-- and intermediate--energy
neutrino--hadron NC processes not considered in previous studies. In
particular, we focus on the use of neutrinos to probe nucleon strangeness
content, noting the relative advantages or disadvantages this offers
when compared with PV electron scattering. We also pay particular
attention to three low-- and intermediate--energy neutrino NC experiments: the
recently completed Brookhaven (BNL) $\nu$--p/$\nubar$--p experiment [Ahr87],
the KARMEN experiment involving inelastic neutrino excitation of a discrete
state in $^{12}$C [KAR92]
and the LSND experiment at LAMPF [Lou89]
currently in progress.

As with PV electron scattering, it
is important when reviewing neutrino scattering prospects to keep
considerations of doability firmly in mind. For the energies and momentum
transfers of interest here, typical cross sections $d\sigma/d Q^2$ range
from $\sim 10^{-38}\to 10^{-40}\ \hbox{cm}^2\ (\hbox{GeV/c})^{-2}$. We shall
return in Sect.~V.A.2 to discuss the existing or planned neutrino
facilities (see Table~5.3 for characteristics of these facilities). Typically
the fluxes of neutrinos or antineutrinos obtained are about
10$^7$~cm$^{-2}$~s$^{-1}$, occasionally somewhat more or somewhat less.
(We shall not discuss the subject of reactor antineutrino physics where
considerably larger fluxes and much smaller cross sections than those
encountered below are relevant --- see Ref.~[Don79a] for further treatment of
this subject.) To
set the scale of neutrino--nuclear physics let us assume a target with
10$^{31}$ protons (typically a target+detector will have about 10\%
hydrogenic protons) and consequently would weigh about 200 tons ---
for example, as does the LSND detector discussed in Sect.~V.C.5. This yields
an effective neutrino--proton luminosity of 10$^{38}$~cm$^{-2}$~s$^{-1}$.
For a typical $\nu$--$p$ total cross section of 10$^{-40}$~cm$^2$ [Don83] this
yields an event rate of 10$^{-2}$~s$^{-1}$.  To accumulate 10$^4$ events and
thus reach the level of 1\% statistical error needed to make such measurements
relevant as probes of
strangeness or non--standard physics would take 278~hr.  Naturally the
running time will scale with the flux and consequently will vary with the
nature of the neutrino--producing facility (see Table~5.3).  Additionally,
the $\nu$--$p$ total cross section varies with energy from about 0.2 $\times$
the above number for beam stop neutrino sources to a few $\times$ the above
number for DIF facilities [Don83].  Accordingly, the typical running time
for 1\% statistics
goes from a few hundred to a few thousand hours. The level of precision
realistically achievable depends on kinematics and neutrino species ($\nu$
or $\nubar$). Similarly, the sensitivity of cross sections to different
form factors or electroweak couplings is also highly dependent  on the
same factors. Consequently, any discussion of the interpretation of
neutrino cross sections must take into account considerations of experimental
doability.

\goodbreak
\bigskip
\noindent IV.J.1.\quad GENERAL FEATURES
\medskip

        The summary of accessible kinematic regimes for neutrino beams
given in Table 5.3 points to an important difference between neutrino
and electron scattering experiments. In the latter instance,
one may tune $\epsilon$ and $\theta$ to define a rather narrow
kinematic window and thereby perform Rosenbluth--type separations and, in
so--doing, highlight
contributions from different pieces of the hadronic NC. Neutrino
scattering experiments, on the other hand, integrate over a range of energies
about some peak energy as determined by the neutrino spectrum $\phi(\epsilon)$.
The corresponding inclusive observable is the energy--integrated differential
cross section
$$
{\overline{d\sigma\over dQ^2}} = \int_{\epsilon_1}^{\epsilon_2}\phi(\epsilon)
{d\sigma(\epsilon)\over dQ^2}\ \ , \eqno\nexteq\nameeq\Edsignu
$$
where $d\sigma(\epsilon)/dQ^2$ is the energy--dependent $\nu$ or $\nubar$
differential cross section
and $(\epsilon_1, \epsilon_2)$ define the range of
available beam energy for a given facility (see Table 5.3). For a given value
of $Q^2$, then, the integral in Eq.~(\Edsignu) effectively integrates over
the lepton scattering angle, thereby precluding the possibility of making
a Rosenbluth--type separation of the inclusive cross section. Consequently,
the strategy appropriate to
neutrino scattering experiments is to consider measurements at a variety of
facilities, thereby accessing different
$\epsilon$ and $Q^2$ regimes, in order to emphasize
in Eq.~(\Edsignu) the dependence on different form factors or electroweak
couplings. Eq.~(\Edsignu) also points to the importance of one's knowledge
of the neutrino spectrum $\phi(\epsilon)$ in the interpretation of the measured
cross section.

When studying electron scattering or charge--changing neutrino/antineutrino
reactions it is possible to detect the final--state charged lepton and
therefore inclusive cross sections can be measured.  In contrast, for
neutrino/antineutrino scattering it is not practical to detect the scattered
lepton and hence some hadronic signature must be found to know that
a scattering event has occurred. This means that a {\it different kind of
inclusive cross section} must be studied: the scattered neutrino is not
detected (and thus the cross section is in general inclusive in the
leptonic sector), whereas some hadronic final--state particle detection is
involved (and thus the cross section is {\it not} inclusive in the hadronic
sector). In other words, the measurement is more like a coincidence
(semi--inclusive) electron scattering experiment where the scattered electron
is not detected, {\it viz.,\/} more like nuclear photoreactions.

In the case of {\it elastic} neutrino scattering, for example,
one typically measures the nuclear recoil having kinetic energy $T_\sst{R}$
and angle
$\theta_\sst{R}$ relative to the incident beam direction, as discussed in
Sect.~III.E.3 (see Eqs.~(\Ekinconvx)). To set the scale for the subject of
the next subsection, in Table~4.1 we give the accessible ranges in
$Q^2$, $\tau$ and $T_\sst{N}$ and their
equivalents in $\theta_\sst{N}$ and $\theta$ for a range of values of
$\epsilon$ available at different facilities ranging from low--energies
(beamstop facilities) to medium--energies (decay--in--flight (DIF)
facilities) for neutrino--nucleon elastic scattering.
Following the discussion of $\nu$--N elastic scattering in Sect.~IV.J.2,
in Sects.~IV.J.3 and IV.J.4 we
return to consider elastic scattering from the deuteron and from spin--0
nuclei such as $^4$He, respectively.

\midinsert
$$\hbox{\vbox{\offinterlineskip
\def\strut{\hbox{\vrule height 15pt depth 10pt width 0pt}}
\hrule
\halign{
\strut\vrule#\tabskip 0.2cm&
\hfil$#$\hfil&
\vrule#&
\hfil$#$\hfil&
\vrule#&
\hfil$#$\hfil&
\vrule#&
\hfil$#$\hfil&
\vrule#&
\hfil$#$\hfil&
\vrule#&
\hfil$#$\hfil&
\vrule#\tabskip 0.0in\cr
& \multispan{11}{\hfil\bf TABLE 4.1\hfil} & \cr\noalign{\hrule}
& \hbox{Neutrino energy} &&\theta\ \hbox{range}  && \theta_\sst{N}
\ \hbox{range}&&|Q^2|\ \hbox{range} && \tau && \hbox{Recoil energy} & \cr
& \hbox{(MeV)} && && &&\hbox{(GeV/c)}^2 && && \hbox{MeV} &\cr
\noalign{\hrule}
& 35 && 0^\circ \rightarrow 180^\circ && 90^\circ \rightarrow 0^\circ
&& 0 \rightarrow 0.00046&& 0 \rightarrow 0.0013 &&
0 \rightarrow 2.43 &\cr
& 50 && 0^\circ \rightarrow 180^\circ && 90^\circ \rightarrow 0^\circ
&& 0 \rightarrow 0.0090&& 0 \rightarrow 0.0026 &&
0 \rightarrow 4.81 &\cr
& 100 && 0^\circ \rightarrow 180^\circ && 90^\circ \rightarrow 0^\circ
&& 0 \rightarrow 0.033&& 0 \rightarrow 0.0093 &&
0 \rightarrow 17.6 &\cr
& 200 && 0^\circ \rightarrow 180^\circ && 90^\circ \rightarrow 0^\circ
&& 0 \rightarrow 0.112&& 0 \rightarrow 0.032 &&
0 \rightarrow 59.7 &\cr
& 500 && 0^\circ \rightarrow 180^\circ && 90^\circ \rightarrow 0^\circ
&& 0 \rightarrow 0.484&& 0 \rightarrow 0.137 &&
0 \rightarrow 257.9 &\cr
& 1000 && 0^\circ \rightarrow 180^\circ && 90^\circ \rightarrow 0^\circ
&& 0 \rightarrow 1.278&& 0 \rightarrow 0.362 &&
0 \rightarrow 680.5 &\cr
\noalign{\hrule}}}}
$$
\smallskip
\baselineskip 10pt
{\ninerm
\noindent\narrower {\bf Table 4.1} \quad Achievable $|Q^2|$, $\tau$,
recoil energy ($T_\sst{N}$), together with equivalent
recoil nucleon scattering angle ($\theta_\sst{N}$) and (undetected)
neutrino scattering angle ($\theta$) for various neutrino energies,
$\epsilon$ (see
also Table 5.3).  The target is assumed to be a nucleon at rest.
\smallskip}
\endinsert

\baselineskip 12pt plus 1pt minus 1pt

In the case of inelastic neutrino or antineutrino scattering alternatives
to detecting the recoiling target nucleus can be sought.  For example, as
discussed in more detail in Sect.~IV.J.5, it is possible to excite a discrete
state in a nucleus via inelastic neutrino scattering and then detect the
decay of that state, say by the emission of a photon.  In a sense such
measurements involve ``nothing in'' and ``nothing out'' except the emission
(when the neutrino--producing beam is on) of a photon.  Of course, other
final--state hadronic signals may be sought (emission of an $\alpha$--particle,
ejection of a proton or neutron, {\it etc.}).  Some of these are
discussed in Ref.~[Don79a] and the references contained therein.  We shall
return in Sect.~IV.J.5 to consider only a few cases to illustrate the basic
nature of such neutrino--nuclear studies.

Let us note for the present that even the above classes of measurements are
not always so cleanly separated when real experiments are contemplated.  For
instance,
for purposes of interpreting elastic $\nu$--N and $\nubar$--N cross
section measurements, one faces a third level of complication not encountered
in elastic PV $\evec p$ scattering. Typically, the elastic neutrino--nucleon
cross section must be extracted from a measurement of the
quasielastic $A(\nu, N)A'$ cross
section, where $A$ and $A'$ denote  target and daughter nuclei. In the
BNL $\nu$--p/$\nubar$--p experiment, for example, 80\% of the \lq\lq elastic"
events actually involved scattering from protons bound in $^{12}$C nuclei.
The remaining 20\% resulted from scattering from $^1$H nuclei in the CH$_2$
target. Extraction of the elastic $\nu$--N cross section from
the $A(\nu, N)A'$ reaction generally introduces more theoretical uncertainty
than would enter the extraction from the quasielastic inclusive $A(\nu, \nu')X$
reaction, were a measurement of the latter possible.
Interpretation of the former process, which is semi--inclusive,
requires knowledge of the final--state interactions between the outgoing
nucleon and residual nucleus. In contrast, the inclusive
quasielastic measurement
involves detection of the outgoing lepton (assuming such were possible
--- as it is for the charge--changing neutrino and antineutrino reactions
where a charged lepton is produced), and
no knowledge of the hadronic final--state interactions is needed in the
interpretation. An indication of the scale of theoretical uncertainties
in the interpretation of $A(\nu, N)A'$ reactions might be obtained from
analyses of quasielastic
$A(e, e'p)A'$ coincidence cross sections. Typically the final--state
interactions play a very important role in interpreting such electromagnetic
coincidence cross sections, but play a much less significant part in
determining the inclusive $A(e,e')$ cross section.  Quantifying this
statement requires a case--by--case analysis and lies beyond the context of
the present review. However, one can get some sense of the importance of
final--state interactions by noting that for coincidence electron
scattering the DWIA (where final--state interactions are included) differs
from the PWIA (where they are not) typically by factors of about 0.6--0.7
(see, {\it e.g.,\/} [Fru84]), while for inclusive
scattering much smaller modifications from final--state interaction effects
are usually found.  We shall return briefly to this point in the next section
when discussing the ideas put forward by the authors of Ref.~[Gar92].

\goodbreak
\bigskip
\noindent IV.J.2.\quad ELASTIC NEUTRINO--NUCLEON SCATTERING
\medskip

        The aforementioned theoretical challenges notwithstanding, measurements
of elastic and low--lying inelastic neutrino cross sections offer a number of
potential advantages. We focus first on elastic neutrino--nucleon scattering,
for which the primary attraction is its sensitivity to $\gtilan$. Unlike
its contribution to $\alr(\evec N)$, which is suppressed by $\gve=-1+
4\sstw$ $\approx -0.092$, the $\gtilan$ contribution to $d\sigma^{\nu
(\nubar)}/dQ^2$
receives no such suppression. Moreover, the interpretation $\gtilan$, as
determined by neutrino scattering, is theoretically less ambiguous than
in the case of PV electron scattering. Recall from Eq.~(\EGtilde) that
one has $\gtilan = \xiateo\gathree\tau_3+\xiatez\gaeight +\xiaz\GAS$,
where $\gathree$ and $\gaeight$ may be determined from neutron and hyperon
semileptonic decays, respectively, and where the $\xi_\sst{A}^{(a)}$
are renormalized electroweak axial--vector NC couplings determined by
the underlying gauge theory. As noted earlier, $\xiatez$ vanishes at
tree--level in the Standard Model, but becomes nonzero once electroweak
radiative corrections are included. For PV electron scattering, these
corrections, which enter all three of the $\xi_\sst{A}^{(a)}$, are
enhanced over the generic $\alpha/4\pi$ scale and contain theoretical
hadronic uncertainties on the same scale. For neutrino scattering, on
the other hand, the corrections are significantly smaller and more
reliably calculable. Consequently, an extraction of $\gtilan$ from the
elastic neutrino--nucleon cross section offers a theoretically \lq\lq cleaner"
probe of, {\it e.g.}, axial--vector nucleon strangeness, than does the
corresponding determination from $\alr(\evec p)$.

        In addition to its dependence on $\GAS$, the energy--integrated
elastic $\nu$--p ($\nu$--n) cross section displays a non--negligible dependence
on other quantities of interest, such as the electroweak couplings $\xivp$
($\xivn$), the axial--vector dipole mass parameter (see Eq.~(\GDaxial)), and
the vector current strangeness form factors (see Sect.~III.C).
For low--energy, low momentum transfer scattering, one may expand the
cross section in Eq.~(\Edsignup) to leading order in $\epsilon/\mn$ and
$\tau$ and obtain [Gar92]
$$
\eqalign{ {d\sigma\over dQ^2}(\epsilon)&={G_\mu^2\over 8\pi}\biggl[
	(1+\rho^2) (\gtilan)^2 + (1-\rho^2) (\gtilen)^2 \cr
        &\quad\quad - 4\rho\sqrt{\tau}
	\Bigl\{{1\over 2}\Bigl((\gtilan)^2+(\gtilen)^2\Bigr)
	\pm\gtilan\gtilmn\Bigr\}\biggr]\ \ , \cr}
\eqno\nexteq\nameeq\Edsignulow
$$
where
$$
\rho\equiv \sqrt{\tau}\bigl({\mn\over\epsilon}\Bigr) = \sqrt{|Q^2|/4\epsilon^2}
=\sqrt{1+\tau}\cos\theta_\sst{N} -\sqrt{\tau}
\eqno\nexteq
$$
so that $0\leq \rho \leq [1+2\epsilon/\mn]^{-1/2}$.
{}From Eq.~(\Edsignulow) one may observe the relative importance of the
axial--vector form factor in comparison with contributions from others,
especially for target protons. In this case, the NC electric form factor
of Eq.~(\EGtilde) is suppressed due to the $\xivp=(1-4\sstw)$ coefficient
of $\GEp$ and the leading $Q^2$--dependence of $\GEn$ and $\GES$ as they
enter $\gtilep$. The contribution from $\gtilmn$ to Eq.~(\Edsignulow) is
suppressed by $\sqrt{\tau}$ relative to the leading $\gtilan$ contribution.
Hence, one expects $\gtilan$ to dominate the cross section at low--energy
and low momentum transfer [Don74]. The situation for target neutrons is
somewhat
different, since the leading contribution to $\gtilenn$ is $\xivn\GEp$,
which is not suppressed.

        To illustrate the sensitivities of the elastic $\nu$--p and
$\nubar$--p differential cross sections in various kinematic regimes, we give
in
Table~4.2 the fractional shift in $d\sigma^{\nu (\nubar)}(\epsilon)/dQ^2$
due to variations in several parameters of interest. The higher
energy and momentum transfer results correspond to the BNL kinematics,
while those for lower energies and $|Q^2|$ are appropriate to the LSND
measurement. Although the BNL experiment integrated over beam energy from
0.2 to 5.0 GeV, the spectrum $\phi(\epsilon)$ was peaked in the neighborhood
of 1 GeV. One expects the sensitivities at this energy, then, to represent
rather fairly the sensitivity of the energy--integrated cross section.

\midinsert
\baselineskip 17pt plus 1pt minus 1pt
$$\hbox{\vbox{\offinterlineskip
\def\strut{\hbox{\vrule height 14pt depth 10pt width 0pt}}
\hrule
\halign{
\strut\vrule#\tabskip 0.2cm&
\hfil$#$\hfil&
\vrule#&
\hfil$#$\hfil&
\vrule#&
\hfil$#$\hfil&
\vrule#&
\hfil$#$\hfil&
\vrule#&
\hfil$#$\hfil&
\vrule#\tabskip 0.0in\cr
& \multispan9{\hfil\bf TABLE 4.2\hfil} & \cr\noalign{\hrule}
& (\epsilon, |Q^2|) && \hbox{Parameter} && \hbox{range} &&
\hbox{variation in} && \hbox{variation in} & \cr
& && && && d\sigma^{\nu}(\epsilon)/dQ^2\ (\%) &&
d\sigma^{\nubar}(\epsilon)/dQ^2
\ (\%) &\cr
\noalign{\hrule}
&(1.0, 0.5) && \xivp && 0.1003\to 0.0821
&&0.7\to -0.6 &&-0.6\to 0.5 &\cr
&(1.0, 0.5) && \mustr &&-0.2\to 0.2 &&6\to -5 &&-4\to 5 &\cr
&(1.0, 0.5) && \rhostr &&-1.4\to 1.4 &&0.3\to -0.7 &&0.6\to -2 &\cr
&(1.0, 0.5) && \eta_s &&-0.12\pm 0.07 &&14\mp 8 &&25\mp 15 &\cr
&(1.0, 0.5) && \lambda_\sst{D}^\sst{A} && 3.12\to 3.52 &&5.7\to -5.3
&&10.1\to -9.0 &\cr
\noalign{\hrule}
&(1.0, 1.0) && \xivp && 0.1003\to 0.0821
&&0.8\to -0.7 &&-2.1\to 2.2 &\cr
&(1.0, 1.0) && \mustr && -0.2\to 0.2 &&6\to -6 &&-15\to 18&\cr
&(1.0, 1.0) && \rhostr &&-1.4\to 1.4 &&0.2\to 0.6 &&1.3\to 5.1&\cr
&(1.0, 1.0) && \eta_s &&-0.12\pm 0.07 &&13\mp 7 &&40\mp 25 &\cr
&(1.0, 1.0) && \lambda_\sst{D}^\sst{A} &&3.12\to 3.52 &&8.2\to -7.2 &&
24.5\to -19.9 &\cr
\noalign{\hrule}
&(0.15, 0.05) && \xivp && 0.1003\to 0.0821
&&0.4\to -0.3 &&-0.5\to 0.5 &\cr
&(0.15, 0.05) && \mustr &&-0.2\to 0.2 && 3\to -3 &&-4\to 4 &\cr
&(0.15, 0.05) && \rhostr && -1.4\to 1.4&&0.07\to 0 &&0.05\to -0.03 &\cr
&(0.15, 0.05) && \eta_s &&-0.12\pm 0.07 &&17\mp 10 &&25\mp 15 &\cr
&(0.15, 0.05) && \lambda_\sst{D}^\sst{A} &&3.12\to 3.52 &&0.9\to -0.9 &&
1.3\to -1.3&\cr
\noalign{\hrule}}}}
$$
\smallskip
\baselineskip 10pt
{\ninerm
\noindent\narrower {\bf Table 4.2} \quad Variation in elastic
$\nu$--p and $\nubar$--p differential cross section associated with
uncertainties in electroweak couplings and form factor parameters (units:
$\epsilon$ [GeV], $|Q^2|$ [(GeV/c)$^2$]). The
final two columns give the percent deviation from nominal Standard Model,
no--strange predictions for the cross sections. The central value of
the axial--vector dipole parameter $\lambda_\sst{D}^\sst{A} =3.32 $
corresponds to
$M_\sst{A} = 1.032$ GeV. Variations due to $\GES$ uncertainty were computed
assuming $\lams=0$.
\smallskip}
\endinsert

\baselineskip 12pt plus 1pt minus 1pt

        From the entries in Table~4.2, one may observe immediately the
level of precision required for a meaningful electroweak test. In all but
a few cases, measurements of $d\sigma(\epsilon)/dQ^2$ to better than
1\% precision are needed to determine $\xivp$ to 10\% (the precision assumed
in deriving the corresponding bands in Fig.~2.4). Since the cross sections
were determined to $\sim 10\% $ precision in the BNL experiment, while the
LSND expectation is for a 10\%
determination of the cross section, significantly higher precision
experiments would be required at existing or future facilities to
make a meaningful extraction of $\xivp$. For the form factors, on the other
hand, significant constraints could be achieved under realistic experimental
conditions. In the case of $\GMS$, for example, a modest improvement in
statistics over the BNL experiment could lead to a determination of $\mustr$
at a level comparable to the projected
SAMPLE constraints. In contrast, the sensitivity
to $\GES$ appears to be well below the observable level. Given this
insensitivity, we have not included in Table~4.2 the shifts associated
with variations in $\lams$.
The cross sections display the greatest sensitivity to variations in
$\GAS$, assuming its current level of uncertainty. As the
sensitivity to $\lambda_\sst{D}^\sst{A}$ makes clear, it is desirable to
attempt
a determination of $\eta_s$ at lower energy and momentum transfer. At the
BNL kinematics, for example, the variation associated with the present
uncertainty in $\lambda_\sst{D}^\sst{A}$ is comparable to that associated with
the error in $\GAS$ (using the BNL results), whereas at the LSND kinematics,
the $\lambda_\sst{D}^\sst{A}$--variation is an order--of--magnitude smaller
than that associated with $\eta_s$. To improve upon the BNL $\GAS$ constraints,
however, a measurement of $d\sigma^{\nu (\nubar)}(\epsilon)/dQ^2$ at LSND
kinematics would need to be carried out with better precision than presently
projected. We note in passing that an analysis similar to the above was
carried out in Ref.~[Bei91a]. In the latter work, the shift in
$d\sigma^{\nu(\nubar)}(\epsilon)/dQ^2$ from the zero--strangeness predictions
are given, assuming the predictions of Ref.~[Jaf89] for the vector current
form factors and the EMC value for $\eta_s$ (see Table~2.3).

        Any extraction of constraints on form factors and couplings must,
of course, account for simultaneous uncertainties in the parameters of
interest. These correlations are somewhat more complicated than in the
case of PV electron scattering, where the NC form factors and electroweak
couplings enter $\alr$ linearly. In contrast, the differential $\nu$--N
and $\nubar$--N cross sections are quadratic in these quantities, so that
the correlations define generalized conic sections in a multi--dimensional
parameter space. To illustrate, we consider elastic $\nu$--p and $\nubar$--p
measurements at the BNL kinematics. The BNL experiment was originally
analyzed assuming that all vector strangeness form factors vanish identically.
A reanalysis of the data allowing for nonzero $\GES$ and $\GMS$ would require
detailed knowledge of the neutrino energy spectra, normalizations,
systematic error correlations, as well as other systematics including
charge--changing and quasielastic cross sections for calibration, {\it etc.}
Such a reanalysis has in fact recently been carried out by the authors of
Ref.~[Gar93], but we wish to examine here, in a more general way, the
sensitivity of these data to the strangeness form factors and electroweak
couplings.

        To this end, we generate \lq\lq fictitious" data points by
assuming the Standard Model with {\it no} strangeness in the nucleon and
add arbitrary random statistical errors of ${\cal O}$(10\%) (roughly the
same magnitude as the actual BNL errors) to $d\sigma/d Q^2$ for the same
kinematic points as in the BNL experiment. Performing a $\chi^2$--minimization
on this fictitious data set to fit different input parameters then yields an
effective uncertainty for the parameters considered. This procedure affords
at least an estimate of the {\it scale} of the uncertainties and correlations
that a more complete reanalysis would provide. At the same time, this
procedure neglects systematic errors, so that the uncertainty estimates
provided below should be considered underestimates.

        We chose as parameters to be varied $\xivp$, $\eta_s$, $\mustr$,
$\rhostr$, and $M_A$. For simplicity, we did not consider additional
uncertainties associated with the $\lambda_\sst{E,M}^{(s)}$. Setting
all but one of the parameters to their nominal Standard Model values and
allowing the last to vary, we take as the statistical uncertainty for a
one--parameter fit the range in this parameter which keeps $\chi^2$ within one
unit of its minimum. Allowing a second parameter to vary leads to a 67\%
confidence contour whose extrema define the correlated uncertainties.
For example, allowing only $\xivp$ to vary gives about a 44\% uncertainty
in this parameter\footnote{*}{Recall that a 10\% uncertainty in $\xivp$
corresponds roughly to a 1\% uncertainty in $\sstw$.}. Allowing $\eta_s$
to vary as well gives $\delta\xivp/\xivp\approx 0.66$ at 67\% confidence
(see Fig.~4.13).
A three--parameter fit in which $M_A$ is also
allowed to vary leads to an 88\% uncertainty in $\xivp$. From these results
one would conclude that a future medium--energy $\nu$--p/$\nubar$--p elastic
scattering experiment would have to be performed with significantly
higher precision than obtained in the BNL experiment in order to constrain
new physics at a level competitive with atomic PV or prospective PV
electron scattering experiments.

The situation with regard to the strangeness form factor constraints is
somewhat more hopeful. In Table~4.3 we summarize these prospective
constraints, allowing for correlated uncertainties among the various form
factors assuming $\xivp$ takes on its Standard Model value.

\midinsert
$$\hbox{\vbox{\offinterlineskip
\def\strut{\hbox{\vrule height 15pt depth 10pt width 0pt}}
\hrule
\halign{
\strut\vrule#\tabskip 0.2cm&
\hfil$#$\hfil&
\vrule#&
\hfil$#$\hfil&
\vrule#&
\hfil$#$\hfil&
\vrule#&
\hfil$#$\hfil&
\vrule#\tabskip 0.0in\cr
& \multispan7{\hfil\bf TABLE 4.3\hfil} & \cr\noalign{\hrule}
& \hbox{Case} && \hbox{Varied Parameters} && \hbox{Experimental
Uncertainty} &&\hbox{Constraints}&\cr\noalign{\hrule}
& \hbox{BNL} && M_A, \eta_s && \pm 10\%\ \hbox{in $\nu ,\nubar$ cross sections}
&&\delta\eta_s=\pm 0.12 &\cr
& && && && \delta M_A = 0.1 &\cr
\noalign{\hrule}
& \hbox{BNL} && M_A, \eta_s, \mustr && \pm 10\%\ \hbox{in $\nu ,\nubar$
cross sections} &&\delta\eta_s=\pm 0.16 &\cr
& && && && \delta M_A = 0.1 &\cr
& && && && \delta\mustr = 0.2  &\cr
\noalign{\hrule}
& \hbox{BNL} && \rhostr, \mustr, \eta_s && \pm 10\%\ \hbox{in $\nu ,\nubar$
cross sections} &&\delta\rhostr=\pm 5.0 &\cr
& && && && \delta\mustr = \pm 0.3&\cr
& && && && \delta\eta_s = \pm 0.06&\cr
\noalign{\hrule}
&\hbox{LSND}  && \eta_s  && \pm 18\%\ \hbox{in $\nu$ cross section}
&& \delta\eta_s=\pm 0.1 &\cr
\noalign{\hrule}}}}
$$
\smallskip
\baselineskip 10pt
{\ninerm
\noindent\narrower {\bf Table 4.3} \quad Multi--parameter fits to
\lq\lq fictitious" BNL and LSND data, assuming $\xivp$ at its
Standard Model value.
\smallskip}
\endinsert

\baselineskip 12pt plus 1pt minus 1pt

{}From the three--parameter fit in Table~4.3, we observe that the
constraint $\delta\mustr=\pm 0.3$ is nearly as stringent as that expected
from the SAMPLE experiment ($\delta\mustr\approx \pm 0.2$).
The recent reanalysis of the actual BNL data by [Gar93] gives error
estimates quite compatible with the above numbers, if somewhat larger due
to their inclusion of the (systematic) normalization uncertainties.
In contrast,
the uncertainty in $\rhostr$ is nearly three times less
stringent than would be achievable from a series of $\alr(\evec p)$
measurements alone. The constraint on $\eta_s$, achieved
for fixed $M_A$,  could not be approached by PV electron scattering, given
the large axial--vector radiative correction uncertainties arising in the
latter.
experiment, where the impact of
Since the $\eta_s$ uncertainty increases in a two--parameter
$M_A$--$\eta_s$ fit, it is desirable to carry out a determination of
$\GAS$ at the lower--energy and lower--$|Q^2|$ of the LSND experiment, where
the impact of $M_A$ uncertainties on the differential cross section is
negligible (see Table~4.2). However, the projected 20\% statistical
error for the LSND determination of the cross section corresponds to an
uncertainty in $\eta_s$ only slightly better than the uncertainty taken
from the fictitious BNL data allowing for the correlation with $M_A$
[Bei91a].

        In order to achieve or surpass the aforementioned form factor
constraints, improvements in both experimental precision and the reliability
of theoretical modeling would be required. Prospects for experimental
progress are discussed in Sect.~V. As far as theoretical analysis is
concerned, contributions from final--state hadronic interactions render the
extraction of $d\sigma(\nu N)/dQ^2$ from the $A(\nu, N)A'$ cross section
theoretically problematic, as discussed above.  It may prove advantageous to
make use of existing analyses of $A(e,e'N)A'$ reactions to ``calibrate'' the
hadronic physics and thus, at least to some extent, to remove the
final--state interaction effects as uncertainties.  Given the possibility of
high--precision measurements in the future, more theoretical work on these
issues will have to be undertaken.

        In order to minimize the impact of some of these ambiguities,
the authors of Ref.~[Gar92] have proposed measuring the ratio $R(\epsilon)$
of proton to neutron yields in quasielastic nucleon knockout.
Assuming the static approximation
(initial nucleon at rest) and neglecting final--state interactions, one has
$$
R(\epsilon)={\int_{|Q^2_1|}^{|Q^2_2|}\Bigl\{d\sigma^{\nu p}(\epsilon)/dQ^2
\Bigr\}dQ^2\over
\int_{|Q^2_1|}^{|Q^2_2|}\Bigl\{d\sigma^{\nu
n}(\epsilon)/dQ^2\Bigr\}dQ^2}\ \ . \eqno\nexteq\nameeq\Ergara
$$
These authors compute $R$
for $\epsilon = 200$ MeV and detection of outgoing nucleons in
the energy range $50 \leq T_\sst{N} \leq 59.7$ MeV. Assuming quasifree
kinematics and taking the initial nucleon to be at rest,
this spread corresponds to a range in $|Q^2|$ of
$0.094 \leq |Q^2| \leq 0.112$ (\gevoc)$^2$ or, from Eq.~(\Ekinconvx),
a range in forward angles of $23.5^\circ\geq\theta_\sst{N}\geq 0^\circ$.
The resultant value of $R$ depends on the strangeness
parameters $\mustr$ and $\eta_s$ as
$$
R\approx 1-3.16\eta_s-0.362\mustr+0.073\eta_s^2+0.024\eta_s\mustr\ \ ,
\eqno\nexteq\nameeq\Ergarb
$$
where in accordance with Table~4.2 the dependence on uncertainties in
$M_A$ has been neglected. The authors also point out that $R$ is
insensitive to uncertainties in neutrino flux, since the same flux is
used in extracting both proton and neutron knockout cross sections
and then forming the ratio. Furthermore,
$R(\epsilon)$ carries only a gentle dependence on $\epsilon$ for
$160\leq\epsilon\leq 240$ MeV, so that one need not possess precise
knowledge of $\phi(\epsilon)$ in order to interpret this ratio.
One might
also hope that at these kinematics, the final--state interactions for
outgoing protons and neutrons behave similarly, so that their impact on
$R$ would be much smaller than on the individual cross sections. Perhaps
the most important feature of Eq.~(\Ergarb) is its large sensitivity to
$\eta_s$. Indeed, a 10\% determination of $R$ could constraint $\eta_s$
to $\delta\eta_s=\pm 0.03$, while the error induced in $\eta_s$
from $\delta\mustr$ assuming the SAMPLE projections would be roughly
$\delta\eta_s=\pm 0.02$. Hence, further analysis of the impact of
final--state interactions on the interpretation of $R$, as well as exploration
of experimental feasibility, appears to be warranted.

\goodbreak
\bigskip
\noindent IV.J.3.\quad ELASTIC SCATTERING FROM $^4\hbox{He}$
\medskip

The basic formalism for elastic neutrino and antineutrino scattering from
spin--0 nuclei has been introduced in Sect.~III.E.3.  In Ref.~[Don83] a
selection of targets (all with spin--0 except $^2$H, which was also
studied in that work --- see the following discussions in Sect.~IV.J.4) was
investigated with an eye to the problem of elastic scattering. The general
characteristics of coherent scattering involve a recoil kinetic energy
which falls with mass number as $A^{-1}$, but a cross section which increases
with the square of a combination of the proton (atomic) and neutron numbers,
$(Z \xi_V^p + N \xi_V^n)^2$, as discussed in Sect.~III.D.1.  Thus, the
challenge
for such studies is to measure small cross sections (albeit, much larger than
single--particle transition cross sections due to the coherence) by
detecting very low energy recoils.
In the present
subsection we shall only briefly touch upon a single example of this class
of reactions, {\it viz.,\/} elastic scattering from $^4$He.  No evaluation
of the do--ability of using such a target will be attempted here, although
it should be remarked that the fact that $^4$He is a scintillator could
make such a target/detector promising.

Typical conditions for use with DIF neutrino beams are represented in
Table~4.4 (taken from Ref.~[Don83]).

$$\hbox{\vbox{\offinterlineskip
\def\strut{\hbox{\vrule height 15pt depth 10pt width 0pt}}
\hrule
\halign{
\strut\vrule#\tabskip 0.2cm&
\hfil$#$\hfil&
\vrule#&
\hfil$#$\hfil&
\vrule#&
\hfil$#$\hfil&
\vrule#&
\hfil$#$\hfil&
\vrule#\tabskip 0.0in\cr
& \multispan7{\hfil\bf TABLE 4.4\hfil} & \cr\noalign{\hrule}
& \hbox{Quantity} &&\epsilon = 100\ \hbox{MeV}  && \epsilon = 150\ \hbox{MeV}
&& \epsilon = 200\ \hbox{MeV}  & \cr
\noalign{\hrule}
& \theta_\sst{R}^{\rm peak} && 62^\circ && 64^\circ && 69^\circ &\cr
& |Q^2|\ \hbox{(GeV/c)}^2 && 0.0084 && 0.016 && 0.018 &\cr
& \tau_\sst{N} && 0.0024 && 0.0045 && 0.0052 &\cr
& q\ \hbox{(MeV/c)} && 92 && 126 && 136 &\cr
& T_\sst{R}\ \hbox{(MeV)} && 1.13 && 2.13 && 2.48 &\cr
& d\sigma/d\Omega_\sst{R}\ (\times 10^{-41}\ \hbox{cm}^2\ \hbox{sr}^{-1}) &&
2.6 && 4.6 && 6.6 &\cr
\noalign{\hrule}}}}
$$
\smallskip
\baselineskip 10pt
{\ninerm
\noindent\narrower {\bf Table 4.4} \quad Elastic neutrino scattering
from $^4$He at intermediate energies.  For given values of the neutrino
energy, $\epsilon$, the quantities in the table are given for recoil
angles corresponding to the peak in the differential cross section.
The dimensionless momentum transfer $\tau_\sst{N}$ is calculated using
the nucleon mass ($\equiv |Q^2|/4\mns$) rather than the mass of $^4$He.
\smallskip}

\baselineskip 12pt plus 1pt minus 1pt

Clearly, even for a target as light
as $^4$He, the recoil energies are rather small for detection in a
large--active--volume detector.  The quantity of interest is that
which multiplies the electromagnetic form factor in Eq.~(\Ehefourneu),
namely
$$
X\equiv\sqrt{3}\xi_V^{T=0}[1+\Gamma(q)]+\xi_V^{(0)}F_{C0}(s)/F_{C0}(T=0)
\ \ . \eqno\nexteq
$$
Estimating the form factor ratio in last term by
$$
{{F_{C0}(s)}\over{F_{C0}(T=0)}}={{G_E^{(s)}}\over{G_E^{T=0}}}
\cong 2\rho_s\tau\ \ , \eqno\nexteq
$$
setting $\xi_E^{(s)}$ to unity, since the momentum transfer is very small,
and for the isospin--mixing factor writing approximately
$$
\Gamma(q)\cong \Gamma_0 \tau + {\cal O}(\tau^2) \ \ , \eqno\nexteq
$$
where from Ref.~[Don89] one has that $\Gamma_0\approx 0.04$. At
tree--level, the entire
multiplying factor above then has the form
$$
X=-4\sin^2 \theta_\sst{W} \Bigl[ 1 + \gamma_0\tau + {\cal O}(\tau^2)
\Bigr] \ \ , \eqno\nexteq\nameeq\Factorx
$$
where tree--level coupling have been employed and where $\gamma_0=
\Gamma_0+\rho_s/2\sin^2 \theta_\sst{W}$.  For $|\rho_s|$ set to the
value 2 [Jaf89] one finds that $|\gamma_0|\cong 4.4$ with only a 1\%
correction from $\Gamma_0$.  The term $\gamma_0 \tau$ then contributes
about 1--2\% compared to the leading term in $X$ for the conditions in
Table~4.4 and it thus provides only a small modification of the
basic $4\sin^2 \theta_\sst{W}$ dependence in Eq.~(\Factorx).  For the
conditions studied here the problem is similar to that of attempting a
Standard Model test with PV elastic electron scattering at low--energies
(see Sect.~IV.B.2).  Naturally, as in the discussions in Sect.~IV.B,
at higher--energies the influence of both the isospin--mixing term and the
electric strangeness term is larger; the possibility exists that the
clarity of the nuclear recoil as a signature for elastic scattering might
make neutrino scattering advantageous for studies of hadronic structure in
much the same way that is expected for PV electron scattering.  Given
sufficiently high--quality determinations from both types of lepton
scattering measurements, it would then be possible to address the issue of
how the radiative corrections differ for the various flavors of leptons.

\goodbreak
\bigskip
\noindent IV.J.4.\quad ELASTIC NEUTRINO--DEUTERON SCATTERING
\medskip

Experimental considerations make it less likely that elastic $\nu$--D
scattering will prove useful as a probe of nucleon strangeness or
non--standard physics than will either elastic neutrino--nucleon
or elastic PV $^2\hbox{H}(\evec, e')$ scattering.
Nevertheless, elastic $\nu$--D scattering
illustrates the special issues one encounters when seeking to study
nucleon structure or electroweak physics with elastic neutrino
scattering from nuclei with nonzero spin.

Apart from experimental considerations, the interpretation of
elastic $\nu$--D
measurements involves complications not present in $\nu$--N or PV electron
scattering. In particular, both the nuclear and nucleon form factors
contribute to $d\sigma(\nu D)/dQ^2$. In contrast, by a suitable choice
of target and kinematics, much of the nuclear physics may be eliminated
from $\alr$, which involves a ratio of hadronic response functions. In the
case of neutrino scattering, one might similarly attempt to minimize
one's sensitivity to nuclear physics by studying the $\nu$--D/$\nubar$--D
asymmetry [Hen91, Fre92]
$$
\delnnb\>=\>{{d\sigma/ dQ^2}\Bigr\vert_{\nu\sst{D}}-{d\sigma/ dQ^2}\Bigr
\vert_{\nubar\sst{D}}\over {d\sigma/ dQ^2}\Bigr\vert_{\nu\sst{D}}+
{d\sigma/ dQ^2}\Bigr\vert_{\nubar\sst{D}}}\ \ . \eqno\nexteq\nameeq\EDnn
$$

With these considerations in mind, we consider briefly the interpretation
of the elastic $\nu$--D cross section, paying particular attention to the
similarities and differences with elastic $\nu$--N and PV elastic
$^2\hbox{H}(\evec, e')$ scattering.
To that end, we convert from a discussion
of $d\sigma/dQ^2$ to $d\sigma/d\Omega$ (see Sect.~III.E.3) and consider
different regimes in
$\nu$ scattering angle, even though a recoil deuteron rather than
outgoing neutrino would be detected.
Following the formalism outlined in Sect.~III.E.3 (see Eqs.~(\Eneutcr) and
(\Eneusign)), we
write the quantity
appearing in ${{d\sigma}\over{d\Omega}}({\nu D})$ as
$$ {\tilde F}_D^2 (q,\theta) \equiv 4{\tilde F}^2 (q,\theta) \equiv
R_{(1)} + R_{(2)} + R_{(3)} + R_{(4)}\ \ ,
\eqno\nexteq
$$
where the response functions have the forms
$$
\eqalign{
R_{(1)}\ =\ & 3 (\xi_V^{T=0})^2 \Bigl\{v_L
	\left(F_{C0}^2(T=0)+F_{C2}^2(T=0)\right)
        + v_T F_{M1}^2(T=0)\Bigr\} \cr
R_{(2)}\ =\ & 2\sqrt3 \xi_V^{T=0}\xi_V^{(0)}
            \Bigl\{v_L\left(F_{C0}(T=0)F_{C0}(s)
                       +F_{C2}(T=0)F_{C2}(s)\right) \cr
                &\qquad\qquad\qquad + v_TF_{M1}(T=0)F_{M1}(s) \Bigr\} \cr
R_{(3)}\ =\ & \pm 2v_{T'}\sqrt3\xi_V^{T=0} F_{M1}(T=0)
           \Bigl\{ \xi_A^{T=0} F_{E1_5}(8)
                        + \xi_A^{(0)}F_{E1_5}(s)\Bigr\}\cr
R_{(4)}\ =\ & v_T\Bigl\{ \xi_A^{T=0} F_{E1_5}(8) +
                \xi_A^{(0)} F_{E1_5}(s)\Bigr\}^2\cr
            &\qquad\qquad\qquad + (\xi_V^{(0)})^2\Bigl\{v_L\left( F_{C0}^2(s)
                                +     F_{C2}^2(s) \right)
                + v_T F_{M1}^2(s)  \Bigr\} \cr
            &\qquad\qquad\qquad \pm 2v_{T'} \xi_V^{(0)} F_{M1}(s)
                        \Bigl\{\xi_A^{T=0} F_{E1_5}(8)
                          +\xi_A^{(0)} F_{E1_5}(s)\Bigr\} \cr
           }\eqno\nexteq
$$
and where the form factors are defined in Sect.~III.D.1.

The response functions $R_{(i)} (i=1,\ldots , 4)$ are the $\nu$--D
analogs of the $\Delta_{(i)} (i=1,\ldots , 3)$ in $\alr(\evec D)$.
$\rone$ contains the leading vector current contribution in the absence
of strangeness; $\rtwo$ depends linearly on the strangeness vector
current form factors; and $\rthree$ contains the product of vector and
axial--vector form factors. Although the combinations of form factors
entering the $R_{(i)}$ are, for the most part,  identical to those
appearing in the numerators of the corresponding $\Delta_{(i)}$, a few
differences exist: (i) the combination $v_\sst{L}
(\fcz^2+\fct^2)+v_\sst{T}\fmo^2$ appearing in $\rone$ is canceled
from $\delone$ by an identical quantity appearing in the denominator of
$W^\sst{\rm (PV)}/F^2$; no such cancellation occurs for $\nu$--D scattering:
(ii) the AM form factor $F_{E1_5}(AM)$ contained in
$\delthree$ does not enter $\rthree$ since the neutrino has no
EM charge; and (iii) the
$R_{(i)}$ depend on products of weak neutral current amplitudes rather
than on the weak--electromagnetic interference product which governs the
$\evec D$ asymmetry. Consequently, the $R_{(i)}$ are bilinear in weak
neutral current couplings, whereas the $\Delta_{(i)}$ depend on them
linearly.

The terms in $\rfour$ have no analog in $\alr(\evec D)$. They arise
because $\dsigomega(\nu D)$ is proportional to a product of weak,
neutral current amplitudes. Since $\rfour$ is second--order in the small
strangeness form factors and
$\xia^{(8)}+\coeff{2}{\sqrt{3}}\xia^{(0)}$ axial--vector coupling, one might
expect it to be negligible in comparison with the other $R_{(i)}$.

In principle, ${{d\sigma}\over{d\Omega}}({\nu D})$ can be quite
sensitive to vector strangeness nucleon form factors,  as is
$A_{LR}(\vec e D)$.  In practice, however, kinematics conspire to make
the neutrino experiments more difficult.  To see why,
consider first the magnetic
contribution, $F_{M1}(s)$. Its contribution
becomes most significant at
backward angles, where momentum transfer is largest.
However, the deuteron body form factors defined in Eq.~(\Efmuld) fall
off rapidly with momentum transfer, thereby suppressing the
backward--angle cross section.  For example, the magnetic term
$\dmm(Q^2)$, whose contribution dominates $F_{M1}(s)$,  is down by two
orders--of--magnitude from its static value $\dmm(0)$, for
backward--angle scattering with incident energies of only 500 MeV.  For
the single--nucleon form factor, on the other hand,
on the other hand, the suppression is only 0.25 at this energy. By going to
lower--energy neutrino beams, one may mitigate this suppression to some
extent. At low--$|Q^2|$, however, $F_{M1}(s)$ also vanishes
linearly with momentum transfer. As a result, optimal neutrino energies,
balancing these two constraints, are roughly several hundred MeV. For these
kinematics, one then encounters experimental difficulties due to the
extremely low resulting recoil deuteron energy. In contrast, no
such difficulties arise in a measurement of $\alr(^2\hbox{H})$,
which involves detection of the outgoing lepton. Moreover, at backward
angles, the deuteron body magnetic form factor essentially cancels from
the PV asymmetry (see Sect.~IV.C.1), so that the large two--body
suppression does not appear in the observable of interest.

At forward angles and low momentum transfer, the $\nu$--D cross section goes
like
$$
\eqalign{
d\sigma/d\Omega (\theta \rightarrow 0^\circ, q\ {\rm small}) &\propto
        3(\xi_V^{T=0})^2\left(F_{C0}^2(T=0) + F_{C2}^2(T=0)\right) \cr
	&\qquad\qquad +{\textstyle{ 1\over2 }}
        \left(\xi_A^{T=0}F_{E1_5}(8) +
                        \xi_A^{(0)}F_{E1_5}(s)\right)^2
        +{\cal O}(q^2)\ \ , \cr}
\eqno\nexteq
$$
where the {\cal O}$(q^2)$ terms include contributions from the
magnetic ($F_{M1}(T=0)$) and strangeness charge ($F_{C0,C2}(s)$)
multipoles. In comparison, the PV asymmetry goes like
$\sqrt3{\xi_V^{T=0}} + \hcal{O}(q^2)$. In
both cases, the strange--quark
vector current contributions are
suppressed by factors of $|Q^2/m_\sst{N}^2|$.\footnote{*}{At small
momentum transfers,
$F_{M1}^2(T=0)$ is explicitly proportional to $q^2$;
$F_{C0}(s)$, being proportional to $\tau$ because the deuteron has no net
strangeness, is likewise small.}  At forward angles, however, the
electromagnetic cross
section gets very large, which keeps the FOM
for PV electron scattering fairly
high [Pol90]. Thus, ignoring systematic errors, one could
hope to measure $\alr(^2\hbox{H})$ with sufficient precision to
be sensitive to $Q^2$--suppressed terms, in analogy with the single
nucleon case.  For the $\nu$--D cross section, however, there is no Mott
divergence, and thus nothing to improve the statistics at forward
angles.  Considering high--energy forward--angle $\nu$--D scattering
leads to only a marginal improvement in sensitivity. When kinematic
coefficients of
order $\tau$ are large enough to allow significant strange quark
contributions, the deuteron body form factor is again very small.

The strangeness axial--vector current contributes to the differential
neutrino cross section via $R_{(3)}$ and $R_{(4)}$. In comparing the
sensitivities of $d\sigma(\nu D)/dQ^2$ and $\alr(^2\hbox{H})$ to
the axial--vector form factor, we reiterate that the interpretation of a
neutrino
measurement in terms of axial--vector strangeness is much less ambiguous than
in the case of PV electron scattering. Both the interference
term $R_{(3)}$ and its analog $\Delta_{(3)}$ in $\alr(^2\hbox{H})$
are suppressed at forward angles. The axial--vector contribution to $R_{(3)}$,
however, carries no $\gve$ suppression factor so that its backward--angle
contribution can be more significant for $\nu$--D scattering than
for PV $^2\hbox{H}(\evec, e')$ scattering. The axial--vector contribution
via $R_{(4)}$ has no analog in $\alr(^2\hbox{H})$. This term has no
explicit forward--angle kinematic suppression, but is still quadratic in
the presumably small axial--vector isoscalar form factor. To use this term to
limit the axial--vector isoscalar form factor to less than or about
$\pm 0.15$ would require
absolute forward $\nu$--D cross section measurements at below the
10\% level. In the backward--angle limit, the axial--vector contributions from
interference and
quadratic terms can be formally combined by
$$
\eqalign{
{d\sigma\over d\Omega}\,{\textstyle{({\nu\atop \bar\nu})}}
\,(\theta\rightarrow 180^\circ)
        \ &\propto\
        \Bigl(\left[\sqrt3\xi_V^{T=0}F_{M1}(T=0)+
                        \xi_V^{(0)}F_{M1}(s)\right] \cr
              &\qquad\qquad \pm \left[\xi_A^{T=0}F_{E1_5}(8) +
                   \xi_A^{(0)}F_{E1_5}(s)\right]\Bigr)^2\ \ . \cr}
\eqno\nexteq
$$
At moderate energies, (where the cross sections are extremely small,
as discussed above) the axial--vector contributions contribute
comparably to the strangeness magnetic form factors. Since
$F_{M1}\propto (q/m_\sst{N})$ for low momentum transfer, however,
the backward--angle cross section is much more sensitive to the axial--vector
terms at low--energies,  becoming essentially directly proportional
to the square of the axial--vector form factor.  There is also a range of
energies where the vector and axial--vector terms become comparable. In this
kinematic
regime, the difference between the $\nu$--D and $\nubar$--D cross sections
can be large, thereby enhancing the sensitivity of $\Delta_{\nu\nubar}$ to
nucleon strangeness.

        Finally, we observe that the use of elastic $\nu$--D scattering to
probe physics beyond the Standard Model through an extraction of the
$\xi_\sst{V,A}$ from the differential cross section is more problematic
than with other processes discussed in this review. The reason is the
additional level of nuclear physics uncertainties not encountered in the
interpretation of the elastic $\nu$--N/$\nubar$--N cross sections. The
impact of these uncertainties on $\alr$ can be mitigated by a suitable
choice of target since the asymmetry involves a ratio of nuclear response
functions.  The extent to which an observable such as $\Delta_{\nu\nubar}$
can reduce the sensitivity of an electroweak test to nuclear physics
uncertainties remains to be analyzed. At present, elastic $\nu$--D scattering
seems most suited to probing the structure of the nucleon.

\goodbreak
\bigskip
\noindent IV.J.5.\quad INELASTIC NEUTRINO SCATTERING
\medskip

We end our discussion of special cases with brief remarks about a few
selected inelastic neutrino scattering transitions where knowledge of
the cross sections is expected to help in determining the single--nucleon
weak--interaction form factors.  Of particular interest in this regard is
the NC transition between the ground state of $^{12}$C and the $1^+1$,
15.11 MeV excited state in the same nucleus, together with the
charge--changing (CC) transitions to the analog ground states of $^{12}$B
and $^{12}$N.  From our summary of the formalism in Sect.~III.E.3 we see
that the NC multipole matrix elements that enter include ${\tilde F}_{M1}$,
${\tilde F}_{C1_5}$, ${\tilde F}_{L1_5}$ and ${\tilde F}_{E1_5}$, together
with their CC analogs. For the states involved only isovector currents occur.
This system is much--studied theoretically (see, for
example, Refs.~[OCo72, Don73, Don74, Don76, Don79b]) and recently has been the
focus of several $^{12}$C$(\nu_e,e^-)^{12}$N(g.s.), charge--changing
[All90, KAR92] and $^{12}$C$(\nu,\nu')^{12}$C(15.11), neutral--current [KAR91]
experimental investigations.
As emphasized in Refs.~[Don74, Don76, Don79b], at low--energies
the allowed multipole is of the Gamow--Teller form and consequently the
nuclear matrix element involved in $\beta$--decay is closely related to
the dominant contribution entering in charge--changing neutrino reactions
and neutral--current neutrino scattering.  For instance, to the extent
that the long wavelength limit (LWL)
can be taken one may write the NC cross section in terms of the $^{12}$N
and $^{12}$B
$\beta$--decay rates, $\omega_{\beta}^{\pm}$, respectively
(see Eq.~(4.13) in [Don79b]):
$$
\sigma_{\nu,\nu'}^{LWL}(J=T=1) = {3\over 4} \pi^2 G_\mu^2 \mn^2
\Bigl[ {{\nu-\omega}\over\mn} \Bigr]^2 (\xi_\sst{A}^{T=1})^2
\times {{\omega_{\beta}^{\pm}(J=1)}\over{G_\mu^2 \cos^2 \theta_c
f^{\pm} (W_0^{\pm})^5}}
\ \ , \eqno\nexteq
$$
where $W_0^{\pm}$ is the energy of the decay, $\theta_c$ is the Cabibbo
angle and $f^{\pm}$ is a dimensionless phase--space integral
[Don79b].  Using the average of the $^{12}$B
and $^{12}$N $\beta$--decay rates, this yields (Eq.~(4.53) in [Don79b])
$$
\sigma_{\nu,\nu'}^{LWL}(J=T=1) = 2.7\times10^{-39}\ {\rm cm}^2\
\Bigl[ {{\nu-\omega}\over\mn} \Bigr]^2 (\xi_\sst{A}^{T=1})^2
\ \ . \eqno\nexteq
$$

Of course, these simple relationships are not exact, since as noted above
there are more nuclear matrix elements than one to consider (and these
enter differently in the
various electroweak processes); in addition, the momentum transfer
dependence of the nuclear form factors must at some level be taken into
account.  For this well--studied $A=12$ system, as discussed in depth in
Ref.~[Don79b], it is possible to constrain
the nuclear one--body density matrix elements (see Sect.~III.D.2) to a very
high degree using
electron scattering, $\gamma$--decay, $\beta^{\pm}$--decay and
$\mu^-$--capture information. Consequently, it is not necessary to resort to
the strict LWL model. It is reassuring, however, that the simple model does
yield an excellent approximation to the more sophisticated result (typically
being in error by only about 10--15\% at low--energies, although
somewhat worse at intermediate--energies) and, on the basis of the
refinements that go into the latter, it is expected that for this special case
the confidence level of the theoretical predictions for the
neutrino--induced reactions should be better than 10\%.

Indeed, the
recent CC results [All90, KAR92] strengthen this expectation, since rather good
agreement is found with the predictions: from early modeling [Don73]
the integrated cross section for a beamstop facility was found to be
0.94$\times 10^{-41}$ cm$^2$ (see also Refs.~[Fuk88, Min89]) while
experiments have
yielded 1.05$\pm$0.10(stat)$\pm$0.10(syst)$\times 10^{-41}$ cm$^2$ [All90] and
0.81$\pm$0.09(stat)$\pm$0.075(syst)$\times 10^{-41}$ cm$^2$ [KAR92].  With
further running it might prove possible to lower the total experimental
uncertainty to below 10\% [KAR91, KAR92], commensurate with the estimated level
of the model uncertainty.

The situation for NC neutrino--excitation of the 15.11 MeV level in
$^{12}$C followed by $\gamma$--decay back to the ground state is
similarly quite satisfactory.  The theoretical modeling [Don74, Don76,
Don79a] updated to a modern value for $\sstw$ yields
0.97$\times 10^{-41}$ cm$^2$ for the integrated beamstop $\nu_e\ +\
{\bar\nu}_\mu$ cross section (see also Refs.~[Ber79, Fuk88, Kol92] which
predict very
similar numbers), to be compared with the experimental value
1.08$\pm$0.51(stat)$\pm$0.11(syst)$\times 10^{-41}$ cm$^2$ [KAR91].  With
extended running it is expected here as well that the total experimental
uncertainty
may be reduced to below 10\% [KAR91, KAR92].  In fact, a theory--to--experiment
comparison of the ratio of NC to CC integrated cross sections might
ultimately provide the best--determined quantity, both from the experimental
point of view where flux normalization uncertainties may be minimized,
as well as from the theoretical point of view where the
ratio is expected to have only a very weak dependence on the nuclear
modeling.

Of particular interest for the NC measurements in which it is the sum of the
$\nu_e$ and ${\bar\nu}_\mu$ scattering cross sections that is
determined is the fact that the VA--interference term changes in
going from particle-- to antiparticle--scattering (see Eq.~(\Eneusign))
and consequently largely cancels in taking the sum (it does not
exactly cancel, since the beamstop spectra for $\nu_e$ and ${\bar\nu}_\mu$,
while rather similar, are not exactly the same --- see Fig.~5.1a).  In fact the
predictions for the quantity $\delta <\!\sigma_{\nu_e}
+\sigma_{{\bar\nu}_\mu}\!>/\delta\sstw$ yield the very small value 0.06,
reflecting the weak dependence on the vector currents.  Since the
summed result is almost completely determined by the isovector
axial--vector form factor and is quadratic in this quantity (see the
above discussions on neutrino reactions), this suggests that
extended experimental measurements in the $A=12$ system could lead to
a determination of $\tilde G_A^{T=1}(0)$ with about 5\% uncertainty.
Coupling such a result with a high--precision
determination of the equivalent quantity from PV electron scattering
where the radiative
corrections are expected to be rather different would provide
a very interesting test of the underlying weak--interaction model.

Let us end this section with a brief discussion of another interesting
case drawn from Ref.~[Don79a].  The same motivation of studying
M1/Gamow--Teller NC excitations can be extended to include {\it isoscalar}
transitions, as emphasized by the authors of Refs.~[Don74, Don76].
However, in this case there are no $\beta$--decay rates to use to
(largely) determine the nuclear matrix element; instead one may use
the corresponding $\gamma$--decay rate $\Gamma_{\gamma}(J=1,T=0)$
to write (see Eq.~(4.17) in [Don79a]):
$$
\eqalign{
\sigma_{\nu,\nu'}^{LWL}(J=1,T=0) = {3\over{4\alpha}} \pi^2 G_\mu^2 \mn^4
&\Bigl( {{2J_0+1}\over{2J_1+1}} \Bigr)
\Bigl[ {{\nu-\omega}\over\mn} \Bigr]^2
\Bigl({{\xi_\sst{A}^{T=0} G_\sst{A}^{(8)}+
\xi_\sst{A}^{(0)} G_\sst{A}^{(s)}}\over{F_2^{T=0}}} \Bigr)_{LWL}^2 \cr
&\times {1\over{\omega^3}} \Gamma_{\gamma}(J=1,T=0)
\ \ , \cr}\eqno\nexteq
$$
where the ground state has angular momentum $J_0$ while the excited state
has $J_1$ and $|J_1-J_0|\le 1$. Specifically, let us consider the $1^+0$
ground and $2^+0$, 7.028 MeV excited states in $^{14}$N (see Eq.~(4.51) in
[Don79a] and associated discussion).
Keeping only the axial--vector
strangeness form factor and using Eq.~(\Egalstra) one obtains
$$
\sigma_{\nu,\nu'}^{LWL}(J=1,T=0) = 2.5\times 10^{-38}\ {\rm cm}^2\
\Bigl[ {{\nu-\omega}\over\mn} \Bigr]^2 (\eta_s)^2
\ \ . \eqno\nexteq
$$
Given the value $\eta_s\approx -0.12$ in Table~4.2 this yields roughly
1/4 the number of photons for this isoscalar $^{14}$N case as obtained for the
above isovector $^{12}$C case (per neutrino per unit time for equal masses of
carbon and nitrogen) and consequently may prove feasible for future
experiments.

Other nuclear transitions are discussed in Ref.~[Don79a] for NC neutrino
scattering of reactor anti--neutrinos and beamstop and DIF
intermediate--energy neutrinos ($\nu_e$,
$\nu_\mu$ and ${\bar\nu}_\mu$ --- see Sect.~V.A).  The results given there
can usually be extended quite easily to the incorporate strangeness content
that was ignored when that previous review was written.

\vfil\eject

%

\secnum=5
\neweq

%

\def\qvecsq{\vec q^{\mkern2mu\raise1pt\hbox{$\scriptstyle2$}}}

\def\xiva{{\xi_\sst{V}^{(a)}}}

\def\evec{{\vec e}}
\def\sigpin{{\Sigma_{\pi\sst{N}}}}

\def\snw{\sin^2\theta_\sst{W}}

\def\rapp{\hbox{${\lower.40ex\hbox{$>$}\atop\raise.20ex\hbox
{$\sim$}}$}}
\newbox\tallstrutbox
\setbox\tallstrutbox=\hbox{\vrule height 12 pt depth 4pt
width 0pt}
\def\tallstrut{\relax\ifmmode\copy\tallstrutbox\else\unhcopy
\tallstrutbox\fi}
\def\pmb#1{\setbox0=\hbox{$#1$}%
\kern-.025em\copy0\kern-\wd0
\kern.05em\copy0\kern-\wd0
\kern-.025em\raise.0433em\box0 }

\newbox\esbox
\setbox\esbox=\hbox{\vrule height20pt depth10pt width0pt}
\def\es{\relax\ifmmode\copy\esbox\else\unhcopy\esbox\fi}

\newbox\headstrutbox
\setbox\headstrutbox=\hbox{\vrule height20pt depth10pt width0pt}
\def\headstrut{\relax\ifmmode\copy\headstrutbox\else\unhcopy\headstrutbox\fi}

\noindent {\bf V.\quad EXPERIMENTAL CONSIDERATIONS}
\bigskip

In this section we discuss some of the experimental considerations that
enter into treatments of parity--violating electron scattering and
neutrino scattering from
nucleons and nuclei. Our aims are several: we wish to summarize the
general considerations that occur in experimental investigations of
these classes of reactions to provide the reader with an assessment of
what issues are likely to determine the future course of the field
(Sect.~V.A); we then turn in Sect.~V.B to overviews of past studies of electron
and neutrino scattering for the purpose of learning about the hadronic
neutral current --- here sufficient detail is given in several cases to
permit an appreciation of the level of difficulty involved in both classes
of reactions; finally, in Sect.~V.C we summarize the present and proposed
worldwide experimental program as we are aware of it and attempt to
indicate where future initiatives in this general area of research may lie.

\bigskip
\noindent{\bf V.A.\quad General Considerations}
\bigskip

We begin with a discussion of the general experimental
considerations that pertain when studying PV electron scattering
(Sect.~V.A.1) or
neutrino scattering (Sect.~V.A.2).

\bigskip
\noindent V.A.1.\quad ELECTRON SCATTERING FACILITIES

Electron accelerators which have been used for PV electron
scattering studies span the energy
range from a few hundred MeV to tens of GeV.  Modern facilities are designed to
produce intense beams of polarized electrons and the newest machines will also
provide CW beams. High--duty--factor beams, with greatly reduced instantaneous
counting rates (more than two orders--of--magnitude), would make possible for
the first time the capability of using particle--counting techniques.
Both particle
discrimination and background rejection would be significantly enhanced.

There are many experimental challenges associated with carrying out
PV experiments at the 1\% level using high--energy electrons.
These involve limitations on
luminosity, beam polarization, detector acceptance, resolution and the ability
to control systematic errors.  The relative importance of the different
factors is a strongly physics--dependent issue.
The luminosity is given by
$$
\eqalign{{\cal L} &= {{I \rho N_0}\over A} \cr
&\cong 4\times 10^{38}\ {{I \rho }\over A } \ \ , \cr}
$$
where the luminosity is in units of cm$^{-2}$~s$^{-1}$, the average current $I$
is in units of 100~$\mu$A, the target density $\rho$ is in g~cm$^{-2}$ and $A$
is the target mass number.
Essentially all PV experiments are designed to run at maximum luminosity in the
interest of running time and cost.  For accelerators in the one to few GeV
range (see Table~5.1), polarized--beam intensities of 100~$\mu$A are
reasonable. For a CW
accelerator the laser requirements are modest, while for pulsed machines very
high--powered lasers are required to produce the polarized beams and such
accelerators are usually not designed to provide much higher beam intensities.
Some lower--energy ($<100$~MeV) electron machines have capabilities in the
ampere range, although these do not have polarized beam capability at present.
Table~5.1 summarizes the beam characteristics of electron accelerators which
are designed to deliver polarized beams.  Both the MIT/Bates and
NIKHEF\footnote{*}{NIKHEF does not
currently have a polarized injector, although
plans exist to install one.} facilities
have recently been upgraded with the addition of stretcher/storage rings which
make possible essentially CW extracted beams, as well as very high current
internal stored beams.  Polarized beams will be available in
both operating modes.

\midinsert
$$\hbox{\vbox{\offinterlineskip
\def\strut{\hbox{\vrule height 10pt depth 5pt width 0pt}}
\def\superstrut{\hbox{\vrule height 17pt depth 10pt width 0pt}}
\hrule
\halign{
\strut\vrule#\tabskip 0.05in
&#\hfil
&\vrule#
&\hfil#\hfil
&\vrule#
&\hfil#\hfil
&\vrule#
&\hfil#\hfil
&\vrule#\tabskip 0.0in\cr
&\multispan7\hfil\headstrut{\bf TABLE 5.1 }
\hfil & \cr\noalign{\hrule}
\superstrut& LABORATORY && $\matrix{\hbox{ENERGY RANGE}\cr\hbox{(GeV)}\cr}$ &&
$\matrix{\hbox{BEAM}\cr\hbox{CURRENT}\cr}$ && DUTY FACTOR &
\cr\noalign{\hrule}
& MIT/Bates && && && & \cr
&\quad Pulsed && 0.1 -- 1.0 && 100$\mu$A && 0.01 & \cr
&\quad CW Extracted && 0.3 -- 1.0 && \phantom{1}50$\mu$A && CW & \cr
&\quad CW Internal && 0.3 -- 1.0 && \phantom{1}80mA && CW &
\cr\noalign{\hrule}
& CEBAF(design) && 0.5 -- 6.0 && 200$\mu$A && CW & \cr\noalign{\hrule}
& Mainz && 0.2 -- 0.9 && 100$\mu$A && CW & \cr\noalign{\hrule}
& NIKHEF && && && & \cr
&\quad Pulsed && 0.1 -- 0.9 && \phantom{1}65$\mu$A && 0.001 & \cr
&\quad CW Extracted && 0.3 -- 0.9 && \phantom{1}65$\mu$A && CW & \cr
&\quad CW Internal && 0.3 -- 0.9 && 200mA && CW & \cr\noalign{\hrule}
& SLAC && && && & \cr
&\quad End Station A (Present) && 1 -- 23 && \phantom{11}10$\mu$A && $2\times
10^{-4}$ & \cr
&\quad End Station A (Upgrade) && 1 -- 50 && 100 nA &&
$10^{-5}$ & \cr\noalign{\hrule}}}}$$
\smallskip
\baselineskip 10pt
{\ninerm
\noindent\narrower
{\bf Table~5.1.} \quad Properties of existing and planned electron
accelerators.
\smallskip}
\endinsert

\baselineskip 12pt plus 1pt minus 1pt

A very practical luminosity--related consideration is the power--handling
capability of the scattering target.  It is possible to construct liquid
hydrogen, deuterium and helium targets which can dissipate $\sim 1$~kW of
beam power.  Certain solid targets, such as carbon and tungsten, have
somewhat higher power--handling capabilities, although
the target thickness is in general limited to
approximately (5--10) g/cm$^2$ of material. Other considerations that limit
target thickness involve radiative effects which can limit the resolution
in the scattered electron energy and affect the accuracy of the results. In
particular, bremsstrahlung effects set limits which are comparable to the
target's maximum power--handling capabilities.
\bigskip
\goodbreak
\noindent \undertext{ Polarized Electron Sources}

Polarized beams are a relatively new capability for electron accelerators.
They now provide many important physics opportunities connected with
electroweak studies and with experiments designed to exploit spin degrees of
freedom in electron scattering.  In addition to the PV experiments discussed
here, major efforts are underway to measure $G_E^n$ and various
observables involving nucleon and nuclear structure.

The measurement of small asymmetries, in the range of $10^{-
4}-10^{-6}$, places severe demands on the minimal
requirements that a polarized source must achieve to make
such experiments practical.  The most important of these are
high intensity (up to hundreds of mA peak current for pulsed accelerators)
and high
polarization.  Other factors being equal, the figure--of--merit in such
experiments is proportional to
P$_e{}^2$ $A_{LR}^2 \sigma$,
where P$_e$ is the beam
polarization, $\alr$ is the asymmetry to be measured and $\sigma$ is the
differential cross section (see Sect.~III.E.2).  In order to control
systematic effects,  rapid and precise polarization reversal is
essential and it must leave all other beam parameters
unchanged.

Although various techniques have been used to construct polarized electron
sources, only the Li atomic beam and photoemission sources have been used as
injectors in accelerators to date and essentially all sources now in
use on electron machines are based on photoemission from GaAs.
Photoemission from semiconductors as a source of polarized electrons was first
proposed by Garwin, Pierce and Siegmann [Gar74] and by Lampel and
Weisbuch [Lam75].  These crystals can provide the
very high peak currents required by the older low--duty--factor machines and
allow for rapid polarization reversal and stability demanded by
PV experiments. A source was constructed by Sinclair and
co--workers for the SLAC PV experiment of
Prescott {\it et al.\/} [Pre79], and most electron accelerator sources
now in use are variants of this early SLAC design.

The operation of a GaAs source is relatively simple. Polarized electrons are
produced in the conduction band of the semiconductor crystal by shining
circularly--polarized laser light of an appropriate wavelength on the material.
To allow the electrons to escape from the crystal, the surface is covered
with a very thin layer of alkali metal and oxidants.  This lowers the
work--function of the surface so that the energy of an electron in the vacuum
is
lower than it is in the bulk material (a condition known as negative electron
affinity).

The theoretical maximum polarization possible with a bulk
GaAs crystal is limited
by level degeneracies  to 50\%.  Polarizations achieved in practice are in the
range of 25 to 43\%.  Since the FOM is proportional to the square
of the polarization, almost an order--of--magnitude improvement is still
possible.  The sign of the electron polarization is simply changed by
reversal of the circular polarization of the light and devices are available
for achieving this on time scales shorter than a microsecond.

During the past decade, much research activity has been directed at
understanding why working sources have a polarization less than the theoretical
maximum of 50\% and finding approaches for achieving much higher
polarizations. Significant progress has occurred.  A recent review by Cardman
summarizes these latest efforts [Car92].
The most promising research is directed towards approaches which would remove
the
degeneracy in GaAs and allow for much higher polarizations.  These approaches
depend on altering the nature of the crystal structure by: (1) applying
stress to the crystal; (2) constructing artificial structures with
varying band gap energies; and (3) finding new crystals where the
degeneracy is absent due to lack of symmetry.

Experiments with strained crystals have been carried out by Nakanishi {\it et
al.\/} [Nak91].  In their approach a layer of GaAs is grown on a surface of
G$_a$P$_x$As$_{1-x}$ (with $x\sim 0.17$) which has a lattice mismatch of
0.6\%.  This stressed crystal has shown electron polarizations as high as
83\% [Mar92].
The quantum efficiency has however been reduced by an order--of--magnitude in
comparison with using bulk GaAs.  This may not be a problem for CW
accelerators, but could be a serious limitation for the older pulsed
machines.

Multi--layered heterostructures, such as alternating GaAs and
Al$_x$Ga$_{1-x}$As, have also been constructed in another approach in
which the band
degeneracy is broken by the introduction of a periodic variation in the band
energy.  A KEK/Nagoya/NEC group [Omo91]
has investigated this technique and achieved a
polarization of 71\%.  The quantum efficiency is, however, almost two
orders--of--magnitude below bulk GaAs.

The third approach to higher polarization involves the use of direct band gap
materials which do not have the valence band degeneracy of GaAs;
one class of such materials is the chalcopyrites.  These materials could
in principle provide both high polarization and high quantum
efficiency and several groups are now investigating these possibilities.


An important consideration in the operation of polarized sources, in addition
to high polarization, is source lifetime.  At present the lifetimes of typical
accelerator photocathode sources are in the range of a few hundred hours,
after which time
the crystal surface must be cleaned and reactivated.  Most evidence
indicates that lifetime is for the most part related to vacuum quality.  Since
the vacuum requirements are extremely high, efforts are also under way to find
less sensitive materials and/or techniques for continuously cleaning the
crystal surface.

There has been much progress in our ability to generate and use intense
polarized electron beams since the first PV experiment at SLAC.  The recent
developments in source technology show possibilities for substantial
improvement in polarization for the future.
\bigskip
\goodbreak
\noindent \undertext{ Detectors}

Detectors that have been used and/or proposed for PV experiments
have acceptances from a few millisteradians to more than a steradian (see
Table~5.2).  The
small--acceptance devices are usually magnetic spectrometers with relatively
good
momentum resolution ($<0.1\%$), whereas the large--acceptance devices involve
the use of \v{C}erenkov counters
optimized for the detection of minimum--ionizing particles.  The latter have
essentially no momentum resolution, since the thresholds are usually set very
low.  There have also been some novel ideas for
systems in the range of (50--200)
msr which have reasonable momentum resolution.  One such system uses a
multi--sector toroidal magnet configuration as proposed for an experiment in
Hall~C at CEBAF.  As discussed in Sect.~V.C.5, the possibility of using a
large solenoid as a detector has
also been studied and looks quite promising, particularly at relatively low
energies ($<$1~GeV).  All experiments until now have used detection of
the scattered electrons.  The toroidal system, mentioned above, is designed to
look at the flux of recoil protons instead, which potentially has some
advantages for experiments where the scattered electron is at an unfavorably
far forward angle.

\midinsert
$$\hbox{\vbox{\offinterlineskip
\def\strut{\hbox{\vrule height 10pt depth 5pt width 0pt}}
\def\superstrut{\hbox{\vrule height 17pt depth 10pt width 0pt}}
\hrule
\halign{
\strut\vrule#\tabskip 0.022in&
#\hfil &
\vrule#&
\hfil#\hfil &
\vrule#&
\hfil#\hfil &
\vrule#&
\hfil#\hfil &
\vrule#&
\hfil#\hfil &
\vrule#\tabskip 0.0in\cr
&\multispan{9}\hfil\headstrut{\bf TABLE 5.2 }
\hfil & \cr\noalign{\hrule}
\superstrut& && Type && $\matrix{\hbox{Acceptance}\cr\hbox{(msr)}\cr}$ &&
$\matrix{\hbox{Momentum}\cr\hbox{Resolution}\cr(\%)\cr}$ &&
$\matrix{\hbox{Scattering}\cr\hbox{Angle}\cr}$ & \cr\noalign{\hrule}
& SLAC && && && && & \cr
&\quad 1977 && Magnetic && 0.5 && 20 && $4^\circ$ & \cr
&\quad End Station A && && && && & \cr
&\qquad\quad (Elastic) && Magnetic (2) && 0.6 && 0.5 && 6.5$^\circ$ & \cr
&\qquad\quad  &&  && 3 && 0.1 && 67.9$^\circ$ (p) & \cr
&\qquad\quad (DIS)     && Magnetic (2) && 1 && 3 && 3.75$^\circ$ &
\cr\noalign{\hrule}
& Mainz               && && && && & \cr
& && \v{C}erenkov && 2500 && --- && $115^\circ-145^\circ$ & \cr
& && Calorimeter && 2200 && 4 && $35^\circ -50^\circ$ & \cr\noalign{\hrule}
& MIT/Bates               && && && && & \cr
\superstrut
&\quad ${}^{12}$C && Magnetic (2) && 20 && 6 && $35^\circ$ & \cr
\superstrut
&\quad SAMPLE && \v{C}erenkov && 2000 && --- && $135^\circ - 160^\circ$ &
\cr
&\quad Solenoid && Magnetic && 600 && 1 && Forward & \cr
& && && 2000 && 1 && Backward &
\cr\noalign{\hrule}
& CEBAF  && && && && & \cr
\superstrut
&\quad Hall A && Magnetic (2) && 16 && .01 && $12.5^\circ$ & \cr
\superstrut &\quad $G^0$ && $\matrix{\hbox{Toroidal}
\cr\hbox{Spectrometer}\cr}$ && 870 && 10 && $60^\circ -77^\circ$ (p) & \cr
\superstrut &\quad && (reversed) && 500 && 10 && $108^\circ$ &
\cr\noalign{\hrule}}}}$$
\smallskip
\baselineskip 10pt
{\ninerm
\noindent\narrower
{\bf Table~5.2.} \quad Properties of parity--violating electron scattering
detector systems (existing and proposed).
\smallskip}
\endinsert

\baselineskip 12pt plus 1pt minus 1pt

Basic physics considerations set the criteria for the required momentum
resolution.  Experiments on hydrogen have the least restrictions: in this
case inelasticities in excess of $m_\pi$ must be accounted for in precise
measurements.  For nuclear scattering the issue involves contributions due to
inelastic levels or electrodisintegration.  Elastic scattering on
deuterium requires a resolution of better than 2.2~MeV, for ${}^{12}$C less
than 4.4~MeV and in the case of ${}^4$He this is relaxed somewhat to
20.1 MeV.  Depending on the incident electron energy, this corresponds to
momentum resolutions of a few percent to less than 0.1\%.  Magnetic devices
provide the only practical means for achieving such resolutions.  In addition,
for experiments at the highest luminosity, where event--by--event tracking is
not possible, the detector must have intrinsic momentum and angular
focusing properties as well.

A monoenergetic beam of electrons passing through a target loses energy at
the rate of $\sim 2.2$~MeV cm$^2$/gm and develops an energy--spread at
approximately 1/4 this rate.  For example, a target of
5~gm/cm$^2$ will develop an energy--spread of approximately 2.5~MeV due to
ionization straggling.  This sets an obvious limit on maximum target thickness
due to these considerations alone.  This limit is, however, comparable to
that due to the maximum level that is practical in handling the
dissipated power.

Each experiment has to identify and solve a number of problems connected with
systematic uncertainties.  Most important are those associated
with beam--related helicity--correlated differences.  Beam parameters such as
energy, intensity, position and direction can have major effects.  The
MIT/Bates ${}^{12}$C results [Sou90a] show that asymmetry
uncertainties at the level of
$2\times 10^{-8}$ are achievable in an electron scattering measurement and
in polarized proton scattering at SIN [Kis87]
the systematic errors were of comparable
size.  It is expected that it should be possible to improve
on these results with the modern CW electron accelerators.

There have been suggestions, at different times, that PV studies
might be feasible using polarized targets or recoil polarimetry.  Practical
considerations, however, make this highly unlikely given present technology.
High--density polarized targets are available as cryogenic gas cells (${}^3$He)
and solids (${}^1$H, ${}^2$H).  In the case of polarized ${}^3$He a density of
$10^{22}$~cm$^{-2}$ can be contemplated.  At a beam current of 100~$\mu$A,
which could present some difficulty for maintaining target polarization, the
maximum luminosity is only $\sim 10^{37}$~cm$^{-2}$~s$^{-1}$.  This is nearly
two orders--of--magnitude less than that achieved in the MIT/Bates ${}^{12}$C
experiment.  In the case of polarized solid targets the situation is even
worse.  State--of--the--art solid cryogenic polarized ${}^2$H targets can
reach densities of $10^{23}$ cm$^{-2}$ at high polarization [Bur93, Day93].
The beam current must, however, be
limited to less than 100 $n$A in order to maintain target
polarization.  The resulting
luminosity is only  a few times $10^{34}$~cm$^{-2}$~s$^{-1}$.  These targets,
in addition, usually have associated impurities which could affect the
interpretation of the results.  Polarization reversal is also more
difficult and the frequency would be much reduced in comparison to reversing
beam electron polarization.  These considerations are of great importance
in the control of systematic errors. Furthermore, while
we have not evaluated in any detail the case of recoil electron polarimetry,
it would
appear that involving a second scattering with its attendant cross
section, analyzing power and efficiency would not be favorable.


Three PV electron scattering experiments of the type described
in this section have been completed to date. These will be discussed in
more detail in Sect.~V.B, followed in Sect.~V.C by a description of the
several new experiments which have been proposed or are currently in
progress.

\bigskip
\goodbreak
\noindent V.A.2.\quad NEUTRINO SCATTERING FACILITIES

Neutrino beams for medium--energy physics research are
in use now at the Los Alamos Meson Physics Facility (LAMPF) and at the
Rutherford Appleton Laboratory (ISIS).  Future facilities are planned,
for the KAON project in Vancouver and
for a Pulsed Lepton Source (PLS) at LAMPF.
These neutrino beams can be put into two generic classes. In the beamstop
source, neutrinos result predominantly from the production and
subsequent decay at rest of single positive pions.  The decay sequence
$\pi^{+}\rightarrow\mu^{+} \nu_{\mu}$~ is followed by
$\mu^{+}\rightarrow e^{+} \nu_{e}\overline{\nu}_{\mu}$, which produces equal
numbers of $\nu_{\mu}, \nu_{e}$, and $\overline{\nu}_{\mu}$ neutrinos
with up to 52.8 MeV kinetic energy.  In the decay--in--flight (DIF)
source, a decay region following a pion--production target allows energetic
$\nu_{\mu}(\overline{\nu}_{\mu})$
neutrinos to be produced from the decay of high--energy $\pi^{+}(\pi^{-})$.


A brief survey of the characteristics of present
and future medium--energy
neutrino beams is displayed in Table~5.3.  The spectrum of a beamstop source,
as shown in Fig.~5.1a, consists of a monochromatic $\nu_{\mu}$~ line at
29.8 MeV from the two--body decay at rest of the $\pi^{+}$
and the well--determined $\nu_{e},\overline{\nu}_{\mu}$~
Michel spectrum, with average energy about 35 MeV,
from the three--body decay at rest of the $\mu^{+}$.  Nuclear
absorption of negative pions and muons typically reduces $\overline{\nu}_e$
contamination from $\mu^-$ decays at rest to less than
$5 \times 10^{-4}$ of the $\mu^+$ decay flux.
Neutrinos from a beamstop source can be separated in time for suitably short
proton pulses.  As shown in Table~5.3, both the ISIS (100 ns)
and the proposed PLS--LAMPF (270 ns) time structures are short
compared to the muon lifetime, and so allow separation of $\nu_{\mu}$
from $\nu_{e}$~ and $\overline{\nu}_{\mu}$ neutrinos.  This is a
particularly useful feature in neutral current measurements.

The more energetic $\nu_{\mu}$ spectra from DIF sources are
shown in Figs.~5.1b and 5.1c for a modified LAMPF beamstop and for the future
KAON facility.  The LAMPF neutrino flux, produced by an 800 MeV proton beam,
extends from 60 to 250 MeV, while the KAON neutrino flux, produced from 30
GeV protons, would peak at about 1 GeV and extend to 6 GeV.  Charge selectivity
in the pion focusing elements of the future KAON and PLS--LAMPF
beams will enhance the
$\nu_{\mu}(\overline{\nu}_{\mu})$ beam while decreasing the
$\overline{\nu}_{\mu}(\nu_{\mu})$ contamination.

Knowledge of the absolute neutrino flux is very difficult to obtain
experimentally and is thus usually determined through Monte Carlo
simulations and calibrations.  For the rather complicated geometries of the
LAMPF and ISIS beamstops, absolute values to $\pm 7$\% accuracy were
obtained through Monte Carlo calculations [Bur90] of the stopping
$\pi^{+}$/proton
ratio, as calibrated by an experimental measurement of that ratio in a
simplified, instrumented beamstop [All89].  As an example of the
accuracy of DIF sources, the BNL neutrino beam intensity
has been quoted [Ahr85]
as $\pm 12$\% by calibration with measured charged--current $\nu_\mu$~ events.

\midinsert
\baselineskip 12pt
$$\hbox{\vbox{\offinterlineskip
\def\strut{\hbox{\vrule height 8.5pt depth 3.5pt width 0pt}}
\def\superstrut{\hbox{\vrule height 15pt depth 10pt width 0pt}}
\hrule
\halign{
\strut\vrule#\tabskip 0.1in&
\hfil#\hfil&
\hfil#\hfil&
\vrule#&
\hfil$#$\hfil&
\hfil$#$\hfil&
\hfil$#$\hfil&
\hfil$#$\hfil&
\vrule#\tabskip 0.0in\cr
& \multispan7\hfil\headstrut{\bf TABLE 5.3 }\hfil & \cr\noalign{\hrule}
\superstrut& Laboratory & $\matrix{\hbox{Neutrino}\cr\hbox{Source}\cr}$ &&
\matrix{\hbox{Neutrino}\cr\hbox{Type}\cr} & \matrix{\hbox{Energy
Range}\cr\hbox{(MeV)}\cr} & \matrix{\hbox{Time}\cr\hbox{Structure}\cr} &
\matrix{\hbox{Flux }\phi\cr \nu\hbox{cm}^{-2}\hbox{s}^{-1} \cr} &
\cr\noalign{\hrule}
& &     &&         &        &    &                & \cr
& LAMPF & beamstop && \nu_e,\,\nu_\mu,\,\bar{\nu}_\mu & 0-53 &
500\,\mu\hbox{s} & 4.2\times 10^7 & \cr
& & DIF && \nu_\mu & 30-250 & '' & 1.1\times 10^6 & \cr
& &     &&         &        &    &                & \cr
& ISIS & beamstop && \nu_e,\,\nu_\mu,\,\bar{\nu}_\mu & 0-53 & 100\,\hbox{ns}
& 1.3\times 10^6 & \cr
& &     &&         &        &    &                & \cr
& KAON & DIF && \mu_\nu & 100-6000 & 3.6\,\mu\hbox{s} & 2.3\times 10^8 & \cr
& &     &&         &        &    &                & \cr
& PLS-LAMPF & beamstop && \nu_e,\, \nu_\mu,\,\bar{\nu}_\mu & 0-53 &
270\,\hbox{ns} & 1.2\times 10^7 & \cr
& & DIF && \nu_\mu & 60-300 & '' & 7.5\times 10^6 & \cr
& &     &&         &        &    &                & \cr\noalign{\hrule}}}}$$
\smallskip
\baselineskip 10pt
{\ninerm
\noindent\narrower
{\bf Table~5.3.} \quad Characteristics of current and future neutrino sources.
The decay--in--flight sources (DIF) have a decay region following the pion
production target.  The time structure is that of the incident
proton beam.  Neutrino fluxes are calculated for representative
geometries for the different facilities.
\smallskip}
\endinsert

\baselineskip 12pt plus 1pt minus 1pt

Investigation of hadronic neutral currents is in progress at
ISIS and at LAMPF.  The KARMEN collaboration has observed [Bod91]
the reaction ${}^{12}\hbox{C}(\nu,\nu'){}^{12}\hbox{C}^*(15.11\,\hbox{MeV})$,
 for combined $\nu_{e},\overline{\nu}_{\mu}$~ neutrinos
from the ISIS beamstop source (see also Sect.~IV.J.5).  The LSND
collaboration at LAMPF is proceeding to
measure both this reaction and the
$p(\nu_\mu,\nu_\mu)p$  elastic scattering cross section at
low momentum transfer [Lu90].  Elastic
scattering from a nucleus, $A(\nu_\mu,\nu_\mu)$,
has been considered for the proposed LAMPF--PLS facility [Don83].

\vfil\eject
\noindent{\bf V.B.\quad Past Studies of Electron and Neutrino Scattering}
\medskip

\def\rapp{\hbox{${\lower.40ex\hbox{$>$}\atop\raise.20ex\hbox
{$\sim$}}$}}
\newbox\tallstrutbox
\setbox\tallstrutbox=\hbox{\vrule height 12 pt depth 4pt
width 0pt}
\def\tallstrut{\relax\ifmmode\copy\tallstrutbox\else\unhcopy
\tallstrutbox\fi}
\def\pmb#1{\setbox0=\hbox{$#1$}%
\kern-.025em\copy0\kern-\wd0
\kern.05em\copy0\kern-\wd0
\kern-.025em\raise.0433em\box0 }

\bigskip
\noindent V.B.1.\quad DEEP--INELASTIC SCATTERING FROM DEUTERIUM (SLAC)

The first scattering experiment to investigate PV
effects due to neutral currents in
electromagnetic interactions was carried out at SLAC in the
late 1970's [Pre79].  It involved the measurement of an asymmetry in
deep--inelastic scattering (incoherent scattering from individual quarks) of
longitudinally--polarized electrons from a deuterium target at energies in the
range (16--22) GeV. These results provided one of the important early
verifications for the predictions of the gauge theories and for the
Weinberg--Salam--Glashow SU(2)$_L\times$U(1)$_Y$
theory (Standard Model) of the
weak and electromagnetic interactions in particular.

The scale of the predicted PV asymmetry for the SLAC kinematics
was small ($\sim 10^{-4} |Q^2|$ (GeV/c$)^2$) and as a consequence
presented some very
difficult technical problems.  It required the development of an
intense polarized electron source, beam control and monitoring, together
with counting
techniques capable of achieving the required sensitivity and precision.
These developments provided a technical base and model for the experiments at
intermediate--energies which were to be carried out later at
Mainz and MIT/Bates.
\bigskip
\goodbreak
\noindent \undertext{ Polarized Electron Source}

The polarized electron sources used as injectors in linear
accelerators, to date, operate on the principle of photoemission from a GaAs
crystal. At SLAC the longitudinally--polarized electrons were extracted
following photoexcitation from the J=3/2 valence band to the J=1/2
conduction band
using circularly--polarized 710 nm light from a flashlamp--pumped dye laser.
The theoretical maximum polarization achievable using this technique is
$P_e = 0.50$, although in actual use
the beam polarization as measured is more typically
$P_e = 0.40$.  A schematic of the original SLAC source is shown in Fig.~5.2.
The Mainz and MIT/Bates polarized sources as well as that planned
for CEBAF are
modern versions of very similar design.  As described in Sect.~V.A,
research is currently underway on developing
crystal structures which could yield polarizations
as high as $P_e = 0.80$.

The electron helicity was reversed rapidly by
reversing the circular polarization of the laser light.
A typical arrangement of optical elements used to achieve
this is shown in Fig.~5.3.  The light is linearly--polarized
in a Glan--Thomson calcite prism and then circular polarization is
achieved by means of a Pockels cell operating as a
quarter--wave plate.  The cell has a birefringence which is linear
with the applied electric field allowing the polarization to
be controlled by the application of a voltage prior to each
beam pulse.  Switching the polarity of the voltage reverses the
helicity of the photons and the helicity of the electrons, allowing
reversal rates up to 1 kHz to be achieved.

In order to isolate asymmetry contributions due to systematic effects (for
example, such as those connected with changes in beam parameters)
from the PV asymmetries, it is important to have an independent
control of the electron helicity.  At SLAC,
a slow helicity--reversal technique was
implemented by rotating the Glan--Thomson prism through $90^\circ$
using a movable holder.

\bigskip
\goodbreak
\noindent \undertext{ Beam and Instrumentation}

The electron accelerator was operated at 120 Hz with beam pulse
lengths of 1.5 $\mu$s.  Currents in the range (1--4)$ \times
10^{11} e^-$/pulse were accelerated and delivered to the
target. A major objective of the experiment was to eliminate and
control the sources of helicity--correlated systematic
errors.  Beam parameters were measured and controlled to
keep the corrections to the measured asymmetry small.  The
measured beam parameters included average polarization,
energy, current, position and angle at the target.

Beam polarization was measured using M{\o}ller scattering in which
the longitudinally--polarized electrons were elastically
scattered from polarized target electrons.  The target was a
thin Fe foil located in a magnetizing external field.  The
scattering cross section is large and for relativistic
energies the analyzing power is a maximum at
90$^\circ$ in the electron--electron center--of--mass.  Since the
scattering process is precisely described by QED, the beam
polarization could be measured to an accuracy of $\pm5$\%,
limited mainly by the uncertainty in background subtraction.

The beam--monitoring system was crucial for controlling beam
position and direction at the target.  Resonant microwave
cavities, with a node in the response which could be located on the beam axis,
were used to measure the beam position.  Pairs of monitors,
sensitive to horizontal and vertical displacements, were
located at 2 m and 50 m upstream of the target.  It was
possible to achieve pulse--to--pulse measurements of beam
position, averaged over the pulse, with resolutions better
than 10 $\mu$m and corresponding angle resolutions better
than $\pm$0.3 $\mu$radians.

Beam currents were measured (0.02\% resolution) on a
pulse--to--pulse basis by means of non--intercepting toroid monitors.
The charge per pulse was used to normalize the detector flux
signals as well as the beam position monitors.  Beam energy
was measured pulse--to--pulse by means of a microwave cavity
position monitor at a location in the beamline with nonzero
dispersion.  The relative energy sensitivity was $\pm 0.01\%$.

The beamline optics were typically nulled in order to minimize
sensitivity to helicity--correlated systematics.  Beam
energy, position, and angle were continuously monitored and
stabilized using computer--controlled feedback loops.
\bigskip
\goodbreak
\noindent \undertext{ Target and Spectrometer}

The incident electrons were scattered in a 30 cm long liquid
deuterium target and momentum--analyzed in a magnetic
spectrometer.  The spectrometer was operated at a
scattering angle of $4^\circ$ and had a very large momentum
acceptance.  Angular acceptances were $\Delta\theta =
\pm 7.5$ mrad and $\Delta\phi = \pm 16.6$ mrad.
%
Two counters were used to detect the scattered electrons:
the first was a gaseous $N_2$ \v{C}erenkov counter operated
at atmospheric pressure in which a spherical mirror collected and
focused the \v{C}erenkov light to an off--axis phototube.
This \v{C}erenkov counter was backed by a shower counter
constructed from nine radiation lengths of lead glass in which
\v{C}erenkov light produced in the lead glass was collected by
an array of phototubes.
A third counter was located behind a thick shielding wall
which absorbed the electron shower.  This counter was
designed to measure the background due to $\pi$'s, $\mu$'s
and K's and the associated asymmetry.  The experimental
arrangement was very clean and the measured contribution of
the background to the final asymmetries was less than 1\%.

A statistical precision of $10^{-5}$ requires
$10^7$ beam pulses or approximately one day at 120 Hz and therefore
a flux--counting technique was developed to deal with the
very high rates of scattered particles.  At
the SLAC kinematics each 1.5 $\mu$s beam pulse yielded
$10^3$ scattered electrons.  The resulting anode currents
for each detector were integrated, digitized and stored to
tape. Normalizing the flux to the incident charge yields a
quantity proportional to the acceptance averaged cross section.
Since background discrimination is not possible in such a flux
counting measurement, the kinematics and experiment
were designed to minimize the contribution of unwanted processes.

\bigskip
\goodbreak
\noindent \undertext{ Results}

The data were collected in runs of a few hours (1--3)
duration.  During each run there was random pulse--to--pulse
reversal of beam helicity using the Pockels cell.  Between
runs, two other methods were employed to achieve independent
helicity reversal.  A rotation of the calcite polarizing
prism, in front of the Pockels cell, resulted in
a rotation of the plane of linear polarization and beam
polarization reversal.
Some data were also
taken with the prism at the $45^\circ$ (unpolarized beam)
position.

The second method of spin reversal was based on the
precession of the electron spin relative to its direction,
due to the electron's anomalous magnetic moment ($g-2$
precession).  The electron beam was transported with a net
deflection of $24.5^\circ$ between accelerator and target.
A beam energy change of 3.237 GeV results in a helicity
reversal.  The experiment was run at four energies in the
range (16.2--22.2) GeV resulting in approximately two spin
reversals due to precession.
Figure~5.4
shows the results for the measured
asymmetries for each of the different energies.  The kinematics for the
scattering process were adjusted by correcting the asymmetry for the varying
$Q^2$.

The above methods for spin reversal were designed to identify and
measure false asymmetries due to systematic effects.  These
independent spin--reversal techniques allow for the
experimental cancellation of helicity--correlated effects that the
source may have on the beam parameters.  It is important
however, that the level of such cancellations or
corrections be kept small relative to the measured physics
asymmetry.

The SLAC data for the two highest--energy points yielded an
experimental asymmetry
$\alr/|Q^2| = (-9.5 \pm 1.6) \times 10^{-5}$ (GeV/c$)^{-2}$ (deuterium).
The data point is at the average value of $y= 1-
\epsilon^\prime/\epsilon = 0.21$ and the average value of $|Q^2| = 1.6$
(GeV/c$)^2$ [Pre78].  The quoted error is a linear combination of
equal statistical and systematic contributions.  The main
systematic uncertainties come from the measurement of beam
polarization, $P_e\ (\sim 5\%)$ and helicity--correlated
differences in beam parameters $(\sim 3\%)$.  The beam
parameter errors are summarized in Table~5.4.  As expected,
the energy differences were the most important.
Normalization corrections have also been applied to the data
for $\pi^-$ background (2\%) and radiative effects (3\%).

\midinsert
$$
\hbox{\vbox{\offinterlineskip
\def\strut{\hbox{\vrule height 10pt depth 5pt width 0pt}}
\def\superstrut{\hbox{\vrule height 17pt depth 10pt width 0pt}}
\hrule
\halign{\strut\vrule#&\quad#\quad\hfil
&\vrule#
&\hfil\quad#\quad\hfil
&\vrule#\cr
&\multispan3\hfil\headstrut{\bf TABLE 5.4 }
\hfil & \cr\noalign{\hrule}
\superstrut&Parameter&&$\Delta \alr/|Q^2|$&
\cr\noalign{\hrule}
& E$_0$&&$-0.37 \times 10^{-5}$ &\cr
& Q&&$-0.03\times 10^{-5}$ &\cr
& X$_t$&&$+0.04\times 10^{-5}$ &\cr
& Y$_t$&&$-0.02\times 10^{-5}$ &\cr
& $\theta_x$&&$\ \ {} 0.00$ &\cr
& $\theta_y$&&$+0.01\times 10^{-5}$ &\cr
&  &&  &\cr
& \ \ \ \ Total &&$-0.37 \times 10^{-5}$ &
\cr\noalign{\hrule}}}}$$
\smallskip
\baselineskip 10pt
{\ninerm
\noindent\narrower
{\bf Table~5.4.} \quad Systematic errors for the SLAC experiment [Pre79].
\smallskip}
\endinsert

\baselineskip 12pt plus 1pt minus 1pt

In
Fig.~5.5
the measured asymmetry is shown as a function of
$y$ (see Ref.~[Pre79]).  The
results are clearly consistent with the Standard
Model and were used to extract a best fit for the weak mixing angle:
$\sin^2\theta_W
= 0.224 \pm 0.020$.  The best fit for the model--independent
parameters is $a_1= -(9.7 \pm 2.6) \times 10^{-5}$ and
$a_2 = \ (4.9 \pm 8.1) \times 10^{-5}$, where $a_i
=(9G_\mu/20\sqrt{2}\pi\alpha)\tilde a_i$ with $\tilde a_i$ defined in
Eq.~(\disasymb).
A less accurate measurement was also done using a liquid hydrogen
target at 19.4 GeV.  The experimental asymmetry was
$\alr/|Q^2| = (-9.7 \pm 2.7) \times 10^{-5}$ (GeV/c$)^{-2}$ (hydrogen).
The quoted error includes both the statistical
and systematic uncertainties.  One expects a slightly
smaller asymmetry for the proton due to the different mix of
quarks.
\bigskip
\goodbreak
\noindent V.B.2.\quad QUASIELASTIC SCATTERING FROM BERYLLIUM (MAINZ)

The PV asymmetries observed in the scattering
of high--energy longitudinally--polarized electrons from deuterium at SLAC and
in atomic physics measurements at
low--energies were sensitive mainly to the hadronic vector
current.  These two measurements are essentially orthogonally dependent on the
linear couplings ${\tilde \alpha}$ and ${\tilde \gamma}$ of Table~3.2.
Contributions due to the hadronic axial--vector currents were
too small to provide a measure of the corresponding coupling
constants ${\tilde \beta}$ and ${\tilde \delta}$.

A recent PV experiment completed at Mainz [Hei89],
involved the scattering of medium--energy electrons at
backward angles.  The measurement involved the inclusive
scattering of polarized electrons of energy $\epsilon = 300$ MeV
from $^9$Be at an average scattering angle of ${\bar \theta}
= 130^\circ$.  At these kinematics, the cross section is
dominated by quasielastic scattering and the sensitivity to
the hadronic axial--vector current is enhanced, as discussed in Sect.~IV.F.
The expected asymmetry at this momentum transfer is
approximately $10^{-5}$, which is about 10 times smaller than for the SLAC
experiment.  The problems associated with achieving the required statistical
accuracy and the control of systematic errors were severe.
\bigskip
\goodbreak
\noindent \undertext{ Technical Considerations}

The polarized source that was built for this experiment
followed closely the SLAC design.  For injection into the
linac the source produced current pulses of 150 mA peak and
3.5 $\mu$s duration at a repetition rate of 50 Hz.  An average
current of about 7 $\mu$A was delivered to the target.
Proper accelerator operation required injected beam pulses
with a high degree of amplitude stability as well as
flatness.  An optical device, consisting of a transverse
Pockels cell mounted between a pair of Glan--Thomson
polarizers, was incorporated in the laser beamline to
provide automatic control of the pulse--shape.  A short--term
amplitude stability of $\sim 0.1\%$ and flatness within $3
\times 10^{-3}$ were achieved.  Helicity--correlated beam
intensity differences had to be maintained at a level
$<10^{-5}$ ---  this was achieved with an active feedback
system incorporated into the laser beamline optics.
The source lifetime during operation was as long as 200 hours.
A short interruption was required approximately once per day
to change the flashlamps and the laser dye.
As in the SLAC experiment, rapid polarization reversal was
achieved by reversing the voltage applied to the Pockels
cell.  A slow reversal cycle, approximately every 15 min,
was accomplished by rotating the half--wave plate between the
Glan prism and the Pockels cell.

The beam transport line to the target consisted of two
successive oppositely deflecting $90^\circ$ achromatic
bends.  As a consequence, the parallel displacement of the
electron beam leaves the longitudinal polarization
unchanged.  The relatively low beam energy precluded the use
of spin precession for independent spin reversal.

Beam polarization was measured using three different
techniques.  Prior to injection into the linac, a Wien
filter was used to produce transverse polarization.  This
was followed by spin analysis of the Mott--scattered 44 keV
electrons.  The polarization of the 300 MeV beam could also
be measured directly using a M{\o}ller polarimeter
downstream of the first $90^\circ$ achromat.

An on--line Compton polarimeter was used to provide relative
polarization measurements during the experiment.  It
consisted of a magnetized iron absorber between two
ionization chambers. The ratios of the ionization currents
in the two chambers for positive and negative electron
helicity measures the circular polarization of the
bremsstrahlung.  This in turn is proportional to the
longitudinal polarization of the electron beam.  The Compton
polarimeter was calibrated relative to the M{\o}ller
polarimeter.  The beam polarization during the experiment
averaged between 44\% and 50\%.

The beamline was fully instrumented to control and minimize
helicity--correlated differences in beam parameters such as
energy, intensity, position and angle at the target.  The
monitoring system included toroid intensity monitors,
microwave cavity position monitors and steering correctors.

\bigskip
\goodbreak
\noindent \undertext{ Detector System}

Given the small PV asymmetry and the cross section for
quasielastic scattering it was necessary to have a detector system of very
large solid angle to achieve the desired statistical accuracy.  Conventional
magnetic spectrometers were inadequate.
The detector system (Fig.~5.6) consisted of 12 ellipsoidal
mirror gas \v{C}erenkov counters mounted symmetrically about
the beam axis.  Electrons scattered in the angular range
$115^\circ < \theta < 145^\circ$ were detected.  The
\v{C}erenkov light emitted by the electrons travelling through
air between target and mirror is focused onto the
photomultiplier cathode.  The mirrors subtended a solid
angle of 20\% of $4\pi$.

The detector has essentially no momentum resolution.  With
atmospheric pressure air as the \v{C}erenkov medium the
threshold energy for detecting electrons was about 21 MeV.
This provided a low--energy cut--off for the electron spectrum and some
discrimination of backgrounds due to heavier particles.
Each of the mirrors could be tilted remotely in order to scan
the target image over the face of the photomultiplier. This allowed the mirror
orientation to be optimized and the signal--to--background ratio
measured.  Four additional photomultipliers
positioned out of the mirror focus and mounted symmetrically about the
beam axis provided continuous on--line background information for
each counter.  One important background source in this experiment was beam
halo which could be directly scattered in the direction of the
photomultipliers.  Under optimal accelerator tuning conditions the
contribution due to beam halo was kept below 1\%.
\bigskip
\goodbreak
\noindent \undertext{ Measurements}

The scattered rate into each detector during a beam burst
$(\sim 3 \times 10^4/$pulse) was too high for individual
event counting.  Instead, the resulting flux (anode current)
for each detector was integrated, digitized and stored to
tape.  Beam information, charge and position, was similarly
averaged and recorded for each beam burst.
The electron helicity was randomly chosen for a
beam burst.  The following beam pulse was set to have the
opposite helicity.  This process was repeated for succeeding pulse pairs.
In analyzing the data, asymmetries were
formed from such adjacent pulse pairs.  This approach compensated for
unavoidable slow drifts in the accelerator and detector hardware.

Figure~5.7 shows a histogram of pulse--pair asymmetries $(\sim
22,500)$ for a 15 min run at a fixed orientation of the
half--wave plate.  The result shows the expected normal
distribution.  In addition, the width of the distribution is
consistent with the expected number of scattered electrons.
One concludes that contributions due to fluctuations of the beam parameters
have been suppressed to a level below the statistical errors.
\bigskip
\goodbreak
\noindent \undertext{ Results}

There were several sources of systematic false asymmetries
which had to be understood and controlled to a very high
level.  One of these was electronic cross--talk correlated
with the high voltage switching of the Pockels cell.
Careful wiring and grounding of the electronics followed by
detector tests with light diodes showed that the electronic
asymmetry was less than $8 \times 10^{-8}$, which is
insignificant.
M{\o}ller and Mott scattering depends on the polarization of
the electron beam and can give rise to false asymmetries.
The axially symmetric detector system reduced possible Mott
effects to less than $4 \times 10^{-8}$ and M{\o}ller
scattering was estimated to contribute at a level of less
than $8 \times 10^{-8}$.
The most troublesome systematic errors were the helicity--dependent
differences in the beam parameters such as position, angle,
energy, and intensity.  The sensitivity to each of these
parameters was determined experimentally from the normal
jitter of the beam itself.  This jitter was typically 100
times larger than the measured correlated differences in
any of the parameters.

After applying various corrections, the measured corrected
asymmetry was found to be
$A_\sst{LR}^{corr} = (-3.5 \pm 0.7\pm0.2) \times 10^{-6}$,
where the first error is the statistical
uncertainty and the second is the systematic.  The
experimental asymmetry was extracted from $A_\sst{LR}^{corr}$
by normalizing for the dilution due to the background (B/S,
assumed to produce no asymmetry) and the dependence on the
electron polarization $(P_e)$ using the relation
$$A_\sst{LR}^{exp} = A_\sst{LR}^{corr} \ {(1 + B/S) \over P_e} \ .$$
\noindent Using averaged values for $B/S = 0.19$ and $P_e =
0.443$ the final result is
$$A_\sst{LR}^{exp} = (-9.4\pm1.8\pm0.5) \times 10^{-6} \ .$$
In interpreting this result one should note that at least
four distinct scattering processes contributed significantly
to the asymmetry.  These include: quasielastic scattering
(59\%), excitation of the $\Delta$--resonance (6.5\%),
contributions from the ``dip'' region (12.5\%) and
contributions from the radiative tail for low final electron
energies (22\%).  The spectrum was assumed to be cut off by
the 21 MeV detector threshold.

This measured result can be combined with previous
model--independent determinations of ${\tilde \alpha}$ and
${\tilde \gamma}$ [Pre79, Kim81] to obtain a relation between the two
axial--vector coupling constants
$${\tilde \beta} + 0.04 {\tilde \delta} = 0.005 \pm 0.17\ \ . $$
The errors reflect a combination of the errors in the measurement
of ${\tilde \alpha}$ and ${\tilde \gamma}$
and the total relative error of this experiment.
The results for ${\tilde \beta}$ and ${\tilde \delta}$ are
shown plotted in Fig.~5.8 in the standard form involving
the quark coupling constants $C_{2d} = (-{\tilde \beta} +
{\tilde \delta})$ and $C_{2u} = ({\tilde \beta} + {\tilde
\delta})$ together with the results of the SLAC
experiment.  The results of this experiment are consistent
with a vanishing axial--vector interaction $({\tilde \beta} =
{\tilde \delta} = 0)$.  The resulting error band has been reduced by
a factor of three in comparison with earlier work.
\bigskip
\goodbreak
\noindent V.B.3.\quad ELASTIC SCATTERING FROM CARBON (MIT/BATES)

An experiment measuring the PV asymmetry in
the elastic scattering of longitudinally--polarized electrons
from $^{12}$C was successfully completed at MIT/Bates [Sou90a].  The
measurement was motivated by the theoretical work of
Feinberg [Fei75] and Walecka [Wal77], as
discussed in Sect.~IV.B: since $^{12}$C is a
spinless and isoscalar nucleus, the electromagnetic
amplitude is described by a single form factor.  At low--energies where a
phenomenological four--fermion interaction is a good
approximation, the asymmetry may be written as in Eq.~(\Espinz).
The expression for the asymmetry is independent of the form
factors if the effects from strangeness, isospin--mixing and dispersion
corrections (all of which are expected to be small, see Sect.~IV.B) are
ignored.
The goal of the MIT/Bates $^{12}$C experiment was to measure
$\snw$ precisely using a technique which involved
very little theoretical ambiguity in the interpretation of
its result.
The choice of optimal kinematics for elastic scattering from
$^{12}$C was influenced by several important considerations.
To first--order it was important to maximize the FOM.
Signal--to--background
considerations, contamination due to Mott asymmetry and
contributions due to inelastic levels in $^{12}$C also
influenced the choice.
The experiment was designed for a beam energy of 250 MeV, a
scattering angle of about $35^\circ$ giving a momentum
transfer $q = 150$ MeV/c.
With a beam polarization $P_e = 37\%$
the Standard Model predicts an experimental asymmetry
$A_\sst{LR}^{exp} = \alr P_e = 0.70 \times 10^{-6}$.  Such a small
asymmetry places severe demands on the apparatus, in terms
of both achieving sufficient statistical precision and
maintaining adequate control of systematic errors.
\bigskip
\goodbreak
\noindent \undertext{ Beam and Instrumentation}

The apparatus for this measurement consisted of a polarized
electron source, a pair of single--quadrupole spectrometers,
a M{\o}ller polarimeter, precision beam monitors, and a
high--capacity data acquisition system.  A schematic layout
of the experimental apparatus in shown in Fig.~5.9.

The source, which provided an intense beam of
longitudinally--polarized electrons, was built especially for this
experiment.  Similar to the SLAC source, it was based on
photoemission from a GaAs crystal using circularly--polarized
light, but in addition involved a unique three--chamber design which
allowed for greater flexibility to select crystals with high quantum
efficiency,
lifetime and polarization.
The main components of the optical system were a 2W CW
krypton--ion laser and a system of optical shutters and
circular polarizers.  An electro--optical shutter, consisting
of two crossed polarizers, half--wave plate and a Pockels cell modulated the
laser beam to match the 1\% duty factor of the accelerator.  A ``flipper''
system consisting of a half--wave plate, crossed polarizers and a Pockels cell
produced the circularly--polarized light and controlled helicity reversal. The
half--wave plate served as a ``slow'' helicity reversal device.  The two
different orientations of the device reversed the relationship between the
polarity of the potential difference applied to the Pockels cell and the
handedness of the emerging laser light.  This made it possible to change the
overall sign of all downstream helicity--correlated differences including the
electroweak PV asymmetry. A converging lens provided
point--to--point focusing between the Pockels cell and the GaAs crystal and
made the position of the laser
beam on the crystal insensitive to the angular fluctuations
that could be induced in the beam by misalignments or
instability in the operation of the Pockels cell.

The beam delivery system was designed and carefully
instrumented to ensure that the expected small asymmetries
could be accurately measured.  A set of monitors measured
the undesirable effects of helicity reversal.  These included seven current
monitors to measure the intensity, four microwave cavity position monitors in
front of the target to measure the position and angle of the beam and a
position monitor to measure the beam energy at a location in the chicane
where the beam was dispersed in momentum.  All of these measurements were
performed on a pulse--by--pulse basis together with the data--taking.

A steering coil pulsing scheme allowed steering correctors along the
beam transport line to be synchronously adjusted (ramped)
during data acquisition.  The coil current changes were
triggered at a 47 Hz rate, which was chosen to be fast
enough so that the system calibration would be unaffected by
drifts and slow enough to add negligibly to the statistical
fluctuations in the cross section measurements.  The coil--pulsing system was
used for approximately one--third of the data collection time.

The combination of precision monitors and the coil pulsing
scheme allowed for the measurement and correction of
systematic errors.  It was also possible to calibrate the
position monitors and to measure the sensitivity of the
detector system to known displacements of beam energy,
position and angle at the target.  As a consequence the beam
position and direction could be located and maintained
in a way that minimized the sensitivity of the experiment to
beam parameters.

The detector system consisted of a pair of single--quadrupole
spectrometers, UVT--lucite \v{C}erenkov counters, analog and
control electronics and a high--capacity data acquisition
system.  The two spectrometers were fixed on either side of
the beamline at a nominal scattering angle of $35^\circ$.
Each had an acceptance of $\sim$10 msr and a momentum
resolution of $\sim$12 MeV.  Lead collimators at the front,
rear, and midplane of the quadrupoles defined the acceptance
and momentum discrimination of the system.  The focusing was in the
vertical plane.  A bank of 12 \v{C}erenkov detectors directly coupled to
photomultipliers formed the detector array.
\bigskip
\goodbreak
\noindent \undertext{ Measurements}

The electron beam impinged on a 5 g/cm$^2$ segmented carbon
target.  About $10^5$ electrons were detected during each
17 $\mu$s beam burst and the integrated responses over the beam
pulse were recorded by 16--bit ADC's.
A M{\o}ller polarimeter, similar to the SLAC design, was
used to measure the beam polarization during the course of
the experiment.

A major consideration in implementing a data--taking scheme
was the issue of noise: it was important to use a
technique where the noise contributions would be less than the statistical
error per pulse.  To achieve this the
accelerator was operated at a pulse rate of 600 Hz locked to
the 60 Hz line frequency.  The noise associated with the
60 Hz frequency was minimized by dividing the data into ten
separate ``timeslots'' corresponding to the 60 Hz harmonics
and then analyzing the data for each timeslot independently.
The beam helicity was set quasi--randomly for each pulse.
Ten random helicities were chosen, one for each timeslot.
The pattern was complemented for the next ten beam bursts,
and ten asymmetries were computed, each based on a
complementary pulse pair.  This process was repeated every
ten pulse pairs, {\it i.e.,\/} at a 30 Hz rate.
Typical beam currents on target averaged 30--60 $\mu$A.  The
accumulated data amounted to 307 half--hour runs.  With each
timeslot treated independently, this corresponds to 3070
individual ``mini--runs''.  The statistical error for each
mini--run was computed using the variance of the asymmetries.
Approximately 1\% of the data was rejected by loose cuts
which were associated with accelerator malfunctions.  A
histogram of the measured asymmetry for each mini--run
normalized to its statistical error is shown in Fig.~5.10.
The shape, as demonstrated by the solid curve, is Gaussian
with the expected width over more than two decades.

The main objective of the experiment was to measure a cross section which,
apart from the electroweak physics, is
independent of the sign of the helicity.  The goal was to
maintain the level of systematic errors below the
statistical precision.
There were many sources of systematic error.  They included:
beam parameters (intensity, energy, position, direction,
emittance, transverse polarization); hardware (drifts and
noise, crosstalk/pickup, linearity); nuclear physics
(isospin admixtures, inelastic contributions, target purity,
strangeness).  Correlations of beam parameters with
helicity, such as energy, position and intensity were the
most important class of systematic errors.  These were
controlled by minimizing helicity correlations during data
acquisition and corrections were made to the raw asymmetries
using position monitor data during analysis.
A major source of helicity--dependent correlations was the
intensity of the laser light striking the GaAs photocathode.  The resultant
helicity--correlated current changes are coupled by accelerator beam loading
to energy changes.  Since the cross section is a very strong function of the
energy it was necessary to maintain the helicity--dependent current
fluctuations to within
$\Delta I/I < 7 \times 10^{-7}$.  The principal cause of the
intensity correlation was a coupling of a slightly elliptically polarized
laser beam to an asymmetric optical transport system in which the transport
efficiency depends on the orientation of the ellipse axes.
In practice this effect was easily controlled.  It was possible to adjust the
appropriate phase electro--optically
by small changes in the voltages applied to the Pockels cell. A slow feedback
loop was implemented to minimize the
intensity asymmetry.  The result was that the helicity--correlated intensity
asymmetry averaged over the experiment was reduced to about 1 ppm.
\bigskip
\goodbreak
\noindent \undertext{ Results}

During data analysis, it was necessary to make small
correction for systematic asymmetries as measured by the
beam monitors.  The corrected asymmetries were related to the raw
asymmetries by the algorithm
$$A_\sst{LR}^{corr} = A_\sst{LR}^{raw} - \Sigma a_i (\delta M_i)\ \ ,$$
where $A_\sst{LR}^{raw}$ is the measured uncorrected
asymmetry, $\delta M_i$ are the helicity--correlated
differences measured by the beam monitors, and the $a_i$ are
correction coefficients, which are a measure of the
sensitivity of the asymmetry to fluctuations in the beam
parameters.  The last were determined simultaneously with
data--taking by using the coil--pulsing system.  Correlation analysis of the
energy
measurement in the beamline chicane was used to extract the energy--sensitive
coefficient.  An energy vernier (fast phase shifter) on one of the klystron
amplifiers provided a controlled and independent measure of
the sensitivity.  Typical values for the $a_i$ were about
10 ppm/$\mu$m and the averaged helicity--correlated position differences
were typically a fraction of 1 $\mu$m.

Table~5.5 lists all of the corrections (ppm) that were applied to
the experimental asymmetry together with their estimated
uncertainty.  The root--mean--square value of the corrections
due to the position and energy monitor differences for
individual runs was 0.3 ppm.  The average correction for
the entire data sample was only 0.04 ppm.
An important check on the systematics is a measurement of a
null asymmetry.  If the half--wave--plate reversal is
neglected in the data analysis, the measured asymmetry is
$0.04\pm0.14$ ppm.  Similarly the difference in the measured asymmetry
between the two spectrometers is $0.14\pm0.14$ ppm.  This latter result is
important in determining that contributions due to transverse polarizations
are not significant in this experiment.

\midinsert
$$
\vbox{\offinterlineskip
\def\strut{\hbox{\vrule height 10pt depth 5pt width 0pt}}
\def\superstrut{\hbox{\vrule height 17pt depth 10pt width 0pt}}
\def\totalstrut{\hbox{\vrule height 15pt depth 5pt width 0pt}}
\hrule
\halign{
\tabskip 0.1in
\strut                
\vrule#&              
#\hfil&               
\vrule#&              
\hfil#\hfil&          
\vrule#&              
\hfil#\hfil&          
\vrule#               
\tabskip 0.0in\cr
&\multispan5          
\headstrut\hfil{\bf TABLE 5.5}\hfil&\cr
\noalign{\hrule}
\superstrut&Correction&&Value&&Error&\cr
\noalign{\hrule}
& Energy and Position Monitors && $0.04$ && $\pm 0.006$ &\cr
& Electronic Cross--Talk       && $-$    && $\pm 0.001$ &\cr
& Transverse Polarization      && $-$    && $\pm 0.005$ &\cr
& Nonlinearities               && $-$    && $\pm 0.007$ &\cr
& Phase Space                  && $-$    && $\pm 0.006$ &\cr
& Magnetized Iron Background   && $-$    && $\pm 0.010$ &\cr
& \ \ \ \ \ \totalstrut Total  &&        && $\pm 0.016$ &\cr
\noalign{\hrule}}}$$
\smallskip
\baselineskip 10pt
{\ninerm
\noindent\narrower
{\bf Table~5.5.} \quad Systematic errors (in units of ppm) for the
Bates $^{12}$C experiment
[Sou90a].
\smallskip}
\endinsert

\baselineskip 12pt plus 1pt minus 1pt

The result for the measured raw asymmetry is
$A_\sst{LR}^{raw} = 0.56\pm0.14$ ppm and the corrected PV
asymmetry is
$A_\sst{LR}^{corr} = 0.60\pm0.14\pm0.02$ ppm, where the first error is
statistical and the second is systematic.

It was necessary to account for various normalization
factors in extracting a value for $\snw$. These
included the average effective $|Q^2|$, the beam polarization
$P_e$, and the backgrounds due to inelastic nuclear levels
and scattered neutrons.  These factors are summarized in
Table~5.6.  Applying these normalization factors the final result for the
experimental asymmetry is
$A_\sst{LR}^{exp} = (1.65 \pm 0.39 \pm 0.06) \times 10^{-6}$.
The resulting extracted value for $\tilde \gamma$ (see Table~3.2) is
$$\tilde \gamma = 0.136\pm0.032\pm0.009\ \ ,$$
which is consistent with the prediction of the
Standard Model, where
$$\tilde \gamma_{SM} = 0.155\pm0.002\ \ ,$$
using a value of $\snw = 0.233\pm0.002$.

\midinsert
$$
\vbox{\offinterlineskip
\def\strut{\hbox{\vrule height 12pt depth 6pt width 0pt}}
\def\superstrut{\hbox{\vrule height 12pt depth 10pt width 0pt}}
\def\totalstrut{\hbox{\vrule height 12pt depth 6pt width 0pt}}
\tabskip=0pt
\hrule
\halign{
\strut\vrule#&
\quad#\quad\hfil&
\vrule#&
\hfil\quad#\quad\hfil&
\vrule#\cr
&\multispan3\superstrut\hfil{\bf TABLE 5.6}\hfil&\cr
\noalign{\hrule}
& Beam Polarization, $P_e$ && $0.37\pm0.02$ &\cr
& Nuclear Structure        && $1.00\pm0.01$ &\cr
& Background               && $0.98\pm0.02$ &\cr
& \totalstrut $< |Q^2| >/<|Q_\circ^2| >$ && $1.00\pm0.02$ &\cr
}\hrule}
$$
\smallskip
\baselineskip 10pt
{\ninerm
\noindent\narrower
{\bf Table~5.6.} \quad Normalization factors for the Bates $^{12}$C experiment
[Sou90a].
\smallskip}
\endinsert

\baselineskip 12pt plus 1pt minus 1pt

This experiment achieved a statistical precision which
is about five times smaller than that for the most sensitive
previous electron scattering measurement.  Based on these
very promising results, especially the very small systematic
errors, it is clear that significant improvements in a
future $^{12}$C measurement are feasible.  It
is reasonable to imagine large acceptance spectrometers with
factors of 10--30 increased solid angle which together with
substantially longer running time would result in a
statistical error of a few percent.  At this level of
precision theoretical uncertainties may become important.
We will need to understand that the effects of hadronic
contributions to the radiative corrections and isospin--mixing are tractable
at this level before committing to such a major effort.  A significantly
larger contribution, about which there is currently much speculation and a
lot of interest, could come from a large charge radius of strange quarks in
the nucleon (see Sects.~IV.B and V.C).
\bigskip
\goodbreak
\noindent V.B.4.\quad NEUTRINO--PROTON SCATTERING (BNL)

Another class of experiments involves those where
neutrino--proton elastic scattering cross sections are measured. As discussed
in Sect.~IV.J,
these experiments are the best choice for accessing the axial--vector
form factors, both in terms of sensitivity and because the
axial--vector radiative corrections are much smaller than in electron
scattering.
The relevant differential cross section is given in Eqs.~(\Edsignulow).

The most recently completed experiment of this type is that of Ahrens
{\it et al.\/}
[Ahr87], performed at the Brookhaven AGS facility. The 28 GeV proton beam,
incident on a production target, produced pions and kaons, which
decay in--flight into primarily muon neutrinos. Using a magnetic horn
the secondary beams were focused into a decay tunnel,
producing a wide--band beam of either $\nu$ or $\overline\nu$, depending
on the polarity of the horn. The neutrino beams covered a spectrum
of 0.2 to 5 GeV, peaked at $\sim$1 GeV. The contamination from electron
neutrinos was typically 1\%, and from wrong--helicity neutrinos was $\sim$3\%
for the $\nu$ beam and $\sim$9\% for the $\overline\nu$ beam.
In general, the neutrino flux cannot be measured in the neutral current
scattering process, so secondary measurements of quasielastic
charged--current scattering are performed and the data are normalized to
the reactions $\nu_\mu n\rightarrow\mu^-p$ and
$\overline\nu_\mu p\rightarrow\mu^+ n$.

As can be seen in Eq.~(\Edsignulow),
the neutrino cross sections are proportional to
$G_\mu^2$, and so are very small. Thus, it is important to have as
high a luminosity and as large a detector as possible. Typically one
uses an active target. In the case at hand, the target/detector was
comprised of 170 metric tons of liquid scintillator cells. In addition to
the liquid scintillator, each of the 112 cells contained crossed
planes of proportional drift tubes to measure particle tracks.
Particle identification was performed by measuring the energy deposited
in each liquid scintillator plane, and the $\nu$--$p$ signature was
identified by the scattered proton. A shower counter directly
downstream of the main detector was used to contain the electromagnetic
shower from neutrino interactions in the downstream half of the detector.
A muon spectrometer was located behind the shower counter and was
used to measure the neutrino flux via charged--current scattering.
An important consequence of having to use an active target such as
liquid scintillator is that only $20\%$ of the protons in the target
are actually free protons; the other $80\%$ are bound in $^{12}$C.
This introduces a systematic error associated with how well the
neutrino scattering process is described by quasifree scattering.

Much of the background was reduced
by using the time structure of the beam to eliminate events which
arrived out of time with respect to the AGS beam extraction.
In this way the background from inelastic pion production
was removed. Neutron--induced events were also removed with this cut,
and also by reducing the active volume of the detector to
only the central region (an $80\%$ reduction). The remainder of the
backgrounds were modeled by a Monte Carlo calculation.
The overall normalization error in the experiment was $10\%$, and
came predominantly from the Monte Carlo calculation.

The differential cross sections for both neutrinos and
antineutrinos are shown in Fig.~5.11. The data were fit to
Eq.~(\Edsignulow),
averaged over the neutrino energy spectrum. The vector form factors
$F_1$ and $F_2$ were assumed to have a dipole $|Q^2|$--dependence with
$M_V=0.843$ GeV, and both vector strangeness form factors were taken to be
zero.  The axial--vector
form factor was also assumed to have a dipole dependence, and although
there was an allowance for a strange axial--vector contribution through
the parameter $\eta_s$, it was assumed that the $|Q^2|$--dependence of
the strange contribution is the same as the nonstrange part.
The adjustable parameters of the fit were $\sin^2\theta_W$, the
axial--vector dipole mass $M_A$, and $\eta_s$. For the fit shown in
Fig.~5.11,
$\eta_s$ was fixed at 0, and values of $\sin^2\theta_W = 0.220\pm 0.016$
(stat) $\pm 0.03$(syst) and $M_A = 1.06\pm 0.05$ GeV/c$^2$ were
simultaneously extracted. Overall normalization adjustments
of $s_\nu = 1.05$ for $\nu_\mu p$ and $s_{\overline\nu} = 1.09$
for $\overline\nu_\mu p$ was made
due to the 10\% normalization uncertainty in the data.
In a second fit, $M_A$ and $\eta_s$ were
simultaneously fit while $\sin^2\theta_W$ was held
at the then world average of 0.220. A contour plot of $\chi^2$ in
$\eta_s$ and $M_A$ space is shown in Fig.~5.12, and, as stated by
the authors, it is seen that they are highly correlated. Nonetheless,
by fixing $M_A$ at the then world average of 1.03 a value of
$\eta_s = 0.12\pm 0.07$ was extracted from the data, corresponding
to $G_A^{(s)}(0)= -0.15 \pm 0.08$. This result is, in fact, consistent with
the published result of the EMC collaboration [Ash89] (assuming
the momentum scale dependence of the latter is negligible).
It is, however, also consistent with zero at the $2\sigma$ level.

In assessing the results of this experiment, several questions come
to mind which one might hope to address in future experiments.
Firstly, the majority of the
protons detected were actually quasielastically knocked out of $^{12}$C
nuclei. Is it reasonable to assume that the both $M_A$ and $G_A (Q^2)$
are the same for nucleons bound in nuclei as for free nucleons?
At least one author has suggested [Bro88] that medium modifications
might lead to a change in the value of $M_A$ on the order of $5\%$
[Whi90] and, since $M_A$ and $\eta_s$ are so highly correlated, this could
lead to a large change in $\eta_s$. Secondly, is the assumption that
$G_A^{(s)}$ has the same $|Q^2|$--dependence as $G_A^{T=1}$ a good one? It has
been suggested [Ber90] that, within the context of soliton models,
the isoscalar axial--vector dipole mass should be somewhat larger than $M_A$.
Unfortunately,
there is no other experimental information available at
this time. Both of these assumptions could be avoided if an experiment
were to be performed at a lower average value of $|Q^2|$.
A third concern is the assumption that both $F_1^{(s)}(Q^2)$ and
$F_2^{(s)}(Q^2)$
are zero. If one assumes the Jaffe parameterizations, the contributions
from the vector strange form factors at these kinematics are each
about 10--15\% [Bei91a].
At lower--$|Q^2|$ the contribution from $F_1^{(s)}$ will certainly
be smaller; however, $F_2^{(s)}$ is as yet undetermined.

%

A re--analysis of the Brookhaven data has recently been performed [Gar93],
taking into account several of the above considerations. The authors use
more recent world averages of $\sin^2 \hat\theta_W = 0.2325$ and
$M_A = 1.061\pm 0.026$
in the new fit. They also use a nonzero value for the neutron electric
form factor. The vector form factors $F_1^{(s)}$ and $F_2^{(s)}$
were included in the fit and a dipole dependence with dipole mass $M_V$
was assumed. The fit parameters were $M_A$, $\eta_s$, $\mu_s$,
$\langle r_s^2\rangle$
(effectively)\footnote{*}{ In Ref.~[Gar93]
the authors define the parameter $F_1^s = -{1\over 6}\langle r_s^2\rangle$.
In the present notation, this corresponds to
$-{dF_1^{(s)}\over dQ^2}\vert_{Q^2=0}$.}
and the overall normalization parameters $s_\nu$ and $s_{\overline\nu}$.
Five different fits are performed, in which $M_A$, $\eta_s$, $\mu_s$,
and $\langle r_s^2\rangle$ were either held constant or extracted from
the fit.
Two of the five fits attempt to account
for the fact that the majority of the target protons were bound in
carbon by reducing the effective number of protons from 6 to $5.4\pm 0.6$.
Reducing the number of target protons generally has the effect of increasing
the fitted value of $\eta_s$ without having much effect on the extracted
values of the vector form factors. The authors point out that because of the
observed $Q^2$--dependence of the data, a positive slope of $F_1^{(s)}(Q^2)$
is preferred. Their combined fit of vector and axial--vector parameters
yields uncertainties of roughly order $\pm 0.5$ for the strangeness
magnetic moment, and $\pm 0.1$ fm$^2$ for the square of the strangeness
radius. The errors
in the fitted vector form factors (as well as those between the
axial--vector form factor and the axial--vector dipole mass) are
highly correlated.




%
%
The implication is that the BNL data do potentially provide some rough
constraints on the vector, as well as axial--vector, nucleon form factors.
This comes largely from the fact that $\nu$ and $\bar\nu$
cross sections have nicely orthogonal dependences on the vector form
factors. Unfortunately, the large values of $|Q^2|$ in this experiment
introduce a strong dependence on the poorly known $|Q^2|$--variation of
the form factors (in this case, parameterized by only two dipole masses,
$M_V$, and $M_A$.) Thus, the constraints are necessarily
rather weak. But coupled with future data, including
electron scattering and lower--$|Q^2|$ neutrino--proton
scattering, the BNL results should continue to play a role in extracting
the desired strangeness parameters.

\vfil\eject
\noindent{\bf V.C.\quad Present and Proposed Experiments ---
Future Prospectives}
\bigskip

The $^{12}$C experiment described in the last section is an
example of the level of systematic error that can
be achieved at present--day electron accelerators.
Since the completion of the $^{12}$C experiment, accurate measurements
of the mass of the Z--boson in high--energy experiments at CERN, Fermilab
and SLAC have caused a change in focus of PV experiments
at medium--energies. The next generation of experiments are centered
around measurements of the strange matrix elements, relying on
the fact that $\snw$ is now known with sufficient accuracy that
uncertainties due to $\snw$ are much smaller than uncertainties
due to strange--quark contributions. MIT/Bates, CEBAF, Mainz and SLAC all
have proposed extensive programs to measure the strange--quark contributions
to vector matrix elements.
In addition, a new experiment at LAMPF is currently underway to
further investigate the axial--vector strange form factor.
This section will describe and contrast each of the proposals,
as well as provide some discussion of where one might go in the future.

\bigskip
\goodbreak
\noindent{V.C.1.\quad SAMPLE (MIT/Bates)}

The next PV experiment currently being prepared at MIT/Bates [McK89] is
intended to
build on much of the work accomplished in the $^{12}$C experiment.
This experiment will focus on measurement of the strange magnetic
contribution to the proton, specifically $\mu_s$ introduced in Sect.~III.C.
Performing an experiment at backward angles allows one to focus on
the second and third terms of Eq.~(\alrprotpn). The axial--vector term is
further suppressed
by the factor $(1-4\snw)$, so to a first approximation one is left with
an asymmetry proportional to $\tilde G_M^p$.
In the limit of zero momentum transfer, the only unknown parameter is
$\mu_s$. The experiment will make a measurement at a beam
energy of 200 MeV and scattering angles of $135<\theta < 160^\circ$,
which is an average momentum transfer of $|Q^2|=0.1$ (GeV/c)$^2$.
At these kinematics the Standard Model asymmetry is
$8\times 10^{-6}$, which is about five times larger than that
measured in the $^{12}$C experiment, and comparable to the
Mainz experiment. The present goal is to achieve a statistical
accuracy of 5\% with a systematic error of 5\%. This can be achieved
in approximately 500 hours of running with 40 $\mu$A of beam polarized
to 40\%. The resulting overall error on $\mu_s$ will be $\Delta\mu_s=0.22$.

The polarized injector, many of the beamline components and most
of the detector electronics will be the same as those used in the
MIT/Bates $^{12}$C experiment. One added complication that the $^{12}$C
experiment did not have is that the beamline at the target is not at
0$^\circ$ with
respect to the accelerator, so the electron spin will precess away from
longitudinal by $16.5^{\circ}$
by the time it reaches the scattering target. This effect
can be removed by purposely aligning the electron spin away from
longitudinal just after the source but before acceleration. With the
appropriate initial alignment angle, the electron spin will
precess into the longitudinal direction at the target.
This will be accomplished using a Wien filter, a
device with crossed electric and magnetic fields. The beam polarization will
again be measured with a M{\o}ller polarimeter, and in this case it is
important that the polarimeter be accurate enough to detect not only
the large longitudinal M{\o}ller asymmetry, but also the transverse
asymmetry which is seven times smaller.

With the small PV asymmetries anticipated in experiments
of this type, it is desirable to obtain the maximum possible luminosity
from the accelerator. The polarized electron beam will
be incident on a 40 cm long circulating liquid hydrogen target.
The total power deposited in the target by a 40 $\mu$A beam
is 500~W, so sufficient cooling power must be available in
order to keep the target from boiling. A 700 W He gas refrigerator will
be used to cool the target. In addition, any volume of
liquid hydrogen seen by a 15 $\mu$sec beam pulse must be out of the
path of the beam by the next beam pulse or density fluctuations in the
liquid will occur due to boiling. This is achieved by flowing the
hydrogen parallel to the beam at a rate of approximately 10 m/s.
At such a high velocity, the liquid is highly turbulent and a significant
transverse component is also present.

Figure~5.13 shows a layout of the SAMPLE detector and target system.
The target will be viewed by a large array of \v{C}erenkov detectors
which cover $\sim 2$ sr of solid angle. The detector consists of a set
of ten mirrors, each 0.7$\times$0.7 m$^2$, which focus \v{C}erenkov light
produced in air onto ten 8 inch diameter photomultiplier tubes.
This detector system is quite similar to that used in the Mainz
experiment, with the modification that the phototubes must be able
to view the entire 40 cm of target length.
The scattered electron rate into each phototube is $\sim$4000/pulse,
so the yield in each tube is integrated over
the 15 $\mu$sec pulse length.  It is therefore important
that background processes which might cause a signal in the detector
be kept to a minimum. With a beam energy of 200 MeV, it is not
possible to produce pions in the target directly, and the inelastic
scattering rate into the detector is kept reasonably low. The largest
source of background is expected to come from low--energy inelastic
scattering from Al in the target plumbing downstream of the target cell.
Because these are mainly low momentum transfer processes,
the main effect on the asymmetry is an overall dilution factor.

As mentioned above, it is the latter two terms in the PV
asymmetry which
are the dominant contributions. The axial--vector term is suppressed because of
the $(1-4\snw)$ term, but it still contributes approximately 20\% (see
Sect.~IV.A).
The weak radiative corrections to the vector term are quite
small ($\sim 3\%$), but the corrections to the axial--vector term are about
20\%
and rather uncertain (see Sect.~III.B). These uncertainties put a natural
limitation on the level of precision one might currently
expect to achieve in extracting $\mu_s$.
It is then interesting to ask
whether it is possible to make some kind of calibration measurement
in order to reduce the
theoretical uncertainties of these effects. In the case of SAMPLE, an
ideal calibration experiment is quasielastic scattering from deuterium
at the same kinematics. As in the case of the Mainz experiment
on $^9$Be and as discussed in Sect.~IV.F, the magnetic strangeness effects in
deuterium are an incoherent sum of
the neutron and the proton and nearly cancel. The axial--vector term
contributes
approximately the same fraction to the asymmetry as in the case of the
proton. Both the asymmetry and the cross section are about 1.5 times
larger than in the proton, which means that only 1/3 as much running time is
needed to achieve the same level of accuracy. If a deviation from the
Standard Model asymmetry is seen in the proton, a measurement on deuterium
would provide an important check on the results.
A complication arising in the case of deuterium is the
fact that one will detect not only electrons from quasielastic scattering,
but also from elastic and threshold inelastic processes. These processes
each contribute about 10\% to the scattering rate. The elastic rate is
particularly troublesome because it is highly sensitive to strange--quark
effects. Fortunately, the elastic and inelastic asymmetries contribute
with opposite sign and roughly cancel each other, leading to an overall
correction factor of $\sim 2\pm 1\%$ at 200 MeV.

\goodbreak
\bigskip
\leftline{V.C.2.\quad MAINZ}

A proposal has recently been submitted to the Mainz Microtron for a
new PV measurement at forward electron angles [Hei93] whose
goal is to measure the contribution of the
strange--quark electric form factor $G_E^{(s)}$ to the asymmetry for elastic
scattering from the proton.
The proposed kinematics would be a beam energy of 855 MeV and an average
scattering angle of
35$^\circ$, corresponding to $\vert Q^2 \vert$ = 0.23 (GeV/c)$^2$.
At these kinematics the PV asymmetry is sensitive to
$(G_E^{(s)} + G_E^n + 0.2 G_M^{(s)})$, and with no strange quarks is
$\sim 7\times 10^{-6}$. Preliminary plans are to build a
calorimeter of six layers of liquid Xenon modules
$5.4 \times 5.4 \times 39$ cm, covering about 2 sr of solid angle.
With the assumed 35 $\mu$A polarized beam current and a 10 cm LH$_2$
target, the data--rate into each detector module would be approximately
2 MHz. The choice of liquid Xenon is driven by the desire for a
fast detector which could count individually scattered electrons at
this rate. On average, a scattered electron would generate an
electromagnetic shower covering an area of $3 \times 3$ modules.
Individual detectors would be
triggered at a signal size corresponding to one--third of the total
elastically scattered electron energy, and signals over the
$3 \times 3$ cluster would be summed together. This would make it
possible to separate elastic scattering from events below the
pion production threshold. A 20 cm polyethylene absorber would be placed
in front of the Xenon modules to reduce background coming from the target.
The PV asymmetry would be measured with a statistical error
of $\sim$4\%, which could be done in 400 hours assuming a beam
polarization of 80\%. To achieve this level of systematic accuracy,
the beam polarization must be measured to 3\% or better. This level
of precision has not currently been achieved and will require
additional development.

\bigskip
\goodbreak
\noindent{V.C.3.\quad CEBAF}

At higher beam energies and forward angles one has the possibility of
obtaining information on the first of the three terms in Eq.~(\alrprotpn).
An explicit extraction of $G_E^{(s)}$ is complicated both by
the neutron electric form factor and by $G_M^{(s)}$, as discussed in
Sect.~IV.
In addition, the uncertainty in the neutron charge form factor
enters directly into the uncertainty in $G_E^{(s)}$. This
may not be such a serious limitation, however, since several experiments
are currently proposed or underway [Mil88, Day89, Mad89, McK91, Ott90,
Pla90, Lun93]
to improve the
limits on $G_E^n$ to $\sim \pm 10\%$. Two proposals for PV ${\vec e}p$
measurements have already been submitted to CEBAF and are conditionally
approved.

\bigskip
\bigskip
\noindent \undertext{ Hall A experiment 91--010}
\bigskip

The first CEBAF proposal [Sou91] is a measurement
at $|Q^2|=0.5$ (GeV/c)$^2$ using the two high--resolution spectrometers in
Hall A.  The pair of spectrometers would be positioned symmetrically on
either side of the beamline at their most
forward scattering angle of 12.5$^\circ$,  yielding a total acceptance
of 16 msr. At these kinematics, the beam energy is 3.5 GeV, and
the calculated asymmetry with no strange--quark effects
is $-2\times 10^{-5}$. With 100 $\mu$A of 50\%
polarized electrons and a 15 cm long LH$_2$ target, a statistical
accuracy of 5\% can be achieved in about 400 hours.
The elastic proton cross section at these kinematics is approximately
700 nb/sr, leading to a total rate into each spectrometer of
$\sim$ 2 MHz. In order to handle these high rates,
the standard detector package of the Hall A spectrometers will be replaced
with a lead glass shower counter and integrating electronics. The
focusing properties of the spectrometer are such that it is possible
to separate elastic scattering events kinematically from other possible
backgrounds by using a reduced area of the spectrometer focal plane.

A single measurement alone would not allow
separation of $G_E^{(s)}$ and $G_M^{(s)}$, but it would determine
if strange quarks play a significant role in the structure of the proton,
modulo the uncertainty in $G_E^n$. If large effects are seen,
a more extended program of measurements at lower--$|Q^2|$
on $^1$H and $^4$He targets is anticipated. A measurement on ${}^4$He
has the advantage of being sensitive only to the electric contribution,
but at low momentum transfer
it represents a significantly greater challenge due to the smaller
asymmetries and lower sensitivity to strange quark effects.

\bigskip
\bigskip
\noindent \undertext{ Hall C experiment 91--017}
\bigskip

A second proposal [Bec91] advocates the construction of a dedicated apparatus
with an open geometry and a large solid angle, which could be used
for both a forward--angle experiment as well as
backward--angle measurements at slightly lower
momentum transfer.
This apparatus, to be placed in Hall C at CEBAF,
would consist of a segmented iron--free toroidal magnet and
an array of scintillators placed along the focal planes of the magnet.
A layout of the apparatus is shown in Fig.~5.14.
The goal of this experiment is to perform several sets of measurements
which would allow a complete separation of the three terms in
Eq.~(\alrprotpn).
At forward electron kinematics, the detector would be sensitive to recoil
protons ($60^\circ<\theta_p <77^\circ$), corresponding to
momentum transfers of $0.1<|Q^2|<0.5$ \gevocsq\ and a Standard Model
asymmetry of $\sim -5\times 10^{-6}$. The entire range of momentum transfer
would be covered in a single measurement, with a solid angle
acceptance of 0.87 sr and a target length acceptance of 20 cm.
The proposed measurements would use 40 $\mu$A of polarized beam,
corresponding to very high counting rates ($\sim$ 10 MHz)
in the detector. The recoil protons would be counted with scalers. Additional
time--of--flight information
would be provided by modifying the CEBAF beam microstructure to
be 32 nsec between beam bunches with increased peak current (the
accelerator design
value is 2 nsec between bunches). This measurement is sensitive to the
combination $G_E^{(s)} + G_E^n + 0.2G_M^{(s)}$, and a 5\% measurement of the
asymmetry could be achieved in approximately 700 hours.

A second set of measurements would involve reversing the orientation
of the spectrometer to detect electrons in the backward direction,
$\theta_e \sim 108^\circ$.
Four measurements, 700 hours each, are proposed. The
beam energies in this configuration
would be 335, 428, 512, and 590 MeV corresponding to momentum transfers of
0.2, 0.3, 0.4, and 0.5 (GeV/c)$^2$.
This set of measurements would be sensitive to a different linear
combination of the strange form factors.
Finally, a 500 hour QE measurement on deuterium would provide some
constraints on the uncertainties associated with the term proportional to
the axial--vector form factor, as discussed in Sect.~IV.F. Combining the
proton detection, electron detection and deuterium
quasielastic measurements would allow the
separation of $G_E^{(s)}$
and $G_M^{(s)}$ over a relatively large range of momentum transfer.
Although a model--independent extraction
of $G_E^{(s)}$ would still be limited by the
uncertainty in $G_E^n$, the SU(3) singlet combination $G_E^{(0)}$
would be determined with good accuracy.

\bigskip
\bigskip
\noindent \undertext{ Hall A experiment 91--004}
\bigskip

As discussed in Sect.~IV.B, another approach to learning something about
strange--quark effects in a
hadronic system is to use PV elastic scattering from
a $J=0$, $T=0$ nucleus. Only the ground--state vector
monopole matrix elements of
the electromagnetic and weak neutral currents occur in this case
and the asymmetry is given by Eq.~(\Ehefour). As
discussed
in Sect.~IV.B, in a light nucleus such as $^4$He, isospin violations
are expected to be quite small and accordingly this case may be the best
to use in looking for effects due
to the presence of strange quarks. Musolf and Donnelly [Mus92a] have argued
that two measurements on a ($0^+0$) nucleus, at low-- and
moderate--$\vert Q^2 \vert$,
might ultimately constrain $G_E^{(s)}$ to a higher level of precision than
that achievable with a measurement on a proton target, since uncertainties
associated with $G_M^{(s)}$, $G_E^n$ and $\tilde G_A^p$ are no longer present.

This third CEBAF proposal [Bei91b] is for a measurement of
$^4$He$(e,e)$ at
$\vert Q^2 \vert =0.6$ (GeV/c)$^2$, again using the Hall A spectrometers.
The target will be $^4$He gas operating at 10K and 70 atm, 15 cm long.
The standard detector package of the spectrometers need not be modified
in this case, since at the highest CEBAF
design luminosity (${\cal L}\sim 3 \times 10^{38}$ cm$^{-2}$~s$^{-1}$) the
counting rate into
each spectrometer is $\sim 2$ kHz. The asymmetry with no strange quarks
is $5\times 10^{-5}$, large on the scale of PV experiments,
thus reducing the demand on the helicity--correlated properties
of the CEBAF beam. This allows for removal of target wall background
since elastically scattered electrons are uniquely identified with
the tracking capabilities of the detector package. Small trigger scintillators
will be installed in the detector package to limit the focal--plane
acceptance of the spectrometers to the region of the elastic peak.

Because of the lower counting rates a good
statistical measure of the asymmetry will take somewhat more running time
than the previous two experiments. However, even a 30\% measure
of the asymmetry may be able to constrain significantly the current
limits of the strange electric contribution to hadronic matter.
This level of precision could be reached with 1000 hours of CEBAF beam,
assuming 40\% beam polarization.

\bigskip
\goodbreak
\leftline{V.C.4.\quad SLAC}

Two proposals were recently developed for measuring PV asymmetries
at very high energies at SLAC.  One proposal suggests the measurement of an
asymmetry in elastic and inelastic electron scattering from the proton at both
high-- and low--$|Q^2|$ (2.0 and 0.5 \gevocsq) [Che92].
The other proposes to measure the deep--inelastic
scattering (DIS) asymmetry on both hydrogen and deuterium at high--$|Q^2|$
(2--9 \gevocsq) [Bos92].
Both experiments would take advantage of high beam energies,
significant $|Q^2|$ and very forward scattering angles with large cross
sections
and asymmetries, resulting in good statistical accuracy.
The DIS experiment
is perhaps the most ambitious of all proposed electron scattering
PV measurements, in terms of the expected accuracy. Both experiments are
an example of future directions in PV electron scattering. Rather than
presenting a discussion of each proposal in detail, a general discussion
is given of both the possibilities and the constraints they might present.

%

\goodbreak
\bigskip
\noindent \undertext{ Proton Elastic and Inelastic Scattering}
\medskip
\nobreak

At beam energies in the 30--50 GeV region it would be possible through
$e$--$p$ elastic scattering to
explore the high--$|Q^2|$ behavior of the strange form factors.
Because of the increasing value of $\tau$, the PV asymmetry becomes
more sensitive to $G_M^{(s)}(Q^2)$ than to $G_E^{(s)}(Q^2)$. Unlike the
lower--energy experiments, possible contamination due to the poorly
known axial--vector form factor and its radiative correction are
negligible (see Sect.~IV.A).

Data could also be acquired simultaneously for excitation of the inelastic
nucleon resonances.  Of great interest is a measurement of the electron
axial--vector/quark vector current isovector coupling for
the $N\rightarrow \Delta$~transition discussed in Sect.~IV.G.  If
the non--resonant background can be neglected, the isovector couplings can be
isolated.  The excitation is also expected to be sensitive to higher--order
electroweak processes and possible contributions from strange quarks.

The biggest difficulty that higher--energy experiments face is the
rapidly falling $e$--$p$ cross section. Although the asymmetries are large,
even modest levels of precision require long periods of running with
a conventional magnetic spectrometer. Nonetheless, these experiments
would be exploring completely unknown territory for which there are
currently few theoretical predictions.

\goodbreak
\bigskip
\noindent \undertext{Deep--Inelastic Scattering}
\medskip
\nobreak

The recently submitted letter of intent [Bos93] for deep--inelastic
scattering is perhaps the most ambitious of all proposed PV experiments
in terms of expected accuracy.
The primary objective of this experiment would be to provide a
precision test of the Standard Model and to determine the
flavor composition of the sea quarks.
The proposed measurement would improve upon the results of the original SLAC
PV experiment by almost an order--of--magnitude.  At the level of
1.5\% uncertainty it would be
comparable to many of the current Standard Model tests.
A planned 1\% atomic PV experiment would achieve approximately
the same uncertainty in $\snw$.  However, the SLAC
and atomic PV experiments are
complementary, since the atomic PV experiment measures $C_{1u} +
C_{1d}$ and
the DIS experiment measures $2C_{1u} - C_{1d}$.
Combining the results would determine $C_{1u}$ and $C_{1d}$
independently to $\pm 0.005$ (see also Sects.~IV.H and IV.I).

It would also be possible to measure the quark
distribution functions.  The overall $x$--dependence of the quark
sea is reasonably well understood, but the flavor composition is not.
At present, the available data from muon,
neutrino, and antineutrino scattering, as well as from the Drell--Yan process,
are insufficient to disentangle the five unknown
quantities $u_{\rm val}$, $d_{\rm val}$, $u_{\rm sea}$, $d_{\rm sea}$ and
$s_{\rm sea}$.  An experiment at SLAC would provide
results for two new observables. By taking
ratios of asymmetries from hydrogen and deuterium targets,
light sea--quark distribution functions would be much more unambiguously
determined.

At beam energies of 30--50 GeV and very forward angles, the PV asymmetry
could be as large as $10^{-3}$. Techniques similar to those used in
previous experiments to control helicity--correlated beam properties
could be used. Count rates would also be high due to the relatively
large deep--inelastic cross sections, so modest beam currents and
target lengths would be sufficient. For the hydrogen/deuterium ratio
measurements, accurate knowledge of the relative target densities would
be required. Absolute densities on the order of 1\%, as well as
1\% knowledge of the relative beam--related density changes, are feasible.
As was demonstrated in the previous SLAC PV experiment, the use of
magnetic spectrometers and segmented shower counters makes it possible
to maintain high count rates while keeping background to a minimum.

The primary technical challenge associated with a precision PV experiment
would be knowledge of the beam polarization at the level of 1\%.
The standard technique used by most experiments is
M{\o}ller scattering from polarized atomic electrons in thin foils.  Analyzing
power is a maximum for scattering at 90$^{\circ}$ in the center--of--mass.
There are several effects which limit the accuracy of this method:
uncertainty in the effective foil polarization, background asymmetries under
the M{\o}ller peak and atomic effects connected with our knowledge of the
atomic
electron polarization.  Significant efforts will be required to understand the
various contributions and reduce the overall systematic errors to the level of
the statistical uncertainties in the experiment itself.  An alternative
approach for determining beam polarization would be to use Compton
scattering which
involves the backscattering of circularly--polarized laser photons from the
beam
of polarized electrons.  The technique is complicated and expensive, but can
probably achieve a systematic error of $< 1\%$ in measuring the beam
polarization.

The two proposals submitted to SLAC are examples of the future directions
one might pursue at a higher--energy electron machine such as that currently
under consideration by the European community [Pro91, EEF93].

\bigskip
\goodbreak
\noindent{V.C.5. SOLENOIDAL SPECTROMETER }

Recent PV experiments have demonstrated that the present
limitation for such measurements is counting statistics.  The next generation
of such studies, with its very ambitious physics goals, will require the
development of large acceptance devices with appropriate momentum resolution.
A novel approach is to use a large solenoid spectrometer: at energies up to
1~GeV this requires a device of up to 4~m diameter, 4~m long and a central
field
of $\sim$2 T.
A recent analysis by Souder {\it et al.\/} [Sou90b]
has shown that there are many
advantages with such an approach.  The device is suitable for both forward--
and
backward--angle measurements.  In the forward direction, for an experiment such
as elastic scattering on ${}^{12}$C, acceptances of up to 600~msr are possible
with adequate resolution to exclude inelastic scattering.  This is more than
an order--of--magnitude larger acceptance than has typically been
available up to now.  At
backward angles, solid angles in excess of 2~sr are possible.  In both cases
the spectrometer can accommodate extended targets such as LH$_2$ and  LD$_2$.
An important advantage of a solenoid is that it makes use of all of the
azimuthal acceptance possible within a given angular range.  It also allows for
the ability to reverse the direction of the magnetic field without any change
in the beam optics.  This could potentially be very important for
understanding systematic effects arising from M{\o}ller scattering backgrounds.

\bigskip
\goodbreak
\leftline {V.C.6.\quad LSND}

The LSND collaboration at LAMPF is currently undertaking a new elastic
neutrino scattering experiment using the A-6 beamline.
The high current (800 $\mu$A) 780 MeV proton beam is
stopped in a water--cooled Cu beam dump, from which
charged pions below 600 MeV are produced. Roughly equal numbers of
$\nu_e$, $\nu_\mu$ and ${\bar\nu}_\mu$ are produced from pion decay
mechanisms at rest. The muon neutrinos result from the process
$\pi^{+}\rightarrow\mu^{+}\nu_\mu$ and are monoenergetic at 27.8 MeV.
The other two species result from
$\mu^{+}\rightarrow e^{+}\nu_e{\bar\nu}_\mu$ and have a distribution
up to a maximum energy of 52.8 MeV. A small percentage of pions
will decay in flight and produce neutrinos with energies in the range of
100--200 MeV. The ratio of ${\bar\nu}_\mu$ from $\pi^{-}$ decay--in--flight
to $\nu_\mu$ from $\pi^{+}$ decay is $\sim 0.2$, and thus the majority of
the higher--energy beam is $\nu_\mu$. An additional decay path at the
beginning of the beamline will enhance the yield of decay--in--flight
neutrinos. The primary goal of the experiment is a continuing study
of neutrino oscillations, for which the decay--at--rest beams are used.

The experiment once again involves an active detector: a cylindrical
tank filled with 200 tons of liquid scintillator, 8.5 m long by 5 m diameter.
High--quality eight inch diameter photomultiplier tubes cover
approximately 25\% of the surface area of the detector. The detector
is surrounded
by a veto shield [Nap89] constructed of planes of Pb and liquid
scintillator to veto events coming from cosmic ray muons and neutrons.
The entire detector is covered with $\sim 2000$ g/cm$^2$ of steel and
earth to provide additional cosmic ray shielding. A water tank
downstream of the detector shields the open end of the detector.

Neutrinos in the 150 MeV range elastically scatter from
the protons in the scintillator (again, 80\% of which are bound in
carbon) with an average momentum transfer of $|Q^2| = 0.07$ \gevocsq.
The recoil protons ($20<E_p<40$ MeV) deposit energy in the scintillator
and are detected. Cosmic ray background events are rejected with the
veto shield and with the good timing resolution of the photomultiplier tubes.
Vertex reconstruction and a 50 cm fiducial cut on the central volume
of the detector provides additional background rejection together with energy
and angle resolution. It is expected that the $\nu_\mu$ elastic
scattering from free protons can be determined to $\sim 20\%$ and the
$\nu_\mu$ cross section from free and bound protons can be measured to
$\sim 10\%$. Because of the very low momentum transfer, the contribution
from $G_E^{(s)}$ is virtually negligible. The remaining ambiguities are
from the strange magnetic moment (which also should be reduced at
lower--$|Q^2|$), and from the nuclear physics uncertainties associated with
quasielastic scattering from $^{12}$C. Using the results from the SAMPLE
experiment discussed above will
resolve the former issue. To address the latter problem the collaboration
has proposed two possible solutions. The first is to use two different
types of liquid scintillator and perform a subtraction to isolate
quasielastic scattering from true elastic neutrino--proton events. A second
proposal is
to make use of the fact that in addition to recoil protons, the LSND
detector will be sensitive to recoiling neutrons. It has been suggested
[Gar92] that the ratio
of quasielastic $(\nu, p)/(\nu ,n)$ from the $^{12}$C in the detector
may provide additional sensitivity to $G_A^{(s)}$ (see also Sect.~IV.J).
Performing such a ratio
measurement should reduce some of the systematic uncertainties associated
with nuclear binding. This technique is also
less sensitive to the uncertainties in the neutrino flux than
($\nu ,p$) elastic cross section measurements.
The nuclear physics uncertainties must still be understood, and are
the subject of future work by the same authors.
The LSND experiment is expected to start production running late in
1993. It is anticipated that in several years of running $G_A^{(s)}(0)$
will be determined to about a factor of two better than was determined
from the BNL experiment.

\vfil\eject

%

\secnum=6
\neweq


\noindent{\bf VI. \quad SUMMARY AND CONCLUSIONS}

        As a result of the foregoing discussion, we hope to have convinced
the reader that intermediate--energy, semileptonic neutral current scattering
offers a promising means for studying a wide array of physics issues. At
the same time, we have endeavored to illustrate the necessity for a
coordinated program of experimental and theoretical work in this field.
Indeed, it should be apparent that a combination of several different
measurements, together with careful and detailed theoretical analysis, is
necessary if one wishes to extract useful new information on physics beyond
the Standard Model, hadronic structure or nuclear dynamics. To this end,
several challenges remain. On the experimental side, one must push the
limits of luminosity ({\it viz.}, beam current) and electron polarization
or neutrino flux in order to
achieve statistical uncertainties on the order of one percent.
Moreover, systematic errors, such as false asymmetries
arising from electron helicity--correlated differences, must be held to the
same level. These considerations impose rather tight restrictions on the
choice of target and kinematics if one is to make meaningful measurements.
{}From a theoretical perspective, one would like a better grasp
of nuclear dispersion corrections, nuclear medium effects in elastic and
quasielastic scattering, and hadronic contributions to electroweak radiative
corrections.

        With these general remarks in mind, we summarize the salient features
of the present and future semileptonic, weak neutral current scattering
program as we view it:

\noindent (1) For the immediate future, intermediate--energy neutrino and
PV electron scattering seem best suited as tools for studying strange--quark
vector and axial--vector form factors of the nucleon. Higher--energy processes,
such as deep inelastic scattering, and lower--energy experiments, such as
atomic PV measurements, cannot access these form factors. In this respect,
intermediate--energy semileptonic scattering is unique.

\noindent (2) Atomic PV experiments are presently more suited to the study of
\lq\lq non--standard" electroweak physics, provided that theoretical nuclear
structure uncertainties can be reduced. To this end, a measurement of the
ground--state neutron distribution in cesium or lead, through a combination
of PV and PC electron scattering experiments, would provide a useful
calibration point for theoretical calculations. Looking somewhat further down
the road, once the nucleon's strange--quark form factors are sufficiently
constrained, PV electron scattering could provide for a set of low-- and
intermediate--energy electroweak tests complementary to those performed with
atomic PV.

\noindent (3) For either type of study (hadron structure or electroweak
physics), neutral current scattering from the proton would not be sufficient,
due to the presence of several unknown or poorly constrained form factors in
the PV asymmetry and neutrino cross section. A program which includes a
combination of scattering from proton and nuclear targets therefore appears
warranted. The components of such a program could include the following:

\item{(a)} A combination of backward--angle $\alr(\evec p)$ measurements,
as in the present SAMPLE and conditionally approved \lq\lq $G^0$" experiments,
with a backward--angle $\alr({\rm QE})$ measurement, could allow one to
constrain separately $\mustr$ and the large and theoretically uncertain
radiative correction, $\rateo(\evec p)$. Alternatively, a series of
backward--angle $\alr(\evec p)$ and elastic $\alr(^2\hbox{H})$ measurements
would accomplish the same purpose. In either case, the resultant limits on
$\mustr$ could be sufficiently tight to test some of the model predictions
discussed in Sect.~II. In the case of quasielastic scattering, a deuterium
target appears to be the most favorable, since one stands the greatest
chance of performing reliable nuclear calculations with this nucleus.

\item{(b)} A combination of low-- and moderate--$|Q^2|$ elastic $\alr(0^+0)$
measurements has the potential to determine $\GES$ at a level sensitive
to most of the model predictions appearing in Table 2.3. Moreover, the
resultant constraints could reduce the $\GES$ uncertainty to a level needed
to make an interesting electroweak test possible with a subsequent
$\alr(0^+0)$ measurement. The $\GES$ constraints obtained from a series of
forward-- and backward--angle $\alr(\evec p)$ alone would be much less
stringent. Nevertheless, measurements made with a proton target would be
an important check on $\alr(0^+0)$ results.

\item{(c)} For purposes of constraining $\GAS$, low--$|Q^2|$ neutrino--nucleon
elastic scattering, as in the LSND experiment, seems better suited than
does PV electron scattering, owing to the large theoretical uncertainties
appearing in the radiative correction $\ratez(\evec p)$ and
the possible large
degree of SU(3)--breaking in the octet axial--vector current matrix elements.
The degree of theoretical uncertainty is much smaller in the case of
$\ratez(\nu p)$. Moreover, since
$\vert\ratez(\nu p)\vert <\!\!< \vert\ratez(\evec p)\vert$, neutrino
scattering determinations of $\GAS(0)$ are much less sensitive to axial--vector
SU(3) breaking. In addition, a determination of the ratio of $\nu p$ and
$\nu n$ yields in QE neutrino scattering could provide for enhanced
sensitivity to $\GAS$ while reducing the experimental and theoretical
systematic uncertainties, and continued studies of selected nuclear
transitions could yield, for example, high--precision determinations of
axial--vector nucleonic couplings.

\item{(d)} Once the strange form factors are constrained with the foregoing
program of measurements, one could consider performing Standard Model tests
with PV electron scattering. In this respect, the most promising targets are
the proton and $(0^+0)$ nuclei. Elastic $\alr$ measurements on these
targets could complement atomic PV measurements as low--energy probes of
physics beyond the Standard Model. An additional possibility is a measurement
of the $N\to\Delta$ PV asymmetry, although for such a measurement to be
competitive, better data on the axial--vector and background contributions
to this asymmetry would be needed.

\noindent (4) Going beyond strangeness form factors and electroweak tests,
one encounters a host of other possibilities for studying hadron--structure
and nuclear dynamics. These possibilities include the following:

\item{(a)} A determination of the axial--vector $N\to\Delta$ matrix element,
which at present is only loosely constrained by neutrino scattering data.
Obtaining an improved determination with PV electron scattering would require
better knowledge of non--resonant background contributions than is presently
available.

\item{(b)} Obtaining a new set of constraints on the conventional picture of
nuclear PV appears within the realm of possibility. Based on some
initial studies, it seems that inelastic $\alr$ measurements at low--energies
($\epsilon\lapp 100$ MeV) would be most ideal. Moreover, given the relatively
broad constraints which presently exist on the
PV meson--nucleon couplings, one can live
with nearly an order--of--magnitude less precision to make meaningful
measurements than is required for strangeness and electroweak studies. In
this regard, PV electron scattering could complement atomic PV determinations
of the nuclear anapole moment as a probe of nuclear PV.

\item{(c)} A determination of the ground--state neutron distribution. A
relatively short ($\sim$ few hundred hours) PV electron scattering experiment
on a heavy nucleus such as lead or cesium could determine $\rho_n(\vec r)$ at
a level significantly beyond the typical present degree of knowledge. Such an
improved determination would provide an important calibration point for
theoretical calculations of $\rho_n(\vec r)$ needed in the analysis of
Standard Model tests using atomic PV measurements with a series of isotopes.

        In comparison with the situation of just a few years ago, it seems
that considerable progress has been made in achieving a comprehensive
understanding of the issues germane to intermediate--energy semi--leptonic
scattering --- especially with regard to the interplay between different types
of physics issues which enter and
the complementarity of different prospective
measurements. Whereas once the future direction of the field appeared rather
murky, one may now envision a rather well--defined program of experiments
and theoretical work. If successfully carried out, such a program could
provide a rich payoff of new insight into the structure of hadrons,
physics beyond the Standard Model and nuclear dynamics.

\vfill
\eject

{\bf ACKNOWLEDGEMENTS}

     It is a pleasure to thank R.~Burman, C.~Williamson, J.~Napolitano,
G.~Dodson, D.~Beck, C.~Wieman, S.~Lamoreaux, P.~Bosted, R.~Lourie, and
A.~Molinari for useful discussions.

\vfil\eject

\centerline{\bf REFERENCES}
\medskip

\item{Abe87} K. Abe {\it et al.}, {\it Phys. Rev. Lett. \bf 58} (1987) 636.

\item{Abe89} K. Abe {\it et al.}, {\it Phys. Rev. Lett. \bf 62} (1989) 1709.

\item{Abe91} F. Abe {\it et al.}, \PRD{43} (1991) 2070.

\item{Abr86} H.~Abramowicz {\it et al.}, \PRL{57} (1986) 298.

\item{Ade85} E.~G. Adelberger and W.~C. Haxton, {\it Ann. Rev. Nucl. Part.
        Sci. \bf 35} (1985) 501. 

\item{Adl68} S.~L. Adler, {\it Ann. Phys. \bf 50} (1968) 189. 

\item{Ahr85} L.~A.~Ahrens {\it et al.\/} {\it Phys.~Rev.~Lett.\/} {\bf 54},
        18 (1985).

\item{Ahr87} L.~A.~Ahrens {\it et al.}, {\it Phys. Rev. \bf D35} (1987) 785.

\item{Ala91} MIT/Bates Experiment 91-09, R. Alarcon and J. van den Brand,
        spokespersons.

\item{Alb88} W. M. Alberico, A. Molinari, T. W. Donnelly, E. L. Kronenberg
        and J. W. Van Orden, {\it Phys. Rev.\/} {\bf C38} (1988) 1801.

\item{Alb90} W.~M.~Alberico, T.~W.~Donnelly, and A.~Molinari, \NPA{512}
(1990) 541.

\item{Alb93a}  W.~M.~Alberico {\it et al.}, {\it Nucl.~Phys.~\bf A} (1993)
to be published.

\item{Alb93b}  W.~M.~Alberico, M.~B.~Barbaro, A.~De Pace, and M.~Molinari,
        {\it Phys.~Lett.~\bf B303} (1993) 5.

\item{Ali92}  J.~Alitti {\it et al.}, {\it Phys.~Lett.~\bf B276} (1992)
        354.

\item{All87} J.~V. Allaby {\it et al.}, {\it Zeit.~f\"ur Phys.
        \bf C36} (1987) 611.

\item{All89} R.~C.~Allen, H.~H.~Chen, M.~E.~Potter, R.~L.~Burman,
        J.~B.~Donahue, D.~A.~Krakauer, R.~L.~Talaga, E.~S.~Smith
        and A.~C.~Dodd,
        {\it Nucl.~Inst.~Meth.\/} {\bf A284} (1989) 347.

\item{All90}R. C. Allen {\it et al.}, {\it Phys. Rev. Lett. \bf 64} (1990)
        1330.

\item{Alt91} G. Altarelli, {\it Helv. Phys. Acta \bf 64} (1991) 761.

\item{Ama87} U. Amaldi, \etal , \PRD{36} (1987) 1385.

\item{Ama91} P. Amandruz {\it et al}, \PRL{66} (1991) 2712.

\item{Aok82} K. Aoki, Z. Hioki, R. Kawabe, M. Konuma, and T. Muta,
        {\it Suppl. Prog. Theo. Phys. \bf 73} (1982) 1. 

\item{Are82} H.~Arenh\"ovel, \NPA{384} (1982) 287.

\item{Are88} H. Arenh\"ovel, W. Leidemann and E. L. Tomusiak, \ZPA{331} (1988)
        123.

\item{Ash89} J. Ashman {\it et al.}, {\it Nucl. Phys. \bf B328} (1989) 1. 

\item{Auf85} S.~Auffret {\it et al.}, {\it Phys.~Rev.~Lett.~\bf 55} (1985)
        1362.

\item{Bec89} D.~H.~Beck, {\it Phys.~Rev.~\bf D39} (1989) 3248.

\item{Bec91} CEBAF proposal PR-91-017, D.H. Beck, spokesperson. 

\item{Bei91a} E.~J. Beise and R.~D. McKeown, {\it Comments in Nucl. Part. Phys.
          \bf 20} (1991) 105. 

\item{Bei91b}CEBAF proposal PR-91-004, E.~J. Beise, spokesperson. 

\item{Bei92} E. J. Beise, in {\it Proceedings of the International Conference
        on the Structure of Baryons and Related Mesons}, M.~Gai, Ed., World
        Scientific (1992) 392.

\item{Ber71} M.~Bernheim, {\it Phys.~Rev.~Lett.~\bf 46} (1971) 402.

\item{Ber79} J.~Bernabeu and P.~Pascual, \NPA{324} (1979) 365.

\item{Ber90} V.~Bernard, N.~Kaiser, and U.-G.~Meissner,
        {\it Phys.~Lett.\/} {\bf B237}, 545 (1990).

\item{Ber91}  W.~Bertozzi, R.~W.~Lourie, and E.~J.~Moniz, in {\it Modern
        Topics in Electron Scattering}, B.~Frois and I.~Sick, Eds.,
        World Scientific (1991) 419.

\item{Bjo69} J.~D.~Bjorken and E.~A.~Paschos, {\it Phys. Rev. \bf 185} (1969)
	1975.

\item{Blo89} C.~Blondell {\it et al.}, CERN report CERN/EP 89-101 (1989).

\item{Blu90} S.~A.~Blundell, W.~R.~Johnson, and J.~Sapirstein, \PRL{65}
        (1990) 1411.

\item{Bod91} B.~Bodmann {\it et al.\/}, {\it Phys.~Lett.\/} {\bf B267},
        321 (1991).

\item{Bos92} SLAC proposal E149, P.~Bosted, contact person.

\item{Bos93} SLAC proposal E149, P.~Bosted, contact person. Note proposal
	status \lq\lq deferred" by EPAC, June 1993.

\item{Bou74} M.~A.~Bouchiat, C.~C.~Bouchiat,
        {\it Phys.~Lett.~\bf B48} (1974) 111;
        {\it J. Phys.~(Paris) \bf 35} (1974) 899;
        {\it J. Phys.~(Paris) \bf 36} (1975) 493.

\item{Bou83} M.~Bourquin {\it et al.\/}, \ZPC{21} (1983) 27. 

\item{Bou86} M.~A.~Bouchiat and L.~Pottier,
        {\it Science \bf 234} (1986) 1203.

\item{Bou91} C.~Bouchiat and C.~A. Piketty, {\it Zeit.~f\"ur Phys.
        \bf C49} (1991) 91.

\item{Bre67} G.~Breit, {\it Rev. Mod. Phys.\/} {\bf 39} (1967) 560.

\item{Bre90} V.~Breton {\it et al.},
        {\it Phys.~Rev.~Lett. \bf 66} (1991) 572.

\item{Bro88}  G.~E.~Brown, C.~B.~Dover, P.~B.~Siegel, and W.~Weise
        {\it Phys.~Rev.~Lett.\/} {\bf 60}, 2723 (1988).

\item{Bur89}  CEBAF Proposal, \lq\lq Electroproduction of the
        $P_{33}$(1232) Resonance", (1989) V.~Burkert and R.~Minehart,
        spokespersons.

\item{Bur90} R.~L.~Burman, M.~E.~Potter and E.~S.~Smith,
        {\it Nucl.~Inst.~Meth.\/} {\bf A291}, 621 (1990).


\item{Bur93} V.~Burkert, private communication.

\item{Cah78} R.~N.~Cahn and F.~J.~Gilman, {\it Phys. Rev. \bf D17} (1978)
        1313.

\item{Car80} L.~S.~Cardman {\it et al.}, {\it Phys.~Lett.~\bf B91} (1980) 203.

\item{Car92} L.~S.~Cardman, {\it Nucl.~Phys.\/} {\bf A546},
       317c (1992).

\item{Che71a} T.~P.~Cheng and R.~Dashen, {\it Phys. Rev. Lett. \bf 26}
        (1971) 594. 

\item{Che71b} M.~Chemtob and M.~Rho, \NPA{163} (1971) 1. 

\item{Che76} T.~P.~Cheng, {\it Phys. Rev. \bf D13} (1976) 2161. 

\item{Che84} T.~P.~Cheng and Ling-Fong Li, {\it Gauge Theory of Elementary
        Particle Physics}, Oxford U. Press (1984).

\item{Che92} SLAC proposal E148, J.~P.~Chen, contact person.

\item{Chu88} MIT/Bates Experiment \#88-10, T. Chupp and A. Bernstein,
spokespersons.

\item{Col78} J.~Collins, F.~Wilczek, and A.~Zee, {\it Phys. Rev. \bf D18}
        (1978) 242. 

\item{Com83} E.~D.~Commins and P.~H.~Bucksbaum, {\it Weak Interactions of
          Leptons and Quarks}, Cambridge U. Press (1983) Chapter 9 and
          references therein.

\item{Con89} M.~Consoli, W.~Hollik, and F.~Jegerlehner in {\it Z Physics
        at LEP}, G.~Altarelli, R.~Kleiss, and C.~Verzegnassi, CERN report
        \# 89-08, Geneva, Switzerland (1989) Vol. 1, p. 7. 

\item{Cot76} J.~Cot\'e, B.~Rouben, R.~de Tourreil, and D.~W.~L.~Sprung,
        \NPA{273} (1976) 269.

\item{Dai91}J. Dai, M. J. Savage, J. Liu, and R. Springer, {\it Phys.~Lett.~\bf
         B271} (1991) 403. 

\item{Day89} CEBAF proposal 89-018, D.~Day, contact person.

\item{Day93} D.~Day, private communication.

\item{deF66}T. de~Forest, Jr. and J. D. Walecka, {\it Adv.
        in Phys.\/} {\bf 15} (1966) 1. 

\item{deF83} T.~de Forest, Jr., \NPA{392} (1983) 232.

\item{Deg89} G.~Degrassi, A.~Sirlin, and W.~J.~Marciano, \PRD{39}
        (1989) 287. 

\item{Des80} B.~Desplanques, J.~F.~Donoghue, and B.~R.~Holstein, {\it Ann.
          Phys. (N.Y.) \bf 124} (1980) 449. 

\item{deT73} R. de~Tourreil and D. W. L. Sprung, {\it Nucl. Phys.\/}
        {\bf A201} (1973) 193.

\item{deT75} R. de~Tourreil, B. Rouben and D. W. L. Sprung,
        {\it Nucl. Phys.\/} {\bf A242} (1975) 445.

\item{Don68} T.~W.~Donnelly, J.~D.~Walecka, I.~Sick, and E.~B.~Hughes,
        {\it Phys.~Rev.~Lett.~\bf 21} (1968) 1196.

\item{Don70} T.~W.~Donnelly, {\it Phys.~Rev.~\bf C1} (1970) 833.

\item{Don73} T.~W.~Donnelly, {\it Phys.~Lett.~\bf 43B} (1973) 93.

\item{Don74} T.~W.~Donnelly, D.~Hitlin, M.~Schwartz, J.~D.~Walecka, and
        S.~J.~Wiesner {\it Phys.~Lett.~\bf 49B} (1974) 8.

\item{Don75} T.~W.~Donnelly and J.~D.~Walecka, {\it Ann. Rev. Nucl. Sci.\/}
         {\bf 25}, (1975) 329. 

\item{Don76} T.~W.~Donnelly and R.~D.~Peccei, {\it Phys.~Lett.~\bf 65B}
        (1976) 196.

\item{Don79a} T.~W.~Donnelly and R.~D.~Peccei, {\it Phys. Rep. \bf 50}
         (1979) 1. 

\item{Don79b} T.~W.~Donnelly and W.~C.~Haxton, {\it At. Data and Nucl.
        Data Tables} {\bf 23} (1979) 103

\item{Don80} T.~W.~Donnelly and W.~C.~Haxton, {\it At. Data and Nucl.
        Data Tables} {\bf 25} (1980) 1

\item{Don83} T.~W.~Donnelly in {\it Proceedings of the Los Alamos Neutrino
        Workshop} (1983).

\item{Don84} T. W. Donnelly and I. Sick, {\it Rev. Mod. Phys.} {\bf 56}
        (1984) 461.

\item{Don85} T. W. Donnelly, {\it Prog. in Part. and Nucl. Phys.} {\bf 13}
(1975) 183.

\item{Don86a} J. F. Donoghue and C. R. Nappi, {\it Phys. Lett. \bf B168}
        (1986) 105. 

\item{Don86b} T.~W.~Donnelly and A.~S.~Raskin, {\it Ann. Phys. \bf 169}
        (1986) 247.

\item{Don88} T.~W.~Donnelly, J.~Dubach, and I.~Sick, \PRC{37} (1988) 2320.

\item{Don89} T.~W.~Donnelly, J.~Dubach and I.~Sick, {\it Nucl. Phys.\/}
{\bf A503} (1989) 589. 

\item{Don92}T. W. Donnelly, M. J. Musolf, W. M. Alberico, M. B. Barbaro,
        A. De Pace and A. Molinari, \NPA{541} (1992) 525. 

\item{Dor89} J. Dorenbosch {\it et al.}, {\it Zeit.~f\"ur Phys.~\bf C41}
        (1989) 567. 

\item{Dre59} S.~D.~Drell and S.~Fubini, {\it Phys. Rev. \bf 113} (1959) 741.

\item{Dub75} J. Dubach, Ph.D. thesis, Stanford University (1975) unpublished.

\item{Dub76} J. Dubach, J. H. Koch, and T. W. Donnelly, \NPA{271} (1976)
     279.

\item{Dub80} J. Dubach, \NPA{340} (1980) 271.

\item{Dub90} D. Dubbers, W. Mampe, and J. D\"ohner, {\it Europhys.
Lett.} {\bf 11} (1990) 195. 

\item{Dum83} O. Dumbrajs \etal, {\it Nucl.~Phys.~\bf B216} (1983) 277.

\item{Dzu86} V.~A.~Dzuba, V.~V.~Flambaum, and I.~B.~Khriplovich,
        {\it Zeit.~f\"ur Phys.~\bf D1} (1986) 243.

\item{EEF93} \lq\lq The EEF Project: Report of the Mainz Workshop",
        J.~Arvieux, E.~de Sanctis, T.~Walcher, and P.~K.~A.~de
        Witt Huberts, Eds., {\it Nuovo Cim.} (1993) to be published.

\item{Ell90} J.~Ellis and G.~L.~Fogli, \PLB{249} (1990) 543. 

\item{Fei75} G.~Feinberg, \PRD{12} (1975) 3575. 

\item{Fet71} A. L. Fetter and J. D. Walecka, {\it Quantum Theory of
Many-Particle Systems}, McGraw-Hill, N.Y., (1971).

\item{Fin91} CEBAF proposal PR-91-010, J.~M.~Finn and P.~A.~Souder,
	spokespersons.

\item{Fla78} J. B. Flanz, {\it et al.}, \PRL{41} (1978) 1642.

\item{Fla79} J. B. Flanz, {\it et al.}, \PRL{43} (1979) 1922.

\item{Fla80} V. V. Flambaum and I. B. Khriplovich, {\it Zh. Eksp. Teor. Fiz.
        \bf 79} (1980) 1656 [{\it Sov. Phys. JETP \bf 52} (1980) 835. 

\item{Fla84}  V.~V.~Flambaum, I.~B.~Khriplovich, and O.~P.~Sushkov,
        {\it Phys. Lett. \bf 146B} (1984) 367. 

\item{For80} E.~N.~Fortson and L.~Wilets, {\it Adv. in Atomic
        and Molec. Phys. \bf 16} (1980) 319.

\item{For84} E.~N.~Fortson and L.~L.~Lewis,
        {\it Phys. Rep. \bf 113} (1984) 289.

\item{For90} E.~N.~Fortson, Y.~Pang, and L.~Wilets, \PRL{65} (1990) 2857. 

\item{Fre90} S. Freedman, {\it Comments in Nucl. Part. Phys.} {\bf 19}
         (1990) 209. 

\item{Fre92} T.~Frederico, E.~M.~Henley, S.~J.~Pollock, and S.~Ying,
        {\it Phys. Rev \bf C46} (1992) 347.

\item{Fri74} J. L. Friar and M. Rosen, {\it Ann.~Phys.~\bf 87} (1974) 289. 

\item{Fri84} J. L. Friar and S. Fallieros, \PRC{29} (1984) 1645.

\item{Fri85} J. L. Friar and W. C. Haxton, \PRC{31} (1985) 2027.

\item{Fro67} R.~F.~Frosch, J.~F.~McCarthy, R.~E.~Rand, and M.~R.~Yearian,
        {\it Phys.~Rev.~\bf 160} (1967) 874.

\item{Fru84} S.~Frullani and J.~Mougey, {\it Adv.~in Nucl.~Phys.~\bf 14}
        (1984).

\item{Fuk88} M.~Fukugita, Y.~Kohyama, and K.~Kubodera, {\it Phys.~Lett.~\bf
        B212} (1988) 139.

\item{Gai84} J.~M.~Gaillard and G.~Sauvage, {\it Ann. Rev. Nuc. Part. Sci.
        \bf 34} (1984) 351. 

\item{Gal71} S. Galster {\it et al.}, {\it Nucl. Phys. \bf B32} (1971) 221. 

\item{Gan72} D.~Ganichot, B.~Grossetete, and D.~B.~Isabelle, \NPA{178}
         (1972) 545.

\item{Gar74} E.~L.~Garwin, D.~T.~Pierce, and H.~C.~Siegmann,
         {\it Helv.~Phys.~Acta\/} {\bf 47} 393 (1974).

\item{Gar92} G.~T.~Garvey, S.~Krewald, E.~Kolbe, and K.~Langacke,
        {\it Phys.~Lett.\/} {\bf B289} (1992) 249.

\item{Gar93} G.~T.~Garvey, W.~C.~Louis, and D.~H.~White, Los Alamos
        preprint LA-UR-93-0037.

\item{Gas91} J. Gasser, H. Leutwyler, and M. E. Sainio, {\it Phys. Lett.
        \bf B253} (1991) 252. 

\item{Gei89} D. Geiregat {\it et al.}, {\it Phys. Lett. \bf B232}
        (1989) 539. 

\item{Gla70}S. L. Glashow, J. Iliopoulos and L. Maiani, \PRD{2} (1970) 1285. 

\item{Gol90} M. Golden and L. Randall, \NPB{361} (1991) 3. 

\item{Got67} K.~Gottfried, {\it Phys.~Rev.~Lett.~\bf 18} (1967) 1174.

\item{Gre69} G. K. Greenhut, {\it Phys. Rev. \bf 184} (1969) 1860. 

\item{Had92} E.~Hadjimichael, G.~I.~Poulis, and T.~W.~Donnelly, {\it
        Phys.~Rev.~\bf C45} (1992) 2666.

\item{Hax89} W.~C.~Haxton, E.~M.~Henley, and M.~J.~Musolf,
        {\it Phys. Rev. Lett.~\bf 63} (1989) 949. 

\item{Hax93} W. C. Haxton, E.M. Henley, and M.J. Musolf, to be published.

\item{Hei89} W. Heil {\it et al.},
          {\it Nucl. Phys. \bf B327} (1989) 1. 

\item{Hei93} Mainz proposal \# A4/1-93, E. Heinen-Konschak {\it et
	al.} collaborators, D. von Harrach, spokesperson.

\item{Hen73} E.~M.~Henley, A.~H.~Huffman, and D.~U.~L.~Yu,
        \PRD{7} (1973) 943. 

\item{Hen79} E.~M.~Henley, W.-Y.~P.~Hwang, and G.~N.~Epstein, \PLB{88}
        (1979) 349. 

\item{Hen82} E.~M.~Henley and W.-Y.~P.~Hwang, \PRC{26} (1982) 2376. 

\item{Hen91} E.~M.~Henley, G.~Krein, S.~J.~Pollock, and A.~G.~Williams,
        {\it Phys. Lett.~\bf B269} (1991) 31.

\item{Hoc73} J.~Hockert, D.~O.~Riska, M.~Gari, and A.~Huffman, \NPA{217}
	(1973) 14.

\item{Hoh76} G. H\"ohler {\it et al.}, \NPB{114} (1976) 505. 

\item{Hol89} B.~R.~Holstein, {\it Weak Interactions in Nuclei}, Princeton
          University Press, Princeton, N.~J., (1989) ch.~5.

\item{Hol90} B.~R.~Holstein in {\it Proceedings of the Caltech Workshop on
        Parity Violation in Electron Scattering}, E.~J.~Beise and
        R.~D.~McKeown, Eds., World Scientific (1990) 27. 

\item{Hor93a} C.~J.~Horowitz, {\it Phys.~Rev.~\bf C47} (1993) 826.

\item{Hor93b} C.~J.~Horowitz and J.~Piekarewicz, {\it Phys.~Rev.}, to
        be published (1993).

\item{Hun76} P.~Q.~Hung and J.~J.~Sakurai, {\it Phys. Lett.
        \bf B63} (1976) 295; see also Table 9.1 of Ref.~[Com83]. 

\item{Hwa81} W.-Y.~P.~Hwang, E.~M.~Henley, and G.~A.~Miller, {\it Ann. Phys.
        \bf 137} (1981) 378. 

\item{Hwa86} W.-Y.~P.~Hwang and T.~W.~Donnelly, {\it Phys.~Rev.~\bf C33}
        (1986) 1381.

\item{Jaf89} R. L. Jaffe, {\it Phys. Lett. \bf B229} (1989) 275. 

\item{Jan72} J.~A.~Jansen, R.~Th.~Peerdeman, and C.~De Vries, \NPA{188}
        (1972) 337.

\item{Jeg89} F.~Jegerlehner, PSI Preprint \# PR-89-23 (1989). 

\item{Jen91} E.~Jenkins and A.~V.~Manohar, {\it Phys. Lett. \bf B255} (1991)
        558. 

\item{Jon80} D.~R.~T.~Jones and S.~T.~Petcov, {\it Phys. Lett. \bf 91B}
        (1980) 137.

\item{Jon89} G.~T.~Jones \etal, {\it Zeit.~f\"ur Phys. \bf C43} (1989) 527.

\item{Jon91} C. E. Jones-Woodward \etal, \PRC{44} (1991) R571.

\item{Kal89}N. Kalantar-Nayestanaki \etal , \PRL{63} (1989) 2032. 

\item{Kap88} D.~B.~Kaplan and A.~Manohar, \NPB{310} (1988) 527. 

\item{Kap92} D.~B.~Kaplan and M.~J.~Savage, UCSD Prinprint \#PTH 92-04 (1992)

\item{KAR91} KARMEN, {\it Phys.~Lett.~\bf B267} (1991) 321.

\item{KAR92} KARMEN, {\it Phys.~Lett.~\bf B280} (1992) 198.

\item{Ken89} D.~C.~Kennedy, B.~W.~Lynn, C.~J.-C.~Im, and R.~G.~Stuart,
\NPB{321}
        (1989) 83. 

\item{Ken90} D.C. Kennedy and P. Langacker, \PRL{65} (1990) 2967;
	\PRL{66} (1991) 395; \PRD{44} (1991) 1591.

\item{Kim81} J.~E.~Kim, P.~Langacker, M.~Levine, and H.~H.~Williams,
        {\it Rev.~Mod.~Phys.\/} {\bf 53}, 211 (1981).

\item{Kin62} T. Kinoshita, {\it J. Math. Phys. \bf 3} (1962) 650. 

\item{Kis87} S.~Kistryn {\it et al.}, \PRL{58} (1987) 1616.

\item{Kit90} T. Kitagaki \etal, {\it Phys.~Rev.~ \bf D42} (1990) 1331,
     and refs therein.

\item{Koe92} W.~Koepf, E.~M.~Henley, and S.~J.~Pollock,
        \PLB{288} (1992) 11. 

\item{Kol92} E.~Kolbe, K.~Langacke, S.~Krewald, and F.-K.~Thielemann,
        \NPA{540} (1992) 599.

\item{Lam75} G.~Lampel and C.~Weisbuch, {\it Solid State Comm.\/}
        {\bf 16} (1975) 877.

\item{Lam93} S. Lamoreaux, private communication.

\item{Lan90}P. Langacker, U. Penn. pre-print \# UPR-0435T (1990) (PASCOS-90).

\item{Lan91} P. Langacker, \PLB{256} (1991) 277.

\item{Lee64} T.~D.~Lee and M.~Nauenberg, \PR{133} (1964) B1549.

\item{Li82} S.-P.~Li, E.~M.~Henley, and W.-Y.~P.~Hwang, {\it Ann. Phys. \bf
143}
     (1982) 372.

\item{Lle72} C.~H.~Llewellyn-Smith, {\it Phys.~Rep.~\bf 3} (1972) 261.

\item{Lou89}LSND collaboration, LAMPF proposal \#1173, Los Alamos National
        Laboratory report LA-UR-89-3764 (1989), W.C. Louis, contact person.

\item{Lou92a} SLAC proposal E148 (1992) R. Lourie, spokesperson.

\item{Lou92b} R. Lourie, private communication.

\item{Lu90} X.~Q.~Lu {\it et al.\/} Los Alamos Report LA-11842-P, Los Alamos
        National Laboratory (1990).

\item{Lun93} A. Lung {\it et al.}, \PRL{70} (1993) 718. 

\item{Mad89} CEBAF proposal 89-005, R.~Madey, contact person.

\item{Mad92} R. Madey {\it et al.}, private communication.

\item{Mar80}W. J. Marciano and A. Sirlin, \PRD{22} (1980) 2695. 

\item{Mar81}W. J. Marciano and A. Sirlin, \PRL{46} (1981) 163. 

\item{Mar83}W. J. Marciano and A. Sirlin, \PRD{27} (1983) 552. 

\item{Mar84} W.~J.~Marciano and A.~Sirlin, \PRD{29} (1984) 75.

\item{Mar90} W.~J.~Marciano and J.~L.~Rosner,
        {\it Phys. Rev. Lett. \bf 65} (1990) 2963. 

\item{Mar93} P. Markowitz {\it et al.}, accepted for publication
	in \PRC{\ } (June 1993).

\item{McC77} J.~S.~McCarthy, I.~Sick, and R.~R.~Whitney,
        {\it Phys.~Rev.~\bf C15} (1977) 1396.

\item{McK89} MIT/Bates proposal 89-06, R.~McKeown and D. H. Beck,
contact people. 

\item{McK91} CEBAF proposal 91-020, R.~McKeown, contact person.

\item{Mil88} MIT/Bates proposal 88-25, R.~Milner, R.~McKeown, T.~Chupp, and
        A.~Bernstein, contact people.

\item{Mil89} MIT/Bates Experiment 89-12, R. Milner and J. van den Brand,
spokespersons.

\item{Min89} S.~L.~Mintz and M.~Pourkaviani, {\it Phys.~Rev.~\bf C40}
        (1989) 2458.

\item{Mo69} L. W. Mo and Y. S. Tsai, {\it Rev. Mod. Phys. \bf
          41} (1969) 205. 

\item{Mon90} C.~Monroe, W.~Swann, H.~Robinson, and C.~E.~Wieman, {\it
        Phys.~Rev.~Lett.~\bf 65} (1990) 1571.

\item{Moo78}  R.~G.~Moorhouse, {\it Electromagnetic Interactions of
Hadrons}, Vol.~1, A.~Donnachie and G.~Shaw, Eds., Plenum Press (1978) 83.

\item{Mus90} M.~J.~Musolf and B.~R.~Holstein,
        {\it Phys. Lett. \bf 242B} (1990) 461. 

\item{Mus91} M.~J.~Musolf and B.~R.~Holstein, \PRD{43} (1991) 2956. 

\item{Mus92a} M.~J.~Musolf and T.~W.~Donnelly, \NPA{546} (1992) 509. 

\item{Mus92b} M.~J.~Musolf in {\it Proceedings of the CEBAF Summer
Workshop}, F.~Gross and R.~Holt, Eds., Newport News, Va.~(1992), 321.

\item{Mus93a} M.~J.~Musolf and M.~Burkardt,
        CEBAF Theory Pre-print \# TH-93-01. 

\item{Mus93b} M.~J.~Musolf, T.~W.~Donnelly, and T.~S.~H.~Lee, work in progress.

\item{Mus93c} M.~J.~Musolf and T.~W.~Donnelly, {\it Zeit.~fur Phys.~\bf C57}
         (1993) 559.

\item{Mus93d} M.~J.~Musolf, S.~J.~Pollock, and M.~D.~Havey, work in
         progress.

\item{Nak91} T.~Nakanishi {\it et al., Phys.~Lett.\/} {\bf A158},
        345 (1991).

\item{Nap89} J.~Napolitano {\it et al.}, {\it Nucl.~Instr.~Meth. \bf A274}
	(1989) 152.

\item{Nap90} J. Napolitano in {\it Proceedings of the Caltech Workshop on
        Parity Violation in Electron Scattering}, E.J. Beise and R.D.
        McKeown, Eds., World Scientific (1990) 90.

\item{Nap91} J.~Napolitano, \PRC{43} (1991) 1473.

\item{Nat82} L.~M.~Nath, K.~Schilcher, and M.~Kretzschmar,
        \PRD{25} (1982) 2300.

\item{Noe88} M.~C.~Noecker, B.~P.~Masterson, and C.~E.~Wieman,
        {\it Phys. Rev. Lett. \bf 61} (1988) 310. 

\item{Nov75} V.~N.~Novikov, and I.~B.~Khriplovich,
        {\it JETP Lett. \bf 22} (1975) 74.

\item{OCo72} J.~S.~O'Connell, T.~W.~Donnelly, and J.~D.~Walecka,
        {\it Phys.~Rev.~\bf C6} (1972) 719.

\item{Omo91} T.~Omori {\it et al., Phys.~Rev.~Lett.\/} {\bf 67},
        3294 (1991).

\item{Ott90} Mainz proposal A1/3-90, E.~Otten, contact person.

\item{Par91}N. W. Park, J. Schechter and H. Weigel, {\it Phys. Rev.\/}
{\bf D43}, 869 (1991). 

\item{Pas73} E. Paschos and L. Wolfenstein, \PRD{7} (1973) 91. 

\item{Pes90} M.~E.~Peskin and T.~Takeuchi, {\it Phys. Rev. Lett. \bf 65}
        (1990) 964. 

\item{Pla90} S. Platchkov {\it et al.}, \NPA{510} (1990) 740. 

\item{Pol87}  S. J. Pollock, Ph. D. Thesis, Stanford University
        (1987) unpublished.

\item{Pol90} S.~J.~Pollock, \PRD{42} (1990) 3010.

\item{Pol92a} S.~J.~Pollock, E.~N.~Fortson, and L.~Wilets,
        {\it Phys.~Rev.~\bf C46} (1992) 2587. 

\item{Pol92b} S.~J.~Pollock and M.~J.~Musolf, to be published.

\item{Pol93} S.~J.~Pollock and M.~J.~Musolf, to be published. 

\item{Pre78} C.~Y.~Prescott {\it et al.}, {\it Phys. Lett. \bf B77}
        (1978) 347. 

\item{Pre79} C.~Y.~Prescott {\it et al.}, {\it Phys. Lett. \bf B84}
        (1979) 524. 

\item{Pro91} {\it Proceedings of the European Workshop on Hadron Physics
        with Electrons Beyond 10 GeV}, B.~Frois and J.-F.~Mathiot,
        Eds., \NPA{532} (1991).

\item{Ras89} A.~S.~Raskin and T.~W.~Donnelly, {\it Ann.~Phys.~\bf 191}
        (1989) 78.

\item{Rei68} R.~V.~Reid, Jr., {\it Ann.~Phys.~\bf 50} (1968) 411.

\item{Rei87} D. Rein, {\it Zeit.~f\"ur Phys. \bf C35} (1987) 43.

\item{Reu82} W.~Reuter, G.~Fricke, K.~Merle, and H.~Miska, {\it
        Phys.~Rev.~\bf C26} (1982) 806.

\item{Ros90} J.~L.~Rosner, \PRD{42} (1990) 3107. 

\item{RPP92} Review of Particle Properties, \PRD{45} (1992) Part II.

\item{Sac62} R.~G.~Sachs, {\it Phys. Rev. \bf 126} (1962) 2256. 

\item{Sal68}A. Salam, in {\it Elementary
        Particle Physics}, ed.~N. Svartholm, Stockholm (1968) 367. 

\item{Sch73} P.~A.~Schreiner and F.~Von Hippel, {\it Nucl. Phys. \bf B58}
        (1973) 333.

\item{Sch88} Andreas Sch\"afer, {\it Phys. Lett. \bf B208} (1988) 175.

\item{Sch90} R.~Schiavilla, V.~R.~Pandharipande, D.~O.~Riska,
         \PRC{41} (1990) 309.

\item{Sea68} R. E. Seamon, K. A. Friedman, G. Breit, R. D. Haracz, J. M. Holt
        and A. Prakash, {\it Phys. Rev.\/} {\bf 165} (1968) 1579.

\item{Ser79} B.~D.~Serot, \NPA{322} (1979) 408. 

\item{Sic70} I.~Sick and J.~S.~McCarthy, \NPA{150} (1970) 631.

\item{Sic91} I.~Sick, personal communication (1991).

\item{Sie37} A.~J.~F.~Siegert, {\it Phys. Rev. \bf 52} (1937) 787.

\item{Sim80} G.~G.~Simon, F.~Borkowski, Ch.~Schmitt, and V.~H.~Walther,
        {\it Z. Naturforsch. \bf 35A} (1980) 1.

\item{Sir80}A. Sirlin, \PRD{22} (1980) 971.

\item{Slo88} T.~Sloan, G.~Smadja, and R.~Voss, {\it Phys.~Rep.~\bf 162}
        (1988) 45.

\item{Sou90a}P.A. Souder, {\it et al.}, {\it Phys. Rev.
Lett. \bf 65} (1990) 694. 

\item{Sou90b} P.~Souder, in {\it Proceedings of the Caltech Workshop on
 Parity--Violating Electron Scattering}, E.~J.~Beise and R.~D.~McKeown,
Eds., World Scientific (1990).

\item{Sou91} CEBAF proposal PR--91--007, P.~Souder and J.M.~Finn,
        contact people.

\item{van91} M.~van der Schaar, {\it Phys.~Rev.~Lett.~\bf 66} (1991) 2855.

\item{Vap88} O.~Vapenikova, {\it Zeit.~f\"ur Phys.~\bf C37} (1988) 251.

\item{Wal75} J.~D.~Walecka in {\it Muon Physics}, Vol. II, V.~W.~Hughes and
        C.~S.~Wu, Eds., Academic Press (1975) 113. 

\item{Wal77} J.~D.~Walecka, \NPA{285} (1977) 349. 

\item{Wei67} S.~Weinberg, \PRL{19} (1967) 1264. 

\item{Whi90} D.~H.~White, in {\it Proceedings of the Caltech
        Workshop on Parity
        Violation in Electron Scattering}, E.~J.~Beise and
        R.~D.~McKeown, Eds., (1990) 224.

\item{Wie93} C. Wieman, private communication.

\item{Wol79}L. Wolfenstein, {\it Phys. Rev. \bf D19} (1979) 3450. 

\item{Zel57}Ia. B. Zeldovich, {\it Zh. Eksp. Teor. Fiz. \bf 33} (1957) 1531
        [{\it Sov. Phys. JETP \bf 6} (1958) 1184]. 

\item{Zel60}Ia. B.~Zeldovich and A.~M.~Perelomov,
        {\it Zh. Eksp. Teor. Fiz. \bf 39} (1960) 1115
        [{\it Sov. Phys. JETP \bf 12} (1961) 777]. 

\item{Zuc71} P.~Zucker,  {\it Phys.~Rev.~\bf D4} (1971) 3350.

\vfil\eject

\centerline{\bf FIGURE CAPTIONS}
\medskip

\def\xivp{{\xi_\sst{V}^p}}

\item{Fig. 2.1} Perturbative contributions to non--valence quark
axial--vector and vector current matrix elements of the nucleon. Loops
involving heavy quarks renormalize light quark current operators via
gluon exchange (curly lines).
\medskip
\item{Fig. 2.2} Non--perturbative contributions to strange--quark vector
and axial--vector matrix elements of the nucleon. Virtual $s\bar s$ pair
with low--momenta ($p\lapp\Lambda_{\rm QCD}$) interacts with valence quarks
to form strange hadronic intermediate states ({\it i.e.} $K\Lambda$,
$\phi,\ldots$).
\medskip
\item{Fig. 2.3} Experimental constraints on parity--violating meson--nucleon
couplings entering meson--exchange model for the PV NN interaction [Hol89].
In the text, $f_\pi$ is denoted by $h_{\sst{NN}\pi}$ to avoid confusion with
the pion decay constant. The black circle refers to theoretical \lq\lq best
value" of Ref.~[Des80]. For a detailed description of experimental bands,
see Ref.~[Ade85].
\medskip
\item{Fig. 2.4} Present and prospective low--energy constraints on $S$ and
$T$ parameterization of physics beyond the Standard Model. Short--dashed lines
give present constraints from atomic PV [Noe88, Mar90]. Long--dashed lines
give limits attainable with a 10\% determination of $\xivp$ from a prospective
forward--angle $\alr(\evec p)$ experiment. Solid lines correspond to potential
constraints from a one percent measurement of $\alr(0^+0)$. Dot--dashed lines
indicate possible constraints from a 10\% determination of $\xivteo$
obtainable from
an $\alr(N\to\Delta)$ measurement [Mus92b, Mus93c]. In all cases, common
central
values for $S$ and $T$ are assumed, and only ranges permitted by experimental
and theoretical uncertainties are indicated.
\medskip
\item{Fig. 3.1} Leading--order parity--conserving (a) and parity--violating
(b) amplitudes for scattering of polarized electrons from a nuclear target.
\medskip
\item{Fig. 3.2} Higher--order contributions to amplitude for neutral
current (EM and weak) scattering involving two elementary
fermions (lepton or quark), $f$ and $f'$. Shaded circle represents
radiative processes (loops). (a,c) indicate vertex corrections, (b,d)
represent corrections to propagators, and (e) indicates exchange of two
vector bosons ($V, V' = \gamma, Z^0, W^\pm$).
\medskip
\item{Fig. 3.3} \lq\lq One--quark" electroweak radiative corrections to
lepton--hadron
scattering. One--quark processes refer to scattering involving only
one quark at a time and result in renormalization of the electroweak
couplings $\xiva$ from their tree--level values (see Table 3.2).
\medskip
\item{Fig. 3.4.} Strong interaction corrections to vector boson
propagators. (a) indicates gluon exchange (curly line) between virtual
quark and anti--quark, while in (b) virtual $q\bar q$ pair interacts
strongly with target quark.
\medskip
\item{Fig. 3.5.} Representative Feynman diagrams for \lq\lq many--quark"
electroweak radiative corrections to lepton--nucleon scattering. In contrast
to processes of Fig.~3.3, many--quark processes involve electroweak
interactions between target quarks, leading to the formation of virtual
hadronic intermediate states ($N\pi$, $\rho$, {\it etc.}).
\medskip
\item{Fig. 3.6.} Feynman diagrams used in modelling many--quark electroweak
contributions to $V(e)\otimes A(N)$ scattering amplitude. Pion--loop (a)
and vector meson pole (b) diagrams contain a weak, PV meson nucleon vertex
(crossed circle) and parity--conserving strong and electromagnetic
vertices (open circles), respectively.
\medskip
\item{Fig. 3.7.} Dispersion corrections to lepton--hadron (nucleon
or nucleus) scattering.
Exchange of vector bosons ($V,V'=\gamma, Z^0, W^\pm$) may excite
and de--excite virtual hadronic intermediate states (denoted by double line).
\medskip
\item{Fig. 3.8.} Two--body meson--exchange contributions to nuclear
electromagnetic and weak neutral currents. In (a), an off--shell
photon or $Z^0$ couples to a virtual $N\bar N$ pair, which interacts with
remaining nucleons via meson exchange. (b) and (c)
give currents associated
with exchanged meson and virtual isobar intermediate states, respectively.
\medskip
\item{Fig. 3.9.} Meson--exchange model for nuclear PV contributions to
PV lepton--nucleon scattering. Exchanged mesons interact strongly (open
circle) with one nucleon and weakly (crossed circle) with second nucleon.
PV component of the latter generates PV NN interaction, which mixes
nuclear states. (a,b) illustrate parity--mixing of initial and
final nuclear states, which couple to the final and initial states via
the one--body EM current operator. (c) shows meson--exchange current
contributions, which are required by current conservation.
\medskip
\item{Fig. 3.10.}Figure--of--merit for PV electron scattering
(see Eq.~(\Edoabl)) at $\theta=10^{\circ}$ versus incident electron
energy $\epsilon$.  The various cases presented are discussed in the text.
\medskip
\item{Fig. 3.11.}As for Fig.~3.10, except now at $\theta=30^{\circ}$.
\medskip
\item{Fig. 3.12.}As for Fig.~3.10, except now at $\theta=150^{\circ}$.
\medskip
\item{Fig. 3.13.} Leading--order neutrino and anti--neutrino charged--current
(a,b) and weak neutral current (c,d) scattering from a nuclear target.
\medskip
\item{Fig. 4.1.} Correlation between $\mustr$ and $g_A^{(1)}=-\tilde
G_\sst{A}^{T=1}(0)$
for a backward--angle $\alr(\evec p)$ measurement. Solid bands correspond to
different assumptions for experimental uncertainty in $\alr$. Dashed lines
give error induced in $\mustr$ due to the 20\% uncertainty in $g_A^{(1)}$
resulting from the theoretical uncertainty in $\rateo$. Arrows indicate
constraints on $\mustr$ anticipated from SAMPLE measurement [Bei92].
\medskip
\item{Fig. 4.2.} Fraction of total $\evec p$ asymmetry resulting from
longitudinal ($F_L$) and axial--vector ($F_A$) responses as a function of
scattering angle for different incident energies (given in MeV).
\medskip
\item{Fig. 4.3.} Fractional statistical uncertainty (a) and axial--vector
contribution to $\alr(\evec p)$ (b)
at backward--angles as a function of incident
energy. Luminosity ${\cal L} = 5\times 10^{38} \hbox{cm}^{-2}\hbox{s}^{-1}$,
100\% beam polarization, and 1000 hours of running time assumed in (a).
\medskip
\item{Fig. 4.4.} Correlation between $\mustr$ and dimensionless
strangeness radius, $\rhostr$, in a measurement of the forward--angle
$\evec p$ PV asymmetry. Solid lines correspond to different assumptions
for experimental uncertainty in $\alr$. Dashed lines indicate uncertainty in
$\rhostr$ assuming theoretical uncertainty in $\mustr$ determined from an
\lq\lq ideal" backward--angle $\alr(\evec p)$ measurement. Arrows indicate
expected SAMPLE limits on $\mustr$ [Bei92].
\medskip
\item{Fig. 4.5.} Doability curves for Standard Model test with PV
$\evec p$ scattering. Solid curves give statistical uncertainty in
$\alr(\evec p)$ under same conditions as in Fig. 4.3(a) but with solid
angle $\Delta\Omega=0.01$ sr. Dashed curve indicates maximum statistical
uncertainty allowed for a $\lapp 10\%$ determination of $\xivp$.
\medskip
\item{Fig. 4.6.} Doability curves for $\GES$ determination from
$\alr(^4\hbox{He})$ measurements under two different sets of kinematic
conditions and assumptions about detector. In each case, a luminosity of
${\cal L}[^4\hbox{He}] = 5\times 10^{38}\hbox{cm}^{-2}\hbox{s}^{-1}$,
100\% beam polarization, and 1000 hours of running time are assumed.
Error associated with $\GES$--uncertainty indicated by dashed and
dotted curves
for $\delta\rhostr=1.4$ and 0.7, respectively. In both cases, a value of
$\lamsE=0$ is assumed.
\medskip
\item{Fig. 4.7.} Constraints on $\GES$ obtainable from prospective
PV elastic electron scattering experiments. Dashed lines give constraints
from a series of $\alr(\evec p)$ measurements, including an \lq\lq ideal"
backward--angle measurement and a forward--angle measurement carried out under
the conditions of Ref.~[Nap91]. Dashed--dot and solid bands give constraints
from series of low-- and moderate--$|Q^2|$ measurements of the $^4$He
asymmetry. Conditions of Fig. 4.6(b), with $\theta=30^\circ$, and $\tau$
corresponding to successive minima in the statistical uncertainty in
$\alr$, are assumed. Black dot gives central values for ($\rhostr$,
$\lamsE$) under models (A) and (B) discussed in section IV.B.
\medskip
\item{Fig. 4.8} Doability curves for Standard Model tests with
$\alr(0^+0)$ measurements. Solid curves indicate statistical uncertainty
in $\alr$ assuming ${\cal L}[^4\hbox{He}]=5\times 10^{38}\hbox{cm}^{-2}
\hbox{s}^{-1}$, ${\cal L}[^{12}\hbox{C}]=1.25\times 10^{38}\hbox{cm}^{-2}
\hbox{s}^{-1}$, 100\% beam polarization, 1000 hours of running time, and
solid angles as indicated in the figure. Dashed and dotted curves have
same meaning as in Fig.~4.6. Arrows indicate maximum allowable error
(either experimental or $\GES$--induced) for a $\lapp$ 1\% determination
of $\xivtez$.
\medskip
\item{Fig. 4.9.}The PV asymmetry (upper panels) and structure--dependent
part of the asymmetry defined in Eq.~(\Egtilnos) (lower panels) for
elastic scattering from $^{40}$Ca, $^{48}$Ca and $^{208}$Pb.  Shown
are calculated
results for $\rho_n({\vec{x}})/\rho_p({\vec{x}})=N/Z$ where
${\tilde\Gamma}(q)=0$ (solid lines) and for
$\rho_n({\vec{x}})/\rho_p({\vec{x}})\neq N/Z$ (dashed lines) using density
parameterizations taken from Ref.~[Don89].
\medskip
\item{Fig. 4.10.} Doability curves for $\GMS$ determination with
backward--angle $\alr(^2\hbox{H})$ measurements. Solid curves give
fractional statistical uncertainty in the asymmetry, assuming
${\cal L} = 5\times 10^{38} \hbox{cm}^{-2} \hbox{s}^{-1}$, 100\% beam
polarization, $\Delta\Omega = 1$ sr, and 1000 hours of running time.
Dashed curve gives error induced in $\alr(^2\hbox{H})$ at $\theta =
175^\circ$ by $\GMS$ assuming $\delta\mustr = \pm 0.12$ (corresponding to
\lq\lq ideal" backward angle $\evec p$ determination).
\medskip
\item{Fig. 4.11.} Correlation plots for backward--angle, PV elastic electron
scattering from hydrogen (labelled H) and quasielastic scattering from
carbon (C) and tungsten (W). In the latter two cases, conditions $q=
500$ Mev$/c$, $\theta=150^\circ$, and $\omega=\coeff{1}{2}|Q^2|/\mn$ are
assumed. As in Fig.~4.1, $g_A^{(1)}=-\tilde
G_\sst{A}^{T=1}(0)$.
Bands correspond to $\pm 1\%$ experimental uncertainty in measured
asymmetry.
\medskip
\item{Fig. 4.12.} Hadronic (sea--quark) corrections to Standard Model
values for the $\tilde a_i$ parameters appearing in the deep inelastic
PV $^2$H asymmetry. Dashed line gives $R_1(\hbox{had})$ while solid
line gives $-R_2(\hbox{had})$ as a function of the Bjorken scaling
variable $x$.
\medskip
\item{Fig. 4.13.} $\chi^2$ contour showing 67\%
confidence region in extraction of
$\xivp$ and $\eta_s$ {\it only}, from our fictitious data set.
Systematic errors are ignored, and data is generated as described in
the text (integrated over an incident neutrino spectrum similar to that
of the BNL experiment.) The $\chi^2$ minimum here is about 13.(/14
d.o.f)
\medskip
\item{Fig. 5.1.}Neutrino spectrum from a $\pi$--$\mu$--$e$
beamstop neutrino facility: this arises from the decay
$\pi^+\rightarrow \mu^+\nu_{\mu}$ (0.026 $\mu$s) followed by
$\mu^+\rightarrow e^+ \nu_e {\bar\nu}_{\mu}$ (2.20 $\mu$s).
\medskip
\item{Fig.~5.2.} The SLAC polarized electron source of 1979.
\medskip
\item{Fig.~5.3.} Optical elements of the laser beamline for a SLAC style
polarized electron source.
\medskip
\item{Fig.~5.4.} Measured asymmetry vs.~beam energy in the SLAC experiment.
The beam helicity reverses every 3.237 GeV due to $g-2$ precession.
\medskip
\item{Fig.~5.5.} Measured asymmetry vs.~$y=1-{\epsilon^\prime/\epsilon}$
in the SLAC experiment.
\medskip
\item{Fig.~5.6.} The detector system for the Mainz quasielastic
$^9$Be experiment (see ref.~[Hei89]).
\medskip
\item{Fig.~5.7.} Histogram of pulse--pair asymmetries from the
Mainz experiment.
\medskip
\item{Fig.~5.8.} Limits on ${\tilde \beta}$ and ~${\tilde \delta}$
for the Mainz experiment (dotted lines) compared with data from other
sources.
Hashed lines are from the SLAC data [Pre79], and broken lines represent
data from $\nu$--hadron scattering [Kim81].
\medskip
\item{Fig.~5.9.} Experimental apparatus for the Bates $^{12}$C PV experiment
(see ref.~[Sou90a]).
\medskip
\item{Fig.~5.10.} Histogram of the measured asymmetry in each mini--run
normalized to its statistical error in the $^{12}$C experiment.
\medskip
\item{Fig.~5.11.} Cross section vs.~$|Q^2|$ for $(\nu_\mu,p)$ and
$(\overline\nu_\mu ,p)$ in the Brookhaven experiment [Ahr87].
\medskip
\item{Fig.~5.12.} Contour plot of $\eta$ vs.~$M_A$ taken from Ref.~[Ahr87].
($\eta$ in Ref.~[Ahr87] has the same meaning as $\eta_{s}$ in the
notation of the text.)
\medskip
\item{Fig.~5.13.}Detector and target system for the Bates SAMPLE experiment
(see ref.~[McK89]).
\medskip
\item{Fig.~5.14.} Apparatus for the proposed CEBAF $G^0$ experiment
(see ref.~[Bec91]).

\vfill
\eject
\end

\vfil\eject\end